\documentclass[11pt]{article}
\newif\iffigs\figstrue

\usepackage[title]{appendix}
\usepackage{epsfig,latexsym}
\usepackage{amsmath}
\usepackage{verbatim}
\usepackage{color}
\usepackage{mathrsfs}
\usepackage{amssymb}
\usepackage{physics}
\usepackage{float}

\DeclareMathAlphabet{\mathpzc}{OT1}{pzc}{m}{it}

 \csname
@addtoreset\endcsname{equation}{section}

\def\gz0{\gamma^{0}}

\def\td{\text d}

\def\g{\gamma}

\def\ve{\varepsilon}

\def\k{\kappa}

\def\f{\phi}
\def\vf{\varphi}

\def\cA{{\cal A}}
\def\cB{{\cal B}}

\def\cL{{\cal L}}

\def\cO{{\cal O}}

\def\cV{{\cal V}}

\def\beq{\begin{equation}}
\def\eeq{\end{equation}}
\def\bea{\begin{eqnarray}}
\def\eea{\end{eqnarray}}
\def\ba{\begin{array}}
\def\ea{\end{array}}
\def\bec{\begin{center}}
\def\ec{\end{center}}
\def\ba{\begin{align}}
\def\ena{\end{align}}

\def\12{\frac{1}{2}}

\newcommand{\SO}{\mathop{\rm SO}}

\newcommand{\U}{\mathop{\rm {}U}}
\newcommand{\USp}{\mathop{\rm {}USp}}

\begin{document}

\begin{flushright}
{\today}
\end{flushright}

\vspace{10pt}

\begin{center}


{\Large\sc Integrable Models and Supersymmetry Breaking}\\


\vspace{25pt}
{\sc P.~Pelliconi$^{a,b}$  \ and \ A.~Sagnotti$^{a}$\\[15pt]

{\sl\small
${}^{a}$Scuola Normale Superiore and INFN\\
Piazza dei Cavalieri 7\\ 56126 Pisa \ ITALY \\
\vspace{10pt}
${}^{b}$University of Geneva, Department of Theoretical Physics \\
24 quai Ernest-Ansermet \\ 1214 Gen\`eve 4 \ SWITZERLAND \\
\vspace{10pt}
e-mail: {\small \it pietro.pelliconi@unige.ch, sagnotti@sns.it}}\vspace{10pt}
}

\vspace{40pt} {\sc\large Abstract}\end{center}
\noindent
We elaborate on integrable dynamical systems from scalar--gravity Lagrangians that include the leading dilaton tadpole potentials of broken supersymmetry. In the static Dudas--Mourad compactifications from ten to nine dimensions, which rest on these leading potentials, the string coupling and the space--time curvature become unbounded in some regions of the internal space. On the other hand, the string coupling remains bounded in several corresponding solutions of these integrable models. One can thus identify corrected potential shapes that could grant these features generically when supersymmetry is absent or non--linearly realized. On the other hand, large scalar curvatures remain present in all our examples. However, as in other contexts, the combined effects of the higher--derivative corrections of String Theory could tame them.

\setcounter{page}{1}

\pagebreak

\newpage


\newpage
\baselineskip=20pt
\section{\sc  Introduction}\label{sec:intro}

Despite decades of intensive effort, the key principles of String Theory~\cite{stringtheory} remain largely elusive. With unbroken supersymmetry~\cite{supersymmetry}, convincing arguments link all different ten--dimensional string models to one another and, strikingly, also to the eleven--dimensional supergravity~\cite{supergravity} of Cremmer, Julia and Scherk~\cite{CJS}, within a intriguing picture that is usually dubbed M--theory~\cite{witten1011}. The comparison with low--energy physics, and ultimately with the Standard Model, demands however that supersymmetry be broken, but the response of String Theory to this inevitable feat remains largely mysterious. Typically the breaking of supersymmetry brings along tachyon instabilities, but three distinct string models exist in ten dimensions with no tachyonic modes in their spectra and with supersymmetry absent or non--linearly realized~\cite{susy95,sugimoto,so1616}.  However, supersymmetry breaking is accompanied by the emergence, in the low--energy effective field theory of these models, of an exponential ``tadpole potential'' for the dilaton field $\phi$. This occurs at the (projective) disk level in the non--supersymmetric $\U(32)$ 0'B orientifold~\cite{orientifolds} model of~\cite{susy95} and in the $\USp(32)$ orientifold model of~\cite{sugimoto} with ``brane supersymmetry breaking''~\cite{bsb}, where supersymmetry is present but non--linearly realized~\cite{va, nonlinearsusy}, and at the torus level in the non--supersymmetric $\SO(16)\times \SO(16)$ heterotic model of~\cite{so1616}. The emergent tadpole potentials lack critical points, and therefore ten--dimensional Minkowski space ceases to be a vacuum when they are taken into account.

In sharp contrast with the original Kaluza--Klein setting, where the internal space can be a circle of arbitrary size, the emergent exponential potentials can yield internal intervals of sizes determined by their strengths. These key solutions, which we shall call Dudas--Mourad  vacua~\cite{dm_vacuum},  include however regions where $g_s=e^{\langle \phi\rangle}$ and/or the space--time curvature grow unbounded, but are perturbatively stable~\cite{bms} and lead, strikingly, to finite values for the reduced Planck mass and gauge coupling~\cite{dm_vacuum}. Consequently, these interactions are still present in the resulting nine--dimensional flat spacetimes: even in the presence of tadpole potentials, \emph{the desirable breaking of supersymmetry can thus result in desirable lower--dimensional dynamics}. It is well known that string effective actions receive two series of perturbative string corrections, sized by the curvature in string units and by the string coupling $g_s$, and a host of non--perturbative ones. All these corrections, however, are not known in general, and even the first few terms appear unwieldy.

The authors of~\cite{dm_vacuum} relied on the leading tadpole potentials, which makes it interesting to explore, even in indirect ways, the possible role of these types of corrections. To this end, we shall study attentively nine classes of integrable dynamical systems emerging from scalar--gravity models that were examined in~\cite{fss} in connection with the ``climbing--scalar''~\cite{dm_vacuum,russo,dks2010} Cosmology. The analytic continuation of those results will help us to address a few detailed questions on corresponding spontaneous compactifications. We shall be particularly interested in potential shapes that can grant one or more of the following desirable features:
\begin{itemize}
    \item an internal space of finite size;
    \item a string coupling $g_s$ that is bounded everywhere in it;
    \item finite values for the lower--dimensional Planck mass and gauge coupling.
\end{itemize}
Our analysis will rest on the low--energy effective field theory, and thus ultimately on General Relativity. Within the limitations of this framework, some typical potential shapes will surface nonetheless that can grant one or more of these properties. However, our analysis will also unveil a tension between these demands and the existence of a bounded spacetime curvature.

These types of non--symmetric vacua, where non--trivial profiles are only present in one internal dimension, are admittedly rather simple and special, but are very instructive toy models. The rationale behind the present investigation is precisely that examples of this type, where string corrections would be naturally bounded, may help one build some intuition on corrected string vacua and on the ultimate lesson of~\cite{dm_vacuum} for them.  Aside from this, the solutions that we shall discuss are an interesting set of dynamical options in the presence of gravity, but we do feel that they have a lesson in store. More symmetric vacua resting on the leading tadpole potentials, as in~\cite{ms}, do exist, but they are typically unstable~\cite{bms,gm}, and actually their instability in the presence of broken supersymmetry has become a general tenet within the swampland picture~\cite{swampland}. The encouraging results of~\cite{bms}, where the Dudas--Mourad vacuum was shown to evade this problem, should perhaps be taken as favoring less symmetric configurations leaving behind a flat space, and the present investigation reinforces somehow this feeling, which has been surfacing time and over again in the past. In this respect, one should keep in mind that the lack of internal symmetries and Ricci flatness are also key features of Calabi--Yau spaces~\cite{chsw}, which play a central role in connection with the partial breaking of supersymmetry.

The plan of this paper is the following. In Section~\ref{sec:scalar-gravity} we explain our conventions and elaborate on the desired features of nine--dimensional scalar-gravity vacua. In Section~\ref{sec:compactness} we present the basic equations arising from string--inspired scalar--gravity models, along the lines of~\cite{dm_vacuum}, and discuss in detail the behavior of spatial profiles when the dynamics is dominated by an exponential potential. In the following sections we rely heavily on the results of~\cite{fss}, which provides a catalogue of scalar--gravity models including a palatable family of exact cosmological solutions for a variety of potential shapes. Interestingly, these potential shapes have generically the look of corrected forms of the leading tadpole potential of the orientifold models of~\cite{susy95, sugimoto, bsb}. In Section \ref{sec:models} we present the different classes of integrable models drawn from~\cite{fss}, together with some additional variants, and describe how to solve the resulting equations in the current setting. As in~\cite{dm_vacuum} and~\cite{fss}, proper gauge choices will be instrumental to this end. In Section~\ref{sec:profiles} we present the solutions of the corresponding models and elaborate on the conditions that identify classes of potential shapes complying to one or more of the requests spelled out above. Section~\ref{sec:conclusions} contains some concluding remarks and elaborates on possible future developments along these lines.

\vskip 12pt
\section{\sc Scalar--Gravity Models and String Theory}\label{sec:scalar-gravity}

We use a ``mostly--plus'' signature and work to a large extent in the Einstein--frame, within the class of metrics
\begin{equation}
	\td s^2 \ = \  e^{\frac{2}{9} \, \cA(r)} \, \eta_{\mu\nu} \, \td x^{\mu} \,  \td x^{\nu} \, + \, e^{2 \,  {\cal B}(r)} \, \td r^2 \ .
\label{einstein_metric}
\end{equation}
We denote by $x^{\mu}$, $\mu = 0,\dots, 8$ the spacetime coordinates and by $r$ the tenth, internal, coordinate. The metric (\ref{einstein_metric}) contains warp factors that depend on $r$, and the equations of motion determine the dependence of ${\cal A}$ on this coordinate. On the other hand ${\cal B}$ is a gauge function, which we shall choose on a case--by--case basis, thus selecting $r$ coordinates that simplify the resulting dynamical systems. As in \cite{dm_vacuum}, the dilaton $\phi$ will be here the only bosonic field, aside from the metric, with a non-vanishing vacuum profile. One is clearly demanding a residual Poincar\'e symmetry in the nine--dimensional subspace, but the dilaton and the string coupling
\begin{equation}
	g_s \ = \  e^{ \phi(r)}
\end{equation}
will be free to depend on $r$. Moreover, in all the cases that we shall explore the potential will include an exponential term of the form
\begin{equation}
	V (\f) \ = \  C e^{\frac{3}{2}\,\gamma \, \phi} \ . \label{leading_tadpole}
\end{equation}
The constant $\gamma$ takes two specific values in the leading contributions that present themselves in the non--tachyonic ten--dimensional strings. In detail, $\gamma = 1$ for the $\USp(32)$ orientifold of~\cite{bsb} and for the $\U(32)$ orientifold of~\cite{susy95}, where the contributions arise at the disk/crosscap level, while $\gamma = \frac{5}{3}$ for the $\SO(16) \times \SO(16)$ heterotic model~\cite{so1616}, where the contribution emerges from the torus amplitude.

The Einstein--frame action considered in \cite{dm_vacuum}, inspired by String Theory, was of the form
\begin{equation}
	S \ = \ \frac{1}{2 \kappa_{10}^2} \, \int \td^{10}x \, \sqrt{-g} \, \left[ \, R \, - \, \frac{1}{2} \, (\partial \phi)^2 \, - C \, e^{\gamma \, \phi}  \, + \, \dots \, \right] \ . \label{low_energy_ft}
\end{equation}
Only the first terms displayed above, together with a handful of others, are under control from first principles. However, as we have anticipated, the solutions found in \cite{dm_vacuum} include regions where the string coupling $g_s$ is large, and also regions where the curvature is large in string units, so that string corrections to the low--energy field theory~\eqref{low_energy_ft} are expected to play an important role, and could affect considerably the resulting picture.

The present work can be regarded as an attempt to build some intuition on scenarios that string corrections might unveil in the models of interest, relying on the elegant mathematics of integrable dynamical systems.
We are actually addressing the simplest conceivable option: our targets are corrections to the tadpole potentials in eq.~\eqref{low_energy_ft} that can grant solutions with a bounded $g_s$. A benign setting of this type would make at least part of the higher--order corrections subdominant, and therefore, in our opinion, even this admittedly blind exercise can have some potential lessons in store.

In general, one expects perturbative corrections to Einstein--frame potentials of the form
\begin{equation}
	V(\phi) \ = \ \sum_{b,c,h} c_{\,b,c,h} \ e^{\left(b+c+2 h+\frac{1}{2}\right)\phi} \ ,
\label{perturbative_potential}
\end{equation}
with arbitrary integer values of $b$ and $h$, which count boundaries and handles, and with $c=0,1,2$, which counts crosscaps, together with a host of additional non--perturbative contributions. As we have stressed, even extracting the next-to-leading terms from String Theory is a difficult task, and for all these reasons we are particularly interested in potentials where the leading tadpole of~\eqref{leading_tadpole} is accompanied by terms of this type.

Our starting point will be generalizations of eq.~\eqref{low_energy_ft} including more general potentials $V(\phi)$,
\begin{equation}
	S \ = \ \frac{1}{2 \kappa_{10}^2}\,  \int \td^{10} x \; \sqrt{-g} \left[ R \, - \,  \frac{1}{2} \, g^{\mu\nu}\,  \partial_{\mu} \phi \, \partial_{\nu} \phi \, - \, V(\phi)  \right] \ .
\label{cosm_action}
\end{equation}
For the class of metrics~\eqref{einstein_metric}, these translate into dynamical systems with the reduced action principles
\begin{equation}
 	S \ = \  \int \, dr\; e^{\cA - \cB} \, \left[  \dot{\cA}^2 \, - \, \dot{\vf}^2 \,  - \, 2\, e^{2\cB} \, {\cal V}(\vf)   \right] \ ,
 \label{action}
 \end{equation}
after performing, as in~\cite{dks2010}, the convenient redefinitions
 \begin{equation}
 	\f \ = \ \frac{4}{3}\, \vf \ , \qquad \qquad V(\phi) \ = \ \frac{16}{9}\,{\cal V}(\vf) \ ,
\label{changenormalization}
\end{equation}
which cast them in their simplest form.
The Euler--Lagrange equations of motion for ${\cA}$ and $\vf$ are then
 \begin{eqnarray}
 	 && 2 \, \ddot{\cA} \ + \ \dot{\cA}^2 \ -\ 2\, \dot{\cA}\, \dot{\cB} \ + \ \dot \vf ^2\ + \ 2\,e^{2\cB} \, {\cal V}(\vf) \ = \  0 \ , \nonumber \\
 	 && \ddot \vf \ +\  (\dot{\cA} - \dot{\cB}) \, \dot \vf \ - \  e^{2\cB} \, \cV'(\vf)\ = \  0 \ ,
\label{eqmotionforA}
 \end{eqnarray}
 while the equation for ${\cal B}$, which we shall often call ``Hamiltonian constraint'', becomes
 \beq
 \dot \vf^2  \ - \ 2\,e^{2\cB} \, {\cal V}(\vf) \ = \ \dot{\cA}^2 \ ,
 \label{Ham_const}
 \eeq
and leads to the reduced system
  \begin{align}
 	& \ddot{\cA} \ - \ \dot{\cA} \, \dot{\cB} \ + \  \dot \vf^2 \ = \ 0 \ , \nonumber \\
 	& \ddot \vf + (\dot{\cA} - \dot{\cB}) \, \dot \vf \ - \  e^{2 \cB} \, \cV'(\vf) \ = \  0 \ ,
\label{staticequations}
\end{align}
Notice that, with these redefinitions,
 \begin{equation}
     {\cal V}(\vf) \ = \ {\cal V}_0\,e^{2\vf} \label{V_orientifold}
 \end{equation}
for the two orientifold models of~\cite{susy95} and~\cite{bsb}, for which $\gamma=1$, while
 \begin{equation}
      {\cal V}(\vf) \ = \ {\cal V}_0\,e^{\frac{10}{3}\,\vf} \label{V_heterotic}
 \end{equation}
 for the heterotic model of~\cite{so1616}, for which $\gamma=\frac{5}{3}$: all our examples will contain contributions of the first type, and some will also contain contributions of the second type.

 Notice that eqs.~\eqref{staticequations} are simply solved whenever the scalar $\vf$ takes a constant value $\varphi_0$ that corresponds to a \emph{negative} extremum of the potential. Indeed, the dilaton equation is identically satisfied by such a constant value $\varphi=\varphi_0$, while the other equations reduce to
 \begin{equation}
 	 \ddot{\cA} \ = \   0 \ ,\qquad \qquad \dot{\cA}^2 \ = \  - 2 \cV(\vf_0) \ .
\label{eq_AdS}
\end{equation}
Their solution exists only if ${\cal V}\left(\varphi_0\right)<0$, and is simply
\begin{equation}
	\cA \ = \  \sqrt{\, 2 \, |\cV(\vf_0)|} \ r \ + \ \alpha, \qquad {\cal B} \ = \ 0 \ , 	
\end{equation}
with $\alpha$ an integration constant. However, these types of solutions describe $AdS_{10}$, and are not of interest to us here since they do not describe compactifications to lower dimensions.

We shall be particularly interested in solutions with one or more of the following features:
\begin{itemize}
\item an $r$--direction with a \emph{finite} string--frame length
\begin{equation}
L \ =\	\int \td r \, e^{\cB + \frac \f 4} \  = \  \int \td r \, e^{\cB + \frac \vf 3}\ ;
\label{compact_cond}
\end{equation}
\item a \emph{bounded} string coupling $g_s = e^{\f} = e^{\frac 4 3 \vf}$;
\item finite values for the reduced nine--dimensional Planck mass and the typical gauge coupling
\begin{align}
	M_P^7 \ \propto & \  \int_0^{\infty} \td r \; e^{\frac 7 9 \cA + \cB} \ , \nonumber \\
	\frac{1}{g_{YM}^2} \ \propto & \ \int_0^{\infty} \td r \; e^{\frac 5 9 \cA + \cB + \frac 2 3 \vf } \ .
\label{planckandgauge}
\end{align}
which would grant the corresponding interactions a non--trivial role in the resulting nine--dimensional spacetime;
\item a \emph{bounded} string--frame scalar curvature, which takes the form
\begin{equation}
    R_{(s)} \ = \ - \, 2 \, \, e^{-\frac{2}{3} \vf - 2 \cB} \left[ \, \frac{32}{9} \, \dot \cA^2 \ + \ 3 \, e^{2\cB} \, \left( \cV'(\vf) \ + \ 2 \, \cV(\vf) \right)  \,  \right]
\label{string_curvature}
\end{equation}
after using the equations of motion~\eqref{Ham_const} and \eqref{staticequations}.
\end{itemize}

Eqs.~\eqref{staticequations} are a system of coupled non--linear differential equations, and solving them analytically is not an easy task in general. However, \cite{fss} identified, among other more complicated examples, nine classes of potentials for which the cosmological counterpart of the system of eqs.~\eqref{eqmotionforA} is solvable, more or less explicitly, in closed form. This exemplified a wide number of contexts where the climbing phenomenon of~\cite{fss, dks2010}, the inevitable emergence of cosmological counterparts of eq.~\eqref{einstein_metric} at weak coupling from the initial singularity, occurs. One can also extract from the underlying Mathematics information on static vacua, along the lines stated above. This is the purpose of the present work, and we can now proceed combining, as in~\cite{fss}, different forms of $\cV(\vf)$ with special choices for the gauge function $\cB$.

\section{\sc Dilaton Dynamics, Compactness and Scalar Curvature}\label{sec:compactness}

Identifying potential shapes that can grant an upper bound for the dilaton, and thus for the string coupling, together with a compact internal space, would provide some clues on how String Theory could overcome the limitations of the Dudas--Mourad setup. At the same time, one would be interested in the behavior of the scalar curvature. These features may seem unrelated, but they are actually tightly connected, and depend crucially on the potential $\cV(\vf)$ that drives the dynamics and on the boundary conditions of the corresponding solutions.

For the class of metrics of eq.~(\ref{einstein_metric}), familiar notions drawn from Newtonian mechanics can shed some light on the presence or absence of strong--coupling regions. The Hamiltonian constraint of eq.~(\ref{staticequations}) can indeed be cast in the form
\begin{equation}
	 \frac12 \,\dot \vf^2 \ - \ e^{2 \cB} \, \cV(\vf) \ = \  \frac12 \, \dot{\cA}^2 \ , \label{ham_c}
\end{equation}
which is reminiscent of the energy conservation condition for a Newtonian particle. Here $\displaystyle \frac12 \, \dot{\cA}^2$ plays somehow the role of an $r$--dependent ``total energy'': although it is not known beforehand, it is manifestly non negative, just like the ``kinetic energy'' $\displaystyle \frac12 \,\dot \vf^2$. The remaining contribution is the ``inverted potential'' $- \, e^{2 \cB} \, \cV(\vf)$: the sign reversal is a standard fact in systems depending on a spatial coordinate, as in this case, or on Euclidean time, but has nonetheless notable consequences. In fact, if $\cV(\vf)$ is defined for all $\vf \in \mathbb R$ and is always positive, one expects the solutions to explore the whole potential, which points to the presence of strong--coupling regions. One can thus anticipate that \emph{solutions where $g_s$ has an upper bound ought therefore to emerge in two cases: if the very form of the potential $\cV(\vf)$ places an upper bound $\tilde \vf$ on $\vf$, or if there are regions where $\cV(\vf)$ is negative, which can give rise to potential hills for $- \cV(\vf)$ capable of ``reflecting the particle''.}

In our examples the potential is typically dominated, in interesting regions, by a single exponential, which can be parametrized in the form
\begin{equation}
    \cV(\vf) \ = \ \cV_0 \; e^{2 \, \gamma \, \vf} \ , \label{vgamma}
\end{equation}
and here it will be important to allow both signs for $\gamma$ and for the constant ${\cal V}_0$.
The exact solutions with this class of potentials are relatively simple in the gauges
\beq
2 e^{2{\cal B}}{\cal V}\left(\varphi\right) \ = \ \pm\, 1 \ ,  \label{gammagauge}
\eeq
and allow one to build some intuition on the types of behavior that can emerge in more general cases. We shall be particularly interested in settings that drive the dilaton toward large negative values. The cosmological counterpart of this behavior in an expanding Universe was addressed, for the critical case, in~\cite{dm_vacuum}, and for arbitrary positive values of ${\cal V}_0$ and $\gamma$, in~\cite{russo}. The peculiar \emph{climbing} phenomenon was then identified in this dynamics in~\cite{dks2010}: \emph{with $\gamma \geq 1$, the scalar can only emerge from $\varphi=-\infty$ after the initial singularity, to then ``climb up'' the potential before inverting its motion at a turning point}. Consequently, all these cosmologies entail upper bounds for the string coupling. On the other hand, two distinct options exist for $\gamma< 1$, since the scalar can also emerge from $\varphi=+\infty$, and climb down the potential. In this respect $\gamma = 1$, which corresponds to $e^{\frac{3}{2}\,\phi}$ in ten dimensions, and thus to the leading tadpole term for orientifolds, is the ``critical'' value that separates these two types of behavior.

We can now translate the analysis of~\cite{dks2010} into the present context, where however more interesting options exist for $\dot{\cal A}$, which denotes here the derivative with respect to $r$.
Let us first consider the case in which the potential is dominated by a single exponential, as in eq.~\eqref{vgamma}, with ${\cal V}_0$ \emph{negative}, while allowing for different values of $\gamma$, which we assume initially to be positive. It is natural to begin with this case, since spatial profiles involve potentials that, as we have stressed, are inverted with respect to the cosmological setting. As a result, spatial profiles induced by \emph{negative} exponential potentials satisfy equations that are formally identical to the cosmological solutions with the usual \emph{positive} potentials if the spatial derivative of ${\cal A}$ is also \emph{positive}, up to the replacement of a temporal variable with a spatial one. Therefore, in this case one should recover, verbatim, the results of~\cite{dks2010}. A convenient gauge choice for a potential dominated by a single negative exponential is
\begin{equation}
    2 \, e^{2\cB} \, \cV(\vf) \ = \ - \, 1 \ .
\end{equation}
Allowing for both signs of $\dot{\cal A}$ and $\dot \vf$, the Hamiltonian constraint would then be solved letting
\begin{equation}
    \dot \cA \ = \ \ve_\cA \, \cosh(v) \ , \qquad \qquad \dot \vf \ = \  \sinh(v) \ ,
\end{equation}
where $\ve_\cA = \pm1$, while the corresponding freedom for $\dot \vf$ is already encoded in the sign of $v$. However, one can always work in the region $r>0$, while also restricting the attention to the case $\ve_\cA=1$, since the behavior for decreasing $\cA$ can be recovered from the mirrored evolution toward smaller values of $r$. The equation for $\vf$ then becomes
\begin{equation}
    \dot v \ + \  \sinh(v) \ + \ \gamma \, \cosh(v) \ = \ 0 \ ,
    \label{static_climbing_3}
\end{equation}
and for $0 < \gamma < 1$ one thus finds
\begin{align}
    \dot \varphi \ = & \  \frac{1}{2} \left[\sqrt{\frac{1-\gamma}{1+\gamma}} \, \coth\left( \frac{\sqrt{1-\gamma^{2}}}{2} \, r \right)\ - \ \sqrt{\frac{1+\gamma}{1-\gamma}} \, \tanh \left( \frac{\sqrt{1-\gamma^{2}}}{2} \,  r  \right)\right] \ , \nonumber \\
    \dot \cA \ = & \ \frac{1}{2} \left[\sqrt{\frac{1-\gamma}{1+\gamma}} \, \coth\left( \frac{\sqrt{1-\gamma^{2}}}{2} \, r \right)\ + \ \sqrt{\frac{1+\gamma}{1-\gamma}} \, \tanh \left( \frac{\sqrt{1-\gamma^{2}}}{2} \, r \right)\right] \ ,
    \label{climbing}
\end{align}
or
\begin{align}
    \dot \varphi \ = & \ \frac{1}{2} \left[\sqrt{\frac{1-\gamma}{1+\gamma}} \, \tanh\left( \frac{\sqrt{1-\gamma^{2}}}{2} \, r \right)\ - \ \sqrt{\frac{1+\gamma}{1-\gamma}} \, \coth \left( \frac{\sqrt{1-\gamma^{2}}}{2} \,  r \right)\right] \ , \nonumber \\
    \dot \cA \ = & \ \frac{1}{2}  \left[\sqrt{\frac{1-\gamma}{1+\gamma}} \, \tanh\left( \frac{\sqrt{1-\gamma^{2}}}{2} \,  r \right)\ + \ \sqrt{\frac{1+\gamma}{1-\gamma}} \, \coth \left(\frac{\sqrt{1-\gamma^{2}}}{2} \, r \right)\right] \ . \label{descending}
\end{align}
which are identical to the cosmological solutions in~\cite{russo,dks2010}, up to the replacement of a temporal variable with a spatial one.
As in the cosmological context, eqs.~\eqref{climbing} describe a \emph{climbing} solution, which starts here at weak coupling for $r = 0^+$ and returns to weak coupling as $r \to +\infty$ after attaining an upper bound. On the other hand, eqs.~\eqref{descending} describe a \emph{descending} solution, which starts here at strong coupling for $r = 0^+$ and approaches weak coupling as $r \to +\infty$. Moreover, the limiting behavior of both solutions is the well--known Lucchin--Matarrese attractor~\cite{lm}. The corresponding solutions for negative values of $\gamma$ can be obtained letting $\varphi \to - \varphi$.

As $\gamma \to 1$ only the climbing solution is well behaved and approaches
\begin{align}
    \dot \vf \ = & \ \ \frac{1}{2 \,r} \ - \ \frac{r}{2} \ , \nonumber \\
    \dot \cA \ = & \ \ \frac{1}{2 \,r} \ + \ \frac{r}{2} \ ,
    \label{critical_negative_V}
\end{align}
and finally for $\gamma >1$ there is again only a climbing solution, which reads
\begin{align}
    \dot \vf \ = & \  \frac{1}{2} \left[\sqrt{\frac{\gamma-1}{\gamma+1}} \, \cot \left(  \frac{\sqrt{\gamma^{2}-1}}{2} \,r  \right) \ - \ \sqrt{\frac{\gamma+1}{\gamma-1}} \, \tan \left( \frac{\sqrt{\gamma^{2}-1}}{2} \,r \right)\right] \ , \nonumber \\
    \dot \cA \ = & \ \frac{1}{2} \left[\sqrt{\frac{\gamma-1}{\gamma+1}} \, \cot \left( \frac{\sqrt{\gamma^{2}-1}}{2} \,r  \right) \ + \ \sqrt{\frac{\gamma+1}{\gamma-1}} \, \tan \left( \frac{\sqrt{\gamma^{2}-1}}{2} \,r \right)\right] \ , \label{climbing_2}
\end{align}
where now $0 < r < \frac{\pi}{\sqrt{\gamma^{2}-1}}$. As before, the corresponding solutions for $\gamma$ negative can be obtained changing the sign of $\vf$.

Even for spatial profiles this dynamics has an important consequence: \emph{if ${\cal V}$ is dominated beyond a certain value of $\vf$ by a \emph{negative} exponential potential with $\gamma \geq 1$, the string coupling has a finite upper bound}. On the other hand, for $\gamma \leq -1$ the string coupling would have a finite lower bound, while in the complementary range $-1 < \gamma < 1$ two different options exist, only of which guarantees an upper (or lower) bounds for $g_s$.
\begin{table}[h]
    \centering
    \begin{tabular}{|c|c|c|c|c|c|c|c|}
    \hline
    $\cV_0 < 0$ & type & $e^{\vf}_{\text L}$  & cp$_{\text L}$ & $R_{(s) \text L}$ & $e^{\vf}_{\text R}$ & cp$_{\text R}$ & $R_{(s)\text R}$   \\
    \hline
    $0 < \gamma < 1$ & c & $|r|^{\frac{1}{1 + \gamma}}$ & $0 < \gamma < 1$ & ub & $\exp\left(- \frac{\gamma r}{\sqrt{1 - \gamma^2}} \right)$ & $0 < \gamma < \frac{1}{3}$ & $\frac{1}{3} \leq \g < 1$ \\
    \hline
    $0 < \gamma < 1$ & d & $|r|^{-\frac{1}{1 - \gamma}}$ & $0 < \gamma < 1$ & ub & $\exp\left(-  \frac{\gamma r}{\sqrt{1 - \gamma^2}}  \right)$ & $0 < \gamma < \frac{1}{3}$ & $\frac{1}{3} \leq \g < 1$ \\
    \hline
    $\gamma = 1$ & c & $|r|^{\frac{1}{2}}$ & $\gamma = 1$ & ub & $\exp\left(- \frac{r^2}{4} \right)$ & n--cp & $\gamma = 1$\\
    \hline
    $\gamma > 1$ & c & $|r|^{\frac{1}{\gamma + 1}}$ & $\gamma > 1$ & ub & $| r_{\text{max}} -  r|^{\frac{1}{\gamma - 1}}$ & n--cp & $\gamma > 1$\\
    \hline
    \end{tabular}
    \caption{\small The available options in the exponential potential of eq.~\eqref{vgamma} with $\cV_0 < 0$. The second column identifies the two types of behavior (descending or climbing), which correspond to eq.~\eqref{descending} and to eqs.~\eqref{climbing},~\eqref{critical_negative_V} and~\eqref{climbing_2}. The other columns collect information on the behavior at the two ends of the interval (L and R). The highlighted features are the limiting behaviors of the string coupling, the ranges of $\gamma$ that grant finite contributions to the length of the internal space (n--cp when there are none) and the limiting behavior of the scalar curvature (ub when it is always unbounded).}
    \label{tab:dilaton_dynamics_1}
\end{table}

The length of the internal interval in the string--frame metric is another feature of interest. One can explore the available options following steps similar to what we did for the string coupling, and the interested reader is invited to verify the entries of Table~\ref{tab:dilaton_dynamics_1}.

On the other hand, if $\cV_0>0$ the gauge condition becomes
\begin{equation}
    2 \, e^{2\cB} \, \cV(\vf) \ = \ 1 \ ,
    \label{climbing_gauge}
\end{equation}
and the Hamiltonian constraint implies that $\vf$ and $\cA$ trade their roles, since it is solved by
\begin{equation}
    \dot \cA \ = \  \sinh(v) \ , \qquad \qquad \dot \vf \ = \  \cosh(v) \ .
\end{equation}
One can bypass again all sign ambiguities proceeding as before, which in this case amounts to focusing on solutions that move \emph{inevitably} from weak to strong coupling as $r$ increases. In terms of the variable $v$ the equation of motion for $\vf$ is now
\begin{equation}
    \dot v \ + \ \cosh(v) \ + \ \gamma \, \sinh(v) \ = \ 0 \ .
    \label{dilaton_dynamics_V_neg}
\end{equation}
For $0 < \gamma < 1$ eq.~\eqref{dilaton_dynamics_V_neg} there is now only one type of solution,
\begin{align}
    \dot \vf \ =& \ \frac{1}{2} \left[\sqrt{\frac{1 - \gamma}{1 + \gamma}} \, \cot \left(\frac{\sqrt{1 - \gamma^2}}{2} \, r \right) \ + \  \sqrt{\frac{1 +\gamma}{1 - \gamma}} \, \tan \left( \frac{\sqrt{1 - \gamma^2}}{2} \, r \right)\right] \ , \nonumber \\
    \dot \cA \ =& \ \frac{1}{2}\left[\sqrt{\frac{1 - \gamma}{1 + \gamma}} \, \cot \left(\frac{\sqrt{1 - \gamma^2}}{2} \, r \right) \ - \  \sqrt{\frac{1 +\gamma}{1 - \gamma}} \, \tan \left( \frac{\sqrt{1 - \gamma^2}}{2} \, r \right)\right] \ ,  \label{climbing_A_sub}
\end{align}
which can be considered again in the range $0 < r < \frac{\pi}{\sqrt{1 - \gamma^2}}$.

For $\gamma = 1$ there is also one type of solution,
\begin{align}
    \dot \vf \ = & \ \frac{1}{2r} \ + \ \frac{r}{2} \ , \nonumber \\
    \dot \cA \ = & \ \frac{1}{2r} \ - \ \frac{r}{2} \ , \label{climbing_A_crit}
\end{align}
and finally for $\gamma > 1$ there are two different solutions
\begin{align}
    \dot \varphi \ = & \ \frac{1}{2} \left[\sqrt{\frac{\gamma - 1}{\gamma + 1}} \, \coth\left( \frac{\sqrt{\gamma^{2} - 1}}{2} \, r\right)\ + \ \sqrt{\frac{\gamma + 1}{\gamma - 1}} \, \tanh \left( \frac{\sqrt{\gamma^{2} - 1}}{2} \, r\right)\right] \ , \nonumber \\
    \dot \cA \ = & \ \frac{1}{2}\left[\sqrt{\frac{\gamma - 1}{\gamma + 1}} \, \coth\left(\frac{\sqrt{\gamma^{2} - 1}}{2} \, r\right)\ - \ \sqrt{\frac{\gamma + 1}{\gamma - 1}} \, \tanh \left( \frac{\sqrt{\gamma^{2} - 1}}{2} \, r\right)\right] \ , \label{climbing_A}
\end{align}
and
\begin{align}
    \dot \varphi \ = & \ \frac{1}{2} \left[\sqrt{\frac{\gamma - 1}{\gamma + 1}} \, \tanh\left(\frac{\sqrt{\gamma^{2} - 1}}{2} \, r\right)\ + \ \sqrt{\frac{\gamma + 1}{\gamma - 1}} \, \coth \left( \frac{\sqrt{\gamma^{2} - 1}}{2} \, r\right)\right] \ , \nonumber \\
    \dot \cA \ = & \ \frac{1}{2} \left[\sqrt{\frac{\gamma - 1}{\gamma + 1}} \, \tanh\left(\frac{\sqrt{\gamma^{2} - 1}}{2} \, r\right)\ - \ \sqrt{\frac{\gamma + 1}{\gamma - 1}} \, \coth \left( \frac{\sqrt{\gamma^{2} - 1}}{2} \, r\right)\right] \ , \label{descending_A}
\end{align}
which differ mostly in the behavior of $\cA$. Summarizing, \emph{with a positive exponential potential $\vf$ is bound to reach strong coupling as $r \to \infty$}.

\begin{table}[h]
    \centering
    \begin{tabular}{|c|c|c|c|c|c|c|}
    \hline
    $\cV_0 > 0$ & $e^{\vf}_{\text L}$  & c$_{\text L}$ & $R_{(s) \text L}$ & $e^{\vf}_{\text R}$ & c$_{\text R}$ & $R_{(s)\text R}$   \\
    \hline
    $0 < \gamma < 1$ & $|r|^{\frac{1}{1 + \gamma}}$ & $0 < \gamma < 1$ & ub & $|r_{\text{max}} - r|^{-\frac{1}{1-\gamma}}$ & $0 < \gamma < 1$ & ub \\
    \hline
    $\gamma = 1$ & $|r|^{\frac{1}{2}}$ & $ \gamma = 1$ & ub & $\exp\left(\frac{r^2}{4} \right)$ & $\gamma = 1$ & ub\\
    \hline
    $\gamma > 1$ & $|r|^{\frac{1}{\gamma + 1}}$ & $\gamma > 1$ & ub & $\exp\left( \frac{\gamma r}{\sqrt{\gamma^2 -1 }} \right)$ & $\gamma > 1$ & ub \\
    \hline
    $\gamma > 1$ & $|r|^{\frac{1}{\gamma - 1}}$ & n--cp & $\gamma > 1$ & $\exp\left(\frac{\gamma r}{\sqrt{\gamma^2 - 1}} \right)$ & $\gamma > 1$ & ub \\
    \hline
    \end{tabular}
    \caption{\small The available options in the exponential potential of eq.~\eqref{vgamma} with $\cV_0 > 0$. The various  columns collect information on the behavior at the two ends of the interval (L and R), corresponding to the solutions of eqs.~\eqref{climbing_A_sub},~\eqref{climbing_A_crit},~\eqref{climbing_A} and~\eqref{descending_A}. The highlighted features are the limiting behaviors of the string coupling, the ranges of $\gamma$ that grant finite contributions to the length of the internal space (n--cp when there are none) and the limiting behavior of the scalar curvature (ub when it is always unbounded).}
    \label{tab:dilaton_dynamics_2}
\end{table}
This analysis will prove very valuable in Section~\ref{sec:profiles}, where it will provide a rationale for the qualitative behavior of most solutions.

\section{\sc Scalar--Gravity Integrable Models}\label{sec:models}

In this section we explore the basic setup to build the static counterparts of the simplest cosmological models discussed in~\cite{fss}.
\vskip 12pt
\subsection{\sc Triangular Systems}

\begin{figure}[ht]
	\centering
	\includegraphics[width=0.55\linewidth]{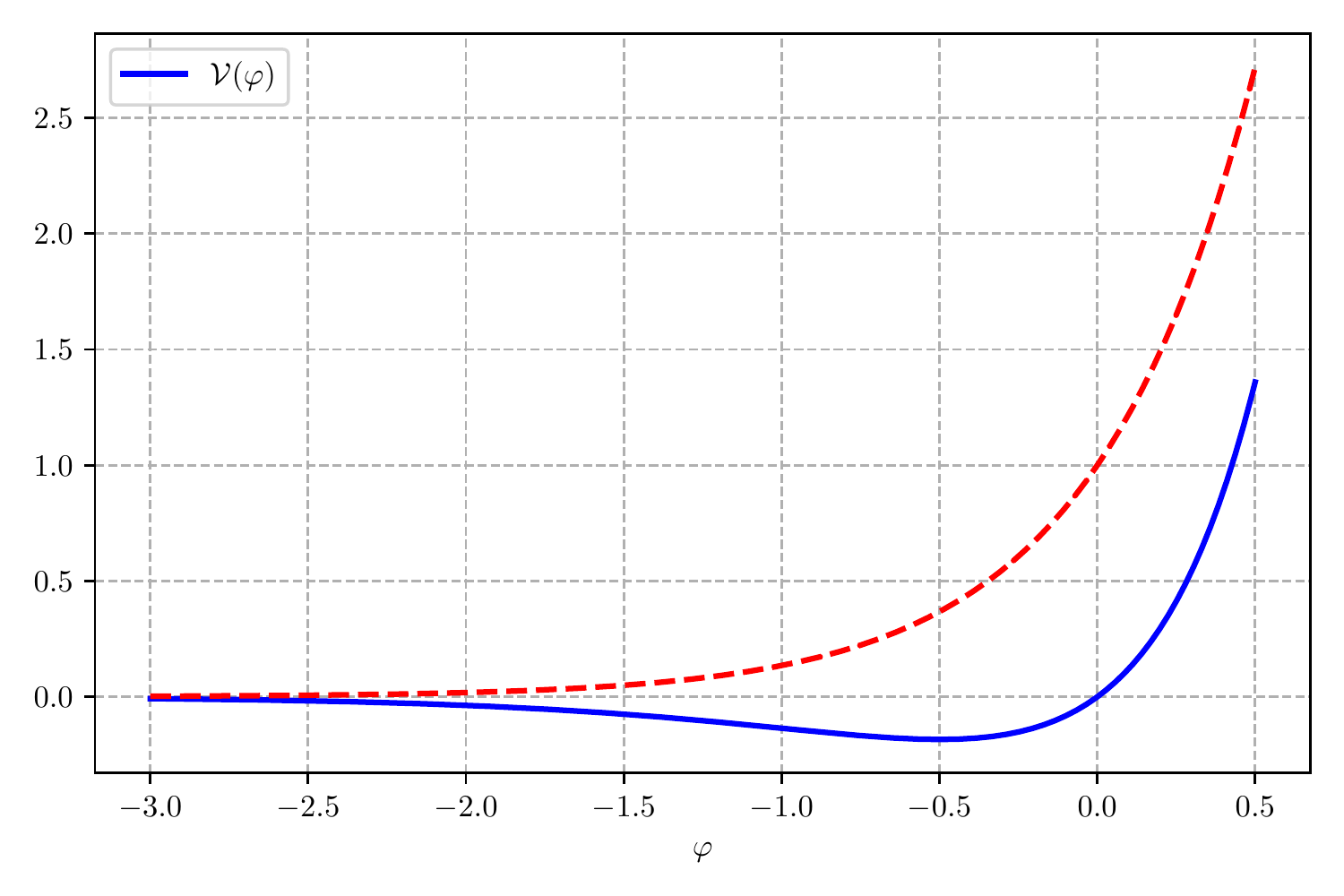}
	\caption{\small The potential of eq.~\eqref{pot_31} with $C = 1$. The dashed potential is the one considered in \cite{dm_vacuum}.}
	\label{pot1}
\end{figure}

\noindent 1). \textbf{The first integrable potential} that we consider is
\begin{equation}
	{\cal V} (\vf) \; = \;  C\,\vf \, e^{2 \vf} \ .
\label{pot_31}
\end{equation}	
One can expand in a power series its string--frame counterpart, and expressing it in terms of the conventional dilaton field $\phi$ gives
\begin{equation}
    \cV_S(\f) \ = \    C \, e^{- \f} \, \sum_{n = 1}^{\infty} \frac{( 1 - e^{- \frac34 \f})^n}{n} \,  \ .
\end{equation}
From a string perspective, the infinitely many contributions have the flavor of corrections to the ``critical'' exponential potential of the orientifold models of~\cite{susy95} and \cite{bsb} that lie beyond the perturbative series. A plot of this potential can be found in fig.~\ref{pot1}, where we have chosen $C>0$, which makes it bounded from below, but as we have seen in Section~\ref{sec:compactness} inverted potentials will be of interest in the following.

Eqs.~\eqref{staticequations} simplify after performing the redefinitions
\begin{equation}
	{\mathcal A}  \ = \    \frac{1}{4} \log|x| + v \ , \qquad \qquad \vf  \ = \ \frac{1}{4} \log|x| \ - \ v \ ,
\label{firstchangepot1}
\end{equation}
and choosing the gauge
\begin{equation}
	{\mathcal B} \ = \  - {\mathcal A} - 2 \vf \ = \  - \frac{3}{4} \log|x| + v \ ,
\label{secondchangepot2}
\end{equation}
the resulting system can be simply solved analytically. Using these new variables the Lagrangian becomes indeed
\begin{equation}
	{\mathcal L} \ = \   \frac12 \ \text{sign}(x) \ \dot x \ \dot v\  +\ C \, \left( -\, \frac 14\,  \log|x| \ + \ v \right),
\end{equation}
and the equations of motion are, for $x  \neq 0$,
\begin{equation}
	\ddot x   \ =  \  2 \  \text{sign}(x) \ C \ , \qquad \qquad \ddot v  \ = \   -\ \text{sign}(x) \ \frac{C}{2x} \ .
\label{primopoteqmotion}
\end{equation}

Systems of this kind are called ``triangular'': one can clearly solve the first of eqs.~\eqref{primopoteqmotion}, and the second then becomes a simple equation with a source term.
Moreover, the Hamiltonian constraint reads
\begin{equation}
	 \dot x \dot v \ = \  - \  \text{sign}(x) \  C \left( \frac12 \, \log|x| \ -\  2\, v \right)\ ,
\end{equation}
and working with absolute values is convenient, since it eliminates the need to restrict the range of validity of the solutions. A closer look, however, reveals that this procedure alters the signs of some constants entering the potentials. For instance, working in the region where $x<0$ is equivalent, insofar as the classical equations of motion are concerned, to flipping the sign of $C$, and thus to considering an inverted potential. Subtleties of this type emerge, in principle, in all cases where redefinitions similar to those in eqs.~\eqref{firstchangepot1} are performed. In order not to be overly pedantic, however, in the following we shall confine ourselves to pointing out when relevant effects of this type emerge.

\noindent 2). \textbf{The second integrable potential} that we consider is
 \begin{equation}
 	\cV(\vf) \ = \  C_1 \, e^{2 \vf} \ +\  C_2  \ , \label{pot_32}
 \end{equation}
which yields another triangular system. One can also consider its string--frame counterpart, and expressing it in terms of the conventional dilaton field $\phi$ gives
\begin{equation}
 	\cV_S(\f) \ = \  C_1 \, e^{- \f} \ +\  C_2 \, e^{-\frac52 \f}  \ ,
 \end{equation}
The term $e^{-\f}$ could be again the leading contribution to the potential in the  orientifold models analyzed in \cite{dm_vacuum}, where $C_1$ would be positive. Here, however, we are also free to elaborate on the behavior of the solutions for negative values of $C_1$. Notice also that a contribution proportional to $C_2$ would lie beyond string perturbation theory.
\begin{figure}[ht]
	\centering
	\includegraphics[width=0.55\linewidth]{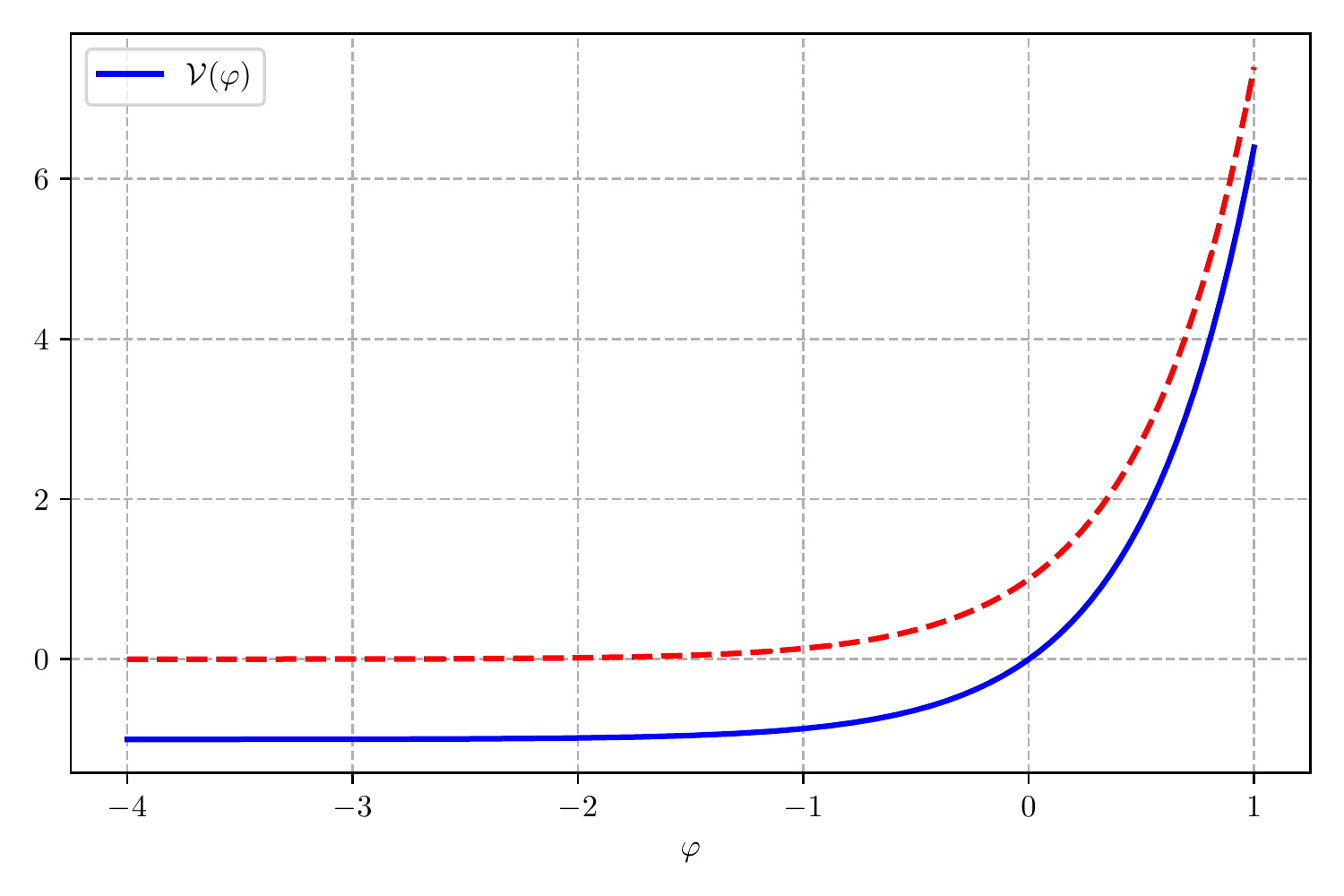}
	\caption{ \small The potential of eq.~\eqref{pot_32} with $C_1 = 1$ and $C_2 = -1$. The dashed potential is the one considered in \cite{dm_vacuum}.}
\end{figure}

For this potential it is convenient to choose the gauge
\begin{equation}
	\mathcal B \ = \   -\, \vf \ ,
\end{equation}
and performing the redefinitions
\begin{equation}
	\mathcal A  \ = \  \frac12 \log(x) + v \ , \qquad \qquad \vf  \ = \  \frac12 \log(x) - v \ , \qquad \qquad \mathcal B  \ = \  v - \frac12 \log(x)\ ,
\end{equation}
the Lagrangian takes the form
\begin{equation}
	\mathcal L \ = \   2 \, \dot x \, \dot v \ -\  2\,  C_1\,  x \ -\  2\,  C_2 \, e^{2v} \ .
\end{equation}
The corresponding equations (\ref{staticequations}) read
\begin{equation}
	\ddot v  \ = \   -\, C_1 \ , \qquad  \ddot x  \ = \ - \, 2 \,  C_2 \, e^{2v} \ , \qquad \dot x \dot v \ = \ -\, C_2 \, e^{2v} \ - \  C_1 \,  x \ . \label{triang_2}
\end{equation}
The first two form again a triangular system, while the third is the Hamiltonian constraint.
The solutions can be considered conveniently also for $x<0$, working with absolute values for the arguments of the logarithm. Proceeding as for the preceding case, a closer look would reveal that the constant $C_2$ in the potential flips sign in regions where $x<0$.

\noindent 3). \textbf{The third integrable potential} that we consider is
\begin{equation}
	\cV(\vf) \ = \   C_1 \, e^{2 \gamma \vf} \ + \ C_2 \, e^{(\gamma + 1) \vf} \ ,
	\label{pot_33}
\end{equation}
whose string--frame counterpart is
\begin{equation}
	\cV_S(\f) \ = \ C_1 \, e^{\frac 12 (3 \gamma - 5)\f} \ + \ C_2 \, e^{\frac 14 (3\gamma - 7) \f} \ ,
\end{equation}
in terms of the conventional dilaton field $\phi$.
This is actually a whole family of potentials, which reduce to the standard one for the orientifold models if $\gamma =  1$, and to the preceding one if $\gamma=  -1$, up to a $\varphi \to - \varphi$ redefinition. Notice also that the choices $\gamma = \frac53$ and $\gamma = \frac73$ include the low--lying potential of the $\SO(16) \times \SO(16)$ heterotic model of~\cite{so1616}.

For $\gamma \neq \pm 1$, combining the gauge choice
\begin{equation}
	\mathcal B \ = \  - \, \gamma \, \vf \ ,
\end{equation}
with the redefinitions
\begin{equation}
	\mathcal A \ = \  \log \left( x^{\frac{1}{1+ \gamma}} \,  y^{\frac{1}{1-\gamma}} \right) \ , \qquad \qquad  \vf \ = \  \log \left( x^{\frac{1}{1 + \gamma}} \, y^{-\frac{1}{1- \gamma}}  \right) \ , \qquad \qquad  \mathcal B \ = \  \log \left( x^{-\frac{\gamma}{1 + \gamma}} \, y^{\frac{\gamma}{1-\gamma}} \right) \ ,
\label{transformations2gammagamma+1}
\end{equation}
reduces the Lagrangian to the convenient form
\begin{equation}
	\mathcal L \ = \  4 \, \dot x \, \dot y \ - \ 2 \, (1-\gamma^2) \, \left[ \, C_1 \, x\, y \ + \ C_2 \, x^{\frac{2}{1 + \gamma}} \, \right] .
\end{equation}
Its equations of motion are
\begin{align}
	\ddot x \ + \ \frac{1 - \gamma^2}{2} \,C_1 \, x \ = & \ 0 \ , \qquad \qquad \ddot y \ +\ \frac{1 - \gamma^2}{2}\, C_1 \, y \ = \  - \,C_2 \, (1 - \gamma) \, x^{\frac{1-\gamma}{1 + \gamma}} \ , \nonumber \\
	\dot x \, \dot y \ = & \ - \, \frac{1 - \gamma^2}{2} \, \left[ \, C_1 \, x\, y\  +\ C_2 \, x^{\frac{2}{1 + \gamma}} \, \right] \ ,
\label{EOM_third_potential}
\end{align}
and the first two form again a triangular system, while the last is the Hamiltonian constraint.

\subsection{\sc Systems Integrable via Quadratures}
\label{Sys_quad}

\noindent We can now turn to a class of systems that can be solved via quadratures.

\noindent 4). \textbf{The fourth integrable potential} that we consider is
 \begin{equation}
 	\cV(\vf) \ = \  C_1 \, e^{2\gamma\vf} \ +\   C_2 \, e^{\frac{2}{\gamma}\vf} \ ,
 	\label{pot_34}
 \end{equation}
 whose string--frame counterpart is
 \begin{equation}
 	\cV_S(\f) \ = \  C_1 \, e^{\frac{3\gamma-5}{2}\f} \ +\   C_2 \, e^{\frac{3 - 5\gamma}{2\gamma} \, \f} \ ,
 \end{equation}
 in terms of the conventional dilaton field $\phi$.
 This is again a whole class of potentials: they comprise two terms that, for general $\gamma$, do not involve integer powers of $g_s$, and can describe the leading contributions of the three tachyon--free models of~\cite{susy95,sugimoto,so1616}. As will become clear in Section \ref{sec:profiles}, the solutions that are more interesting for our current purposes result from two exponential terms of opposite signs. For this reason, up to a shift in $\vf$, we write these potentials in the form
\begin{equation}
	\cV(\vf) \ = \  \lambda \left( \ve_1 \, e^{2\gamma\vf} + \ve_2 \, e^{\frac{2}{\gamma}\vf} \right) \ ,
\label{pot_34_1}
\end{equation}
where $\ve_1 = \pm 1$ and $\ve_2 = \pm 1$, which encompasses a wider range of options than those examined in~\cite{fss}. Notice that the redefinition $\gamma \, \longrightarrow \, \displaystyle \frac{1}{\gamma}$ connects pairs of potentials in this family, and for this reason one can restrict the attention to the range $0 < |\gamma| < 1$, or even to $0 < \gamma < 1$, up to a redefinition $\varphi \to - \varphi$.
\begin{figure}[ht]
\centering
\includegraphics[width=0.55\textwidth]{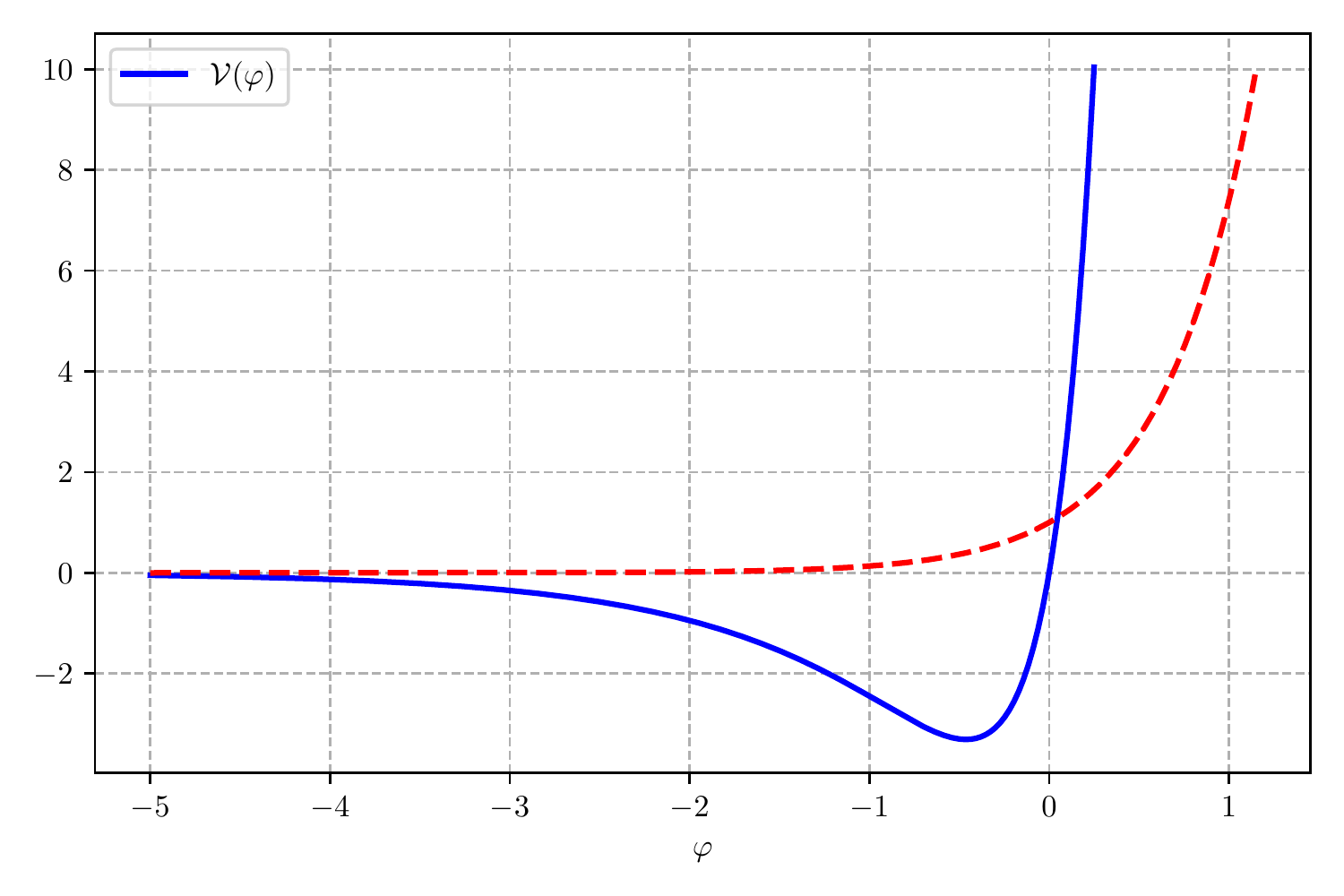}
\caption{ \small  The potential of eq.~\eqref{pot_34_1} for $\gamma =  \frac{1}{2}$, $\lambda = 7$, $\ve_1 = -1$ and $\ve_2 = 1$. The dashed potential is the one considered in \cite{dm_vacuum}}
\end{figure}
With the gauge choice
\begin{equation}
	 \mathcal B \ = \  \mathcal A \ ,
\end{equation}
the Lagrangian takes the form
\begin{equation}
	\mathcal L \ = \  (\, \dot{\mathcal A}^2 \ - \ \dot \vf^2\, )\  -\  2\, \lambda \, e^{2\mathcal A} \, (\, e^{\frac{2}{\gamma} \vf} \ - \ e^{2 \gamma \vf} \,) \ .
\end{equation}
The peculiar structure of kinetic terms and exponents clearly bring to one's mind the Lorentzian boosts of Special Relativity. Hence, if $0 <\gamma < 1$, one is led to define the new variables
\begin{equation}
	\hat{\mathcal A} \ = \  \frac{1}{\sqrt{1 - \gamma^2}}\, (\mathcal A + \gamma \vf) \ , \qquad \qquad  \hat{\vf} \ = \  \frac{1}{\sqrt{1 - \gamma^2}}\, (\vf+ \gamma \mathcal A)  \ ,
\label{lorboost}
\end{equation}
which bring the Lagrangian into the separable form
\begin{equation}
	\mathcal L \ = \  \dot{\hat{\mathcal A}}^2 \ - \ \dot{\hat \vf}^2 \ + \ 2 \, \lambda \,  e^{2 \sqrt{1-\gamma^2} \, \hat{\mathcal A}} \ - \ 2 \, \lambda \, e^{\frac{2}{\gamma} \sqrt{1 - \gamma^2} \,\hat \vf} \ .
\end{equation}
As usual, its two apparently independent equations of motion,
\begin{align}
	& \ddot{\hat{\mathcal A}} \ - \ 2 \, \lambda \, \sqrt{1 - \gamma^2} \, e^{2  \sqrt{1- \gamma^2} \, \hat{\mathcal A}} \ = \  0 \ , \nonumber\\
	& \ddot{\hat{\vf}} \ - \ \frac{2 \lambda}{\gamma} \sqrt{1 - \gamma^2} \, e^{\frac{2}{\gamma} \, \sqrt{1- \gamma^2} \,\hat{\vf}} \ = \  0 \ ,
	\end{align}
are to be supplemented by the Hamiltonian constraint
\begin{equation}
	\dot{\hat{\mathcal A}}^2 \ - \ \dot{\hat \vf}^2 \ = \ 2 \,\lambda \, e^{2 \sqrt{1-\gamma^2} \, \hat{\mathcal A}} \ -\  2\, \lambda \,  e^{\frac{2}{\gamma} \sqrt{1 - \gamma^2} \, \hat \vf} \ ,
\label{equationthirdpotential}
\end{equation}
which links their integration constants.

\begin{figure}[ht]
    \centering
    \includegraphics[width=0.55\linewidth]{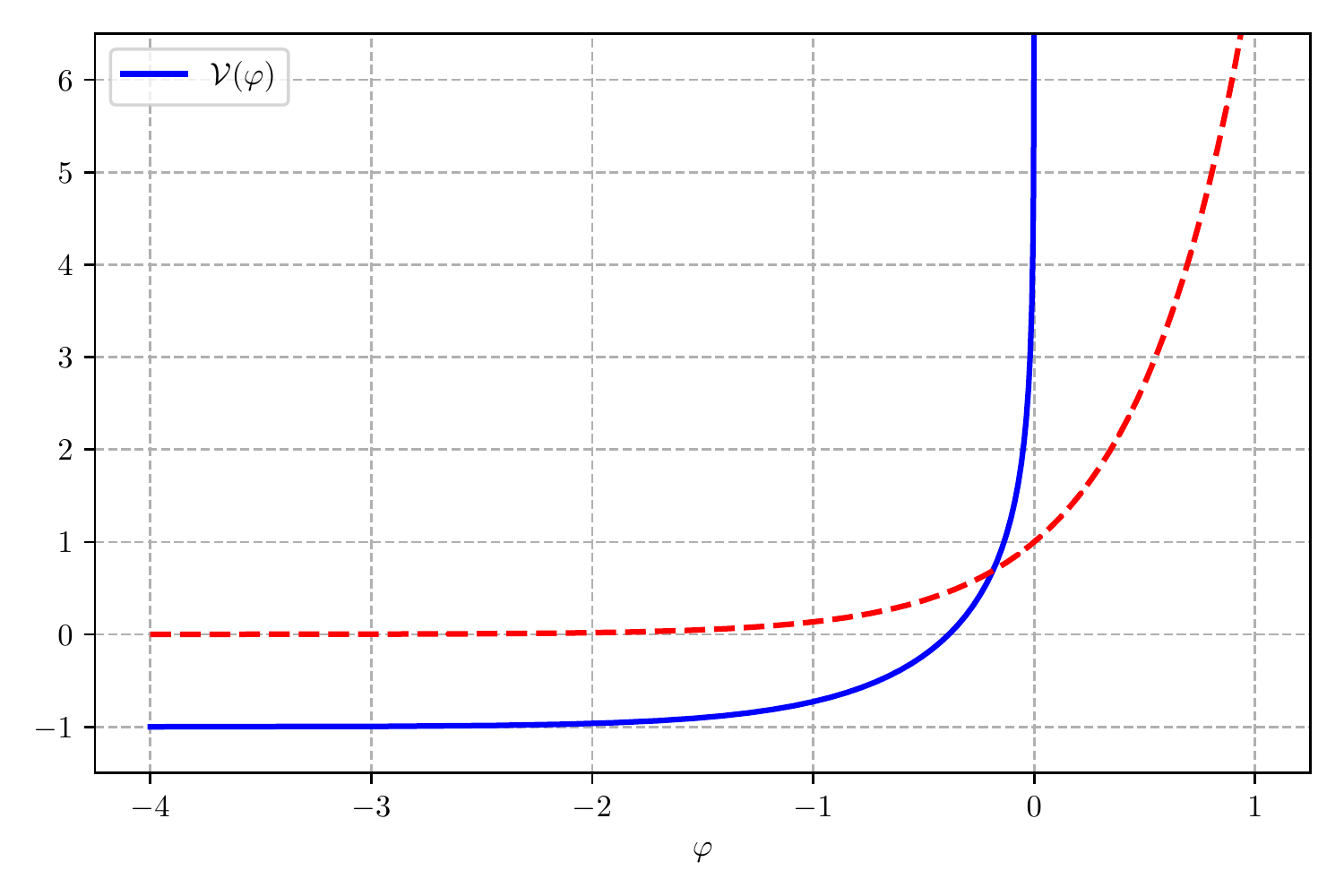}
    \caption{ \small The potential of eq.~\eqref{pot_35} for $C =  1$ and $D =  -1$. This potential has the peculiarity that it is defined only for $\vf < 0$. The dashed potential is the one considered in \cite{dm_vacuum}.}
    \label{logcoth}
\end{figure}

\noindent 5). \textbf{The fifth integrable potential} that we consider,
\begin{equation}
	\cV(\vf) \ = \  C \log \left( - \coth(\vf) \right) + D \ , \label{pot_35}
\end{equation}
is displayed in fig.~\ref{logcoth}.
In this case it is convenient to let
\begin{equation}
	e^{\cA} \ = \  \sqrt{xy} \ , \qquad \qquad e^{\vf} \ = \  \sqrt{\frac y x} \ , \qquad \qquad e^{\cB} \ = \  \sqrt{\frac{1}{xy}} \ , \label{xy_variables}
\end{equation}
so that the gauge choice is
\begin{equation}
	\mathcal B \ = \  - \, \mathcal A \ .
\end{equation}

This potential exhibits a novel feature: it is real only for $\vf < 0$, which sets an upper bound on $g_s$. Interestingly, it includes infinitely many perturbative contributions together with additional terms, all of which conspire to enforce this restriction.
Indeed, in terms of the conventional dilaton field $\phi$, the string--frame counterpart of eq.~\eqref{pot_35} reads
\begin{equation}
    \cV_S(\f) \ = \  C \sum_{s,k = 0}^{\infty} \frac{(-1)^s}{s+1} \, 2^{s+1} \, \binom{k + s}{s} \, e^{\left(\frac32 (k+s) - 1 \right)\, \f} \ + \  D \, e^{- \frac52 \, \f} \ ,
\end{equation}
and we see that it includes the leading orientifold contribution considered in~\cite{dm_vacuum}. The variables of eq.~\eqref{xy_variables} turn the Lagrangian into
\begin{equation}
	\mathcal L \ = \  - \, 4 \, \dot x \, \dot y \ - \ 8 \, C \, \log \left( \frac{x + y}{x - y} \right) \ - \ 8 \, D \ ,
\end{equation}
and letting
\begin{equation}
	x  \ = \  \frac{\xi + \eta}{2} \ , \qquad \qquad y  \ = \   \frac{\xi - \eta}{2} \ ,
\end{equation}
leads finally to
\begin{equation}
	\mathcal L \ = \   \dot \eta^2 \ - \ \dot \xi^2 \ - \ 8\, C \, \log \left( \frac{\xi}{\eta} \right) \ - \ 8 \, D \ ,
\end{equation}
whose equations of motion
\begin{equation}
	\ddot \xi \ = \  - \, \frac{4C}{\xi} \ , \qquad \qquad \ddot \eta \ = \  - \, \frac{4C}{\eta} \ , \qquad \qquad \dot \eta^2 \ - \ \dot \xi^2 \ = \  8 \, C \log \left( \frac{\xi}{\eta} \right) \ + \ 8\, D
\label{EOM_fourth_potential}
\end{equation}
can be solved by quadratures. As usual, the last equation is the Hamiltonian constraint.

\noindent 6). \textbf{The sixth integrable potential} that we consider is
\begin{equation}
	\cV(\vf) \ = \  C \, \cosh(\vf) \ + \ \Lambda \ . \label{pot_36}
\end{equation}
\begin{figure}[ht]
    \centering
    \includegraphics[width=0.55\linewidth]{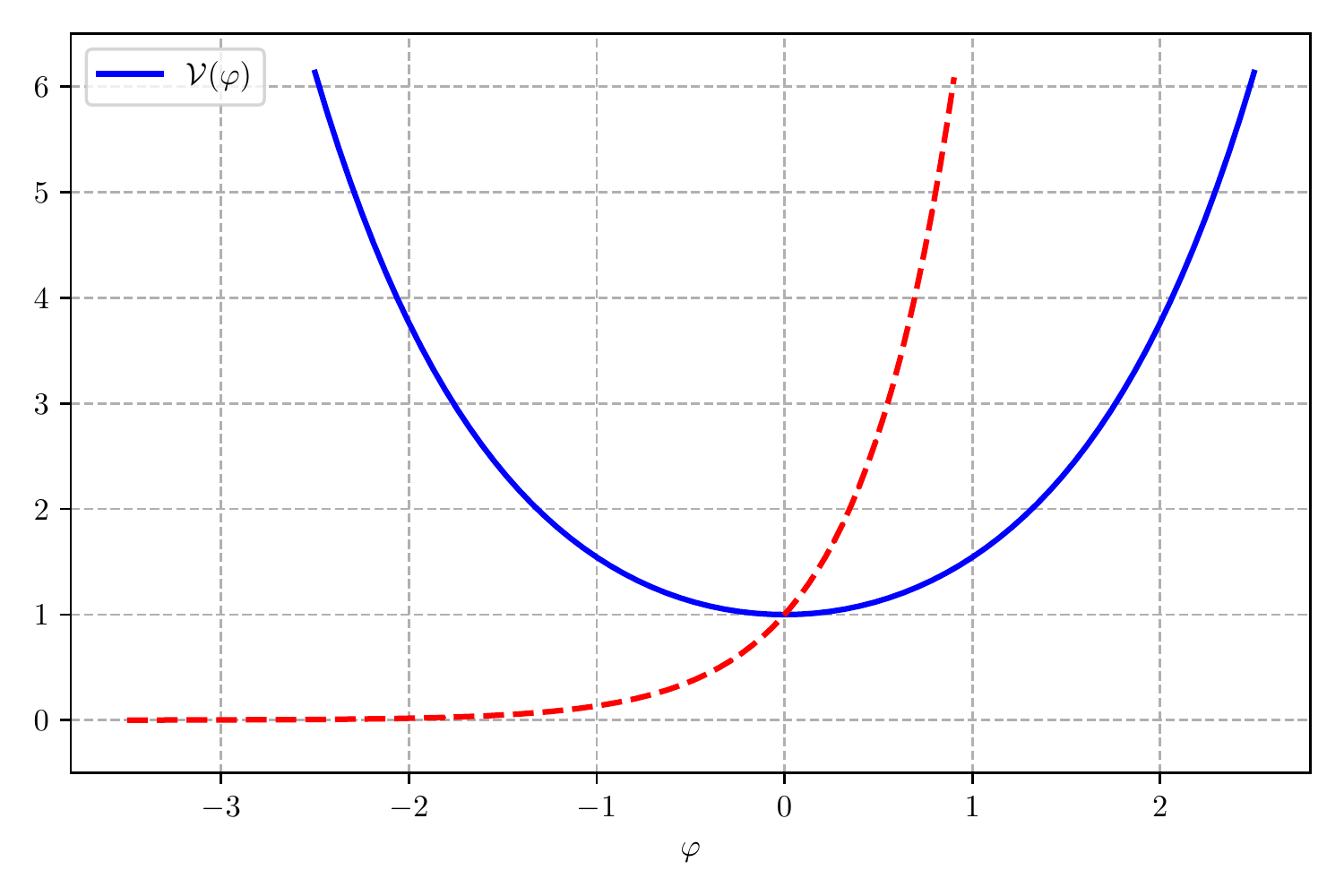}
    \caption{\small The potential of eq.~\eqref{pot_36} for $C =  1$ and $\Lambda =  0$. The dashed potential is the one considered in \cite{dm_vacuum}.}
\end{figure}
Its string--frame counterpart, in terms of the conventional dilaton field $\phi$, is
\begin{equation}
	\cV_S(\f) \ = \  \frac C 2  \left(e^{-\frac74\f} + e^{-\frac{13}4\f} \right) + \Lambda \, e^{-\frac52 \, \f} \ .
\end{equation}
It is now convenient to let
\begin{equation}
	\mathcal A  \ = \  \log(\, x\, y\, ) \ , \qquad \qquad \vf \ = \  \log \left( \frac{x}{y} \right) \ ,
\end{equation}
and to work in the gauge
\begin{equation}
	\mathcal B \ = \  0 \ .
\end{equation}
In terms of the new coordinates $x$ and $y$ the Lagrangian becomes
\begin{equation}
	\mathcal L \ = \  2\, \dot x \, \dot y \ - \ C \, \frac{x^2+y^2}{2} \ - \ \Lambda \,x \, y \ ,
\end{equation}
whose equations of motion are
\begin{equation}
	\ddot x  \ = \  - \, \frac{\Lambda}{2} \, x \ - \ \frac{C}{2} \, y \ , \qquad \ddot y  \ = \   - \, \frac{C}{2} \, x \ - \ \frac{\Lambda}{2} \, y \ , \qquad - \, 2 \, \dot x \, \dot y \ = \   \Lambda \, x \, y \ + \ \frac{C}{2} \, ( \, x^2 + y^2 \, )\ ,
\end{equation}
where the last is the Hamiltonian constraint. One can now decouple the system, letting
\begin{equation}
	\xi \ = \  x \ + \ y \ ,\qquad \qquad \eta \ = \  x \ - \ y ,
\end{equation}
so that the equations of motion become
\begin{equation}
	\ddot \xi \ = \  - \, \left( \frac{\Lambda + C}{2} \right) \, \xi \ , \qquad \qquad \ddot \eta \ = \  - \, \left( \frac{\Lambda - C}{2} \right) \, \eta \ .
\label{EOM_five_potential}
\end{equation}
The first two are again solvable by quadratures, while the last is the Hamiltonian constraint.

\begin{figure}[ht]
	\centering
	\includegraphics[width=0.55\textwidth]{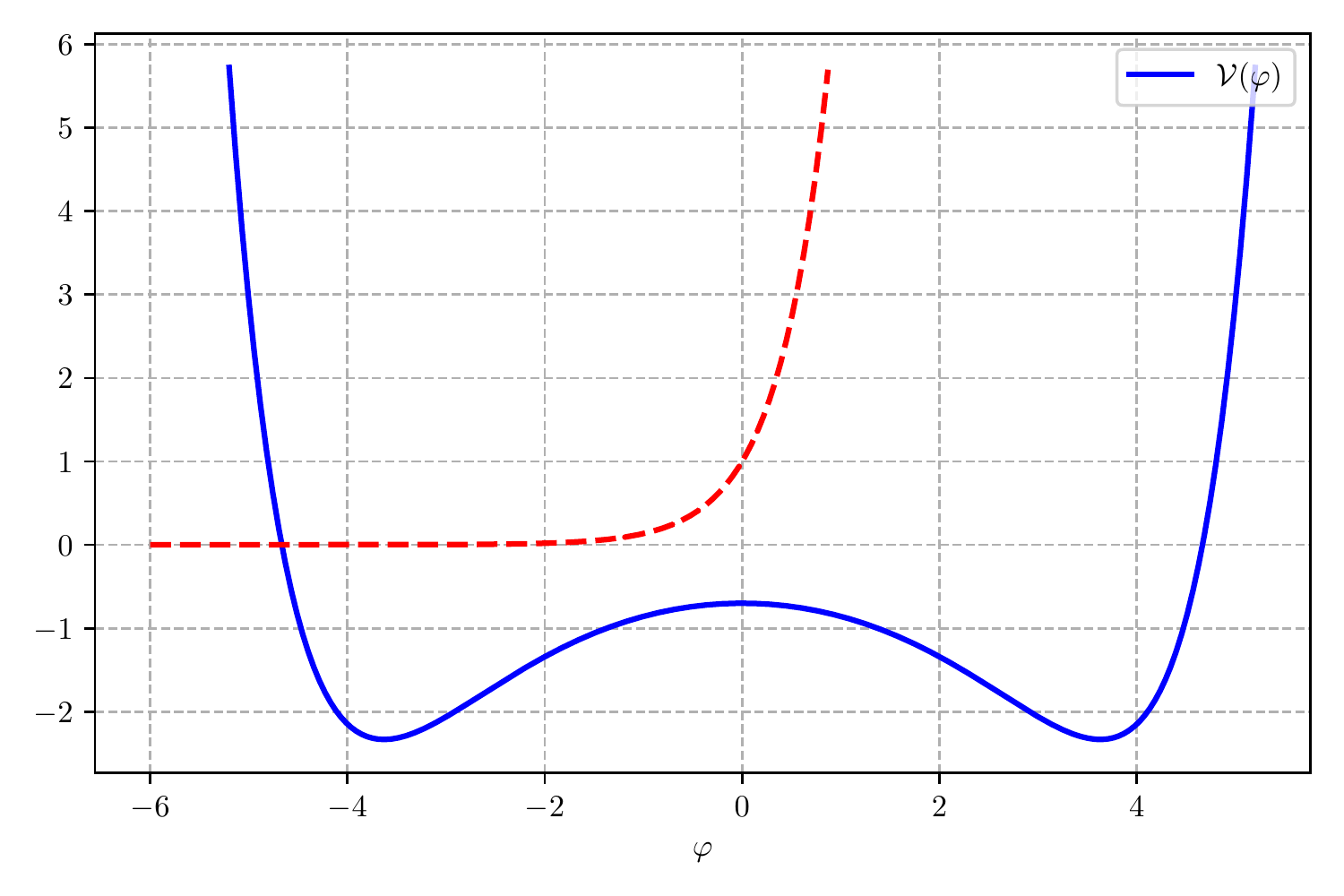}
	\caption{ \small The potential of eq.~\eqref{pot_37} for $C_2 =  1$ and $C_1 =  - \,\frac{7}{10}$. The dashed potential is the one considered in \cite{dm_vacuum}.}
	\label{potcosh4sinh4}
\end{figure}

\noindent 7). \textbf{The seventh integrable potential} that we consider is
\begin{equation}
	\cV (\vf) \   =  \  C_1 \,  \cosh^4\left( \frac{\vf}{3} \right) \ + \  C_2 \, \sinh^4\left( \frac{\vf}{3} \right)\ , \label{pot_37}
\end{equation}
and a particularly interesting option, with $C_1<0$, $C_2>0$, and $C_2 \geq |C_1|$ is displayed in fig.~\ref{potcosh4sinh4}.
Its string--frame counterpart in terms of the conventional dilaton field $\phi$ is
\begin{equation}
	\cV_S (\f) \  = \  \frac{C_1}{16}  \left( e^{-\frac{3}{8} \f} + e^{-\frac{7}{8} \f} \right)^4 \ + \  \frac{C_2}{16} \left( e^{-\frac{3}{8} \f} - e^{-\frac{7}{8} \f} \right)^4\ .
\end{equation}
In this case a convenient gauge choice is
\begin{equation}
	\mathcal B \ = \  \frac{1}{3} \ \mathcal A\ ,
\end{equation}
and the Lagrangian
\begin{equation}
	\cL \ = \  e^{\frac{2}{3} \mathcal A} \, \Big\{ \dot{\mathcal A}^2 \ -\ \dot \vf^2 \ - \ 2 \, e^{\frac{2}{3} \mathcal A} \, \left[ C_1  \, \cosh^4 \left( \frac{\vf}{3} \right)  \ + \  C_2 \, \sinh^4 \left( \frac{\vf}{3} \right) \right] \Big\}\  ,
\end{equation}
is simplified by the redefinitions
\begin{equation}
	e^{\mathcal A} \ = \  ( x\, y)^{\frac{3}{2}} \ , \qquad \qquad e^{ \vf} \ = \  \left( \frac{x}{y} \right)^{\frac{3}{2}} \ , \qquad \qquad e^{\cB} \ = \  (x \, y)^{\frac{3}{2}}\ ,
\end{equation}
which turn it into
\begin{equation}
	\mathcal L \ = \   9 \, \dot x \, \dot y \ - \ 2 \, C_1  \left( \frac{x+y}{2} \right)^4 \ - \ 2\,  C_2  \left( \frac{x - y}{2}\right)^4 \ .
\end{equation}
Finally, letting
\begin{equation}
	\xi  \ = \   \frac{x + y}{2} \ , \qquad \qquad \eta  \ = \   \frac{x - y}{2} \ ,
\end{equation}
leads to the simplest form,
\begin{equation}
	\mathcal L \ = \   9 \, \left( \dot \xi^2 - \dot \eta^2 \right) \ - \ 2 \, C_1 \, \xi^4 \ - \ 2 \, C_2 \, \eta^4 \ ,
\end{equation}
whose equations of motion are
\begin{equation}
	\ddot \xi \ + \ \frac{4}{9} \, C_1 \, \xi^3 \ = \  0 \ , \qquad  \ddot \eta \ - \ \frac{4}{9} \, C_2 \, \eta^3 \ = \  0 \ , \qquad   \dot\eta^2\ - \  \dot \xi^2 \ = \  \frac{2}{9} \, \left[ C_1 \, \xi^4 \, + \, C_2 \, \eta^4 \right] \ .
\label{EOM_six_potential}
\end{equation}
The first two are again solvable by quadratures, while the last is the Hamiltonian constraint.

\begin{figure}[ht]
	\centering
	\includegraphics[width=0.55\textwidth]{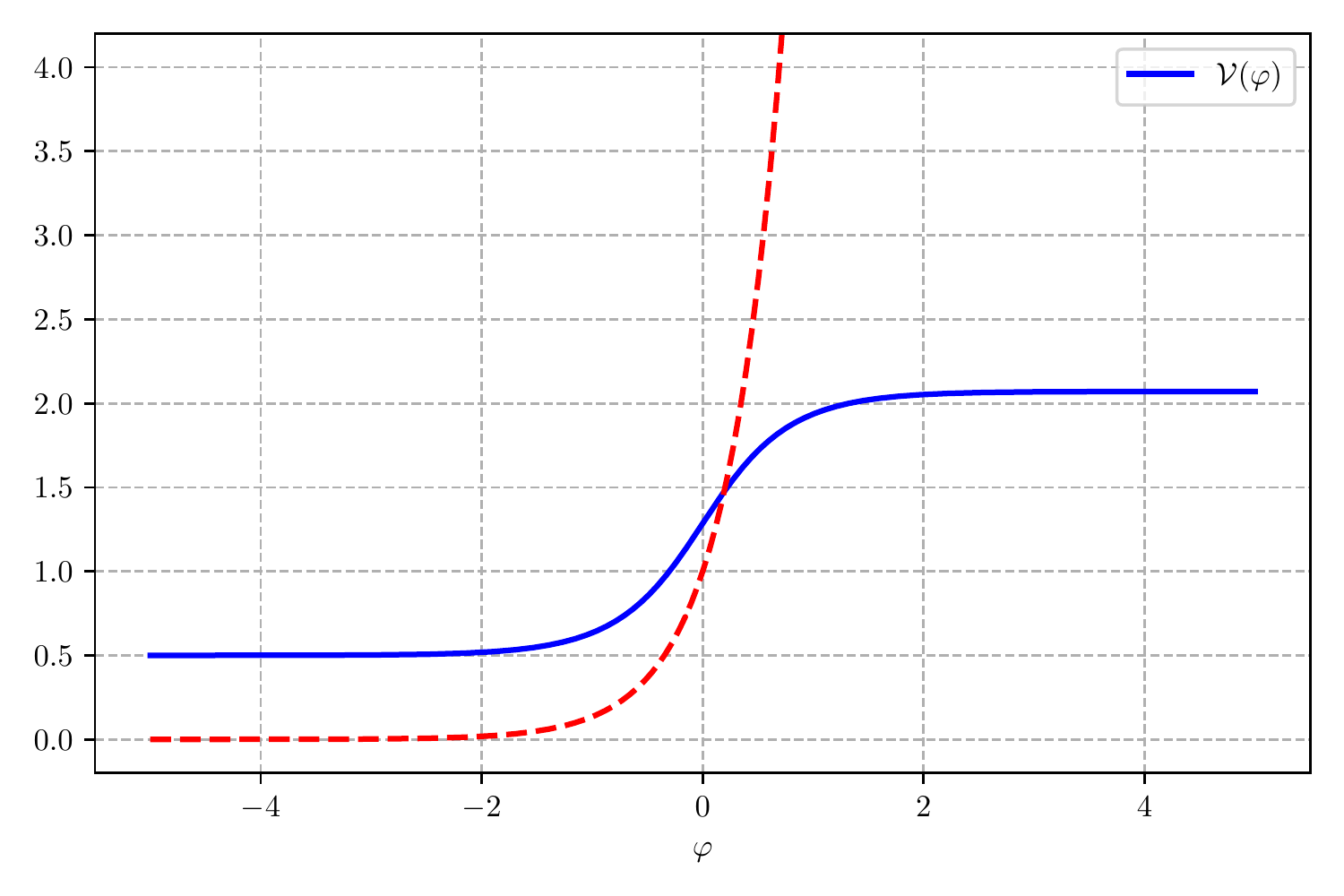}
	\caption{ \small The potential eq.~\eqref{pot_38}, with $C =  1$ and $\Lambda =  \frac 12$. The dashed potential is the one considered in \cite{dm_vacuum}. Notice that, even if the dashed contribution is contained also in \eqref{pot_38}, the higher--order terms make the two potentials qualitatively very different.}
	\label{potarctan}
\end{figure}

\noindent 8). \textbf{The eighth integrable potential} that we consider is
\begin{equation}
	\cV (\vf)  \ = \   \Im \left[  C \, \log \left( \frac{e^{-2 \vf} + i}{e^{-2 \vf} - i} \right)  \ + \ i \, \Lambda \right] \ .
\label{pot_38}
\end{equation}
In principle, the coefficient $C$ could be a complex parameter, but the logarithm is purely imaginary, and one can restrict the attention to real values of $C$.
This potential is equivalent to the step--like function
\begin{equation}
	\cV(\vf) \ = \  2 \, C \arctan \left( e^{2 \vf} \right) \ + \ \Lambda \ ,
\label{arctan}
\end{equation}
which is displayed in fig.~\ref{potarctan}. One can now expand in a power series its string--frame counterpart, and expressing it in terms of the conventional dilaton field $\phi$ gives

\begin{equation}
    \cV_S(\phi) \ =  \   2 \, C \sum_{n=0}^{\infty} \; \frac{(-1)^n}{2n+1}\;  e^{(3n-1)\f} \ + \ \Lambda \, e^{- \frac 5 2 \f} \ .
\end{equation}
Aside from the $\Lambda$ term, which is of non--perturbative flavor, this potential rests on an infinite number of perturbative contributions, starting from the leading one for the orientifolds of~\cite{susy95,sugimoto}.

The change of variables
\begin{equation}
	e^{\cA} \ = \  \sqrt{xy} \ , \qquad \qquad e^{\vf} \ = \  \sqrt{\frac y x} \ , \qquad \qquad e^{\cB} \ = \  \sqrt{\frac{1}{xy}} \ ,
\end{equation}
which also embodies the gauge choice
\begin{equation}
	\mathcal B \ = \  - \mathcal A \ ,
\end{equation}
turns the Lagrangian into
\begin{equation}
	\cL \ = \  4 \, \dot x \, \dot y \ - \ 8 \, \Im \left[ C \, \log \left(\frac{x +iy}{x -iy} \right) \ + \  i \, \Lambda \right] \ .
\end{equation}
Working with the complex variable
\begin{equation}
    z \ = \ x \ + \ i \, y
\label{eight_z}
\end{equation}
in this case one can express the Lagrangian in terms of a complex coordinate and the corresponding velocity, as
\begin{equation}
	\cL \ = \  2 \,  \Im \left[ \, \dot z^2 \ - \ 8 \, C \, \log (z) \ - \ 4 \, i \,  \Lambda \, \right] \ ,
\end{equation}
whose equations of motion are
\begin{equation}
	\ddot z  \ = \ -\,\frac{4 C}{z} \ , \qquad \qquad  \Im \left[ \, \dot z^2 \ + \ 8 \, C \, \log (z) \ + \ 4 \, i \, \Lambda  \,  \right]  \ = \  0 \ .
\end{equation}
The first can be integrated by quadratures in the complex plane, while the second is the usual Hamiltonian constraint.

\noindent 9). \textbf{The ninth integrable potential} that we consider is
\begin{equation}
	\cV (\vf) \ = \  \Im \left[ \,C\, \Big( i + \sinh (2 \gamma \vf) \Big)^{\frac{1}{\gamma} - 1} \right] \ , \label{pot_39}
\end{equation}
so that the Lagrangian becomes
\begin{equation}
	\mathcal L \ = \  e^{\mathcal A - \mathcal B } \, \left\{ \frac{1}{2} \,  \dot{\mathcal A}^2 \ - \ \frac{1}{2} \, \dot \vf^2 \ - \ e^{2 \mathcal B} \, \Im \left[ \, C \, \Big( \, i \ + \ \sinh (2 \gamma \vf) \,  \Big)^{\frac{1}{\gamma} - 1} \, \right]  \right\} \ .
\label{lagrsinhgammameno1}
\end{equation}
\begin{figure}[ht]
	\centering
	\includegraphics[width=0.55\linewidth]{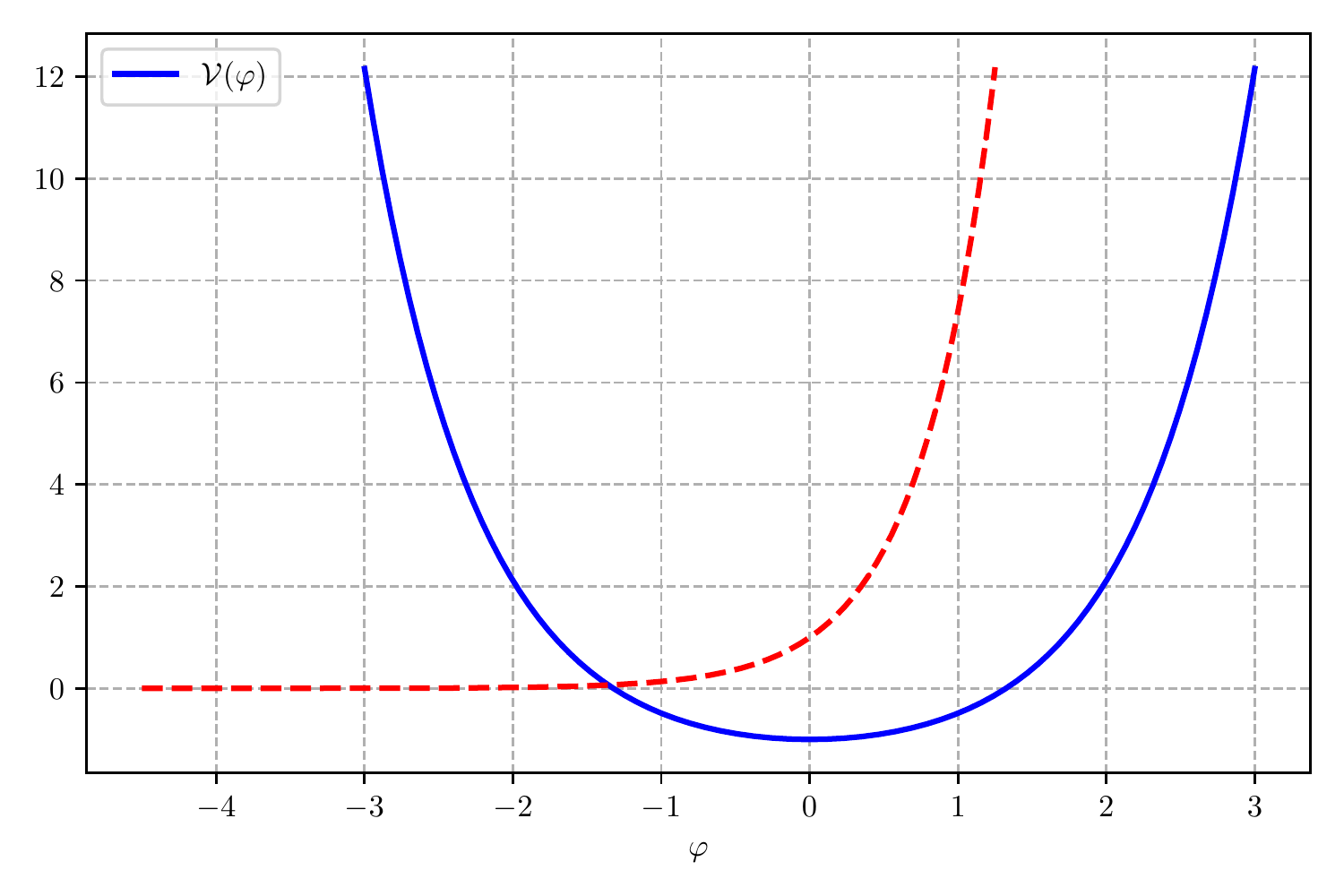}
	\caption{ \small The potential of eq.~\eqref{pot_39} for $\gamma =  \frac13$ and $C =  i$. The dashed potential is the one considered in \cite{dm_vacuum}.}
\end{figure}
This is actually a class of potentials and, resorting to the gauge choice
\begin{equation}
	\mathcal B \ = \  (1 \, - \,  2 \gamma) \, \mathcal A \ ,
\end{equation}
one can introduce the convenient $(x,y)$ variables
\begin{equation}
	e^{ \mathcal A} \ = \  (x\, y)^{\frac{1}{2\gamma}} \ , \qquad \qquad e^{\vf} \, \ = \  \left( \frac{x}{y} \right)^{\frac{1}{2\gamma}} \ , \qquad \qquad e^{\cB} \ = \  (x \, y)^{\frac{1 - 2\gamma}{2 \gamma}} \ ,
\end{equation}
in terms of which
\begin{equation}
	\mathcal L \ = \  \frac{1}{2\gamma^2} \, \dot x \,  \dot y \ - \ \Im \Bigg[ \, \frac{C}{2^{\frac{1}{\gamma} - 1}} \, (x + iy)^{\frac{2}{\gamma} - 2} \, \Bigg] \ .
\end{equation}
This system can be recast in a more compact form introducing the complex variable
\begin{equation}
	z \ = \  x \ + \  i \, y \ ,
\end{equation}
so that, up to an overall factor and a redefinition $\tilde C = \frac{C}{2^{\frac{1}{\gamma} - 2}}$ ,
\begin{equation}
	\mathcal L \ = \  \frac{1}{2}\,\Im \left( \frac{1}{\gamma^2}\, \dot z^2 \ - \ {\tilde C} \, z^{\frac{2}{\gamma} -2}  \right) \ .
\end{equation}
In this fashion, the equations of motion are
\begin{equation}
	\ddot z  \ = \  - \, \gamma \, \left( 1 - \gamma \right) \, \tilde C\, z^{\frac{2}{\gamma} -3} \ , \qquad \qquad \Im \left(
	\frac{1}{\gamma^2}\, \dot z^2 \ + \ {\tilde C} \, z^{\frac{2}{\gamma} -2}  \right) \ = \  0 \ .
\end{equation}
The first can be integrated once more by quadratures in the complex plane, while the second is the Hamiltonian constraint.

\vskip 12pt
\section{\sc  Vacuum Profiles and Bounded String Coupling}\label{sec:profiles}
In this section we analyze the models that we have introduced, paying special attention to cases that grant a bounded string coupling $g_s$, an $r$--direction that has a finite length in the string frame and, in addition, finite values of the nine--dimensional Planck mass and gauge coupling.
We follow the same order as in Section~\ref{sec:models}, so that we begin from the examples that result in triangular systems.
\subsection{\sc Triangular Systems}

\noindent {\sc 1). The Potential of eq.~\eqref{pot_31}}
\begin{figure}[ht]
\centering
\begin{tabular}{cc}
\includegraphics[width=0.4\textwidth]{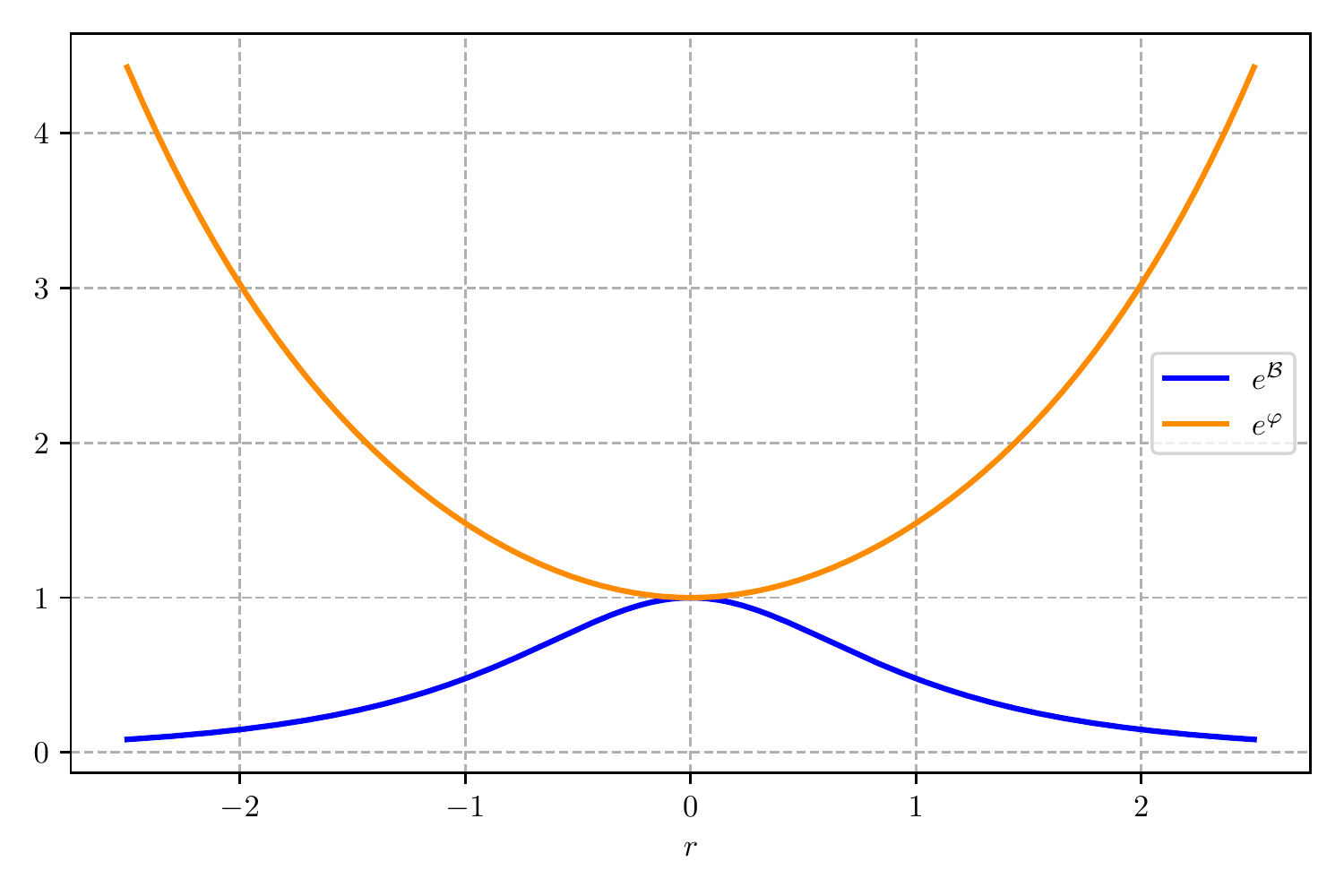} &
\includegraphics[width=0.4\textwidth]{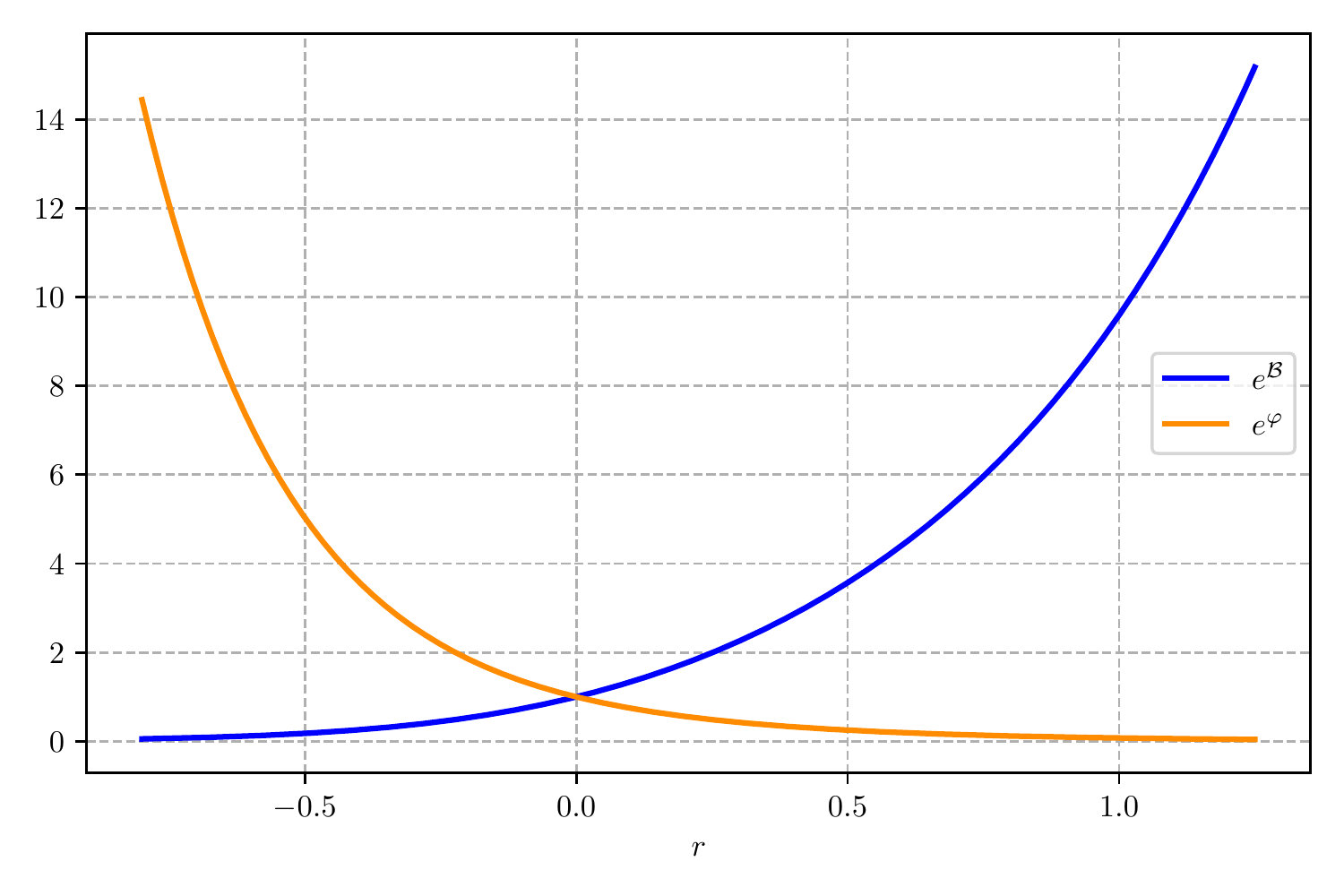} \\
\end{tabular}
\caption{ \small The left panel displays typical solutions for the potential of eq.~\eqref{pot_31_2} corresponding to \emph{positive} values of $D$, with $C =  1$, $a = 1$ and $\alpha =  0$, so that $|\alpha \, a| < \frac{\pi}{4}$. As implied by eq.~\eqref{andamenti}, $g_s$ is unbounded as $r \to \pm \infty$ while $e^{\cB}$ is bounded in both limits. The situation is different in the right panel, where $D$ is again positive and $C =  1$, $a = 1$ and $\alpha =  3$, so that $|\alpha \, a| > \frac{\pi}{4}$. In this case $g_s$ diverges as $r \, \rightarrow \, - \infty$, while $e^{\cB}$ diverges as $r \, \rightarrow \, + \infty$.}
\end{figure}

\noindent For the potential
\begin{equation}
	\cV(\vf) \ = \  C\, \vf \, e^{2 \vf} \ , \label{pot_31_2}
\end{equation}
up to shifts in the coordinate $r$, the general solution for $x(r)$ in eqs.~\eqref{primopoteqmotion} is
\begin{equation}
	x \ = \   C \,(r^2 + D) \ ,
\end{equation}
 where $D \in \mathbb R$ is an integration constant. To begin with, let us choose a positive value for the integration constant, letting $D = a^2$. For $v(r)$ one then needs to solve
\begin{align}
	\ddot v \ = \   - \, \frac{1}{2} \, \frac{1}{r^2 + a^2} \ ,
 \end{align}
so that
 \begin{equation}
 	v \ = \  \alpha \, r \ - \ \frac{r}{2a} \, \arctan\Bigg( \frac{r}{a} \Bigg) \ + \ \frac{1}{4} \, \log \Bigg( \frac{r^2}{a^2} \, + \, 1 \Bigg) \ + \ \beta \ ,
 \end{equation}
where $\alpha$ and $\beta$ are integration constants. The Hamiltonian constraint now sets $\beta = 0$, so that the complete solution reads
 \begin{equation}
 	x \ = \ C (r^2 +a^2) \ , \qquad \qquad v \ = \  \alpha \, r \ - \ \frac{1}{2} \, \frac{r}{a} \, \arctan\Bigg( \frac{r}{a} \Bigg) \ + \ \frac{1}{4} \, \log \Bigg( \frac{r^2}{a^2} \, + \, 1 \Bigg)  \ ,
 \end{equation}
and consequently
 \begin{align}
 	e^{\mathcal A} \ =& \  \sqrt[4]{\frac{|C|}{a^2}} \, \sqrt{r^2 + a^2} \ \exp \left[\alpha \, r  \ - \ \frac{r}{2a}\, \arctan \left(\frac{r}{a} \right) \right] \ , \nonumber \\
	e^{\vf} \ =&\ \sqrt[4]{|C| a^2}  \ \exp \left[ \, - \,  \alpha \, r \ + \ \frac{r}{2a} \, \arctan\left(\frac{r}{a} \right)  \right] \ , \nonumber \\
	e^{\cB} \ =&\ \sqrt[4]{\frac{1}{|C|^3 a^2}}  \ \frac{1}{\sqrt{r^2 + a^2}}\  \exp \left[ \, \alpha \, r \ - \ \frac{r}{2a} \, \arctan\left(\frac{r}{a} \right) \right] \ .
\label{firstsolution}
 \end{align}
Notice that the arguments of the exponential functions behave as
 \begin{align}
 	-\,\alpha \, r \ +\ \frac{r}{2a} \, \arctan\left(\frac{r}{a}\right) \; & \longrightarrow \; -\left( \alpha \, a \ - \ \frac{\pi}{4} \right) \, \frac{r}{a} \qquad \text{for} \qquad r \; \longrightarrow \; + \infty  \ , \nonumber \\
 	-\, \alpha \, r \ +\ \frac{r}{2a} \, \arctan\left(\frac{r}{a}\right) \; & \longrightarrow \; -\left( \alpha \, a \ + \ \frac{\pi}{4} \right) \, \frac{r}{a} \qquad \text{for}\qquad r \; \longrightarrow \; -\infty  \ ,
 	\label{andamenti}
 \end{align}
and therefore \emph{for any choice of $\alpha \, a$ there are strong--coupling regions}.
\begin{figure}[ht]
\centering
\begin{tabular}{cc}
\includegraphics[width=45mm]{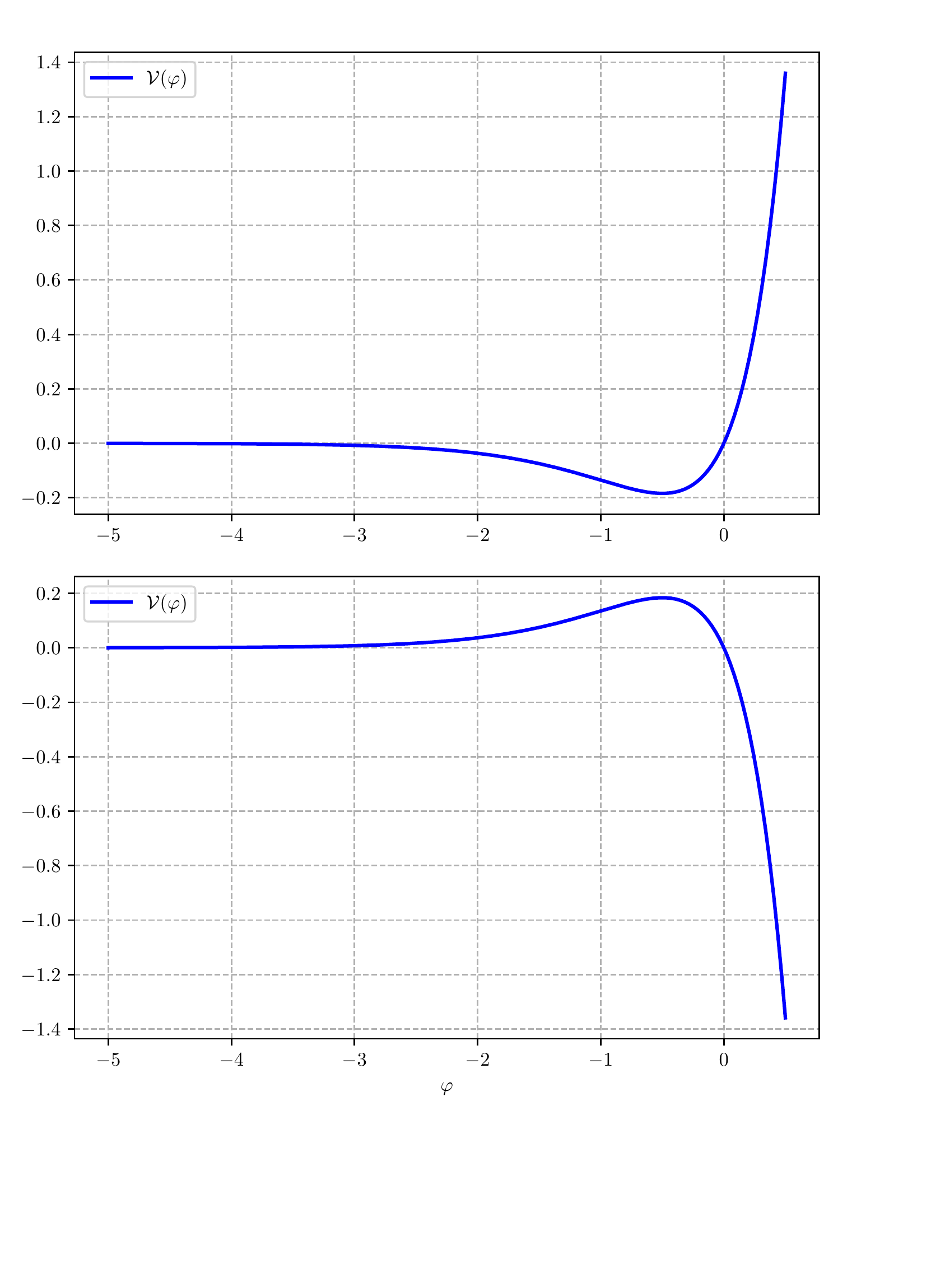} &   \includegraphics[width=0.57\textwidth]{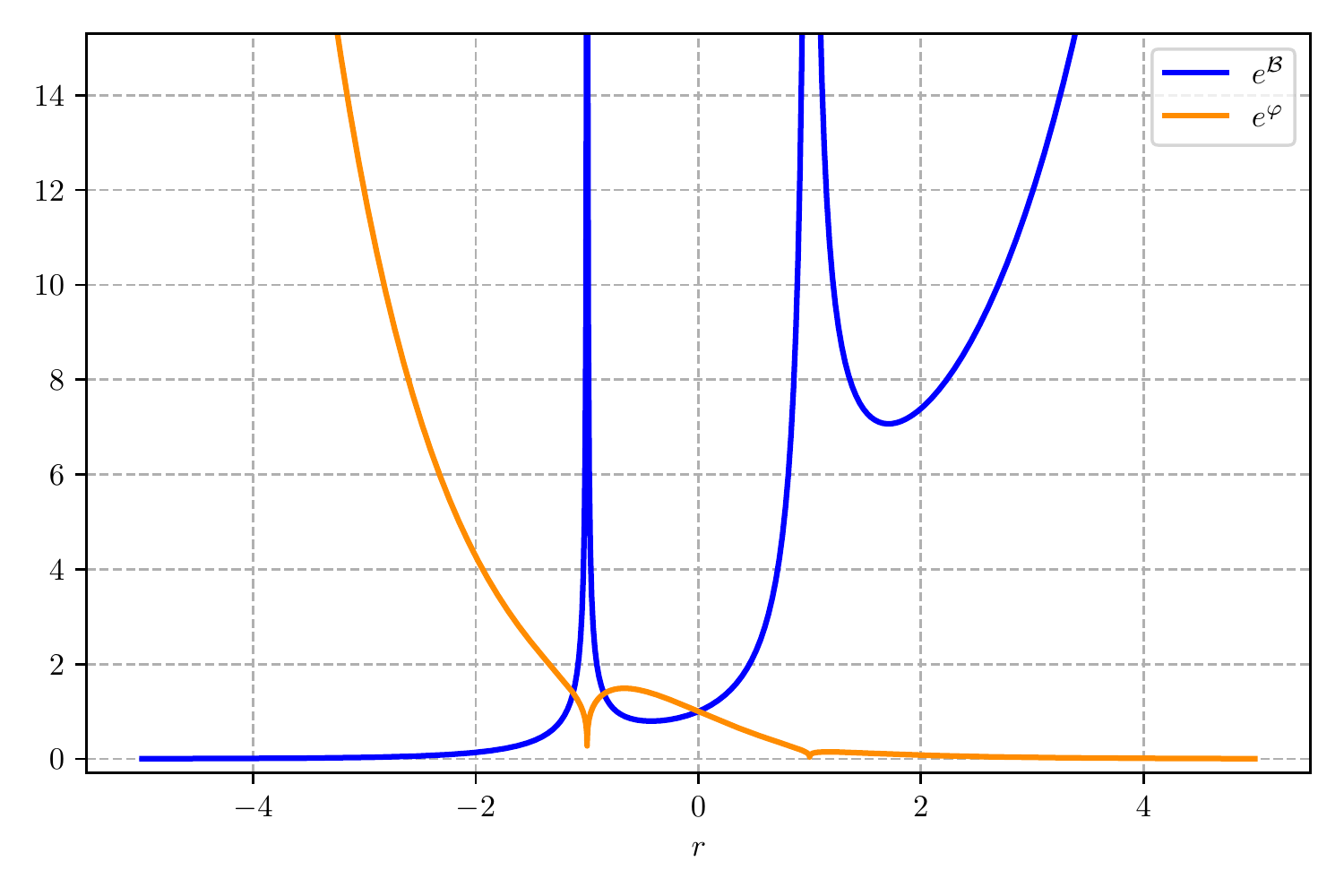}\\
\end{tabular}
\caption{ \small The upper--left panel displays the potential of eq.~\eqref{pot_31_2} for $C>0$, while the lower--left one displays the corresponding inverted potential, which drives the dynamics. The right panel displays $e^{\cB}$ and $e^\vf$ for negative $D$, for the special case $C= 1$, $a =  1$ and $\alpha =  1$. The solution comprises three distinct sectors, and the intermediate one results in a bounded string coupling within an internal interval of finite length.}
\label{D_negative}
\end{figure}

However, one can also allow $D$ to be negative, and letting $D = -a^2$ leads to
\begin{equation}
    x \ = \ C \, (r^2 \, - \, a^2) \ , \qquad v \ = \  \alpha \, r \ -\  \frac{r}{4a} \, \log \left| \frac{r - a}{r + a} \right|\ +\ \frac14 \log|r^2 \, - \, a^2| \ + \ \beta \ .
\end{equation}
The Hamiltonian constraint now fixes $\beta = \frac14 \log(|C|)$, and finally
\begin{align}
 	e^{\mathcal A} \ =& \  \sqrt{|C|} \sqrt{|r^2 \, - \, a^2|} \,\exp \left[\alpha \, r \,  - \, \frac{r}{4a} \log \left| \frac{r - a}{r + a} \right| \right] \ , \nonumber \\
	e^{\vf} \ =& \ \exp \left[- \, \alpha \, r  \, + \, \frac{r}{4a} \, \log \left| \frac{r - a}{r + a} \right| \right] \ , \nonumber \\
	e^{\cB} \ =&\ \frac 1 {\sqrt{|C|}} \, \frac 1 {\sqrt{|r^2 \, - \, a^2|}} \,\exp \left[\alpha \, r \, - \, \frac{r}{4a} \, \log \left| \frac{r - a}{r + a} \right| \right] \ .
 \end{align}
These solutions are displayed in fig.~\ref{D_negative}, but some care is needed to grant them a proper interpretation. To begin with, the dilaton profile exhibits three different regions, which are actually separated by curvature singularities. This indicates that they describe different dynamics, to be considered independently. The inner and right--most regions appear particularly interesting, since the string coupling is bounded there.

However, a closer inspection and the discussion in Section~\ref{sec:models} reveal that the inner portion of the solution actually concerns the inverted potential in the lower--left panel of fig.~\ref{D_negative}, with $C<0$, since $x$ is negative there. The Newtonian analogy is quite useful to guide one's intuition here, provided one takes properly into account that the variable $r$ plays the role of a Euclidean Newtonian time, as we have stressed. As a result, the analogy ought to rely on the \emph{inverted potential}, as is the case in the physics of instantons. One can thus understand qualitatively the inner region referring to the upper--left panel of fig.~\ref{D_negative} (the ``inversion of the inverted potential''), since in this case the sign of the constant $C$ in eq.~\eqref{pot_31_2} flips effectively in regions where $x<0$: the dilaton emerges from large negative values and overcomes the small potential well close to the origin before being reflected by the exponential wall. Alternatively, and equivalently, drawing some parlance from Quantum Mechanics, one might capture this behavior directly looking for ``tunneling regions'' in the actual potential displayed in the lower--left panel. For the outer solutions, the potential is the usual one, and therefore the particle moves in allowed regions of the inverted potential in the lower--left panel, \emph{to the left} of the hill for the region $r>a$, where the coupling is bounded, or \emph{above} the hill for $r<-a$, exploring all possible values of the string coupling. Finally, the interval has a finite length in the first two cases and an infinite length in the last one, and the Planck mass and gauge coupling are finite in the first two cases and infinite in the third. Because of all this, \textbf{the central region is very interesting}, since
\begin{itemize}
    \item the internal $r$--direction is compact;
    \item string coupling $g_s$ is bounded, and vanishes at the ends of the interval;
    \item the 9D Planck mass and gauge coupling are finite.
\end{itemize}
Notice that these results are precisely as expected from the discussion in Section~\ref{sec:compactness}, summarized in Tables~\ref{tab:dilaton_dynamics_1} and~\ref{tab:dilaton_dynamics_2}: the inner region in fig.~\ref{D_negative} refers to the inverted potential, which is dominated by a critical exponential for $\vf>0$, and an upper bound for the string coupling is thus inevitable. In other regions, which refer to the positive potential, weak coupling is possible but is not guaranteed. Similar considerations apply to the length of the internal interval and to the curvature scalar.

\noindent {\sc 2). The Potential of eq.~\eqref{pot_32}}

\noindent Other interesting solutions originate from the class of potentials
\begin{equation}
	\cV(\vf) \ = \   C_1 \,e^{2 \vf} \ + \ C_2 \ .
	\label{pot_32_2}
\end{equation}
Positive values of $C_1$ yield a potential $\cV(\vf)$ that is bounded from below, but a negative value of $C_1$ provides an interesting illustration of what we said in Section~\ref{sec:compactness}. Eqs.~\eqref{triang_2} define again a triangular system, and the general solution for $v$ reads
 \begin{equation}
 	v \ =\  - \,\frac{C_1}{2} \, r^2 \ +\  a \, r \ +\ b \ ,
 \end{equation}
 where $a$ and $b$ are two real integration constants. Up to a shift of $r$ one can set $a=0$, and up to rescalings and a constant shift of the dilaton one can also set $b=0$. Therefore, for the sake of brevity we shall work with the simpler expression
 \begin{equation}
 	v \ = \ - \, \frac{C_1}{2} \, r^2 \ .
 \end{equation}

We can now recall the definition of the \textit{error function},
\begin{equation}
	\erf (s) \ = \  \frac{2}{\sqrt{\pi}} \int_0^s \td t  \; e^{-t^2} \ ,
\label{erf}
\end{equation}
which will recur in the following, and actually its extension to the complex plane will also play a role. The solution found for $v$ leads indeed to
 \begin{equation}
 	\ddot x \ = \  - \  2 \, C_2 \, e^{- \, C_1 \, r^2}
 \end{equation}
and, taking into account the definition of $\erf(x)$, two integrations lead to
\begin{equation}
 	x  \ = \   \alpha \, r \ + \  \beta \ -\ C_2 \, \sqrt{\frac{\pi}{C_1}} \, r \, \erf \left(\sqrt{C_1} r \right) \ - \ \frac{C_2}{C_1} \, e^{- \, C_1 \, r^2} \ ,
\end{equation}
where $\alpha$ and $\beta$ are integration constants. The Hamiltonian constraint demands that $\beta = 0$, so that finally
 \begin{equation}
 	v \ =\ - \,\frac{C_1}{2} \, r^2 \ , \qquad \qquad  x \ = \ \alpha \, r \ - \ C_2 \, \sqrt{\frac{\pi}{C_1}} \, r \, \erf \left(\sqrt{C_1} r \right) \ - \ \frac{C_2}{C_1} \,  e^{-\, C_1 \, r^2}\ .
\label{coord_second}
\end{equation}
Consequently
\begin{align}
	e^{\mathcal A} \, = & \; \; x^{\frac12} \, e^v \; =   \left( \alpha \, r \, \, e^{- C_1 r^2} - \, C_2 \, \sqrt{\frac{\pi}{C_1}} \, r \, \erf \left(\sqrt{C_1} r \right) \, e^{-C_1 r^2}\, - \, \frac{C_2}{C_1} \,  e^{-2 \, C_1 \, r^2} \right)^{\frac12} \ , \nonumber \\
	e^{\vf} \, = & \; x^{\frac12} \, e^{-v} \, =  \left( \alpha \, r \, e^{C_1 r^2} \, - \, C_2 \, \sqrt{\frac{\pi}{C_1}} \, r \, \erf \left(\sqrt{C_1} r \right) \, e^{C_1 r^2} \, - \, \frac{C_2}{C_1}  \right)^{\frac12} \ , \nonumber\\
	e^{\cB} \, = & \; x^{-\frac12} \, e^v \, =  \left( \alpha \, r \, e^{C_1 r^2}\, - \, C_2 \, \sqrt{\frac{\pi}{C_1}} \, r \, \erf \left(\sqrt{C_1} r \right) \, e^{C_1 r^2} \, - \, \frac{C_2}{C_1}  \right)^{-\frac12}  \ .
\label{secondsolution}
\end{align}
If $C_1$ is positive, the potential is the one analyzed in~\cite{dm_vacuum}, with the addition of a constant term that would have a non--perturbative origin in String Theory. In this case one can see that the string coupling becomes unbounded for all possible values of $C_2$. In fact, if $C_2 = 0$ one can assume without loss of generality that $\alpha > 0$, and $e^{\vf}$ becomes unbounded for large positive values of $r$: this is what happens in~\cite{dm_vacuum}, up to our different choice of normalization for ${\cal A}$ and $\varphi$. On the other hand, if $C_2$ is positive the square root in the solution attains positive values for some $r$ only if $|\alpha| > C_2\sqrt{\frac \pi {C_1}}$. One can consider again $\alpha$ positive, without any loss of generality, since negative values can be compensated by a reversal of $r$, and then $e^{\vf}$ diverges for large positive values of $r$. The analysis for $C_2$ negative is very similar, and in this case one is allowed to choose any value of $\alpha$, but the end result is still an unbounded string coupling. In all these cases the nine--dimensional Planck mass and the gauge coupling are finite.
\begin{figure}[ht]
\centering
\begin{tabular}{cc}
\includegraphics[width=45mm]{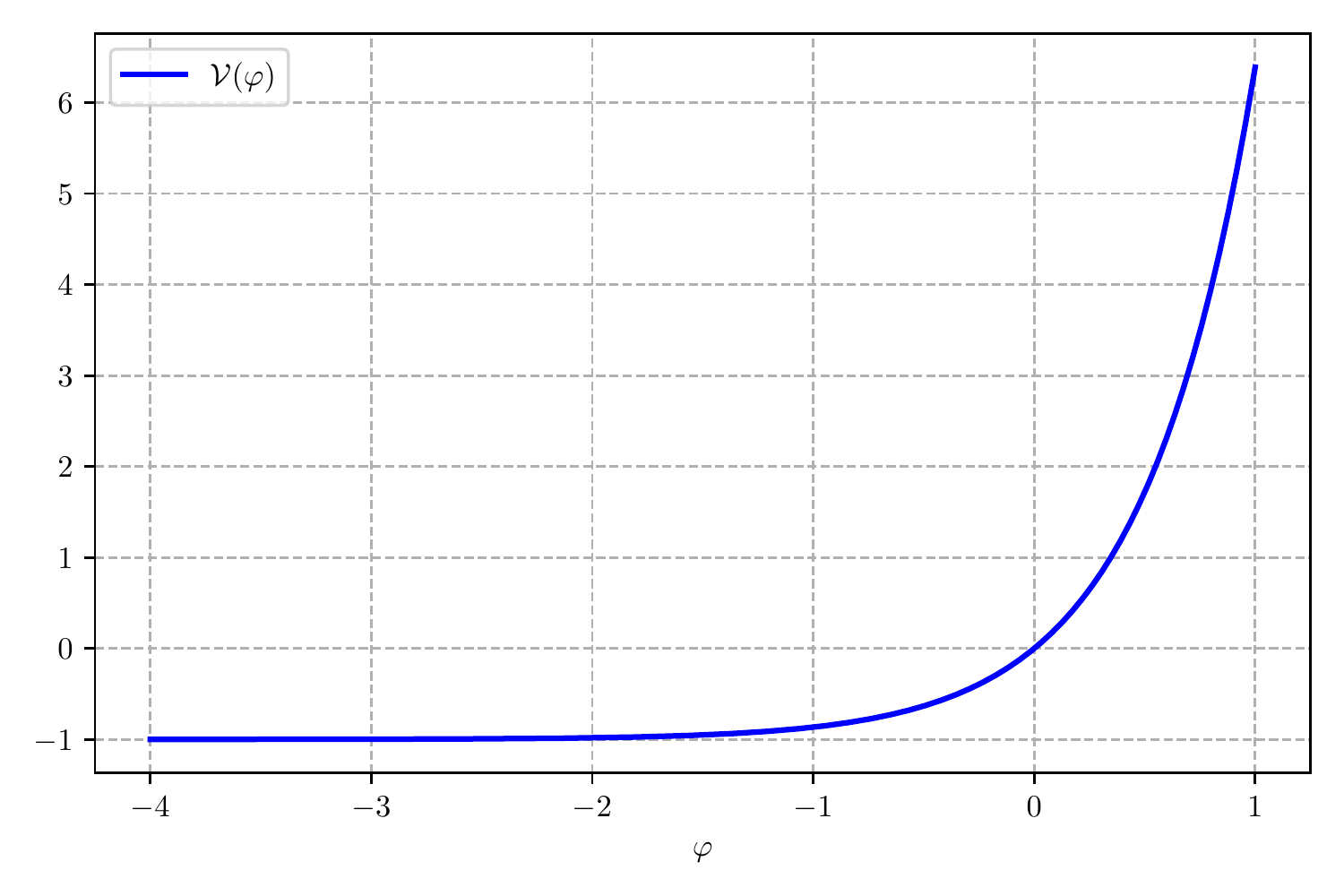} &
\includegraphics[width=45mm]{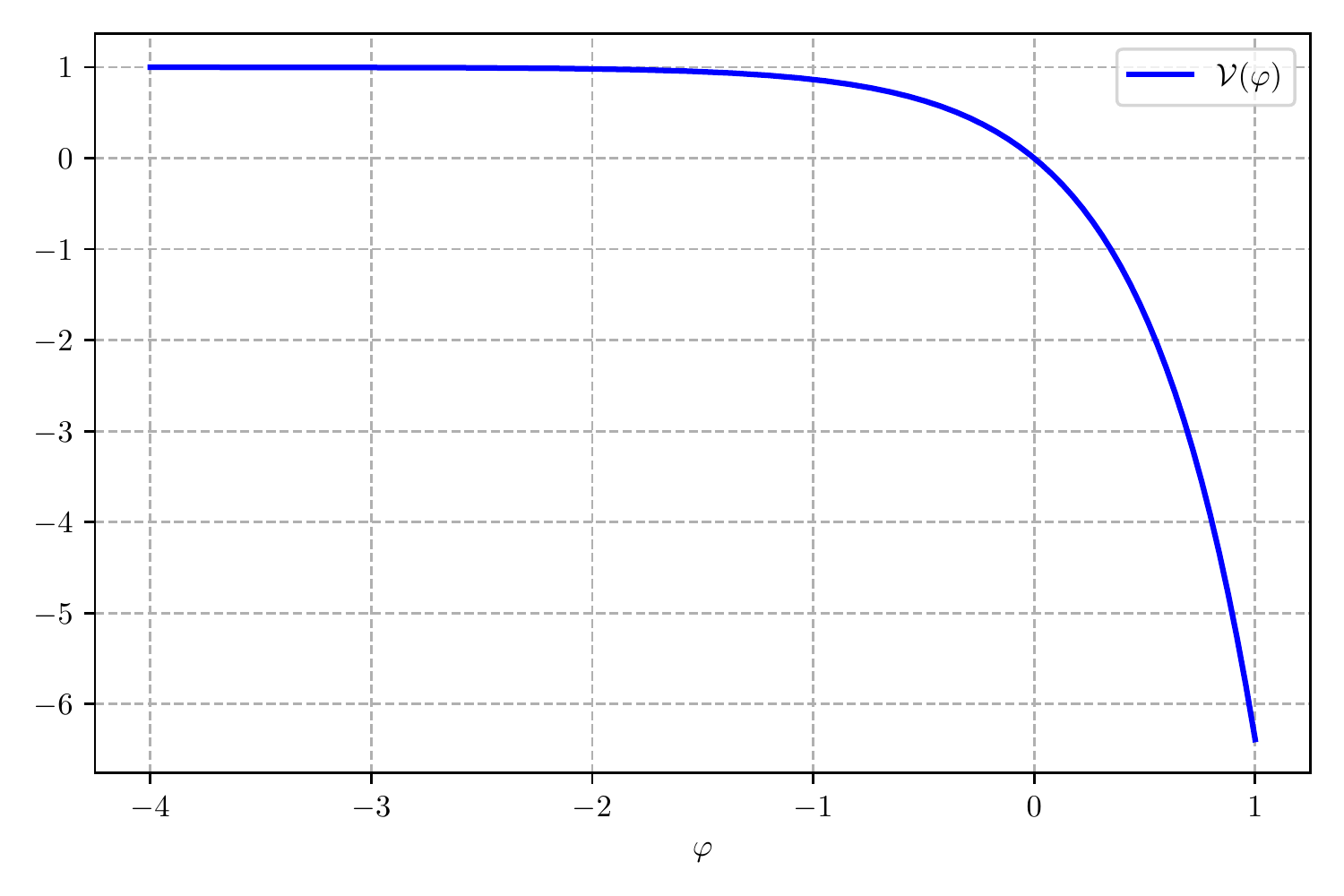} \\
\includegraphics[width=0.4\textwidth]{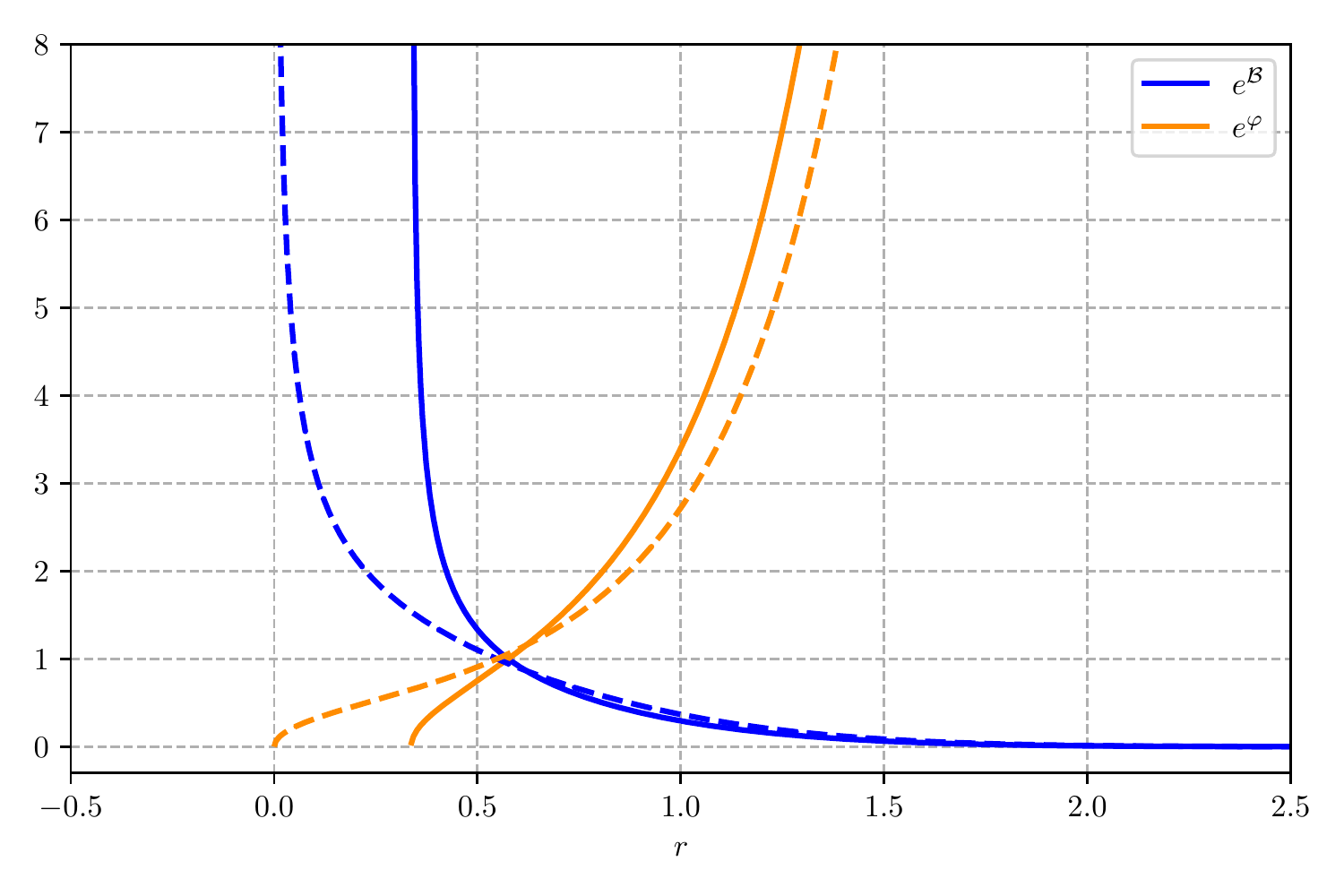} &
\includegraphics[width=0.4\textwidth]{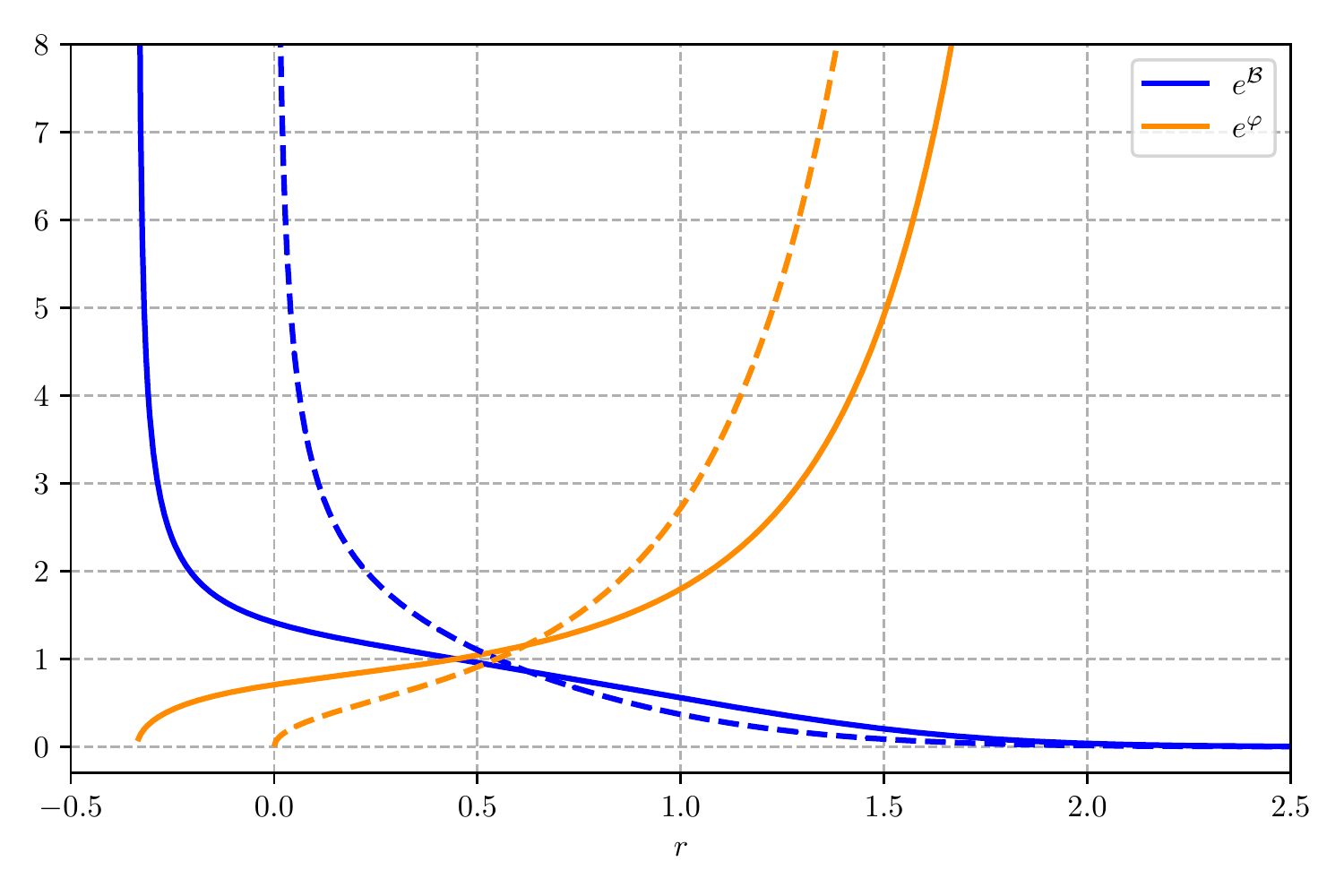} \\
\end{tabular}
\caption{ \small The upper--left panel displays the potential of eq.~\eqref{pot_32_2} for $C_1 > 0$ and $C_2 < 0$, while the upper--right one displays the corresponding inverted potential. The lower left panel displays the solution for $C_1 =  1$, $C_2 = \frac12$ and $\alpha =  1$, so that $|\alpha| > C_2 \sqrt{\frac{\pi}{C_1}}$. As expected $g_s$ is unbounded, and the same is true for $e^{\cB}$. The situation is similar in the lower right panel, where $C_2 =  -\frac12$. In both cases, the dashed lines are the solutions for the low--lying orientifold potential considered in \cite{dm_vacuum}.}
\end{figure}

\begin{figure}[ht]
\centering
\begin{tabular}{cc}
\includegraphics[width=45mm]{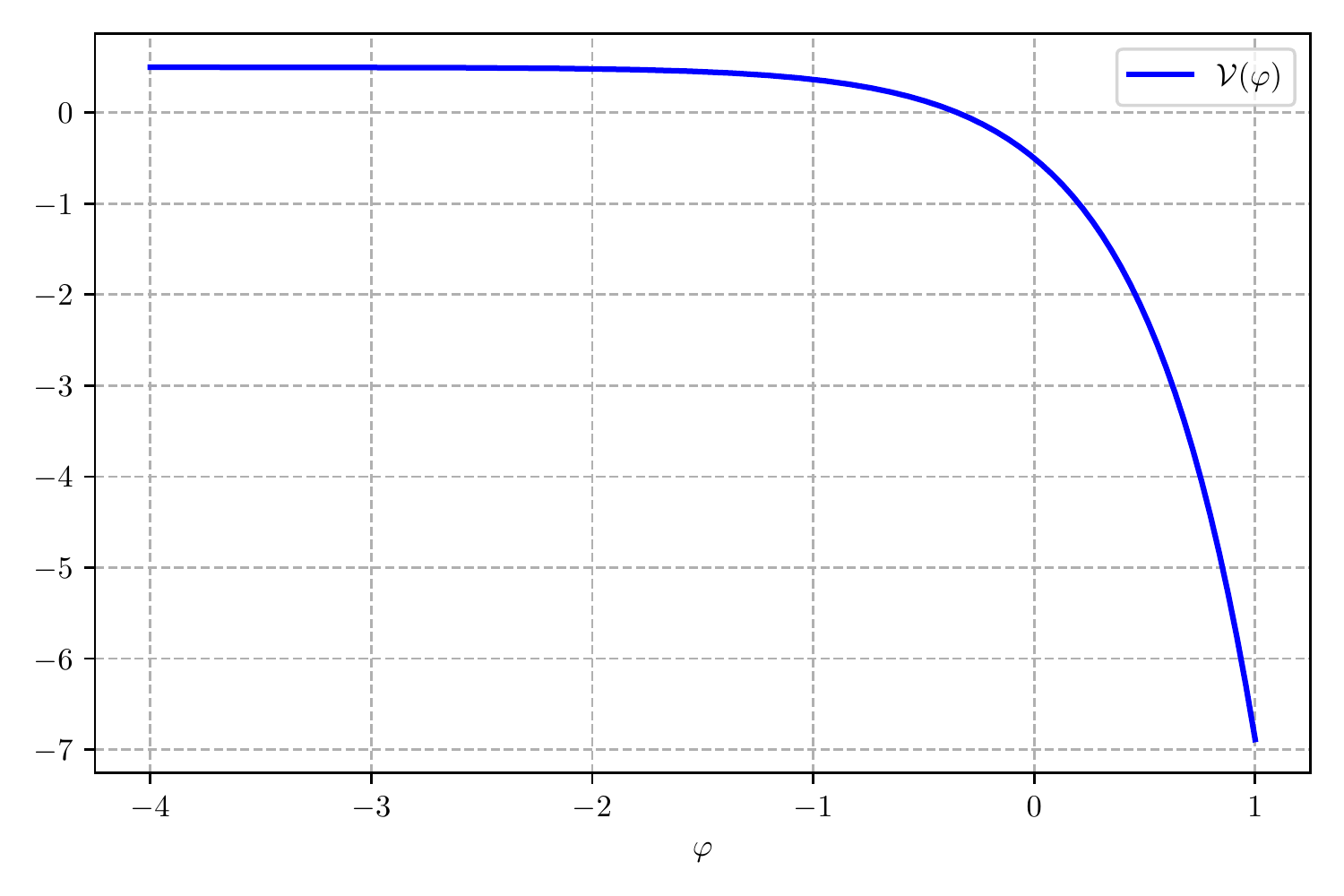} &
\includegraphics[width=45mm]{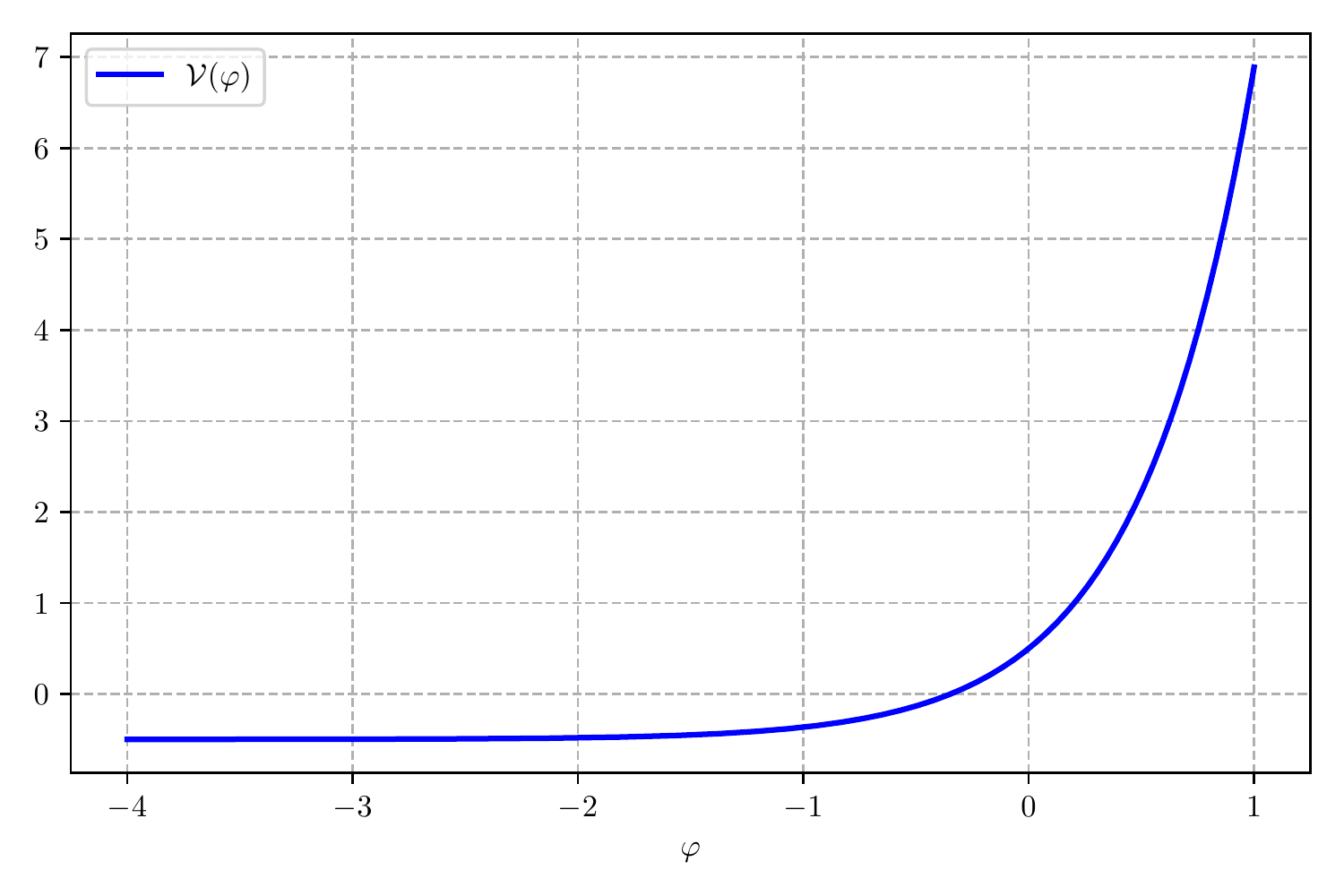} \\
\includegraphics[width=0.4\textwidth]{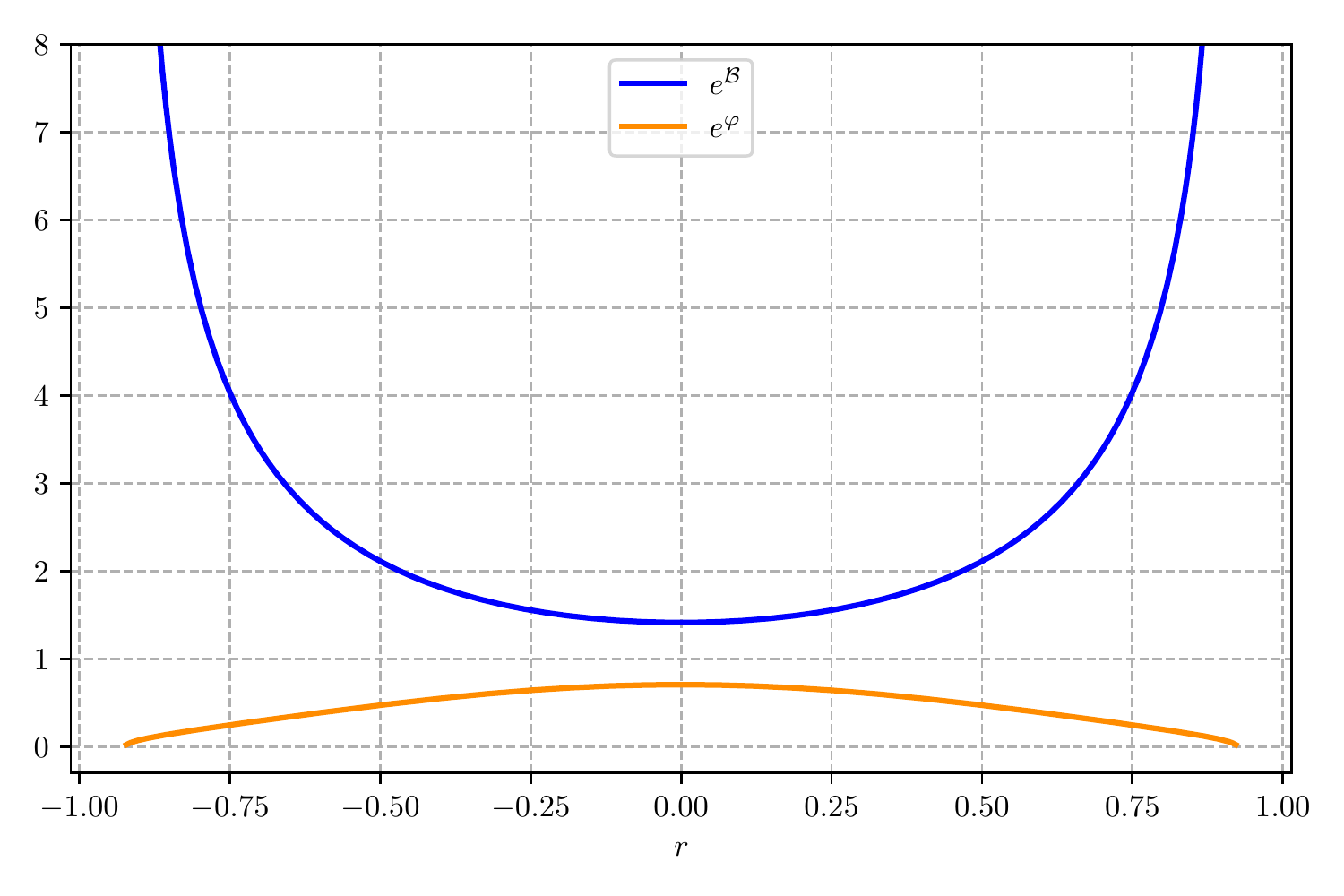} &
\includegraphics[width=0.4\textwidth]{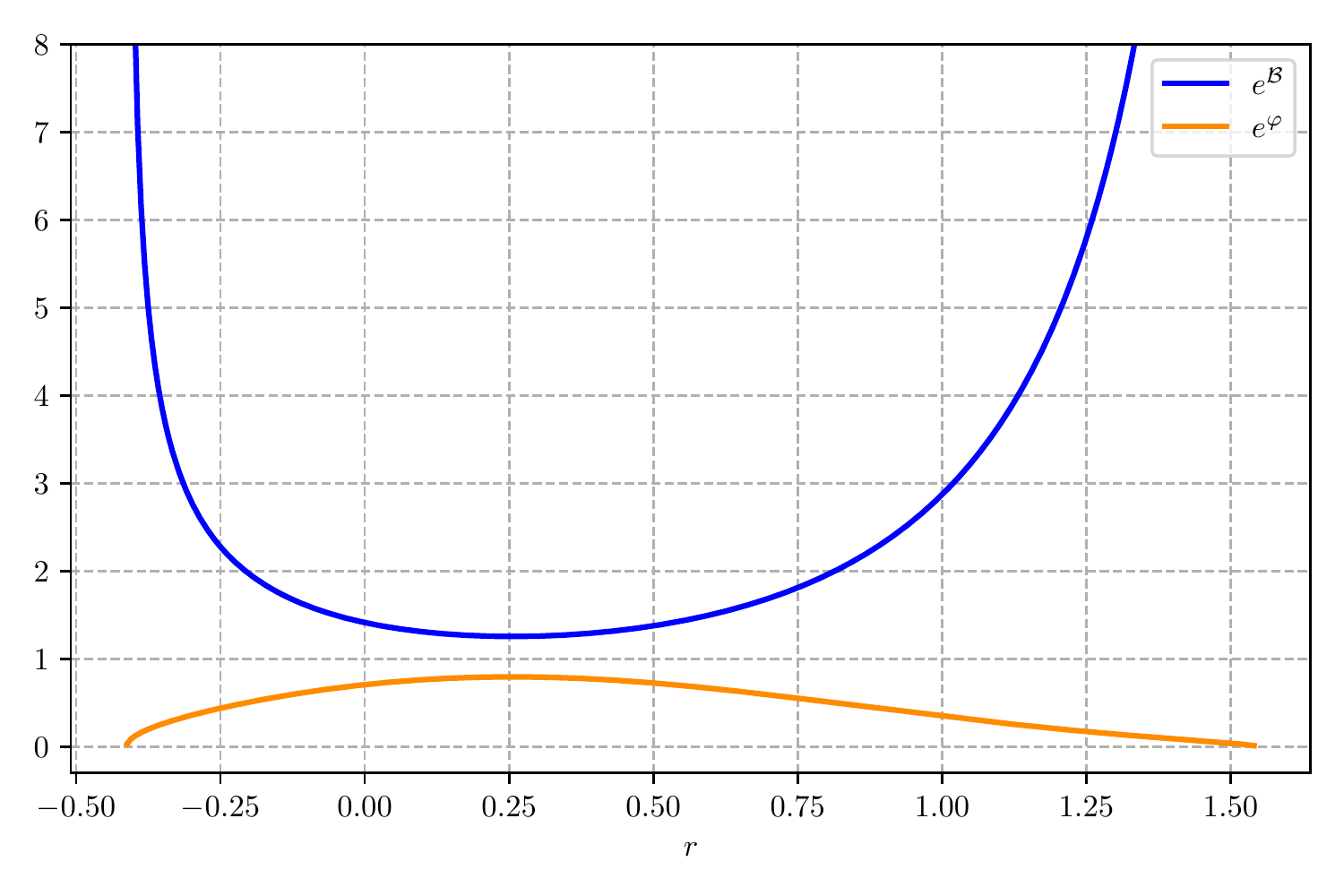} \\
\end{tabular}
\caption{ \small The upper--left panel displays the potential of eq.~\eqref{pot_32_2} for $C_1 < 0$ and $C_2 > 0$, while the upper--right one displays the corresponding inverted potential. The lower left panel shows the behavior of the solution for $C_1 = -1$, $C_2= \frac12$ and $\alpha = 0$. As expected, $g_s$ is bounded and $e^{\cB}$ diverges. In the lower right panel the situation is similar, but the non--vanishing $\alpha = 1$ deforms slightly the solution.}
\label{C2negative1}
\end{figure}

However, if we allow $C_1$ to be negative, differently from \cite{dm_vacuum} (and from the actual starting point in String Theory), the resulting solution can be cast in the form
\begin{align}
	e^{\mathcal A} \ = & \ \; x^{\frac12} \, e^v \ = \  \left( \alpha \, r \,  e^{|C_1|r^2} \,+\, i\,  C_2 \, \sqrt{\frac{\pi}{|C_1|}} \, r \erf \left(i \sqrt{|C_1|} \,r \right) \, e^{|C_1|r^2} \, + \, \frac{C_2}{|C_1|} \, e^{2 |C_1|r^2} \right)^{\frac12}  \ , \nonumber \\
	e^{\vf} \ = & \ x^{\frac12} \, e^{-v} \ = \ \left( \alpha \, r \,e^{-|C_1|r^2} \, + \, i \, C_2 \, \sqrt{\frac{\pi}{|C_1|}} \, r \erf \left(i \sqrt{|C_1|} \,r \right) \, e^{-|C_1|r^2} \, + \, \frac{C_2}{|C_1|}  \right)^{\frac12} \ , \nonumber\\
	e^{\cB} \ = & \ x^{-\frac12} \, e^v \ = \ \left( \alpha\, r \,e^{-|C_1|r^2} \, + \,i \, C_2 \, \sqrt{\frac{\pi}{|C_1|}} \, r \erf \left(i \sqrt{|C_1|} \,r \right) \, e^{-|C_1|r^2} \, + \, \frac{C_2}{|C_1|}  \right)^{-\frac12}  \ ,
\label{secondsolution2}
\end{align}
and \textbf{has some peculiar features}, which were anticipated in Section~\ref{sec:profiles}. For $C_2 > 0$, indeed:
\begin{itemize}
    \item the internal $r$--direction has a finite length;
    \item the string coupling $g_s$ is bounded;
    \item the 9D Planck mass and gauge coupling are finite.
\end{itemize}
However, the string--frame curvature \eqref{string_curvature} is unbounded. One can show that this is the case noting that, for large values of $x$
\begin{equation}
	-i \erf(ix) \ = \  \frac{2}{\sqrt{\pi}} \int_0^x \td s \, e^{s^2} \ \simeq \ \frac{e^{x^2}}{\sqrt{\pi} \, x} \ + \ \frac{ e^{x^2}}{2 \, \sqrt{\pi}\, x^3} \ + \ \cO \left( \frac{e^{x^2}}{x^5}  \right) \ ,
\end{equation}
so that the argument of the square roots in (\ref{secondsolution}) in this limit approaches
\begin{equation}
	\alpha\, r \,e^{-|C_1|r^2} \, + \,i \, C_2 \, \sqrt{\frac{\pi}{|C_1|}} \, r \erf \left(i \sqrt{|C_1|} \,r \right) \, e^{-|C_1|r^2} \, + \, \frac{C_2}{|C_1|} \ \simeq \ \alpha\, r \,e^{-|C_1|\, r^2} \ - \ \frac{C_2}{|C_1|^2 r^2} \ + \ \cO \left( \frac{1}{r^4} \right)  \ .
\end{equation}
For this reason, if $C_2$ is positive, this argument will eventually become negative for large values of $r$. When this happens, the string coupling, which was previously bounded, vanishes and the very range of $r$ ends. Moreover, the gauge choice $\cB =  - \, \vf$ implies that $e^{\cB}$ diverges, but only as an inverse square root. Therefore, the behavior of $e^{\cB + \frac{\vf}{3}}$ results in a compact internal direction. Fig.~\ref{C2negative1} displays a solution of this type.
It is also straightforward to show that the reduced Planck mass and gauge coupling are finite. On the other hand, the behavior of the curvature is slightly more subtle, since it is potentially singular at the two ends of the allowed range of $r$. It will suffice to consider the left end $r=r^*$, where
\begin{equation}
	e^{\vf} \sim \sqrt{r - r^*} \ , \qquad \qquad e^{\cA} \sim \sqrt{r - r^*} \ , \qquad \qquad e^{\cB} \sim \frac{1}{\sqrt{r - r^*}} \ .
\end{equation}
Consequently
\begin{equation}
	e^{-\frac 2 3 \vf - 2 \cB} \sim (r - r^*)^{\frac 2 3}
\end{equation}
and
\begin{equation}
	\dot \cA \sim \frac{1}{r- r^*} \ , \qquad \qquad \ddot \cA \sim \frac{1}{(r- r^*)^2} \ , \qquad \qquad \dot \cB \sim \frac{1}{r- r^*} \ ,
\end{equation}
and the string--frame curvature \eqref{string_curvature} is not bounded. We refrain from discussing explicitly the additional option for the potential, corresponding to negative values for both $C_1$ and $C_2$, since it leads to a non--compact internal space. Once more, the potential is dominated by a critical exponential for $\vf>0$, and consequently these results are precisely along the lines of what we discussed in Section~\ref{sec:compactness}.

\noindent {\sc 3). The potential of eq.~\eqref{pot_33}}

\noindent Up to a shift in $\vf$ and a rescaling of the coordinate $r$, the potential of eq.~\eqref{pot_33},
\begin{equation}
	\cV(\vf) \ = \  C_1 \, e^{2 \gamma \vf} + C_2 \, e^{(\gamma + 1) \vf} \ ,
\end{equation}
can be cast in the form
\begin{equation}
	\cV(\vf) \ = \    \varepsilon_1 e^{2\gamma \vf} + \varepsilon_2 e^{(\gamma+1) \vf} \ ,
	\label{pot_33_2}
\end{equation}
with $\varepsilon_1, \varepsilon_2 = \pm1$. One can find solutions for all values of $\gamma$, but for brevity we shall content ourselves with the region $\gamma > -1$, and thus with cases where at least one exponential is raised to a positive power. These are more directly connected to perturbative String Theory, but cases with two negative powers are also interesting and could be discussed performing a $\vf \to - \vf$ redefinition. In this fashion, peculiar behaviors that set upper bounds on $g_s$ are mapped into others that set lower bounds on it.

\noindent $\boxed{|\gamma| < 1\,, \;\varepsilon_1 =  1\,, \;\varepsilon_2 = 1}$

\noindent With these choices the potential is always positive, and consequently the solutions contain strong--coupling regions.
\begin{figure}[ht]
\centering
\begin{tabular}{cc}
\includegraphics[width=0.4\textwidth]{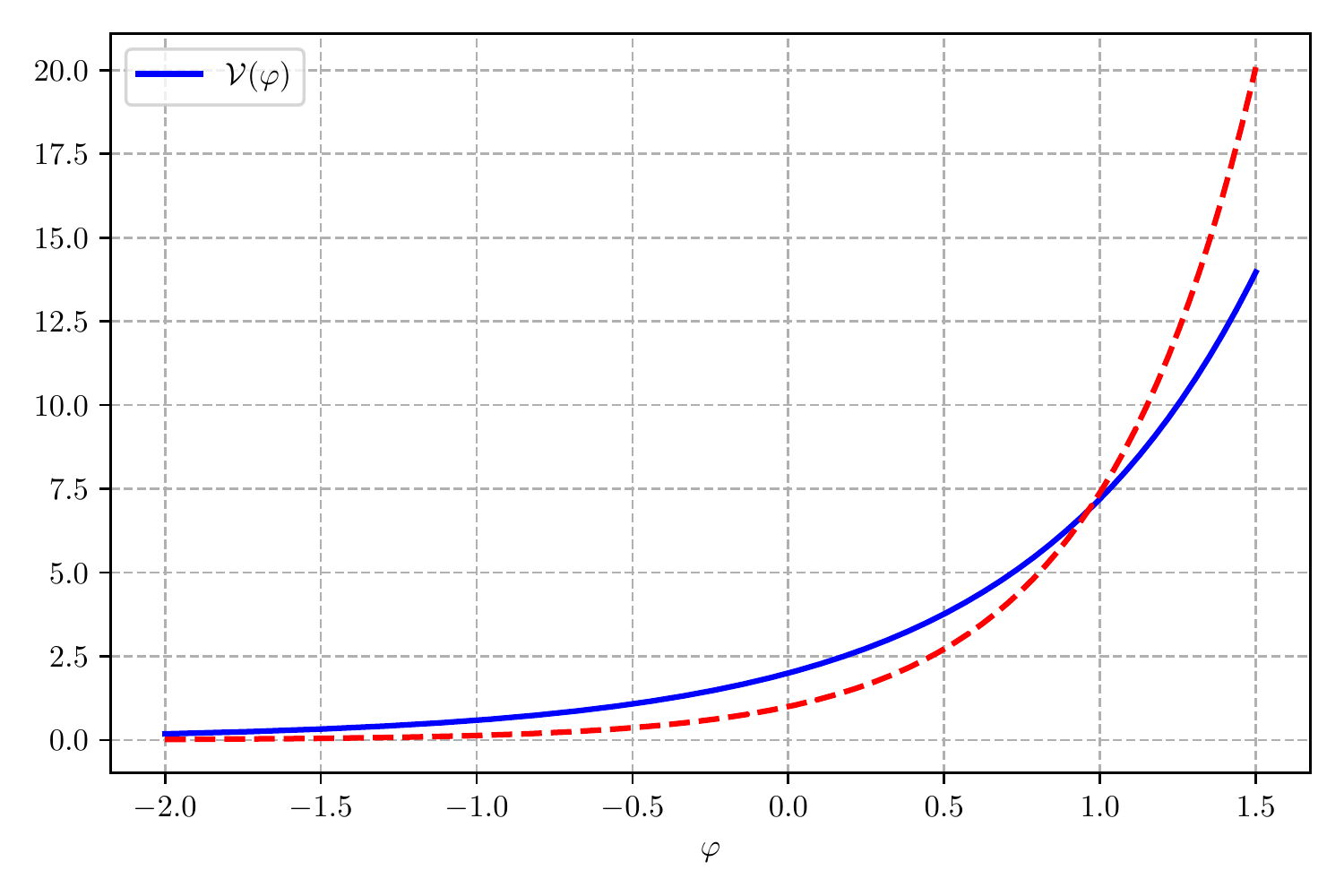} &
\includegraphics[width=0.4\textwidth]{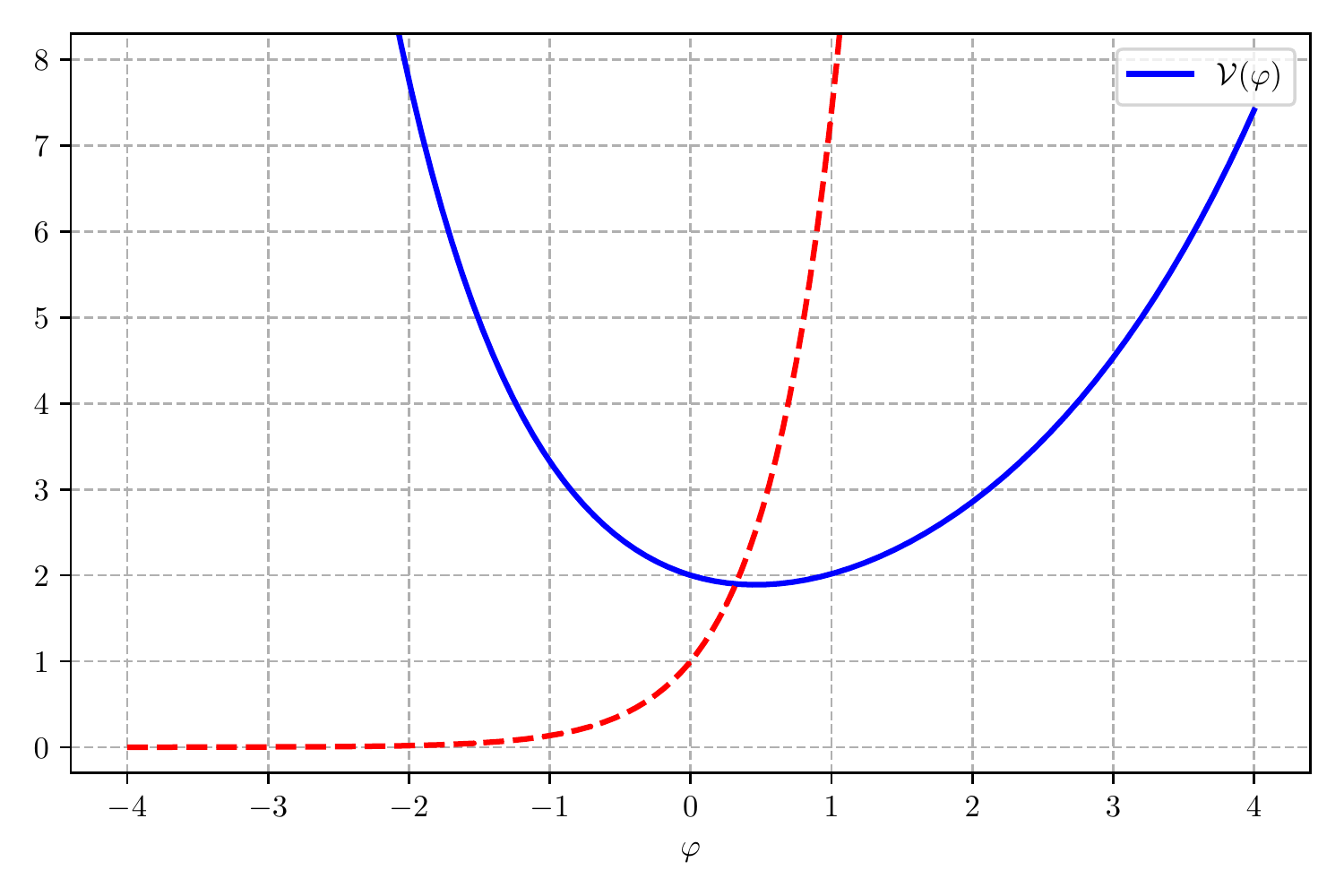} \\
\end{tabular}
\caption{ \small The left panel displays a potential $\cV(\vf)$ of eq.~\eqref{pot_33_2}, with $0 < \gamma < 1$ and $\ve_1 = \ve_2 = 1$, and in particular here $\gamma =  \frac12$. The right panel displays instead a typical potential $\cV(\vf)$ with $-1 < \gamma < 0$, and in particular here $\gamma =  - \frac12$. In both cases, the dashed potentials are the low--lying orientifold contributions considered in \cite{dm_vacuum}.}
\end{figure}
Letting
\begin{equation}
    \Omega^2 =  \displaystyle \frac{1 - \gamma^2}{2} \ ,
\label{omega2}
\end{equation}
the equations of motion reduce to
\begin{equation}
	\ddot x \ +\  \Omega^2 \, x  \ = \  0 \ , \qquad  \ddot y \ + \ \Omega^2 \, y  \ = \  - \, (1-\gamma) \, x^{\frac{1 - \gamma}{1+ \gamma}} \ , \qquad \dot x \, \dot y \ = \  - \, \Omega^2 \, \left[  x \, y \, + \, x^{\frac{2}{1 + \gamma}} \right] \ .
\end{equation}
Up to the usual shift in $r$, and up to some prefactors in $e^{\mathcal A}$ and $e^{\vf}$, the solution for $x$ can be brought to the form
\begin{equation}
	x(r) \ = \  \sin(\Omega r) \ ,
\end{equation}
so that the equation  one needs to solve for $y(r)$ is
\begin{equation}
	\ddot y \ + \ \Omega^2 \, y \ = \  - \, (1-\gamma) \left[ \sin(\Omega r) \right]^{\frac{1 - \gamma}{1 + \gamma}} \ .
\end{equation}
The solution is
\begin{equation}
	y(r) \ = \  \Bigg \{ b + \frac{2 \Omega}{1 + \gamma} \int_0^{r} \td s \; \sin^{\frac{2}{1 + \gamma}} (\Omega s) \Bigg \}  \cos(\Omega r) -  \sin(\Omega r)^{\frac{3 + \gamma}{1 + \gamma}} \ ,
\end{equation}
where we have eliminated one integration constant imposing the Hamiltonian constraint, and this result translates into
\begin{align}
	e^{\mathcal A}  \, = & \;  \sin(\Omega r)^{\frac{1}{1+ \gamma}} \Bigg\{ \Bigg [ b + \frac{2 \Omega}{1 + \gamma} \int_0^{r} \td s \; \sin^{\frac{2}{1 + \gamma}} (\Omega s) \Bigg ] \cos(\Omega r)  - \left[ \sin(\Omega r) \right]^{\frac{3 + \gamma}{1 + \gamma}} \Bigg\}^{\frac{1}{1-\gamma}} \ , \nonumber \\
	e^{\vf} \,  = \;&   \sin(\Omega r)^{\frac{1}{1 + \gamma}} \Bigg\{ \Bigg [ b + \frac{2 \Omega}{1 + \gamma} \int_0^{r} \td s \; \sin^{\frac{2}{1 + \gamma}} (\Omega s) \Bigg ] \cos(\Omega r)  - \left[ \sin(\Omega r) \right]^{\frac{3 + \gamma}{1 + \gamma}}  \Bigg\}^{-\frac{1}{1- \gamma}} \ , \nonumber \\
	e^{\cB} \, = \; &  \sin(\Omega r)^{\frac{-\gamma}{1 + \gamma}} \Bigg\{ \Bigg [ b + \frac{2 \Omega}{1 + \gamma} \int_0^{r} \td s \; \sin^{\frac{2}{1 + \gamma}} (\Omega s) \Bigg ] \cos(\Omega r) - \left[ \sin(\Omega r) \right]^{\frac{3 + \gamma}{1 + \gamma}} \Bigg\}^{\frac{\gamma}{1-\gamma}} \ .
\label{duegammagamma+1soluzione1}
\end{align}
Since $\sin(\Omega r)$ is raised to a real power, we take $r \in \left(0, \frac \pi \Omega \right)$, and $b > 0 $ since $y(0) = b$, but the actual range depends on $b$ and is smaller. Indeed $y\left(\frac \pi {2\Omega} \right) = -1 < 0$, and therefore there is a point $\tilde r \in \left[0, \frac \pi {2\Omega} \right]$ where $y(r)$ vanishes, so that $r \in (0, \tilde r)$. At $\tilde r$, $e^{\vf}$ diverges, as one would have expected from the discussion devoted to the Hamiltonian constraint in Section~\ref{sec:compactness}. These solutions have both strong--coupling and weak--coupling regions.

\noindent $\boxed{|\gamma| < 1\,, \; \varepsilon_1 =  -1\,, \;\varepsilon_2 = 1.}$

\noindent This potential is rather interesting because for $\vf < 0$ it is negative, and therefore it is expected to yield a bounded string coupling $g_s$.
\begin{figure}[ht]
\centering
\begin{tabular}{cc}
\includegraphics[width=0.4\textwidth]{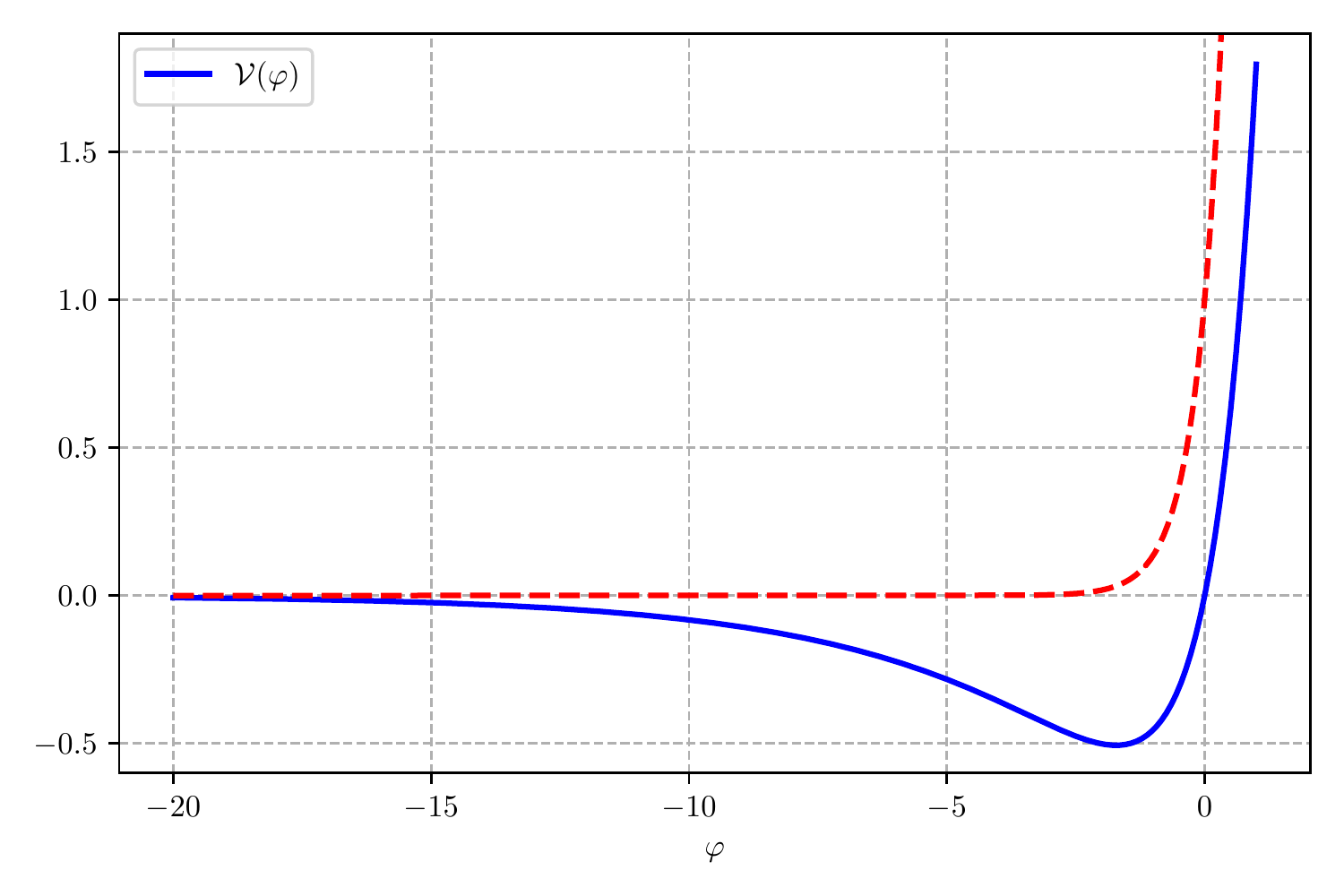} &
\includegraphics[width=0.4\textwidth]{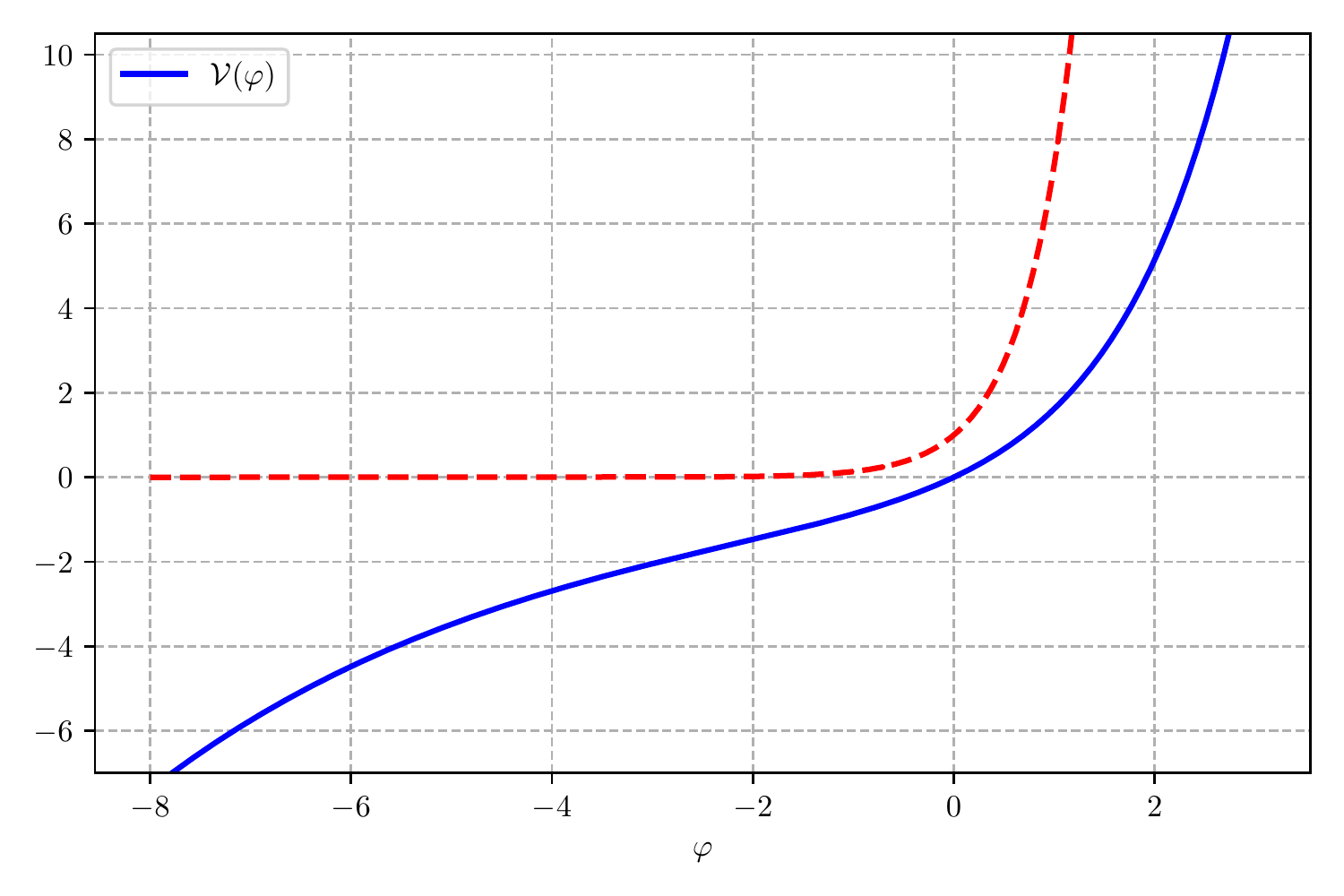} \\
\end{tabular}
\caption{ \small Two potentials of eq.~\eqref{pot_33_2} for $\ve_1 = -1$, $\ve_2 = 1$ and $-1 < \gamma < 1$. In particular, the left panel corresponds to $\gamma = \frac{1}{8}$, while the right one corresponds to $\gamma =  -\frac{1}{8}$. The dashed potential is the one considered in \cite{dm_vacuum}.}
\label{eps-11}
\end{figure}
As before, $\Omega^2$ is still defined by eq.~\eqref{omega2}, and therefore one needs to solve
\begin{equation}
	\ddot x \ - \ \Omega^2 \, x  \ = \  0 \ , \qquad \ddot y \ - \ \Omega^2 \, y  \ = \  - \, (1-\gamma)\, x^{\frac{1 - \gamma}{1+ \gamma}} \ , \qquad \dot x \, \dot y  \ = \  - \, \Omega^2 \left[ - \, x \, y \, + \, x^{\frac{2}{1 + \gamma}} \right] \ .
\end{equation}
For $x(r)$ we choose the solution
\begin{equation}
	x(r) \ = \  \sinh(\Omega r) \ ,
\end{equation}
so that the equation for $y(r)$ is
\begin{equation}
	\ddot y \ - \ \Omega^2 \, y \ = \  - \, (1-\gamma) \left[ \sinh(\Omega r) \right]^{\frac{1 - \gamma}{1 + \gamma}} \ .
\end{equation}
The solution is
\begin{equation}
	y(r) \ = \  \Bigg \{ b +\frac{2 \Omega}{\gamma + 1} \int_0^{r} \td s \; \sinh(\Omega s)^{\frac{2}{1 + \gamma}}  \Bigg \} \cosh(\Omega r)  -\left[\sinh(\Omega r) \right]^{\frac{3 + \gamma}{1 + \gamma}} \ ,
\end{equation}
where we have eliminated one constant of integration imposing the Hamiltonian constraint.

Returning to the original variables,
\begin{align}
	e^{\mathcal A} \, = & \; \sinh(\Omega r)^{\frac{1}{1+ \gamma}} \left\{ \Bigg [ b +\frac{2 \Omega}{\gamma + 1} \int_0^{r} \td s \; \sinh(\Omega s)^{\frac{2}{1 + \gamma}}  \Bigg ] \cosh(\Omega r)  -\left[\sinh(\Omega r) \right]^{\frac{3 + \gamma}{1 + \gamma}} \right\}^{\frac{1}{1-\gamma}} \ , \nonumber \\
	e^{\vf} \, = & \;  \sinh(\Omega r)^{\frac{1}{1 + \gamma}} \left\{ \Bigg [ b +\frac{2 \Omega}{\gamma + 1} \int_0^{r} \td s \; \sinh(\Omega s)^{\frac{2}{1 + \gamma}}  \Bigg ] \cosh(\Omega r)  -\left[\sinh(\Omega r) \right]^{\frac{3 + \gamma}{1 + \gamma}}  \right\}^{-\frac{1}{1- \gamma}} \ , \nonumber \\
	e^{\cB} \, = & \;  \sinh(\Omega r)^{\frac{-\gamma}{1 + \gamma}} \left\{ \Bigg [ b +\frac{2 \Omega}{\gamma + 1} \int_0^{r} \td s \; \sinh(\Omega s)^{\frac{2}{1 + \gamma}} \Bigg ] \cosh(\Omega r)  -\left[\sinh(\Omega r) \right]^{\frac{3 + \gamma}{1 + \gamma}} \right\}^{\frac{\gamma}{1-\gamma}} \ ,
\label{duegammagamma+1soluzione2}
\end{align}
where now $0 < r < +\infty$.
\begin{figure}[ht]
\centering
\begin{tabular}{cc}
\includegraphics[width=45mm]{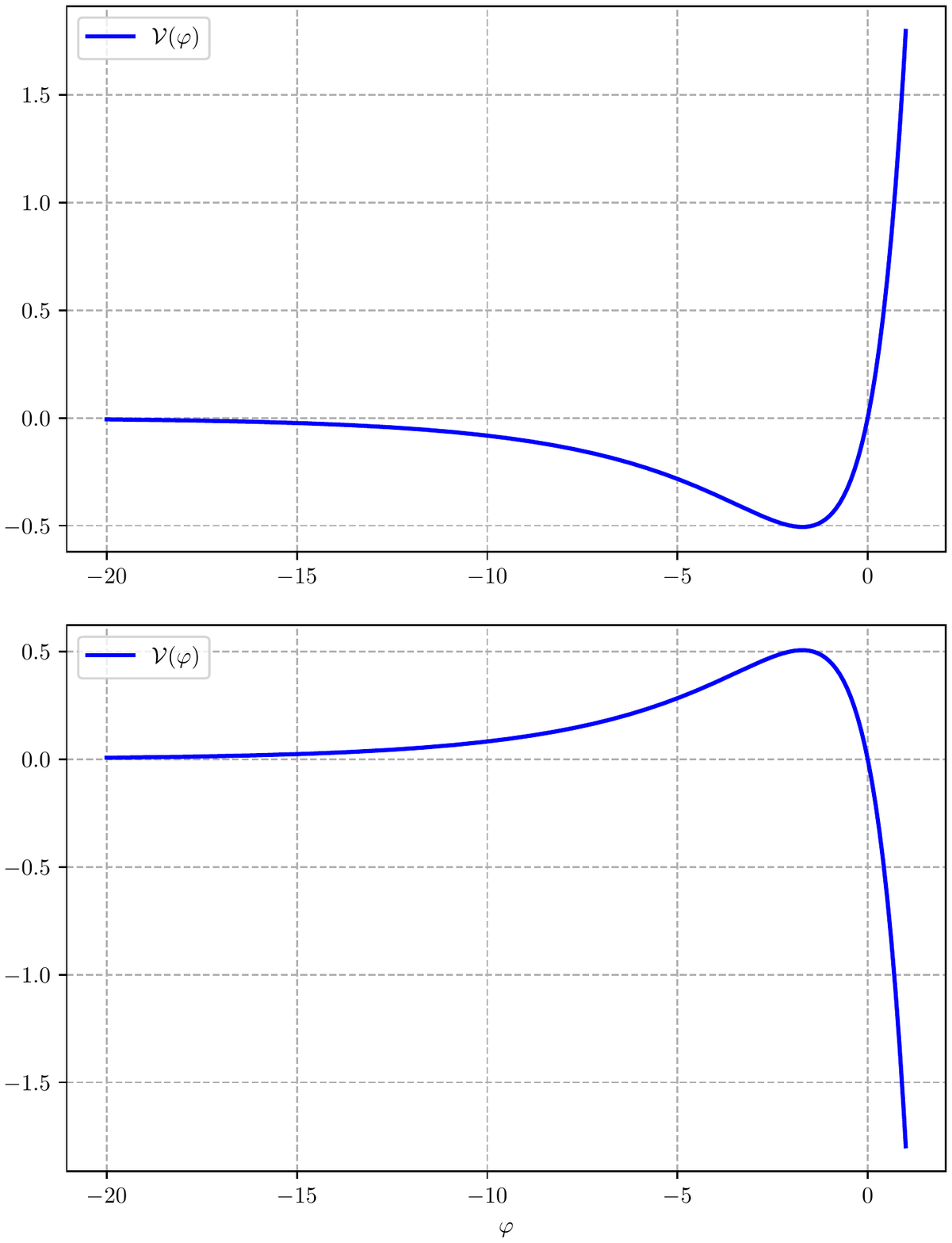} &
\includegraphics[width=0.57\textwidth]{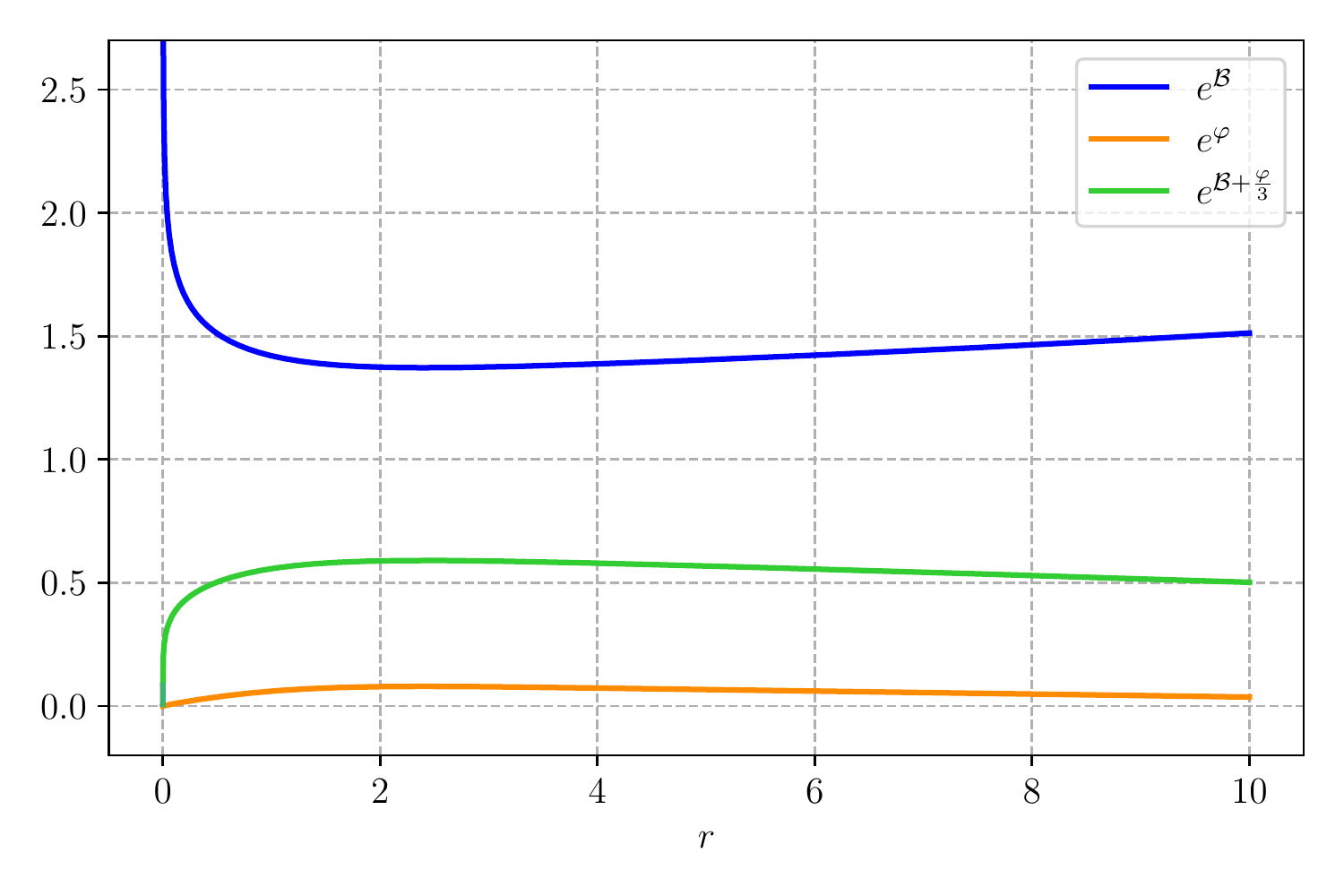} \\
\end{tabular}
\caption{\small Solutions for the potential of eq.~\eqref{pot_33_2} for $0 <\gamma < 1$, $\varepsilon_1 = -1$ and $\varepsilon_2 =  1$. The upper--left panel displays the potential for $\gamma = \frac18$, while the lower--left panel displays the inverted potential. The right panel shows the solution for $b = \frac{137}{18}$, so that the inequality of eq.~\eqref{b_condit} holds. The string coupling is always perturbative, and while $e^{\cB}$ diverges, in the limit of large $r$ the volume form $e^{\cB + \frac{\vf}{3}}$ is an exponential with negative exponent. This solution results in a compact space where the coupling is always perturbative.}
\label{2gamma-gamma+1_6}
\end{figure}
As $r$ approaches zero, $y$ is continuous and $y(0) = b$. On the other hand for $r$ large and $\gamma \neq 0$,
\begin{equation}
	y(r)\ \sim \ \frac{1}{2} \left(b \, - \, \frac{\gamma^2 + \gamma + 2}{\gamma(\gamma + 1)} \; 2^{-\frac{2}{1 + \gamma}}\right)e^{\Omega r} \, + \, \frac{1}{\gamma(\gamma + 1)} \, \frac{e^{\frac{1 - \gamma}{1+\gamma} \Omega r}}{2^{\frac{2}{1+\gamma}}}  \ ,
\label{y_for_gamma_0_1}
\end{equation}
while for $\gamma = 0$
\begin{equation}
	y(r)\ \sim \ \frac{1}{2} \, \Big(b \ - \ \Omega \, r  \Big)e^{\Omega r}  \ .
\end{equation}
Notice that for $0< \gamma < 1$ the first term dominates, while the opposite is true in the complementary range $-1 < \gamma < 0$. In the first case, if
\begin{equation}
    b \ > \ \frac{\gamma^2 + \gamma + 2}{\gamma(\gamma + 1)} \; 2^{-\frac{2}{1 + \gamma}} \ ,
\label{b_condit}
\end{equation}
the solution is defined in the whole semi--axis. The string coupling approaches zero for both small and large values of $r$, while $e^{\cB}$ diverges in both cases. Consequently, the dominant behaviors of the volume form $e^{\cB + \frac \vf 3}$ for $r$ small and large are
\begin{equation}
	e^{\cB + \frac \vf 3} \ \sim \ r^{\frac{1- 3 \gamma}{1 + \gamma}} \qquad r \text{ small} \ , \qquad \qquad e^{\cB + \frac \vf 3} \ \sim \ e^{\frac 2 3 \frac{3 \gamma^2 - \gamma}{1 - \gamma^2} \Omega r} \qquad r \text{ large} \ ,
\end{equation}
so that a bounded $r$--direction obtains if the two conditions
\begin{equation}
	\frac{1- 3 \gamma}{1 + \gamma} > -1 \ , \qquad \qquad \qquad \frac 2 3 \frac{3 \gamma^2 - \gamma}{1 - \gamma^2} < 0
\end{equation}
hold in the range $0 < \gamma < 1$. These inequalities are satisfied if $0 < \gamma < \frac 1 3$, and for this range the internal direction is bounded. However, the reduced Planck mass and gauge coupling (\ref{planckandgauge}) are both infinite, so that these models do not provide interesting compactifications, since the corresponding interactions disappear in nine dimensions. Moreover, in the range $0 < \gamma < \frac{1}{3}$, the string--frame scalar curvature~\eqref{string_curvature} is not bounded both near $r=0$ and for $r$ large. These results are precisely along the lines of what we said in Section~\ref{sec:compactness}.

This solution is displayed in fig.~\ref{2gamma-gamma+1_6}. Notice that in the Einstein frame the internal space would be non compact, since $e^\cB$ diverges for large values of $r$.

In the complementary range $-1 < \gamma < 0$, the solution includes a region of strong coupling. We can see this from the fact that, for this range of $\gamma$, the behavior of $y(r)$ for large $r$ is dominated by the second term of eq.~\eqref{y_for_gamma_0_1}, whose coefficient is negative, which drives $y(r)$ to zero. Where this occurs, the string coupling diverges. In more physical terms, if one inverts the potential in fig.~\ref{eps-11}, one can see that the dilaton is naturally driven toward large values. The discussion in Section~\ref{sec:compactness} indicates that and additional type of behavior is possible, where the string coupling has a lower bound. In practice, one would obtain it starting from a cosh--function.

\noindent $\boxed{|\gamma| < 1\,, \;\varepsilon_1 =  1\,, \;\varepsilon_2 = -1}$

\begin{figure}[ht]
\centering
\begin{tabular}{cc}
\includegraphics[width=0.4\textwidth]{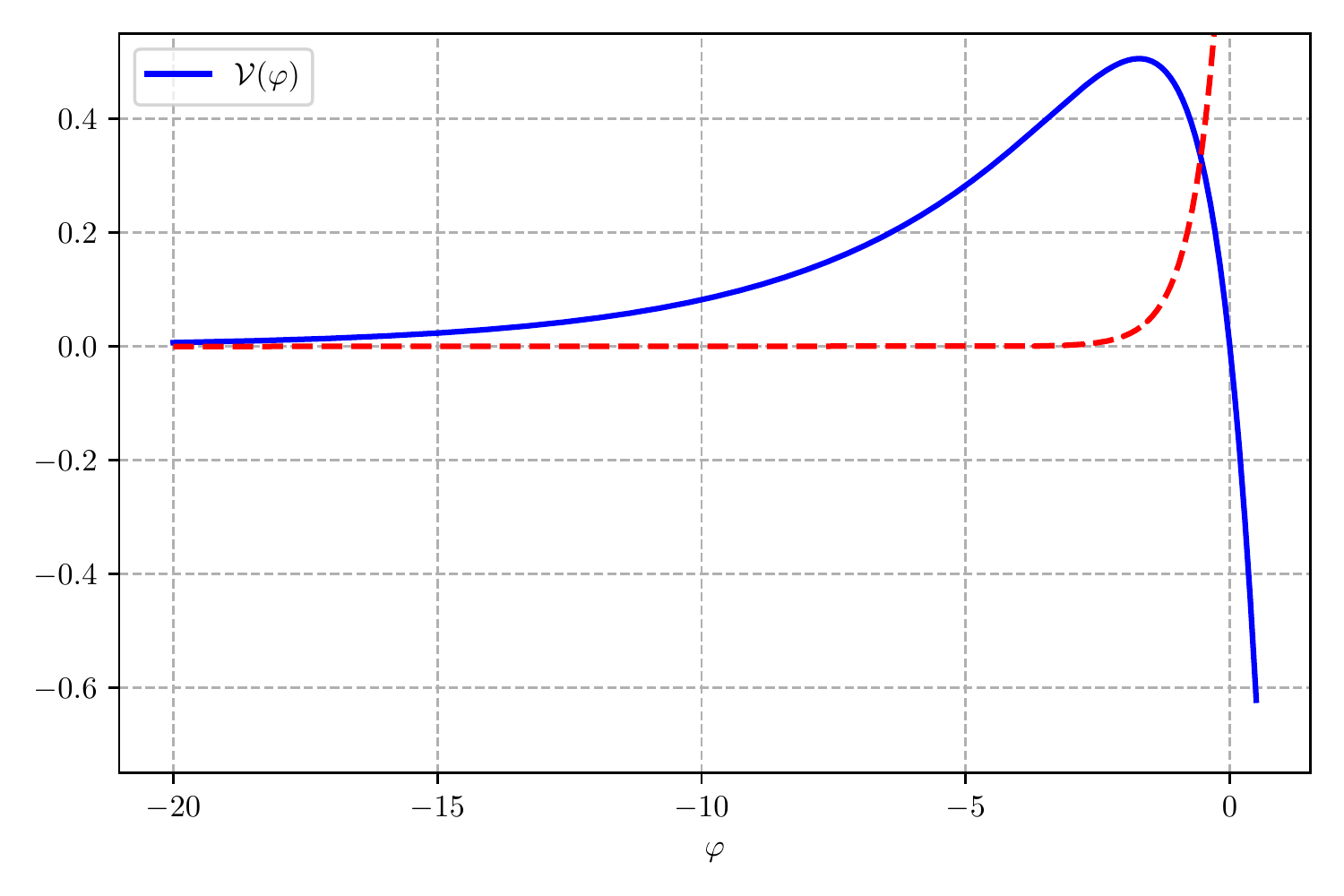} &
\includegraphics[width=0.4\textwidth]{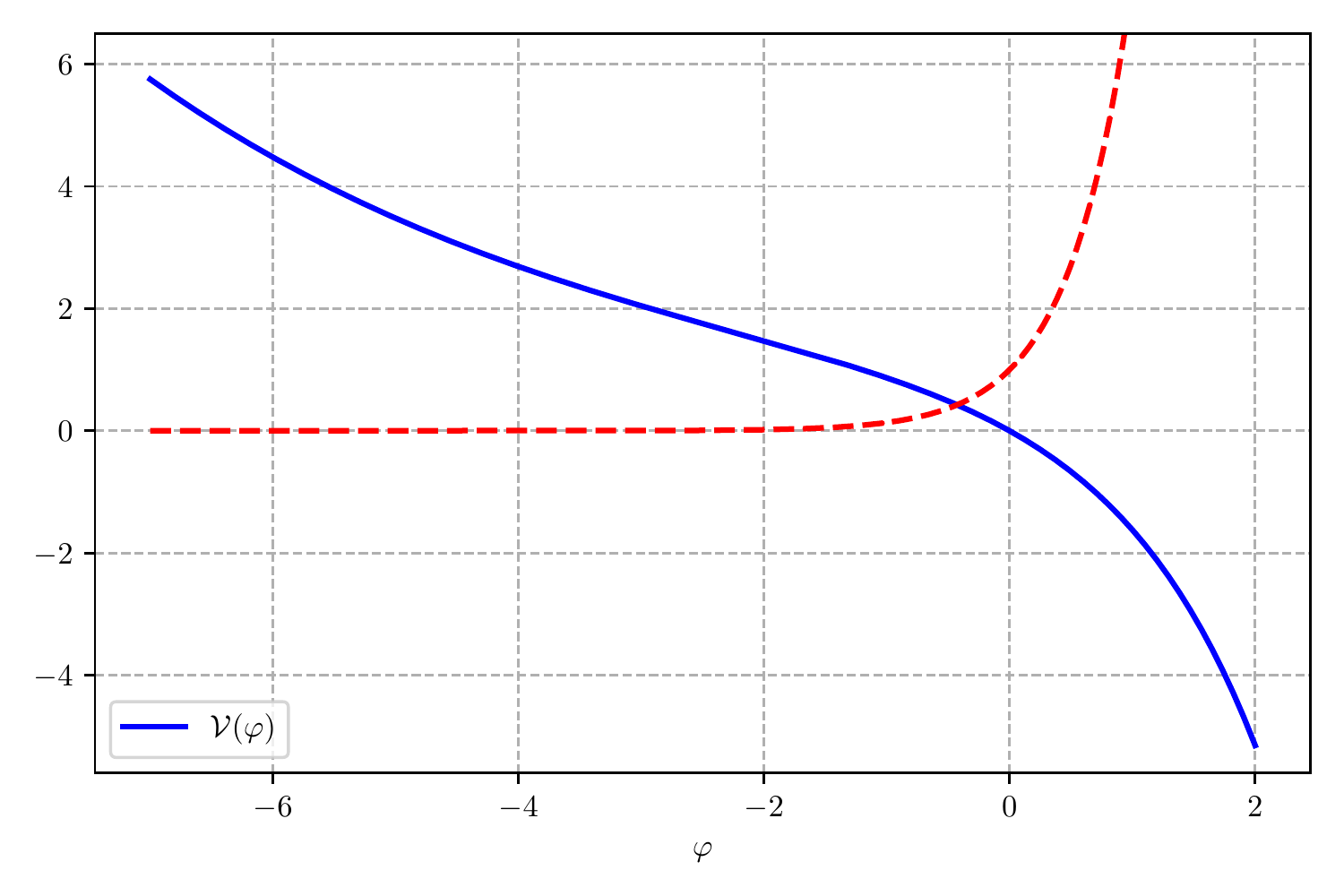} \\
\end{tabular}
\caption{\small The potential of eq.~\eqref{pot_33_2} for $-1 < \gamma < 1$, $\ve_1 = 1$ and $\ve_2 = -1$. In particular, the left panel displays $\gamma =  \frac18$, while the right panel displays $\gamma = - \frac18$. The dashed potential is the one considered in \cite{dm_vacuum}.}
\label{2gamma-gamma+1_7}
\end{figure}

\noindent We can now explore the case $\ve_1 =  1$ and $\ve_2 =  -1$, an example of which is displayed in fig.~\ref{2gamma-gamma+1_7}. Notice that the resulting potentials are unbounded from below.
The solution of eqs.~\eqref{EOM_third_potential} is, up to shifts in the coordinate $r$ and up to an overall factor in $x(r)$, which would reverberate in the other solution,
\begin{align}
    x(r)\ =& \ \sin(\Omega r ) \ , \nonumber \\
    y(r) \ =& \ a \sin(\Omega r ) \ + \ \left[ b - \frac{1 - \gamma}{\Omega} \int_0^r \sin(\Omega s)^{\frac{2}{1 + \gamma}} \td s  \right] \ \cos(\Omega r) + \sin(\Omega r)^{\frac{3 + \gamma}{1 + \gamma}} \ .
\label{sol_unbounded}
\end{align}
$a$ and $b$ are two integration constants, and the Hamiltonian constraint demands that $a =  0$.
\begin{figure}[ht]
\centering
\begin{tabular}{cc}
\includegraphics[width = 45mm]{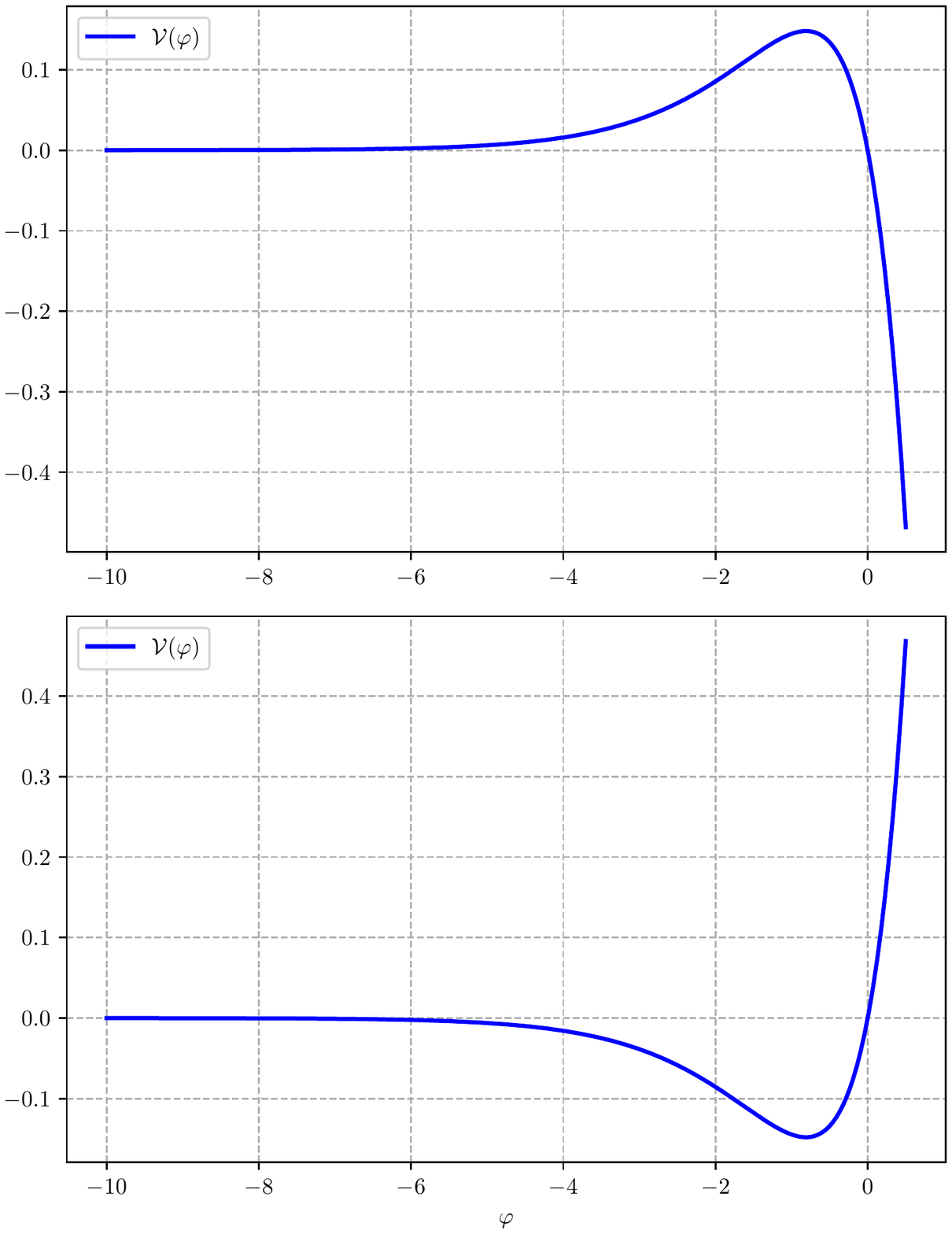} &
\includegraphics[width=0.57\textwidth]{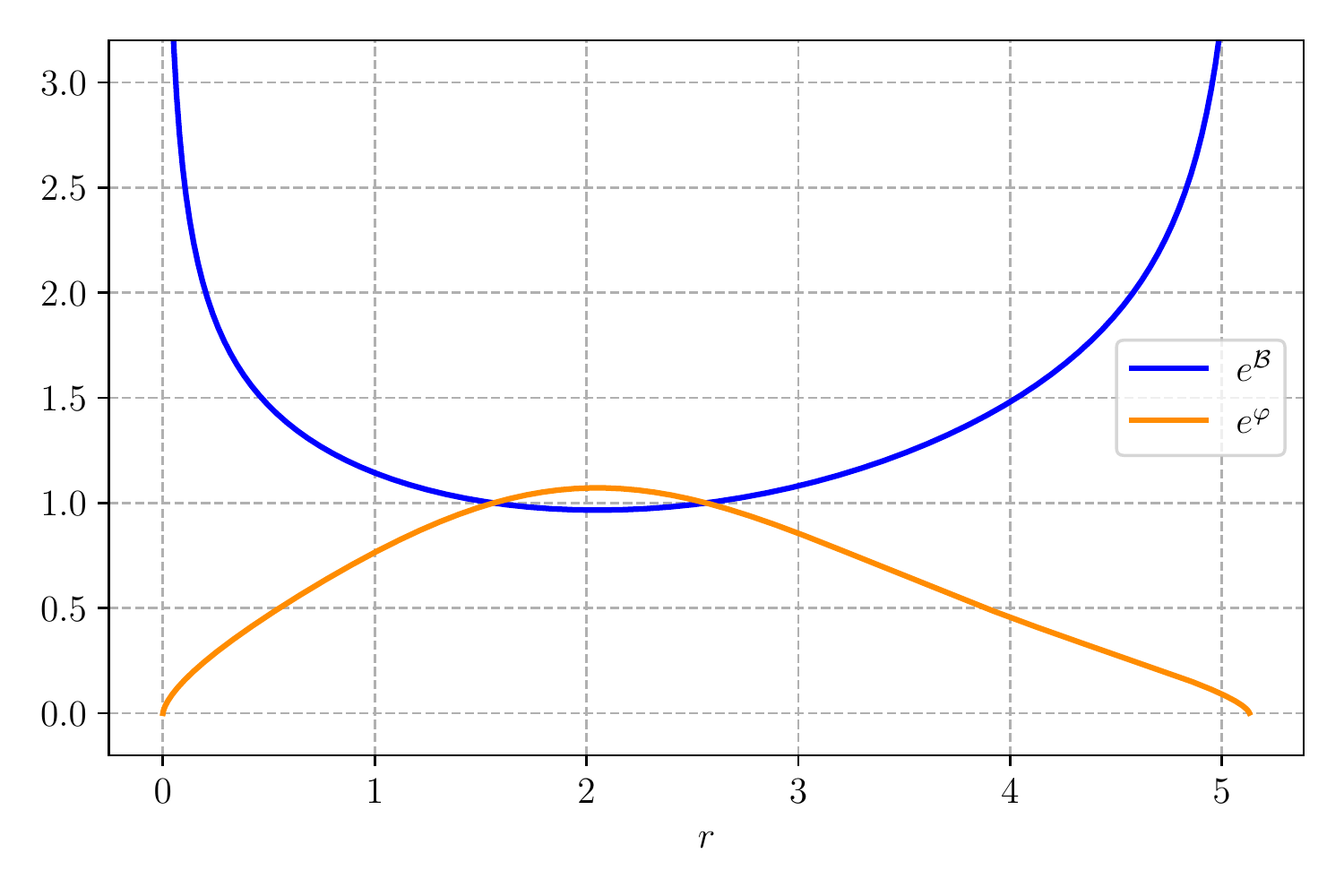}\\
\end{tabular}
\caption{\small The upper--left panel displays the potential of eq.~\eqref{pot_33_2} for $\ve_1 = 1$, $\ve_2 = -1$ and $0 < \gamma < 1$, in particular with $\gamma = \frac{1}{2}$, while the lower--left panel displays the inverted potential. The right panel displays the behavior of $e^{\cB}$ and $e^{\vf}$ for $b = 1$. This choice of $b$ respects \eqref{b_bounded_gs}, and $g_s$ is bounded as expected.}
\label{e11e2-1gmin_1}
\end{figure}

\begin{figure}[ht]
\centering
\begin{tabular}{cc}
\includegraphics[width = 45mm]{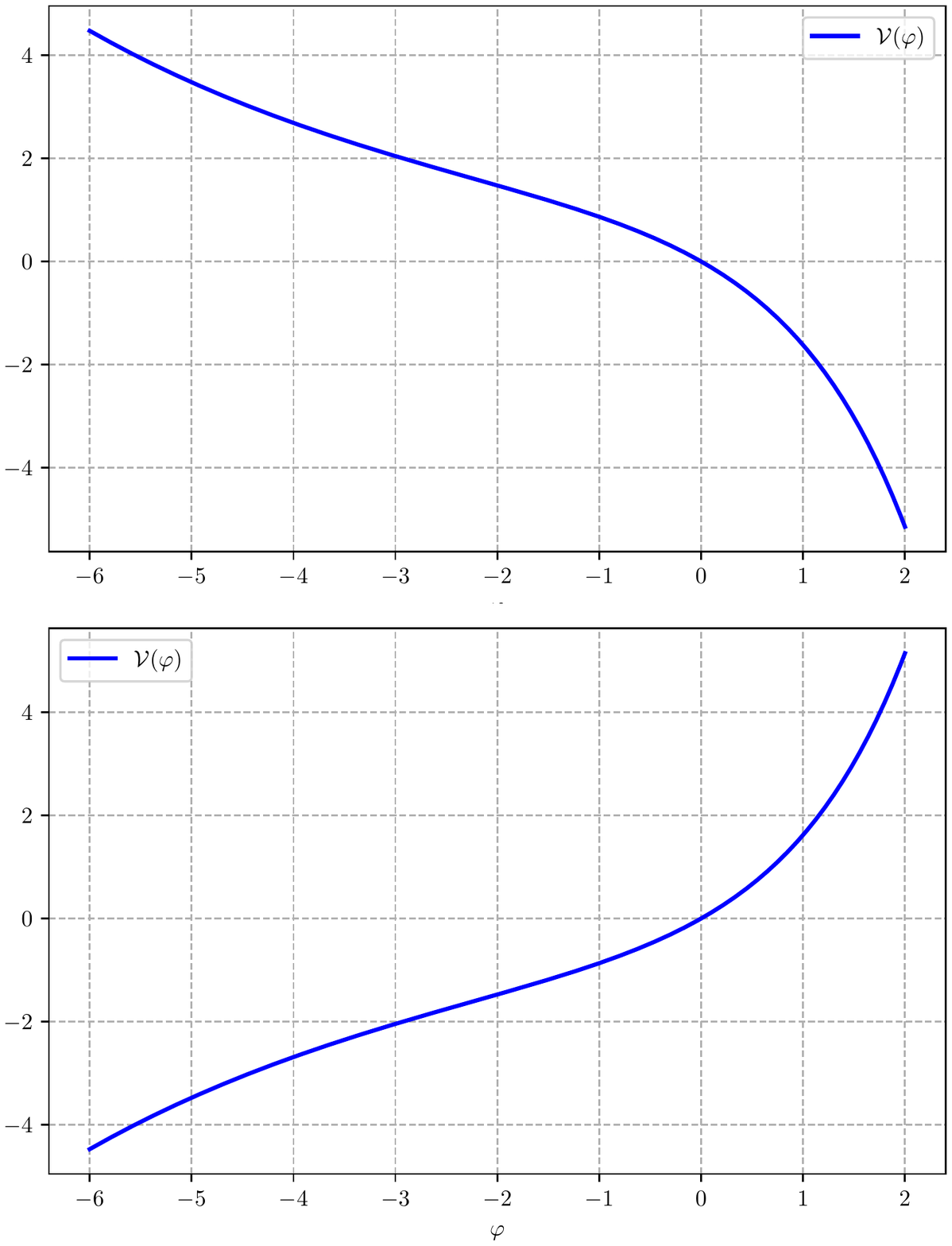} &
\includegraphics[width=0.56\textwidth]{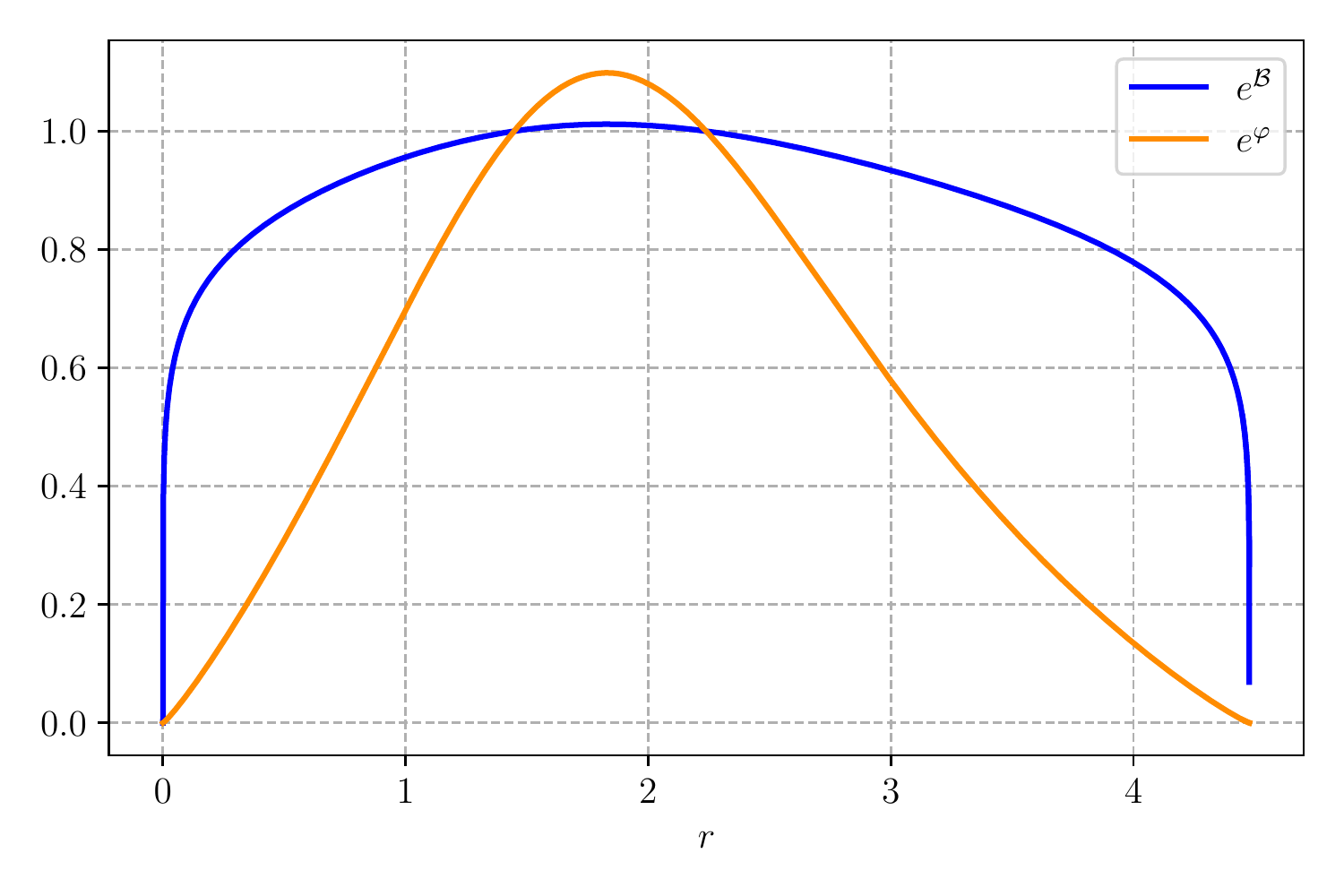} \\
\end{tabular}
\caption{\small The upper--left panel displays the potential of eq.~\eqref{pot_33_2} for $\ve_1 = 1$, $\ve_2 = -1$ and $-1 < \gamma < 0$, in particular $\gamma = -\frac{1}{8}$, while the lower--left panel the inverted potential. The right panel displays the behavior of $e^{\cB}$ and $e^{\vf}$ for $b = 1$. This choice of $b$ respects \eqref{b_bounded_gs}, and $g_s$ is bounded as expected.}
\label{e11e2-1gmin_2}
\end{figure}
Using \eqref{sol_unbounded}, one then obtains
\begin{align}
    e^{\mathcal A} \, = & \; \sin(\Omega r)^{\frac{1}{1+ \gamma}} \left\{ \left[ b - \frac{1 - \gamma}{\Omega} \int_0^r \sin(\Omega s)^{\frac{2}{1 + \gamma}} \td s  \right] \ \cos(\Omega r) + \sin(\Omega r)^{\frac{3 + \gamma}{1 + \gamma}} \right\}^{\frac{1}{1-\gamma}} \ , \nonumber \\
	e^{\vf} \, = & \;  \sin(\Omega r)^{\frac{1}{1 + \gamma}} \left\{ \left[ b - \frac{1 - \gamma}{\Omega} \int_0^r \sin(\Omega s)^{\frac{2}{1 + \gamma}} \td s  \right] \ \cos(\Omega r) + \sin(\Omega r)^{\frac{3 + \gamma}{1 + \gamma}}  \right\}^{-\frac{1}{1- \gamma}} \ , \nonumber \\
	e^{\cB} \, = & \;  \sin(\Omega r)^{\frac{-\gamma}{1 + \gamma}} \left\{ \left[ b - \frac{1 - \gamma}{\Omega} \int_0^r \sin(\Omega s)^{\frac{2}{1 + \gamma}} \td s  \right] \ \cos(\Omega r) + \sin(\Omega r)^{\frac{3 + \gamma}{1 + \gamma}} \right\}^{\frac{\gamma}{1-\gamma}} \ .
\end{align}

The periodicity of these functions, and the condition that $\sin(\Omega r)$ be positive definite, restrict the range of $r$ to the interval $\left(0,\frac{\pi}{\Omega}\right)$. In order to define the function at $r =  0$, however, $b$ must be positive, otherwise a negative $b$ would restrict the interval further, to  $r> r^* > 0$. Moreover, if
\begin{equation}
    b \ > \ \frac{1 - \gamma}{\Omega} \int_0^{\frac{\pi}{\Omega}} \sin(\Omega s)^{\frac{2}{1 + \gamma}} \td s \ ,
\end{equation}
at some point within the interval $\left[0, \frac \pi \Omega \right]$, $y$ vanishes, and $g_s$ diverges there. If instead
\begin{equation}
    0 \ < \ b \ < \ \frac{1 - \gamma}{\Omega} \int_0^{\frac{\pi}{\Omega}} \sin(\Omega s)^{\frac{2}{1 + \gamma}} \td s \ ,
\label{b_bounded_gs}
\end{equation}
$y(r)$ is always positive and $g_s$ is bounded. This type of solution is displayed in figs.~\ref{e11e2-1gmin_1} and \ref{e11e2-1gmin_2}.
When \eqref{b_bounded_gs} holds, \textbf{the solutions are very interesting}, and one can verify that:
\begin{itemize}
    \item the internal $r$--direction is compact;
    \item string coupling $g_s$ is bounded;
    \item the 9D Planck mass and gauge coupling are finite.
\end{itemize}
Even in these examples, the potential is dominated by an exponential for $\vf>0$, and the considerations of Section~\ref{sec:compactness} apply. The potential has $\gamma<1$, but the chosen range for the parameter $b$ selects the climbing behavior.
However, the scalar curvature in string frame~\eqref{string_curvature} is not bounded at both ends of the interval.

\noindent $\boxed{|\gamma| < 1\,, \;\varepsilon_1 =  -1\,, \;\varepsilon_2 = -1}$

\noindent Finally, we consider the case $|\gamma| < 1$, $\ve_1 =   -1$ and  $\ve_2 =  -1$, so that these potentials are again unbounded from below.
\begin{figure}[ht]
\centering
\begin{tabular}{cc}
\includegraphics[width=0.4\textwidth]{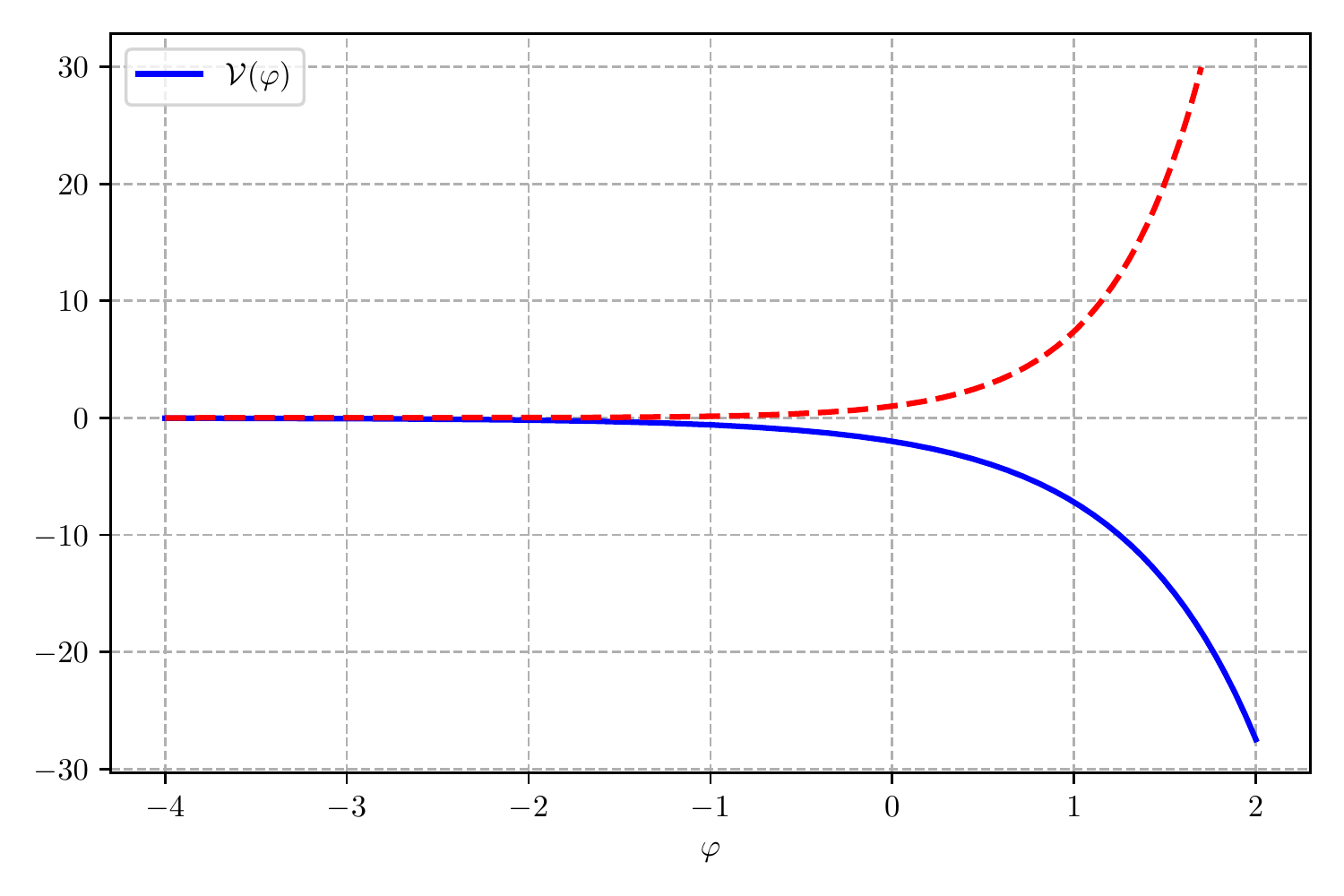} &
\includegraphics[width=0.4\textwidth]{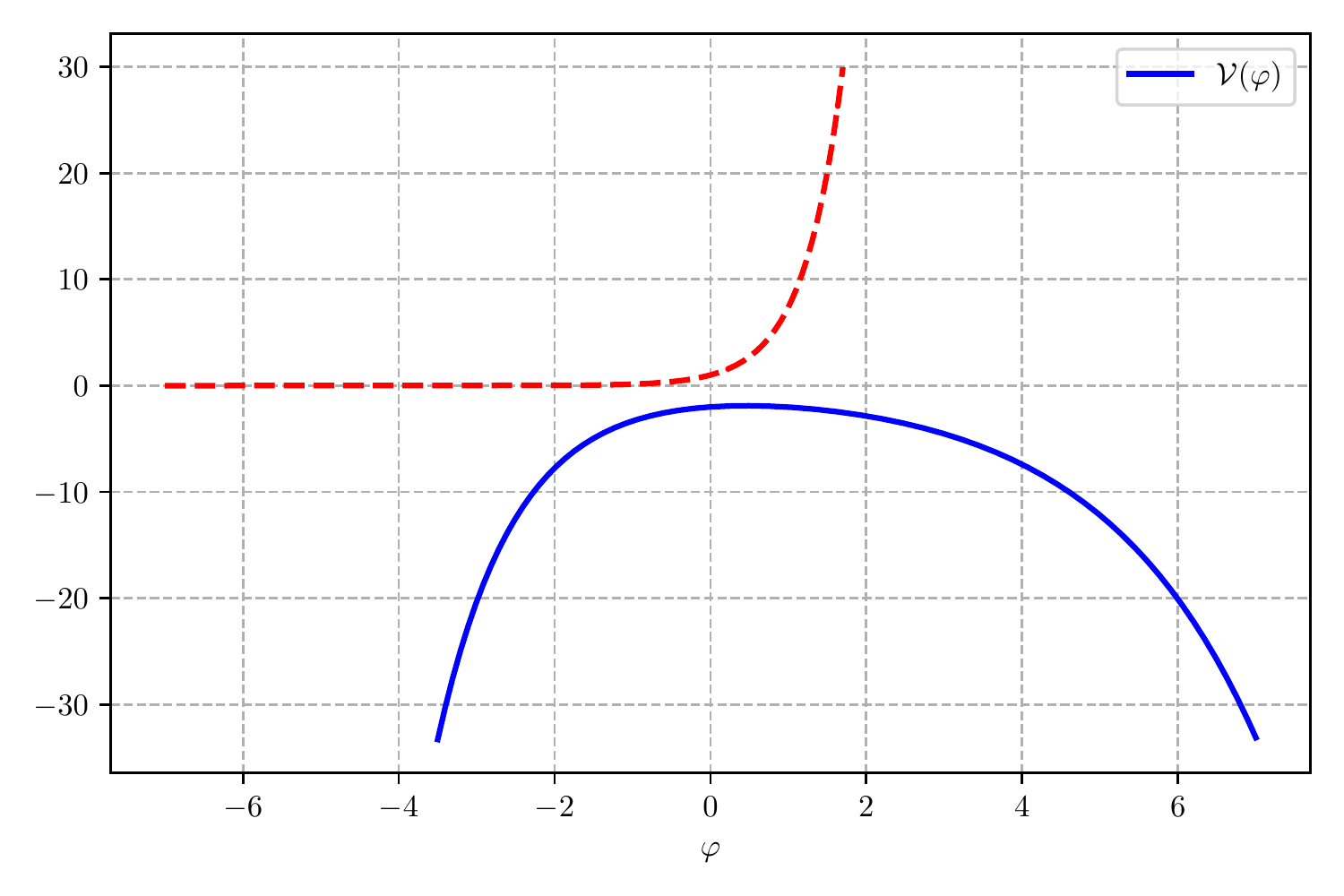} \\
\end{tabular}
\caption{\small The potential of eq.~\eqref{pot_33_2} for $|\gamma| < 1$, $\ve_1 = -1$, and $\ve_2 = -1$. In particular, in the left panel $\gamma = \frac 12$, while in the right one $\gamma = -\frac12$. The dashed potential is the one considered by \cite{dm_vacuum}}
\end{figure}
 As usual,  $\Omega^2$ is defined in eq.~\eqref{omega2}, and a solution of the equations of motion is in this case
\begin{align}
    x \ = \ & \sinh(\Omega r) \ , \nonumber \\
    y \ = \ & \left[ b - \frac{1 - \gamma}{\Omega} \int_0^r \sinh(\Omega s)^{\frac{2}{1 + \gamma}} \ \td s \right] \cosh(\Omega r) + \sinh(\Omega r)^{\frac{3 + \gamma}{1 + \gamma}} \ .
\end{align}
\begin{figure}[ht]
\centering
\begin{tabular}{cc}
\includegraphics[width=45mm]{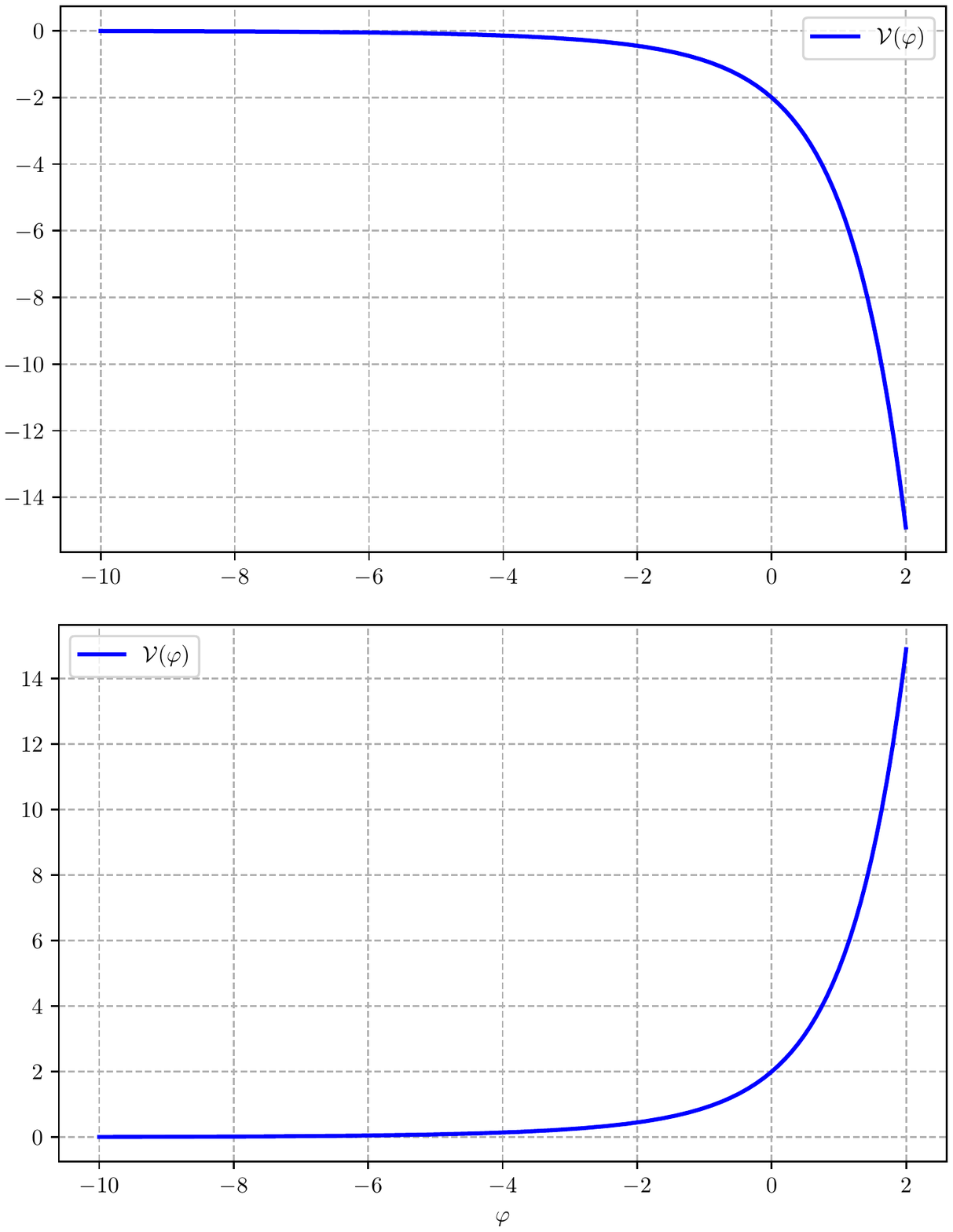} &
\includegraphics[width=0.57\textwidth]{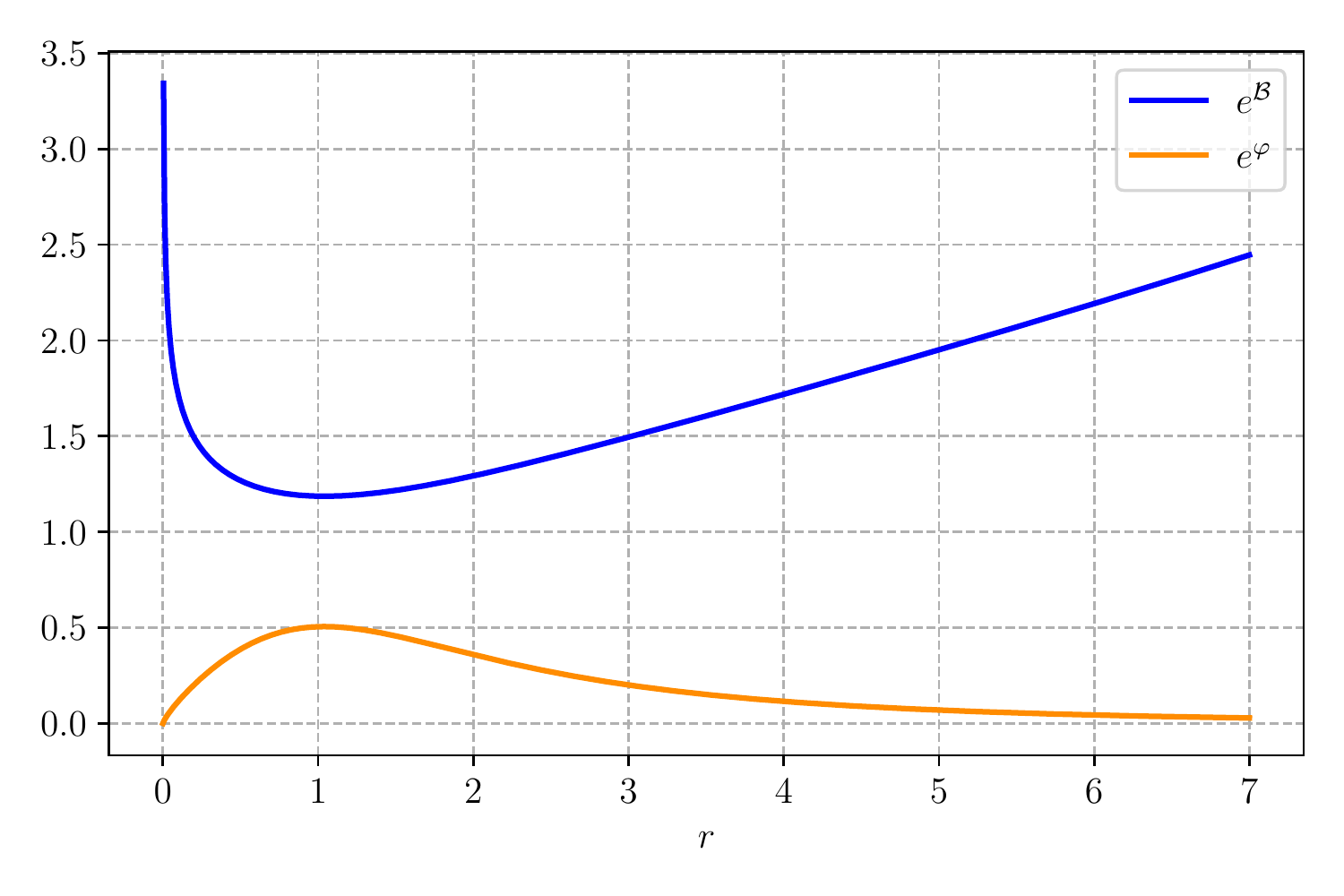}
\end{tabular}
\caption{\small The upper--left panel displays the potential of eq.~\eqref{pot_33_2} for $0 < \gamma < 1$, $\ve_1 = \ve_2 = -1$, in particular $\gamma = \frac{1}{4}$, while the lower--left panel displays the inverted potential. The right panel displays the corresponding solution of $e^{\cB}$ and $e^{\vf}$ for $b = 1$. Notice that for this value of $\gamma$ the $r$--direction is compact.}
\label{e1-1e2-1sol_1}
\end{figure}
\begin{figure}[ht]
\centering
\begin{tabular}{cc}
\includegraphics[width=45mm]{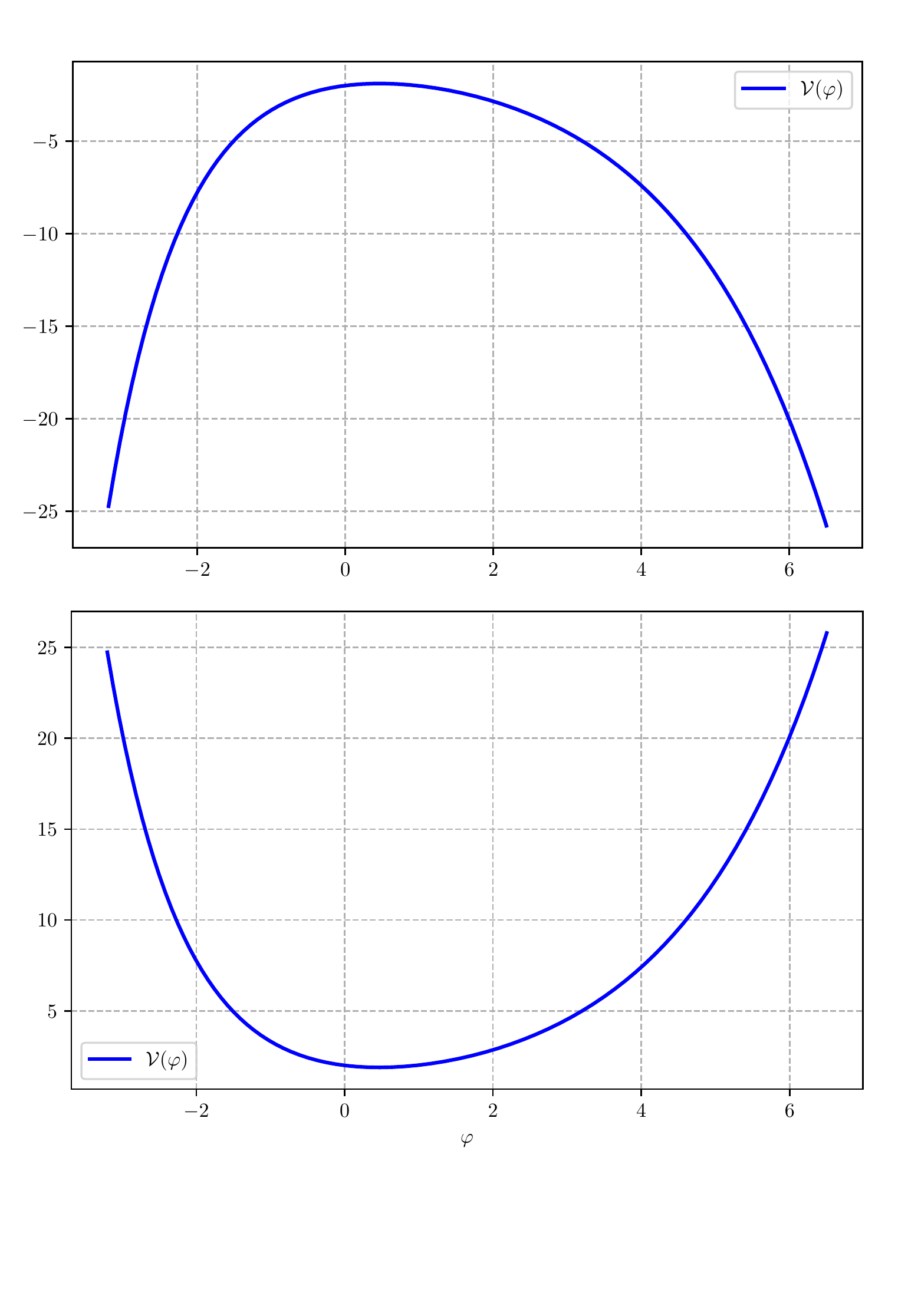} &
\includegraphics[width=0.57\textwidth]{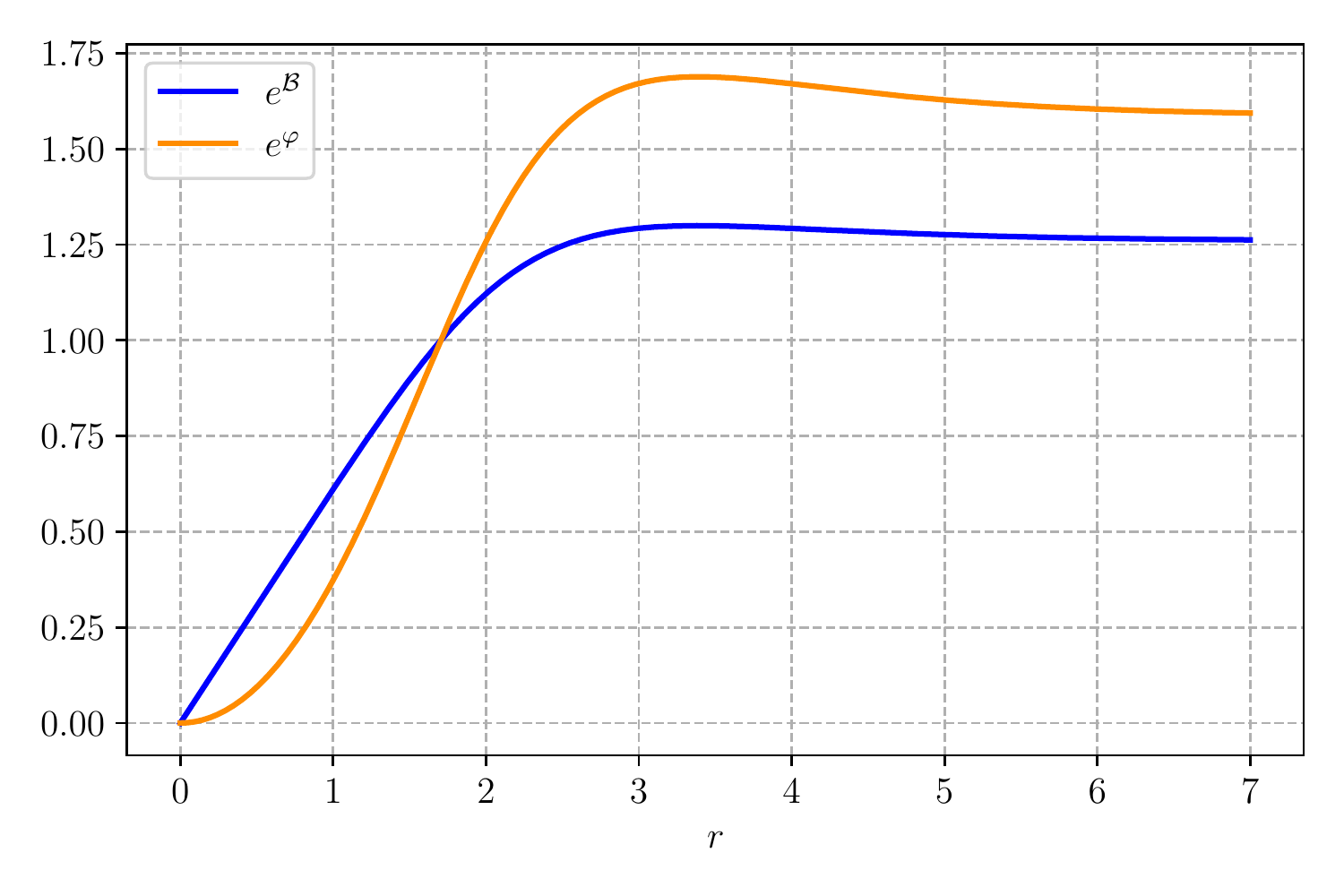}
\end{tabular}
\caption{\small The upper--left panel displays the potential of eq.~\eqref{pot_33_2} for $-1 < \gamma < 0$, $\ve_1 = \ve_2 = -1$, in particular $\gamma = -\frac{1}{2}$, while the lower--left panel displays the inverted potential. The right panel displays the corresponding solutions for $e^{\cB}$ and $e^{\vf}$ for $b = 1$. This case is very peculiar, since both fields become constant for large values of $r$. This behavior makes the string coupling bounded, since the dilaton approaches the negative critical point, but the $r$--direction in not compact.}
\label{e1-1e2-1sol_2}
\end{figure}
The original variables then read
\begin{align}
	e^{\mathcal A} \, = & \;  \sinh(\Omega r)^{\frac{1}{1+ \gamma}} \left\{\left[ b - \frac{1 - \gamma}{\Omega} \int_0^r \sinh(\Omega s)^{\frac{2}{1 + \gamma}} \ \td s \right] \cosh(\Omega r) + \sinh(\Omega r)^{\frac{3 + \gamma}{1 + \gamma}} \right\}^{\frac{1}{1-\gamma}} \ ,  \nonumber \\
	e^{\vf} \, = & \;  \sinh(\Omega r)^{\frac{1}{1 + \gamma}} \left\{ \left[ b - \frac{1 - \gamma}{\Omega} \int_0^r \sinh(\Omega s)^{\frac{2}{1 + \gamma}} \ \td s \right] \cosh(\Omega r) + \sinh(\Omega r)^{\frac{3 + \gamma}{1 + \gamma}} \right\}^{-\frac{1}{1- \gamma}} \ , \nonumber \\
	e^{\cB} \, = & \;  \sinh(\Omega r)^{\frac{-\gamma}{1 + \gamma}} \left\{ \left[ b - \frac{1 - \gamma}{\Omega} \int_0^r \sinh(\Omega s)^{\frac{2}{1 + \gamma}} \ \td s \right] \cosh(\Omega r) + \sinh(\Omega r)^{\frac{3 + \gamma}{1 + \gamma}} \right\}^{\frac{\gamma}{1-\gamma}} \ .
\end{align}

For both positive and negative values of $\gamma$, the solution is defined for $r > 0$, and $g_s$ vanishes at the origin. However, some differences arise in the large--$r$ behavior. For $\gamma > 0$
\begin{equation}
    y(r) \ \sim \ \frac12 \left(b + \frac{\gamma^2 + \gamma + 2}{\gamma(\gamma + 1) }\; 2^{- \frac 2 {\gamma + 1}} \right) \, e^{\Omega r} \ - \ \frac{1}{\gamma(\gamma + 1)}\; 2^{- \frac{2}{1 + \gamma}} e^{\frac{1 - \gamma}{1 + \gamma} \Omega r} \ ,
\end{equation}
and therefore, if
\begin{equation}
    b \ <\  - \ \frac{\gamma^2 + \gamma + 2}{\gamma(\gamma + 1) }\; 2^{- \frac 2 {\gamma + 1}}
\end{equation}
the solution does not exist, since $y(r)$ is always negative. On the other hand, if
\begin{equation}
    b \ >\  - \ \frac{\gamma^2 + \gamma + 2}{\gamma(\gamma + 1) }\; 2^{- \frac 2 {\gamma + 1}} \label{ineq_2}
\end{equation}
the solution does exist, but in order to have a bounded string coupling one has to demand that $b > 0$. In this range $g_s$ tends to zero both near the origin and for large values of $r$. For $b$ negative this solution exhibits a \emph{descending} behavior, while for $b$ positive it exhibits a \emph{climbing} one.

Moreover, the space is compact if $0 < \gamma < \frac{1}{3}$, but the reduced nine--dimensional Planck mass and gauge coupling are infinite, and the scalar curvature of eq.~\eqref{string_curvature} is unbounded. On the other hand, for negative $\gamma$ $g_s$ is bounded at the origin when $b$ is positive, while for large $r$
\begin{equation}
    y(r) \ \sim \ - \ \frac{1}{\gamma(\gamma + 1)}\; 2^{- \frac{2}{1 + \gamma}} e^{\frac{1 - \gamma}{1 + \gamma} \Omega r} \ + \ \frac12 \left(b + \frac{\gamma^2 + \gamma + 2}{\gamma(\gamma + 1) }\; 2^{- \frac 2 {\gamma + 1}} \right) \, e^{\Omega r}  \ .
\end{equation}
This behavior implies that for large $r$ both $e^{\vf}$ and $e^{\cB}$ approach constant values. Therefore, the string coupling is bounded, but the $r$--direction is again non--compact. A solution of this type is displayed in figs.~\ref{e1-1e2-1sol_1} and \ref{e1-1e2-1sol_2}. In the first case (fig.~\ref{e1-1e2-1sol_1}) a single exponential dominates, and the results are consistent with the discussion in Section~\ref{sec:compactness}. On the other hand, in the second case the dilaton settles eventually at the negative critical point of the potential (fig.~\ref{e1-1e2-1sol_2}).

\noindent $\boxed{\gamma > 1 \, , \; \varepsilon_1 =  1\,, \; \varepsilon_2 = \pm 1} \qquad$

\noindent We can now explore another class of potentials with $\gamma > 1$.
\begin{figure}[ht]
\centering
\begin{tabular}{cc}
\includegraphics[width=0.4\textwidth]{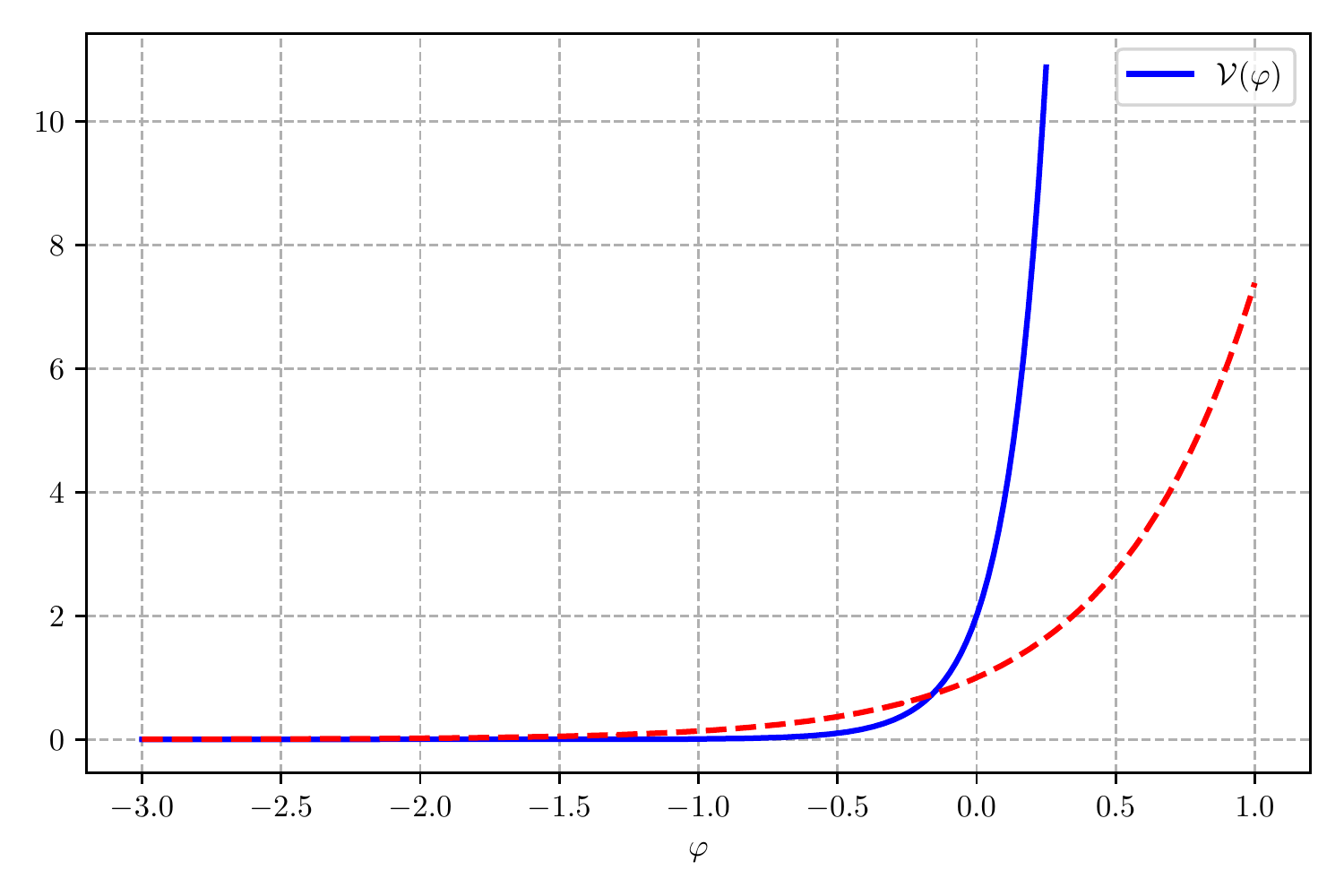} &
\includegraphics[width=0.4\textwidth]{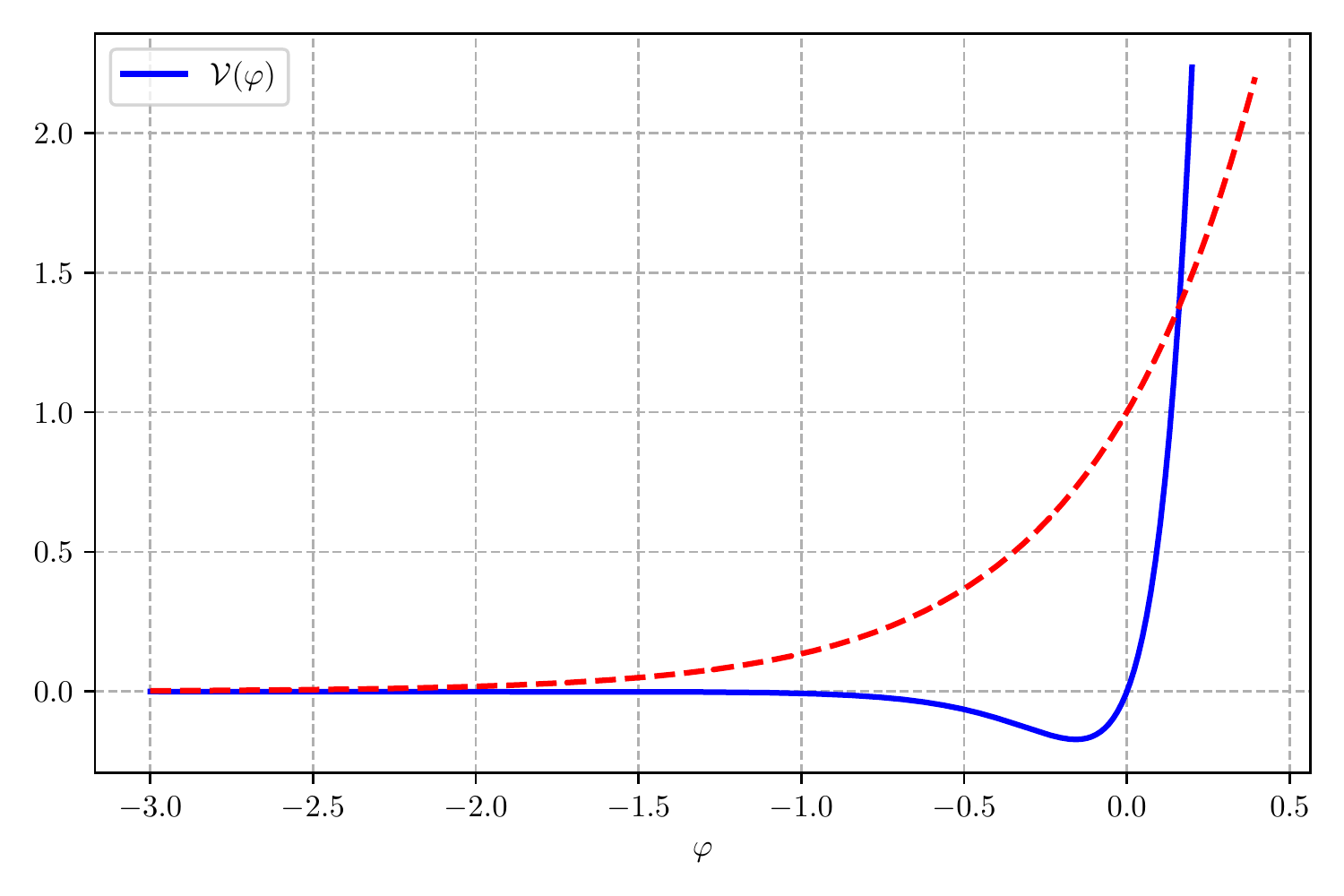} \\
\end{tabular}
\caption{\small The left panel displays the potential of eq.~\eqref{pot_33_2} for $\gamma =4$, $\varepsilon_1 = 1$ and $\varepsilon_2 =  1$. The right one displays the potential for $\varepsilon_2 = -1$ and the same value of $\gamma$. The dashed potential is the one considered by \cite{dm_vacuum}.}
\label{fig:22}
\end{figure}
In this case, letting
\begin{equation}
	\omega^2 \ = \ \frac{\gamma^2 - 1}{2} \ ,
\label{lower_omega}
\end{equation}
the equations of motion read
\begin{equation}
	\ddot x \ - \ \omega^2 \, x \ = \  0 \ , \qquad \qquad \ddot y \ - \ \omega^2 \, y \ = \  \varepsilon_2 \, (\gamma - 1) \, x^{\frac{1 - \gamma}{1+ \gamma}} \ , \qquad \qquad \dot x \, \dot y \ = \  \omega^2 \left[ x\, y \ + \ \varepsilon_2 \, x^{\frac{2}{1 + \gamma}} \right] \ .
\end{equation}
For the first, we choose the solution
\begin{equation}
	x \ = \  \cosh(\omega r) \ ,
\end{equation}
for the purpose of illustration, although it is not the most general one. With this proviso, the $y$--equation reads
\begin{equation}
	\ddot y \ - \ \omega^2 \, y \ = \   \ve_2 \, (\gamma - 1) \, \cosh(\omega r)^{\frac{1 - \gamma}{1+ \gamma}} \ ,
\end{equation}
and, once the Hamiltonian constraint is enforced, its solution
\begin{equation}
	y(r) \ = \ \Bigg \{ a + \frac{2 \, \ve_2 \, \omega}{\gamma + 1} \int_0^{r} \td s \; \cosh(\omega s)^{\frac{2}{\gamma+1}}  \Bigg \} \sinh(\omega r)  - \ve_2 \left[\cosh(\omega r) \right]^{\frac{3 + \gamma}{1 + \gamma}} \ .
\end{equation}
is similar to previous ones.
Returning to the original variables, one finds
\begin{align}
	e^{\mathcal A} \, = & \;  \cosh(\omega r)^{\frac{1}{1+ \gamma}} \left\{\Bigg [ a + \frac{2\, \ve_2 \,\omega}{\gamma + 1} \int_0^{r} \td s \; \cosh(\omega s)^{\frac{2}{\gamma+1}}  \Bigg ] \sinh(\omega r)  - \ve_2 \left[\cosh(\omega r) \right]^{\frac{3 + \gamma}{1 + \gamma}} \right\}^{-\frac{1}{\gamma-1}} \ ,  \nonumber \\
	e^{\vf} \, = & \;  \cosh(\omega r)^{\frac{1}{1 + \gamma}} \left\{ \Bigg [ a + \frac{2 \, \ve_2 \, \omega}{\gamma + 1} \int_0^{r} \td s \; \cosh(\omega s)^{\frac{2}{\gamma+1}}  \Bigg ] \sinh(\omega r)  - \ve_2 \left[\cosh(\omega r) \right]^{\frac{3 + \gamma}{1 + \gamma}} \right\}^{\frac{1}{ \gamma - 1}} \ , \nonumber \\
	e^{\cB} \, = & \;  \cosh(\omega r)^{-\frac{\gamma}{1 + \gamma}} \left\{ \Bigg [ a + \frac{2 \, \ve_2 \, \omega}{\gamma + 1} \int_0^{r} \td s \; \cosh(\omega s)^{\frac{2}{\gamma+1}}  \Bigg ] \sinh(\omega r)  - \ve_2 \left[\cosh(\omega r) \right]^{\frac{3 + \gamma}{1 + \gamma}} \right\}^{-\frac{\gamma}{\gamma-1}} \ .
\label{sixth_solution}
\end{align}
These solutions can be analyzed along the lines of previous cases, and the most interesting features emerge at large $r$. Without loss of generality, one can assume that $a$ be positive, since the solution is invariant for $a \, \rightarrow \, - a$ and $r \, \rightarrow \, -r$. For large positive values of $r$ $y$ behaves as
\begin{equation}
    y(r) \ \sim \ \frac{1}{2} \left( a \, -\,  \ve_2 \, \frac{\gamma^2 + \gamma - 2}{\gamma(\gamma + 1)} \; 2^{- \frac{2}{1 + \gamma}} \right) \, e^{\omega r} \, - \, \frac{\ve_2}{\gamma(\gamma+1)} \; 2^{- \frac{2}{\gamma+1}} \, e^{\frac{1-\gamma}{1+\gamma} \omega r}  \ ,
\end{equation}
and a close inspection reveals that regions of strong coupling are always present. This is a consequence of our choice of a cosh--function to begin with. For $\ve_2<0$, starting from a sinh--function one could exhibit solutions without strong--coupling regions, consistently with the presence of a dip in the right panel of fig.~\ref{fig:22}.

\noindent $\boxed{\gamma > 1 \, , \; \varepsilon_1 =  -1\,, \; \varepsilon_2 \  =  \ \pm 1}$

The potentials with $\ve_1 =  -1$ and $\gamma > 1$ are not bounded from below. However, in view of the discussion in Section~\ref{sec:compactness}, one can anticipate that these examples have some interesting features.
\begin{figure}[ht]
\centering
\begin{tabular}{cc}
\includegraphics[width=0.4\textwidth]{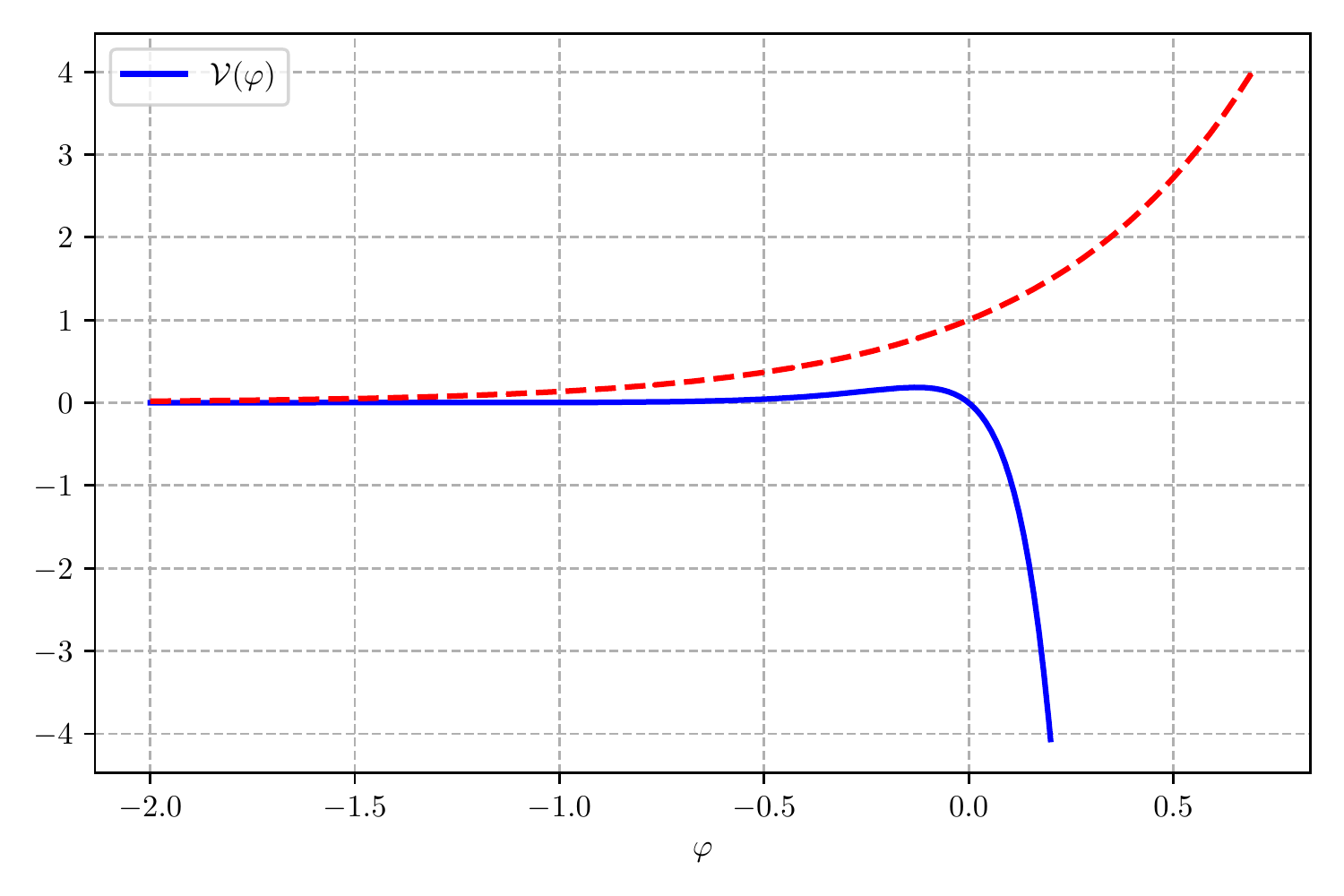} &
\includegraphics[width=0.4\textwidth]{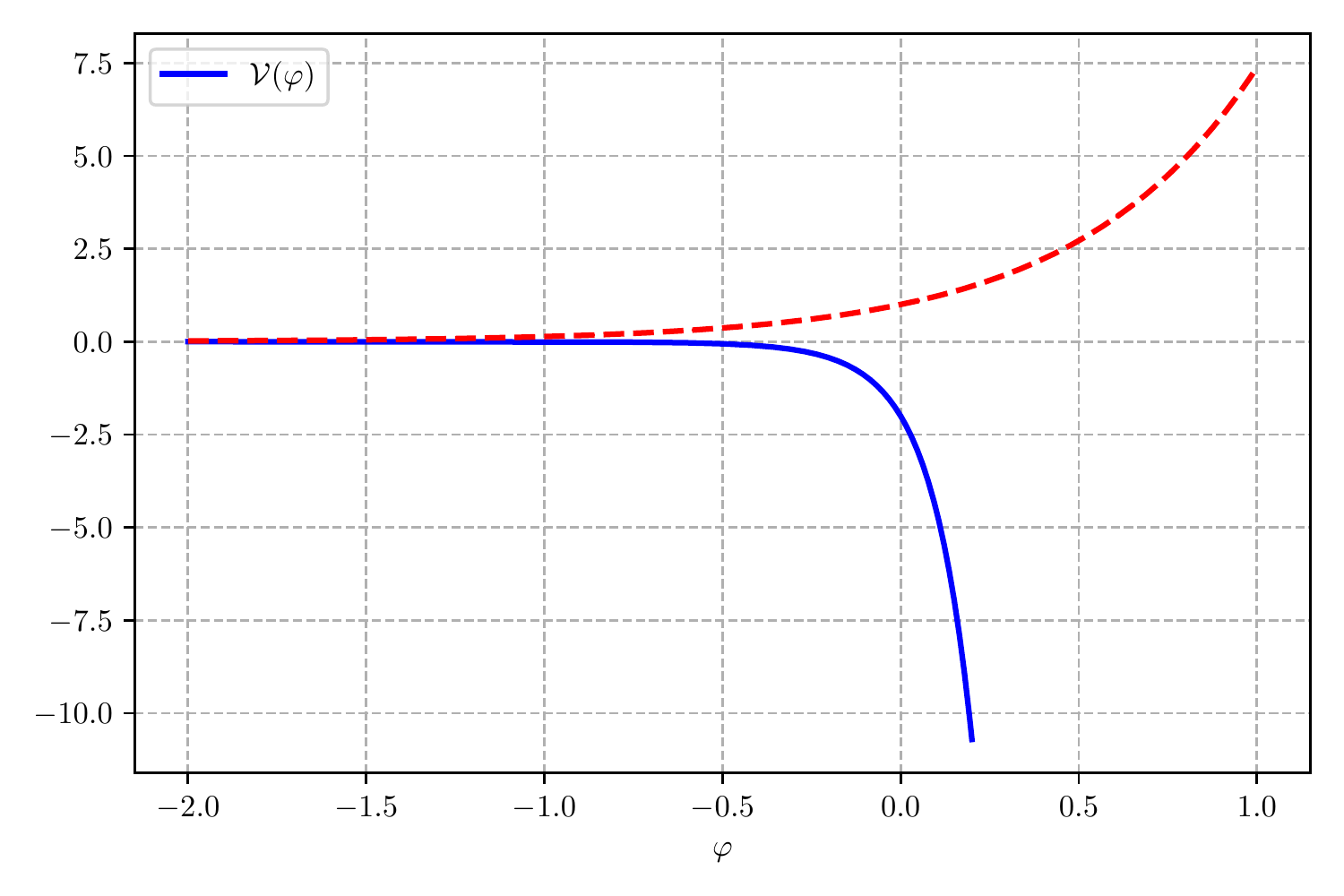} \\
\end{tabular}
\caption{\small The left panel displays the potential of eq.~\eqref{pot_33_2} for $\gamma =5 $, $\varepsilon_1 = -1$ and $\varepsilon_2 =  1$. The right one displays the potential for $\varepsilon_2 = -1$ and the same value of $\gamma$. The dashed potential is the one considered by \cite{dm_vacuum}.}
\end{figure}
Defining $\omega^2$ as in~\eqref{lower_omega}, the solutions read
\begin{align}
    x(r)\ =& \ \sin(\omega r ) \ , \nonumber \\
    y(r) \ =& \ \left[ b - \frac{\ve_2(\gamma - 1)}{\omega} \int_0^r \sin(\omega s)^{\frac{2}{1 + \gamma}} \td s  \right] \ \cos(\omega r) + \ve_2 \sin(\omega r)^{\frac{3 + \gamma}{1 + \gamma}} \ ,
\label{sol_x_y_gammamagg1}
\end{align}
where we have already imposed the Hamiltonian constraint. Returning to the original variables,
\begin{align}
	e^{\mathcal A} \, = & \;  \sin(\omega r )^{\frac{1}{1+ \gamma}} \left\{\left[ b - \frac{\ve_2(\gamma - 1)}{\omega} \int_0^r \sin(\omega s)^{\frac{2}{1 + \gamma}} \td s  \right] \ \cos(\omega r) + \ve_2\sin(\omega r)^{\frac{3 + \gamma}{1 + \gamma}} \right\}^{-\frac{1}{\gamma - 1}} \ ,  \nonumber \\
	e^{\vf} \, = & \; \sin(\omega r )^{\frac{1}{1 + \gamma}} \left\{ \left[ b - \frac{\ve_2(\gamma - 1)}{\omega} \int_0^r \sin(\omega s)^{\frac{2}{1 + \gamma}} \td s  \right] \ \cos(\omega r) + \ve_2 \sin(\omega r)^{\frac{3 + \gamma}{1 + \gamma}} \right\}^{\frac{1}{\gamma - 1}} \ , \nonumber \\
	e^{\cB} \, = & \;  \sin(\omega r )^{-\frac{\gamma}{1 + \gamma}} \left\{ \left[ b - \frac{\ve_2(\gamma - 1)}{\omega} \int_0^r \sin(\omega s)^{\frac{2}{1 + \gamma}} \td s  \right] \ \cos(\omega r) + \ve_2 \sin(\omega r)^{\frac{3 + \gamma}{1 + \gamma}} \right\}^{-\frac{\gamma}{\gamma - 1}} \ .
\end{align}
Let us first consider the case $\ve_2 = 1$. If the inequality
\begin{equation}
    b \ > \ \frac{\gamma - 1}{\omega} \int_0^\frac \pi \omega \sin(\omega s)^{\frac{2}{1 + \gamma}} \td s \ ,
\end{equation}
holds, the solution is defined in the interval $r \in (0, r^*)$, where $y(r^*) = 0$, and the string coupling vanishes at both ends. However, the internal space is not compact, since the volume form $e^{\cB + \frac \vf 3}$ is too singular at $r = r^*$. The behavior is different for
\begin{equation}
    0 \ < \ b \ < \ \frac{\gamma - 1}{\omega} \int_0^\frac \pi \omega \sin(\omega s)^{\frac{2}{1 + \gamma}} \td s \ ,
\label{cond_b}
\end{equation}
since $r \in \left( 0, \frac \pi \omega \right)$ and $y(r)$ in eq.~\eqref{sol_x_y_gammamagg1} never vanishes.
\begin{figure}[ht]
\centering
\begin{tabular}{cc}
\includegraphics[width=45mm]{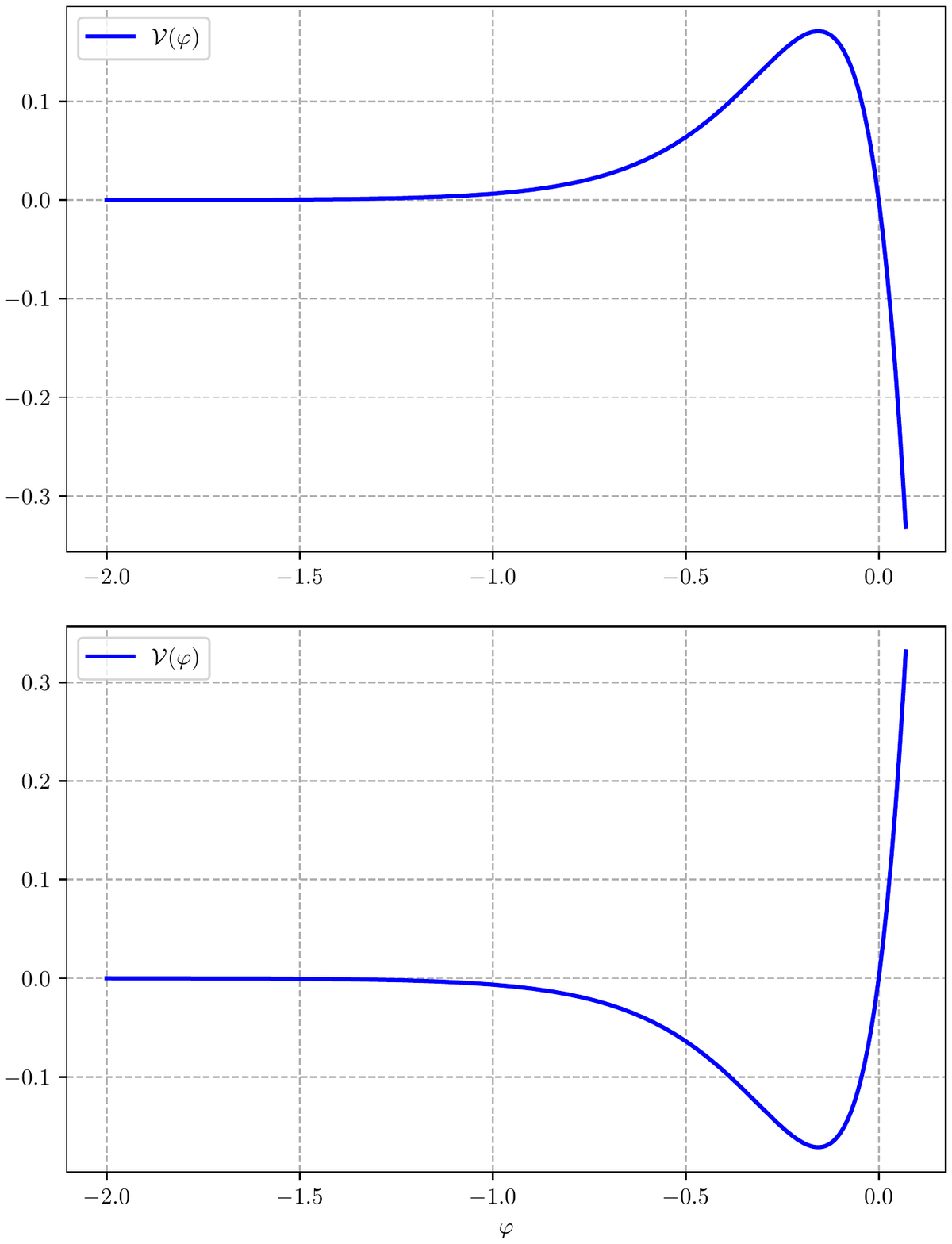} &
\includegraphics[width=0.57\textwidth]{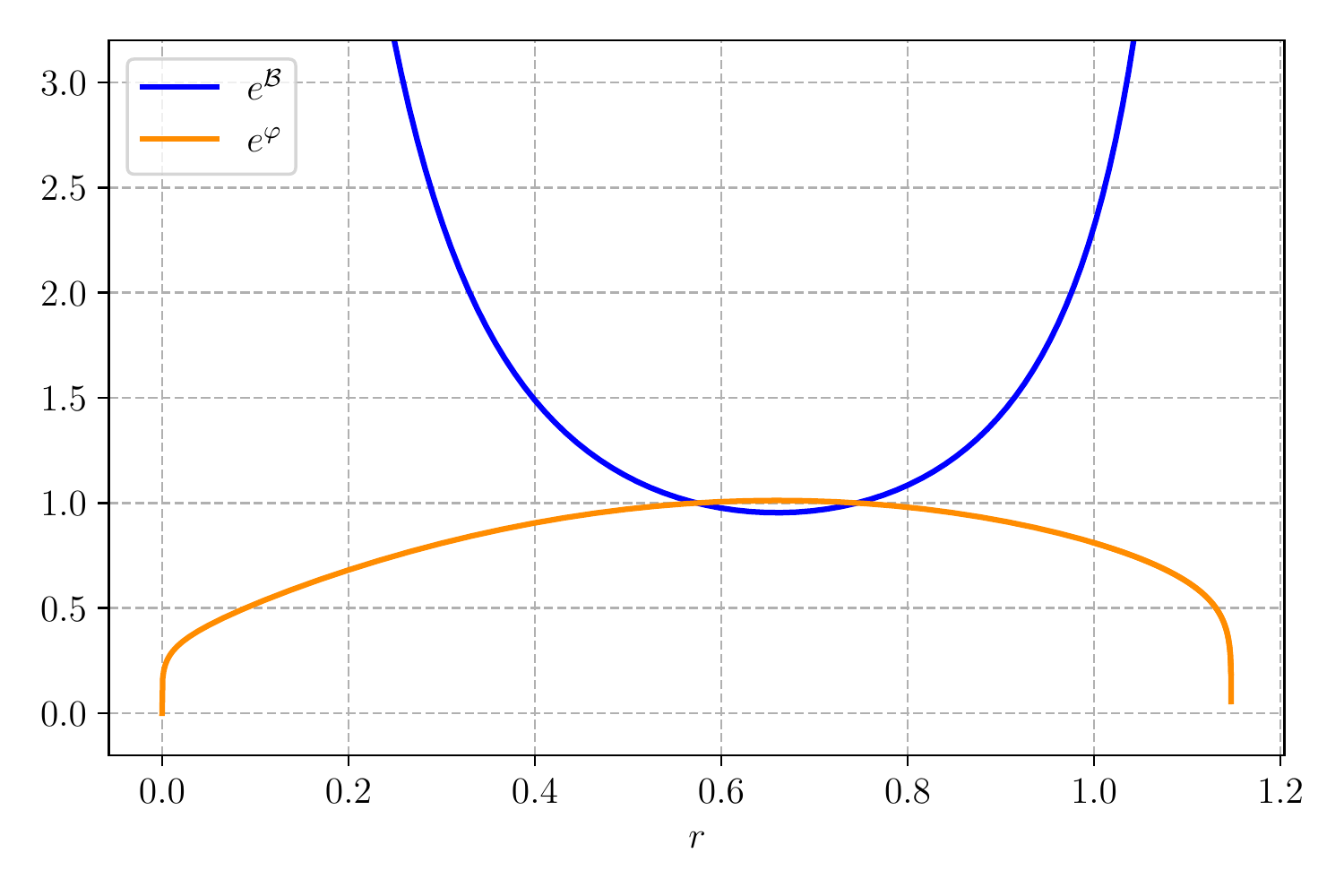}
\end{tabular}
\caption{ \small The upper--left panel displays the potential of eq.~\eqref{pot_33_2} for $\gamma = 4$, $\ve_1 = -1$ and $\ve_2 = 1$, while the lower--left one displays the inverted potential. The right panel shows the solution of $e^{\cB}$ and $e^{\vf}$ for $b = \frac{2}{10}$, so that the condition~\eqref{cond_b} is respected. This solution has several desirable properties, such as a bounded $g_s$, a compact $r$--direction, but yet an unbounded scalar curvature.}
\end{figure}
{\bf This case is very interesting}, since one can verify that:
\begin{itemize}
    \item the string coupling $g_s$ is bounded, and vanishes at the ends of the interval;
    \item the internal $r$--direction is compact;
    \item the 9D Planck mass and gauge coupling are finite.
\end{itemize}
However, the scalar curvature in string frame~\eqref{string_curvature} is not bounded at the ends of the interval.
On the other hand, if $\ve_2 = -1$ the domain of definition is again restricted to $r \in (0, r^*)$, with $r^* < \frac \pi \omega$. In this case $g_s$ is always bounded, but the $r$--direction is not compact.  In this case the potential is dominated for $\vf>0$ by a single exponential with $\gamma>1$, and consequently an upper bound for the string coupling is inevitable, consistently with the discussion in Section~\ref{sec:compactness}.

\subsection{\sc Systems Integrable via Quadratures}

\noindent {\sc 4). The Potential of eq.~\eqref{pot_34}}

\noindent The equations of motion obtained with
\begin{equation}
    \cV (\vf) \ = \  \lambda \left( e^{\frac{2}{\gamma} \vf} \ - \ e^{2 \gamma \vf} \right)
    \label{pot_34_2}
\end{equation}
can be replaced by their first integrals
\begin{equation}
	\dot{\hat{\mathcal A}}^2 \ = \  2 \lambda \left( e^{2 \hat{\mathcal A} \sqrt{1 - \gamma^2}} + C \right) \ , \qquad \qquad \dot{\hat{\vf}}^2 \ = \  2 \lambda \left( e^{\frac{2}{\gamma} \hat{\vf} \sqrt{1 - \gamma^2}} + D\right) \ ,
\end{equation}
where $C$ and $D$ are two integration constants. The Hamiltonian constraint demands that $C = D$, and one can consider a small number of different cases.

If $C = D = 0$ the preceding equations reduce to
\begin{equation}
	\dot{\hat{\mathcal A}}^2 \ = \  2 \lambda \left( e^{2 \hat{\mathcal A} \sqrt{1 - \gamma^2}} \right) \ , \qquad \qquad \dot{\hat{\vf}}^2 \ = \  2 \lambda \left( e^{\frac{2}{\gamma} \hat{\vf} \sqrt{1 - \gamma^2}}\right) \ ,
\end{equation}
where for consistency $\lambda$ must be positive, and the solutions read
\begin{equation}
    e^{\mathcal A} \ = \ \frac{\sqrt{2 \lambda (1 - \gamma^2)}}{\gamma^{\frac{\gamma^2}{1 - \gamma^2}}} \, \frac{|r - r_{\hat \varphi}|^{\frac{\gamma^2}{1 - \gamma^2}}}{|r - r_{\hat{\mathcal A}}|^{\frac{1}{1 - \gamma^2}}} \ , \qquad \qquad e^{\vf} \ = \ \frac{1}{\sqrt{2 \lambda (1 - \gamma^2)}} \, \left|\frac{r - r_{\hat{\mathcal A}}}{r - r_{\hat \varphi}}\right|^{\frac{\gamma}{1 - \gamma^2}} \ .
\end{equation}
Notice that in the limit of large $\left| r \right|$ the string coupling approaches a constant value. Consequently, the argument presented in Section~\ref{sec:compactness} leads one to expect that the internal space be of infinite length. Recalling that in this case the gauge choice is $\mathcal B = \mathcal A$, one can indeed conclude that
\begin{equation}
    e^{\mathcal B} \ \sim \ r^{-1} \ ,
\end{equation}
and the $r$--direction is indeed not compact. Moreover, the string coupling diverges at $r = r_{\hat{\varphi}}$.

If, on the other hand $C \, , \, D \neq 0$, one can conveniently write the first integrals as
\begin{equation}
	\dot{\hat{\mathcal A}}^2 \ = \  2 \lambda \left( e^{2 \hat{\mathcal A} \sqrt{1 - \gamma^2}} + \eta \,e^{2 \hat{\mathcal A}_0 \sqrt{1 - \gamma^2}} \right) \ , \qquad \qquad \dot{\hat{\vf}}^2 \ = \  2 \lambda \left( e^{\frac{2}{\gamma} \hat{\vf} \sqrt{1 - \gamma^2}} + \eta \, e^{\frac{2}{\gamma} \hat{\vf}_0 \sqrt{1 - \gamma^2}}\right) \ , \label{eqs_a_phi_1}
\end{equation}
where $\eta = \pm 1$, while the constants $\hat{\mathcal A}_0$ and $\hat{\vf}_0$ are not independent, but must satisfy
\begin{equation}
    \hat{\mathcal A}_0 = \frac{\hat{\vf}_0}{\gamma}
\end{equation}
in order to respect the Hamiltonian constraint. In the following, we shall explore the possible values of $\eta$ and $\lambda$, and to this end it is convenient to let
\begin{align}
	X  \ = \  & \, (\hat{\mathcal A} - \hat{\mathcal A}_0) \,  \sqrt{1 - \gamma^2} \ , \qquad \qquad  Y \ = \  (\hat{\vf} - \hat{\vf}_0) \, \frac{\sqrt{1 - \gamma^2}}{\gamma} \ , \nonumber \\
	\omega^2 \  = & \  2 \, |\lambda| \, e^{2 \hat{\mathcal A}_0 \sqrt{1 - \gamma^2}} \;\frac{1 - \gamma^2}{\gamma^2} \ ,
\label{XandYchange}
\end{align}
before pausing on various options.

\noindent \framebox{$\lambda > 0 \, , \, \eta = 1$}

\noindent The preceding definitions of eq.~\eqref{XandYchange} turn eqs.~\eqref{eqs_a_phi_1} into
\begin{equation}
	\dot X^2  \ = \  \omega^2 \, \gamma^2 \, ( 1 + e^{2X} ) \ , \qquad \qquad \dot Y^2 \ = \  \omega^2 \, (1 + e^{2Y}) \ ,
\end{equation}
whose solutions are
\begin{equation}
	e^{X} \ = \  \frac{1}{\sinh(\omega \gamma |r - r_{\hat{\mathcal A}}|)} \ , \qquad \qquad e^{Y} \ = \  \frac{1}{\sinh(\omega |r - r_{\hat \vf}|)} \ ,
\end{equation}
and returning to the original variables
\begin{equation}
	e^{\mathcal A} \ = \  e^{\mathcal A_0} \frac{\big[\sinh(\omega|r - r_{\hat\vf}|)  \big]^{\frac{\gamma^2}{1 - \gamma^2}}}{\big[ \sinh(\gamma\omega|r - r_{\hat{\mathcal A}}|)  \big]^{\frac{1}{1 - \gamma^2}}} \ = \ e^{\cB} \ , \qquad \qquad e^{\vf} \ = \  e^{\vf_0} \frac{\big[ \sinh(\gamma\omega|r - r_{\hat{\mathcal A}}|)  \big]^{\frac{\gamma}{1 - \gamma^2}}}{\big[ \sinh(\omega|r - r_{\hat{\vf}}|)  \big]^{\frac{\gamma}{1 - \gamma^2}}} \ .
\label{solgammaunosugamma}
\end{equation}

As we explained in Section~\ref{Sys_quad}, the range $0 < \gamma < 1$ suffices to explore the available options for this family of potentials. The string coupling then vanishes at $r =  r_{\hat{\cA}}$ but diverges at $r = r_{\hat{\vf}}$. Here it will suffice to choose $r_{\hat \cA} > r_{\hat \vf}$, and then $e^{\cB} \, \sim \, e^{\vf} \, \sim \, e^{- \frac{\gamma}{\gamma + 1} \, \omega \, r}$ for large values of $|r|$. All in all, the infinite excursion of $\varphi$ translates into a \emph{finite} contribution to the length of the internal interval from the large--$r$ region. Something different happens, however, at $r = r_{\hat \cA}$, where the string coupling vanishes but the volume form $e^{\cB + \frac{\vf}{3}} \sim |r - r_{\hat{\cA}}|^{\frac{\gamma - 3}{3(1 - \gamma^2)}}$, which results in a \emph{finite} contribution to the internal length only if $0 < \gamma < \frac{1}{3}$ and in an \emph{infinite} one in the complementary range $\frac{1}{3} \leq \gamma < 1$.

\begin{figure}[ht]
\begin{tabular}{cc}
\includegraphics[width=45mm]{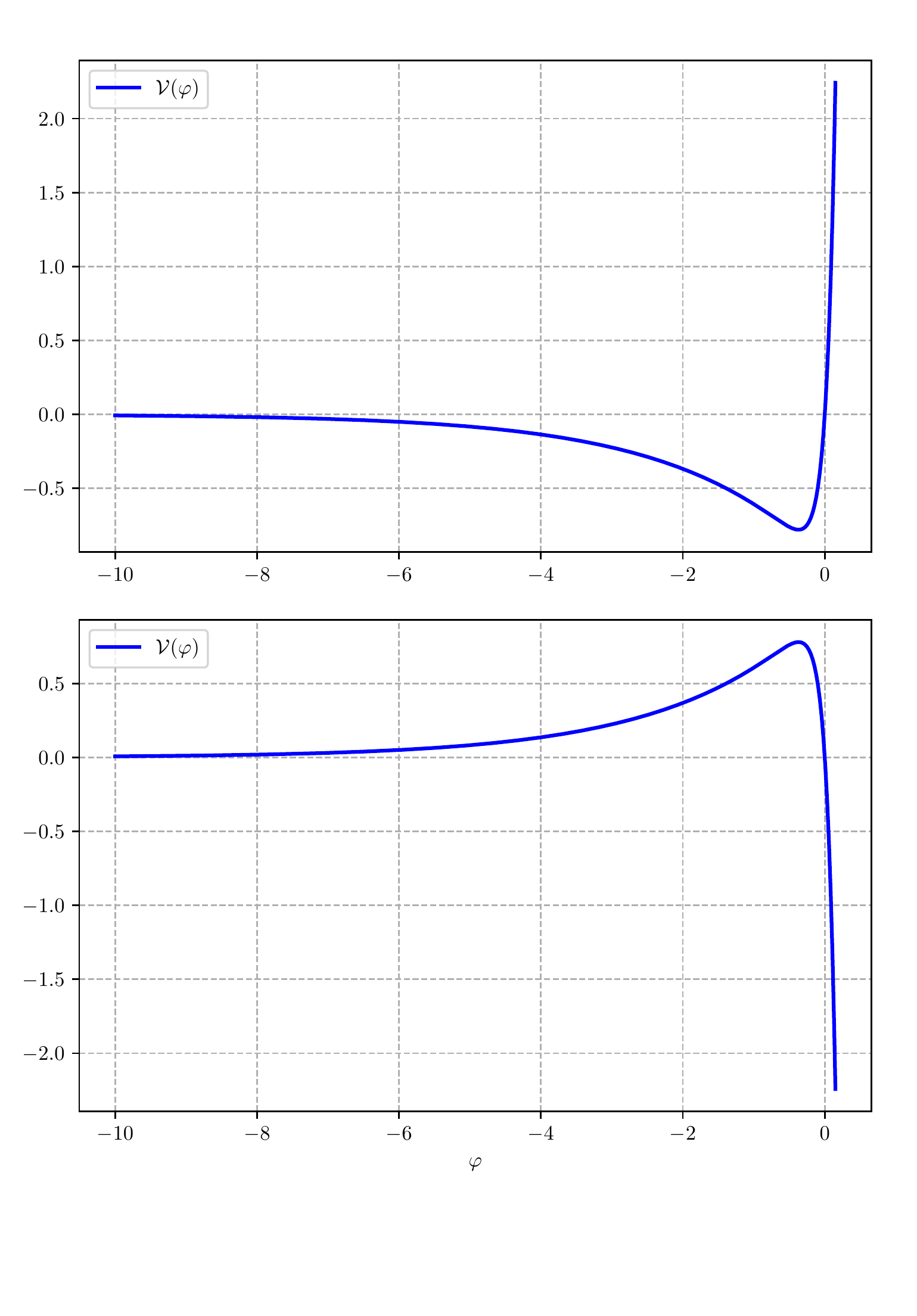} &
\includegraphics[width=0.57\textwidth]{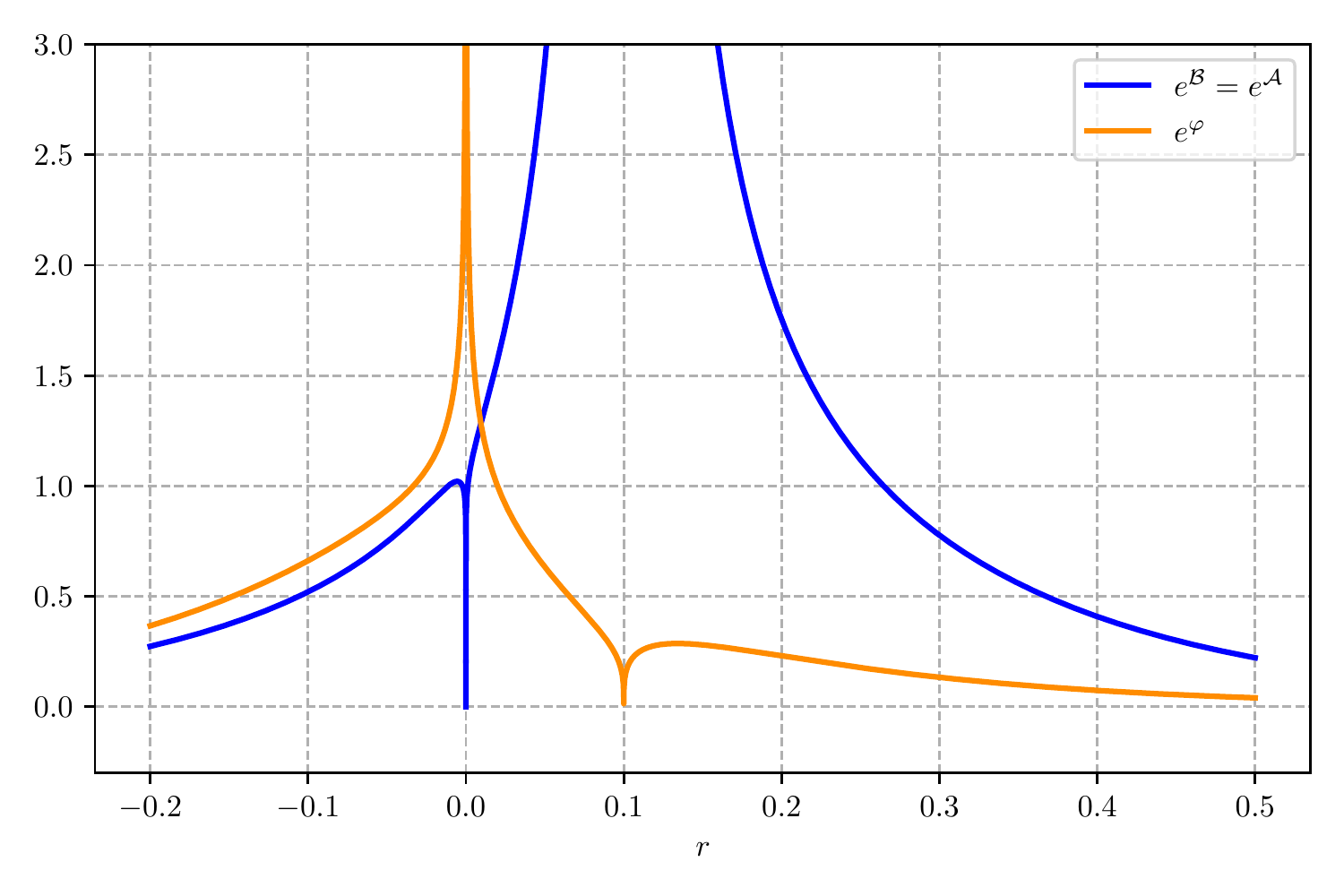}
\end{tabular}
\centering
\caption{ \small The upper--left panel displays the potential of eq.~\eqref{pot_34_2} for $\gamma =  \frac{1}{4}$ and $\lambda =  1$, while the lower--left panel displays the inverted potential. The right panel displays $e^{\cB}$ and $e^{\vf}$ for $\eta = 1$, $r_{\hat{\cA}} = 0.1$, $r_{\hat{\vf}} =  0$ and $\cA_0 =  \vf_0 =  0$.}
\label{gamma1sugamma}
\end{figure}

The scalar curvature has the opposite behavior: it is infinite in the first case and finite in the second. Moreover, it is also unbounded for large values of $r$.
Finally, the reduced Planck mass and gauge coupling \eqref{planckandgauge} are always infinite for this class of models, due to the behavior at $r_{\hat{\cA}}$. This type of solution is displayed in fig.~\ref{gamma1sugamma}. \emph{Curvature singularities have an important effect: the different regions that they separate describe once more distinct vacua!} This type of pattern had already emerged with the first example, but here the different regions do not bring along inverted potentials. Here \textbf{the region $r > r_{\hat \cA}$ is particularly interesting} since it yields finite values for the string coupling and a finite interval length for the whole range $0 < \gamma < \frac{1}{3}$.

\noindent \framebox{$\lambda > 0 \, , \, \eta = -1$}

\noindent In this case the change of variables of eq.~\eqref{XandYchange} turns the equations of motion into
\begin{equation}
	\dot X^2  \ = \  \omega^2 \, \gamma^2 \, ( e^{2X} - 1 ) \ , \qquad \qquad \dot Y^2 \ = \  \omega^2 \, ( e^{2Y} - 1 ) \ ,
\end{equation}
whose solutions are
\begin{equation}
	e^{\mathcal A} \ = \  e^{\mathcal A_0} \frac{\big[ \cos(\omega (r - r_{\hat \vf})) \big]^{\frac{\gamma^2}{1 - \gamma^2}}}{\big[ \cos(\gamma \omega (r - r_{\hat{\mathcal A}}))  \big]^{\frac{1}{1 - \gamma^2}}} \ , \qquad e^{\vf} \ = \  e^{\vf_0} \frac{\big[ \cos(\gamma \omega (r - r_{\hat{\mathcal A}}))  \big]^{\frac{\gamma}{1 - \gamma^2}}}{\big[ \cos(\omega (r - r_{\hat \vf}))  \big]^{\frac{\gamma}{1 - \gamma^2}}} \ .
\end{equation}
This solution has a non-trivial domain of validity if the regions where $\cos(\omega (r - r_{\hat \vf}))$ and \\ $\cos(\gamma \omega (r - r_{\hat {\mathcal A}}))$ are both positive is not empty. Notice that the interval where this is the case for $\cos(\gamma \omega (r - r_{\hat {\mathcal A}}))$ is longer than the corresponding one for $\cos(\omega (r - r_{\hat \vf}))$. As a result, for all choices of integration constants, this class of solutions has a region of strong coupling. In the mechanical analogy, ``the particle overcomes the hill''.

\noindent \framebox{$\lambda < 0 \, , \, \eta = -1$}

\noindent In this case the change of variables of eq.~\eqref{XandYchange} turns the equations of motion into
\begin{equation}
	\dot X^2  \ = \  \omega^2 \, \gamma^2 \, ( 1 - e^{2X} ) \ , \qquad \qquad \dot Y^2 \ = \  \omega^2 \, ( 1 - e^{2Y} ) \ ,
\end{equation}
whose solutions are
\begin{equation}
	e^{\mathcal A} \ = \  e^{\mathcal A_0} \frac{\big[ \cosh(\omega (r - r_{\hat \vf})) \big]^{\frac{\gamma^2}{1 - \gamma^2}}}{\big[ \cosh(\gamma \omega (r - r_{\hat{\mathcal A}}))  \big]^{\frac{1}{1 - \gamma^2}}} \ , \qquad e^{\vf} \ = \  e^{\vf_0} \frac{\big[ \cosh(\gamma \omega (r - r_{\hat{\mathcal A}}))  \big]^{\frac{\gamma}{1 - \gamma^2}}}{\big[ \cosh(\omega (r - r_{\hat \vf}))  \big]^{\frac{\gamma}{1 - \gamma^2}}} \ . \label{fourth_cosh}
\end{equation}
\textbf{This case is very interesting}. The solution is defined for all values of $r$, and for large $|r|$
\begin{equation}
    e^{\vf} \ \sim \  e^{\mathcal A} \ \sim \ e^{- \frac{\gamma}{\gamma + 1} \omega |r|} \ .
\end{equation}
Taking into account the gauge choice $\mathcal B = \mathcal A$, one can conclude that
\begin{itemize}
    \item the internal $r$--direction has finite length;
    \item the string coupling $g_s$ is bounded;
    \item the 9D Planck mass and gauge coupling are finite.
\end{itemize}
\begin{figure}[ht]
\begin{tabular}{cc}
\includegraphics[width=45mm]{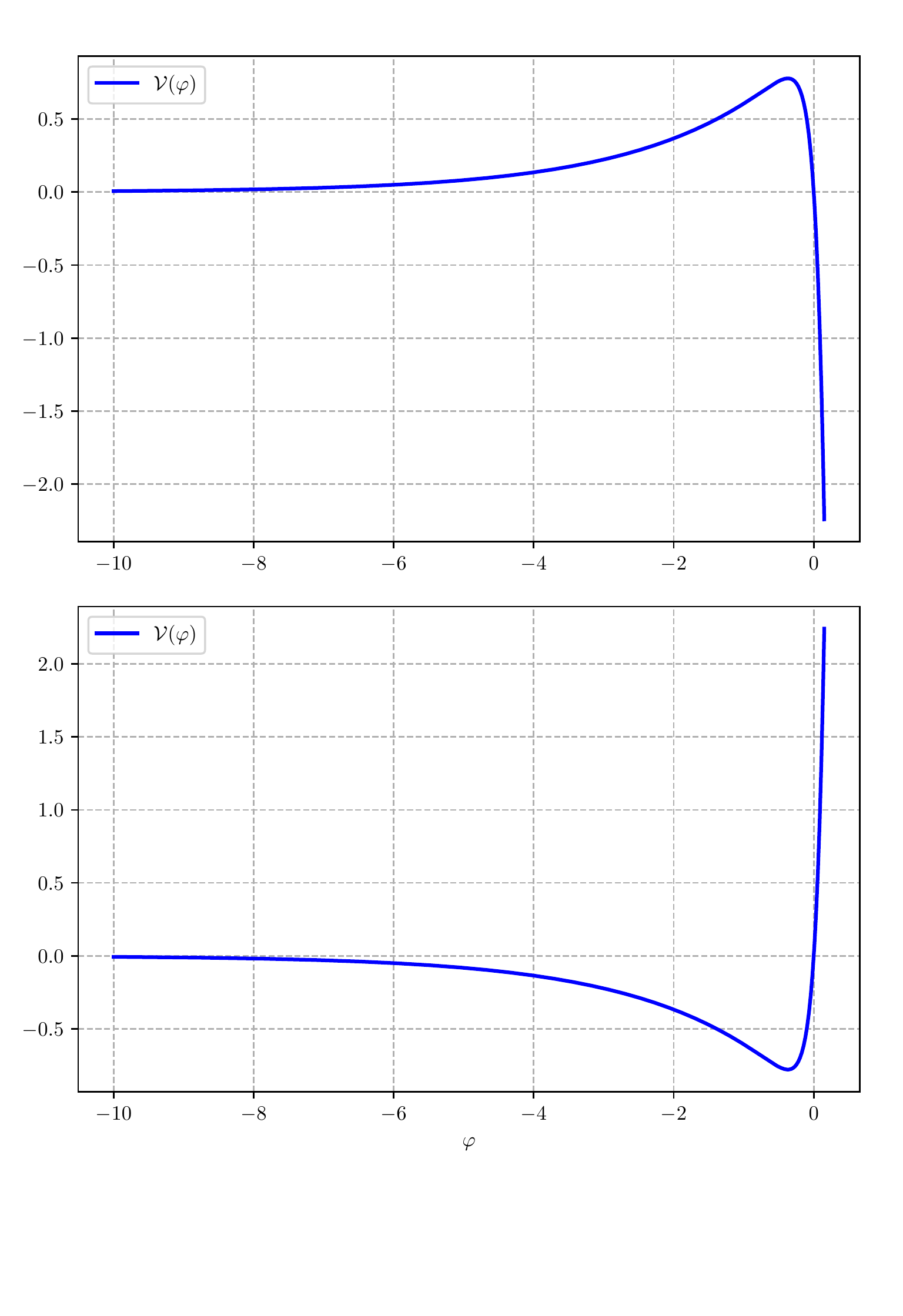} &
\includegraphics[width=0.57\textwidth]{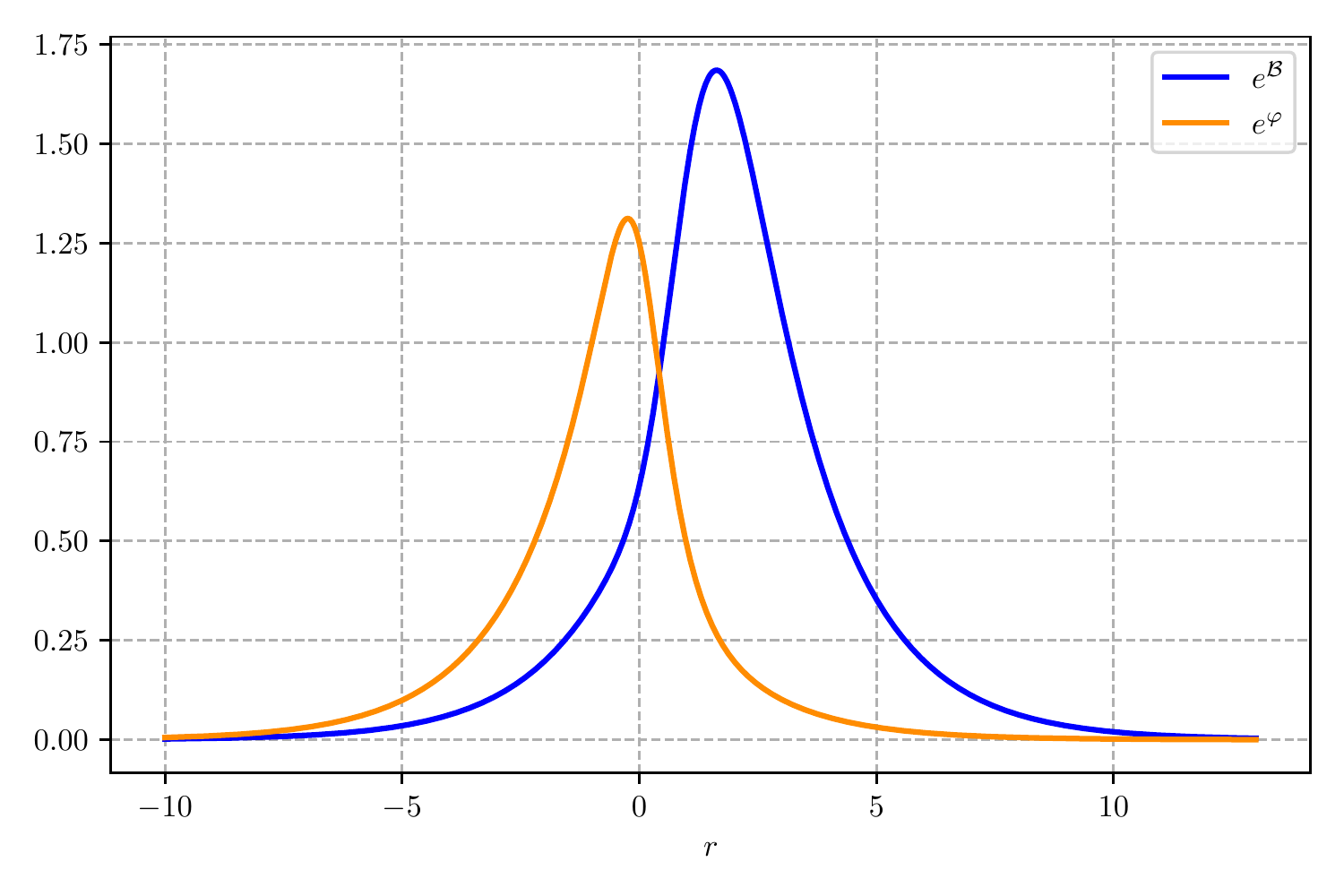}
\end{tabular}
\centering
\caption{ \small The upper--left panel displays the potential of eq.~\eqref{pot_34_2} for $\gamma =  \frac{1}{2}$ and $\lambda = - \frac{1}{2}$, while the lower--left panel displays the inverted potential. The right panel shows $e^{\cB}$ and $e^{\vf}$ for $\eta = -1$, $r_{\hat{\cA}} =  1$, $r_{\hat{\vf}} =  0$ and $\cA_0 =  \vf_0 =  0$.}
\label{fourth_inverted}
\end{figure}
Once more, the string--frame scalar curvature~\eqref{string_curvature} is unbounded for large values of $r$. This is a typical case in which the analysis of Section~\ref{sec:compactness} guarantees directly an upper bound on the string coupling, via the Euclidean counterpart of the climbing mechanism identified in~\cite{dks2010}, since for $\vf \geq 0$ the potential is readily dominated by a negative term with $\gamma > 1$. One the other hand, for $\vf < 0$, the potential is readily dominated by a positive term with $\gamma < 1$. Consequently, Table~\ref{tab:dilaton_dynamics_2} guarantees the compactness of the internal space.

\noindent \framebox{$\lambda < 0 \, , \, \eta = 1$}

\noindent In this case the change of variables of eq.~\eqref{XandYchange} turns the equations of motion into
\begin{equation}
	\dot X^2  \ = \  - \,\omega^2 \, \gamma^2 \, ( 1 + e^{2X} ) \ , \qquad \qquad \dot Y^2 \ = \  - \, \omega^2 \, ( 1 + e^{2Y} ) \ ,
\end{equation}
which cannot be solved over the real numbers, and therefore this choice of the integration constants is unphysical.

\noindent {\sc 5). The Potential of eq.~\eqref{pot_35}}

\noindent The potential
\begin{equation}
	\cV (\vf) \ = \  C \, \log (- \coth(\vf)) + D
\label{pot_35_2}
\end{equation}
is another interesting case.
One can solve the equations of motion (\ref{EOM_fourth_potential}) for $\xi$ and $\eta$, since each entails a ``conserved energy''. This can be shown multiplying the first by $\dot \xi$ and the second by $\dot \eta$, thus obtaining
\begin{equation}
	\dot \xi^2  \ = \  - \, 8 \, C \, \log(\xi)  \ +\  \varepsilon_{\xi} \ , \qquad \qquad \dot \eta^2 \ = \  - \, 8 \, C \, \log(\eta) \ + \ \varepsilon_{\eta} \ ,
\end{equation}
where $\varepsilon_{\xi}$ and $\varepsilon_{\eta}$ are integration constants. After the rescaling $\rho = \sqrt{8C} r$, one thus arrives at the simplified set of equations
\begin{equation}
	\dot \xi^2  \ = \  - \, \log \left( \frac{\xi}{\xi_0} \right) \ , \qquad \qquad \dot \eta^2  \ = \ - \, \log \left( \frac{\eta}{\eta_0} \right) \ .
\label{logcothcharges}
\end{equation}
 where $\log(\xi_0) = \varepsilon_{\xi} / 8C$ and $\log(\eta_0) = \varepsilon_{\eta} / 8C$. The Hamiltonian constraint links $\xi_0$ and $\eta_0$ according to
\begin{equation}
	\log \left( \frac{\eta_0}{\xi_0} \right) \ = \  \frac{D}{C} \quad \longrightarrow \quad \eta_0 \ = \  \xi_0 \, e^{\frac{D}{C}} \ ,
\end{equation}
so that eq.~\eqref{logcothcharges} becomes
\begin{equation}
	\rho \ = \  \pm \, 2 \xi_0 \, \int_{u_0}^{\sqrt{- \log(\frac{\xi}{\xi_0})}} e^{-u^2}  \, \td u
\label{solutionforxi}
\end{equation}
for $\xi$, and one obtains a similar expression for $\eta$. Notice that eq.~\eqref{solutionforxi} should be inverted, an issue that we shall soon return to. However, in terms of the original variables,
\begin{align}
	\rho \, = & \; \pm 2 \xi_0 \int^{\sqrt{-\mathcal A - \log(\frac{2}{\xi_0} \, \cosh(\vf))}}_{u_0} e^{-u^2} \td u \ , \nonumber \\
	\rho \, = & \; \pm 2 \eta_0 \int^{\sqrt{-\mathcal A - \, \log(-\frac{2}{\eta_0} \, \sinh(\vf))}}_{v_0} e^{-v^2} \td v \ .
	\label{integrals}
\end{align}
It is useful to define
\begin{equation}
    f = \displaystyle \frac{\eta_0}{\xi_0} = e^{\frac{D}{C}} \ .
\end{equation}
Moreover, for this system we shall often refer to $\erf^{-1}(x)$, the inverse of $\erf(x)$, whose domain is restricted to $x \in (-1,1)$. In this fashion, with the additional rescaling
\begin{equation}
    \rho \ \longrightarrow \  \sqrt{\pi} \  \xi_0 \ \rho \ ,
\end{equation}
the complete solution reads
\begin{align}
    e^{\mathcal A} \, = & \; \frac{\xi_0}{2} \ \sqrt{ 1 - f^2 \exp \Bigg\{ 2 \bigg[ \erf^{-1} (\rho - \alpha)\bigg]^2 - 2 \bigg[\erf^{-1} \left( \displaystyle \frac{\rho}{f} \right) \bigg]^2   \Bigg\}} \; \; \times \nonumber  \\
	& \qquad \qquad \qquad \qquad \qquad \qquad \qquad \qquad \qquad \exp\left\{ - \left[ \erf^{-1}(\rho - \alpha) \right]^2 \right\} \ , \nonumber \\
	e^{\vf} \, = & \; \sqrt{\frac{1 - f \exp \Bigg\{ \bigg[ \erf^{-1} (\rho - \alpha)\bigg]^2 - \bigg[\erf^{-1} \left( \displaystyle \frac{\rho}{f} \right) \bigg]^2   \Bigg\}}{1 + f \exp \Bigg\{ \bigg[ \erf^{-1} (\rho - \alpha)\bigg]^2 - \bigg[\erf^{-1} \left( \displaystyle \frac{\rho}{f} \right) \bigg]^2   \Bigg\}}} \ , \nonumber \\
	\mathcal B \, = & \;  -\mathcal A \ .
\label{fourth_solution}
\end{align}
Notice that, as expected, in this class of examples \textbf{$g_s$ is manifestly not singular and less than one}.
The integration constant $\alpha$ originates from the integrals of eqs.~\eqref{integrals}. More precisely, two integration constants arise from the two integrals, but one can always absorb one of them shifting the coordinate $\rho$.
In order to analyze the solution~\eqref{fourth_solution}, it is useful to define the function
\begin{equation}
	h(\rho, \alpha) \ = \  1 - f^2 \exp \left\{ 2 \, \bigg[ \erf^{-1} (\rho -\alpha) \bigg]^2 -  2 \, \bigg[ \erf^{-1} \bigg(\frac{\rho}{f} \bigg) \bigg]^2  \right\} \ .
\end{equation}
Moreover, since an upper bound on $g_s$ is guaranteed by the potential, which restricts the dynamics to the region $\varphi \leq 0$, we can move on to ascertain whether or not the $r$--direction can be compact, for some values of $f$ and $\alpha$. The function $\erf^{-1}(x)$ has a limited domain, and consequently, in the preceding expressions, the function $\erf^{-1}\bigg( \displaystyle \frac{\rho}{f} \bigg)$ is defined for $\rho \in (-f, f)$, so that $\erf^{-1}(\rho - \alpha)$ is defined for $\rho \in (-1 + \alpha, 1 + \alpha)$. Consequently, if $(-1 + \alpha, 1 + \alpha) \cap (-f, f)  = \emptyset$, the solution does not exist. We can now analyze different cases of interest.

\noindent \framebox{$f < 1$ , $\alpha = 0$}

\noindent The interval of definition is $\rho \in (-f, f)$, and one has to check whether there are regions where the condition
\begin{equation}
	h(\rho, 0) \ = \  1 - f^2 \exp \left\{ 2 \, \bigg[ \erf^{-1} (\rho) \bigg]^2 -  2 \, \bigg[ \erf^{-1} \bigg(\frac{\rho}{f} \bigg) \bigg]^2  \right\} \ > \ 0
\label{squarerootpositive1}
\end{equation}
is satisfied, since this is the argument of the square roots of (\ref{fourth_solution}). A close inspection reveals that (\ref{squarerootpositive1}) is always satisfied for $\rho \in (-f, f)$, and therefore the square root determining $e^{\cB}$ is always well defined. Something similar applies to $e^{\vf}$, whose square root is a fraction of two factors that never vanish. Therefore, $g_s$ never vanishes as well, and since $f>0$, $g_s < 1$, as we anticipated.
\begin{figure}[ht]
\begin{tabular}{cc}
\includegraphics[width=45mm]{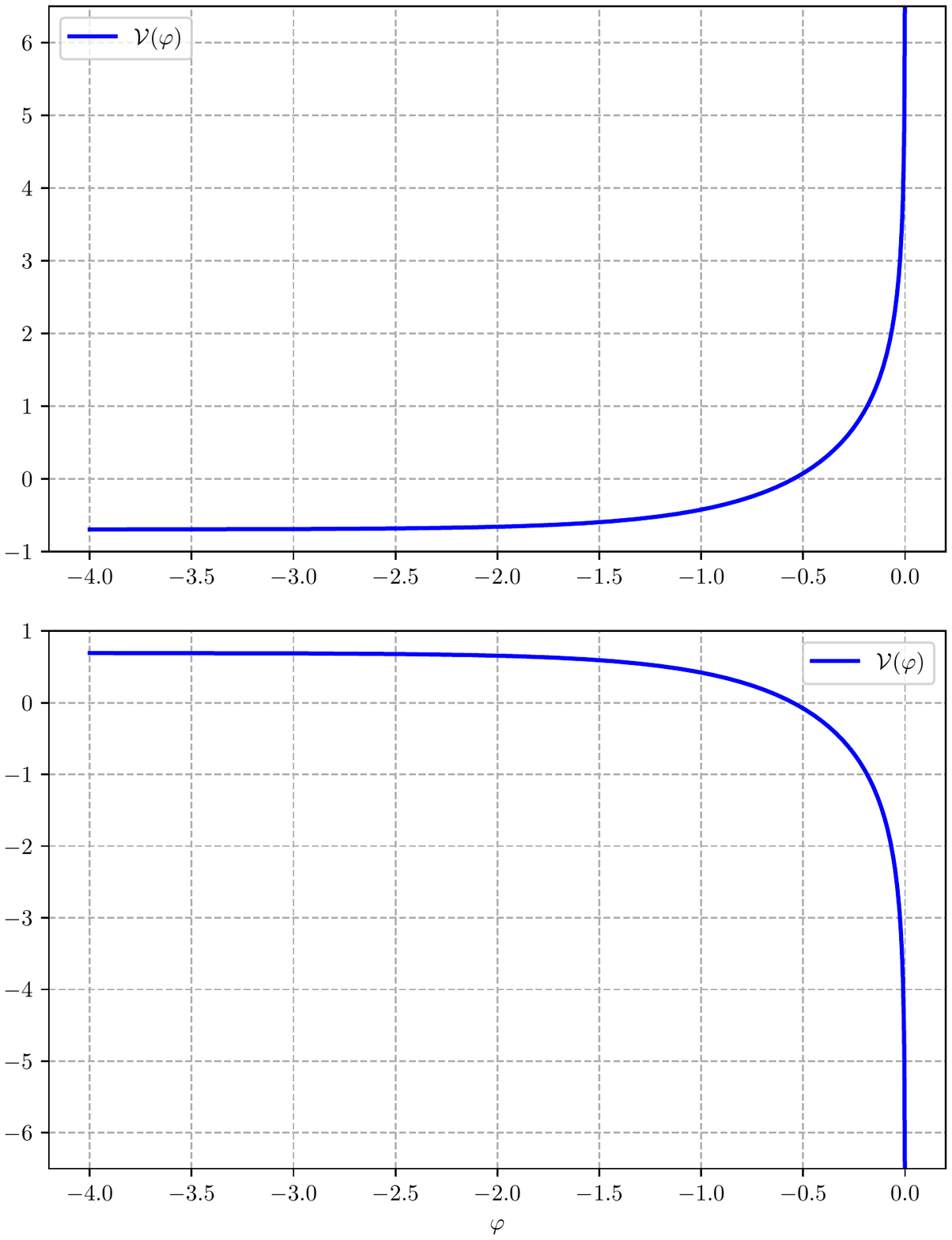} &
\includegraphics[width=0.57\textwidth]{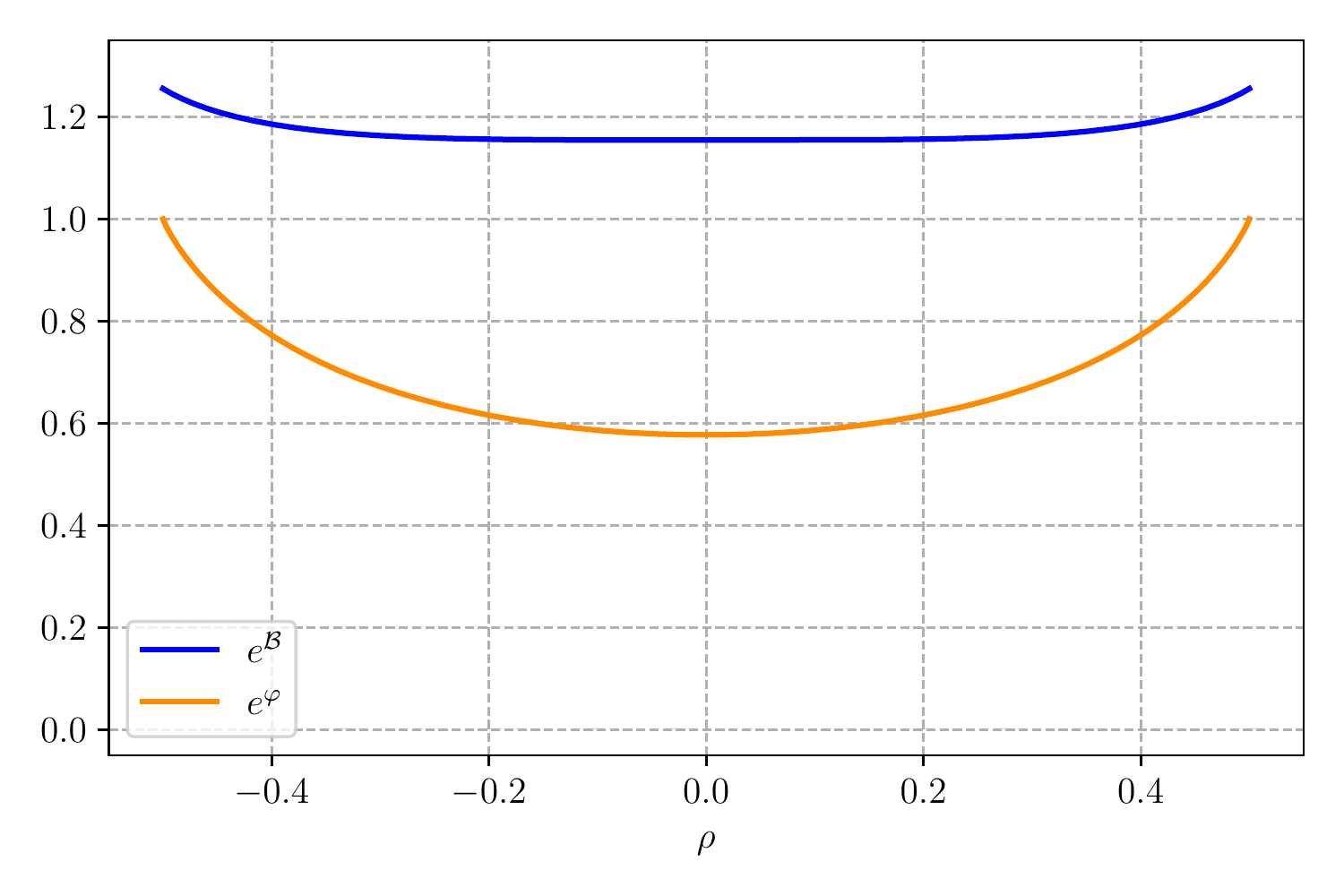}
\end{tabular}
\centering
\caption{ \small  The upper--left panel displays the potential of eq.~\eqref{pot_35_2} for $C = 1$ and $D = - \log(2)$, so that $f = \frac{1}{2}$, while the lower--left panel displays the inverted potential. The right panel displays $e^{\cB}$ and $e^{\vf}$ for $\alpha = 0$ and $\xi_0 = 2$.}
\label{fmin13}
\end{figure}
The $r$--direction is also compact, since $e^{\cB + \frac{\vf}{3}}$ is finite and has to be integrated over a finite domain. Similar considerations apply to the reduced Planck mass and gauge coupling. However, one can verify that the scalar curvature in string frame of eq.~\eqref{string_curvature} diverges at the boundaries.

\noindent \framebox{$f < 1$ , $\alpha \neq 0$}

\begin{figure}[ht]
\centering
\begin{tabular}{cc}
\includegraphics[width=45mm]{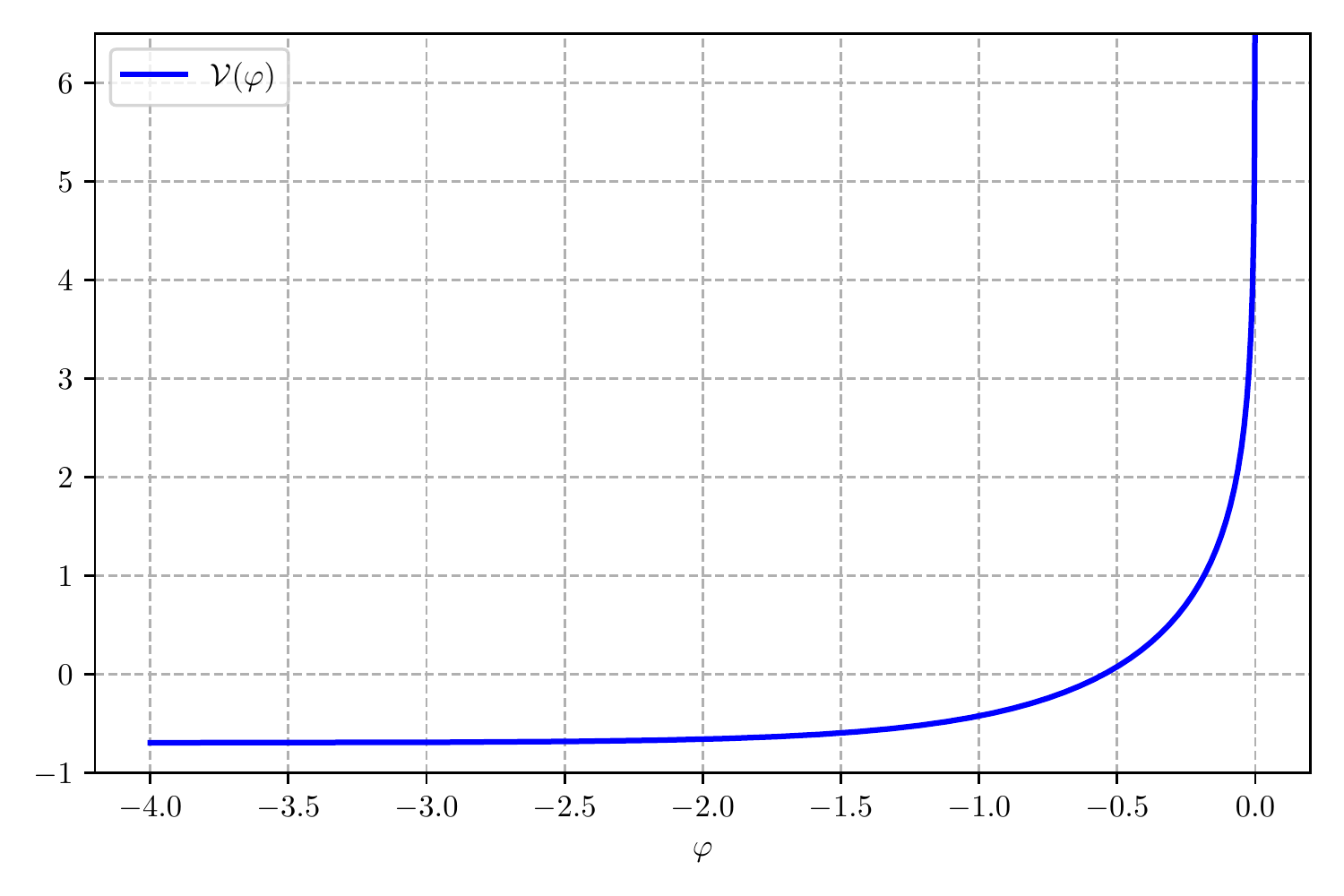} &
\includegraphics[width=45mm]{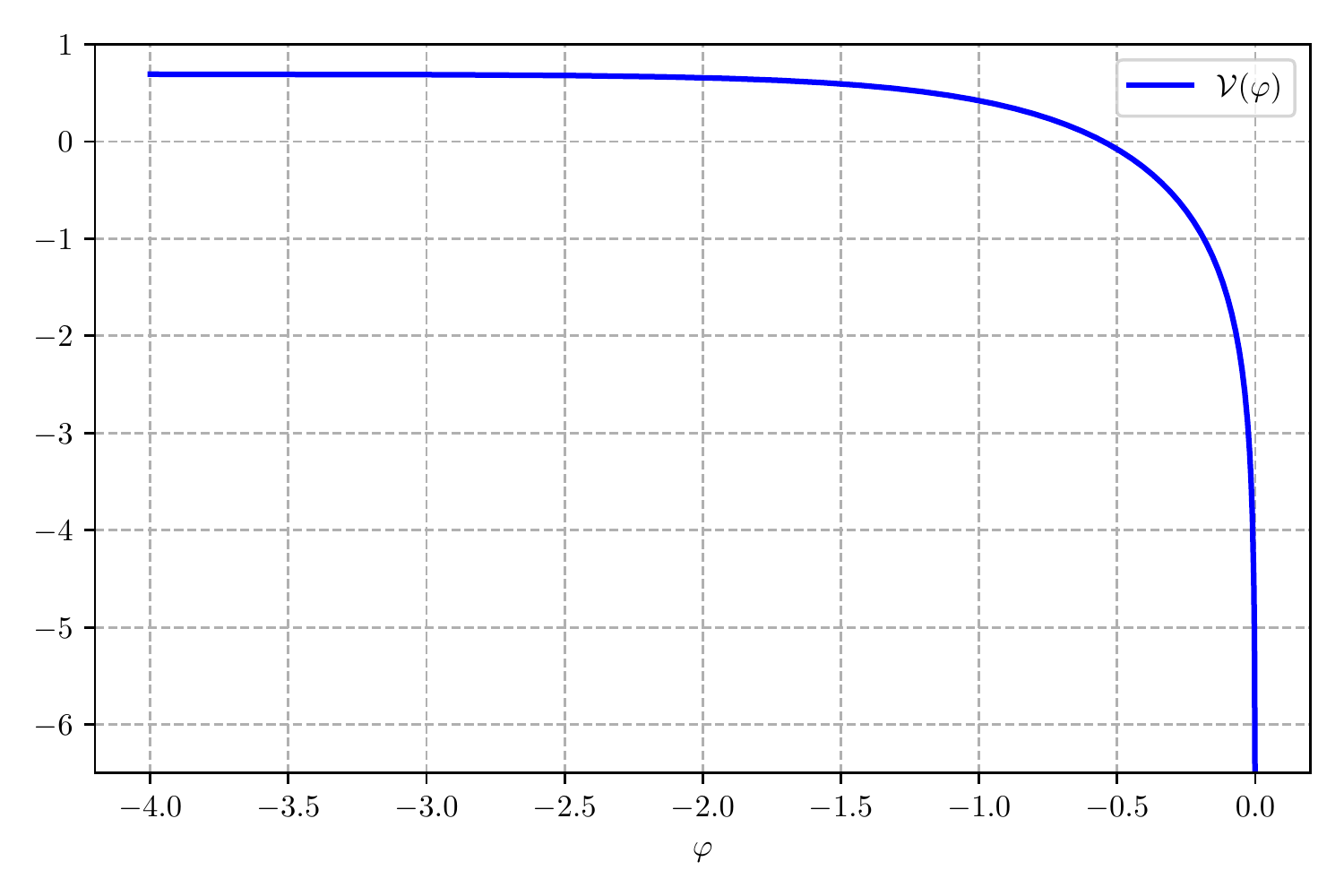} \\
\includegraphics[width=0.4\textwidth]{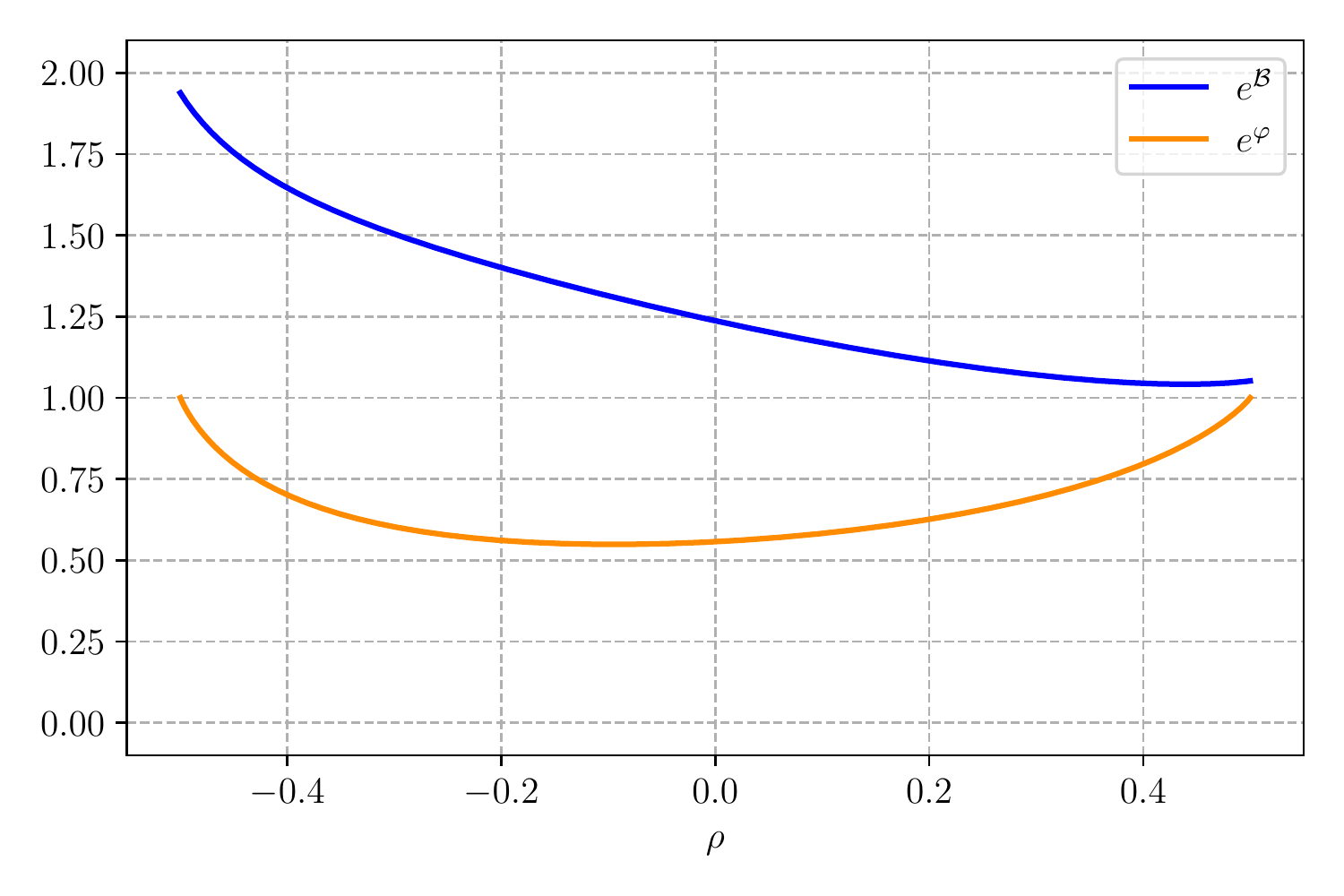} &
\includegraphics[width=0.4\textwidth]{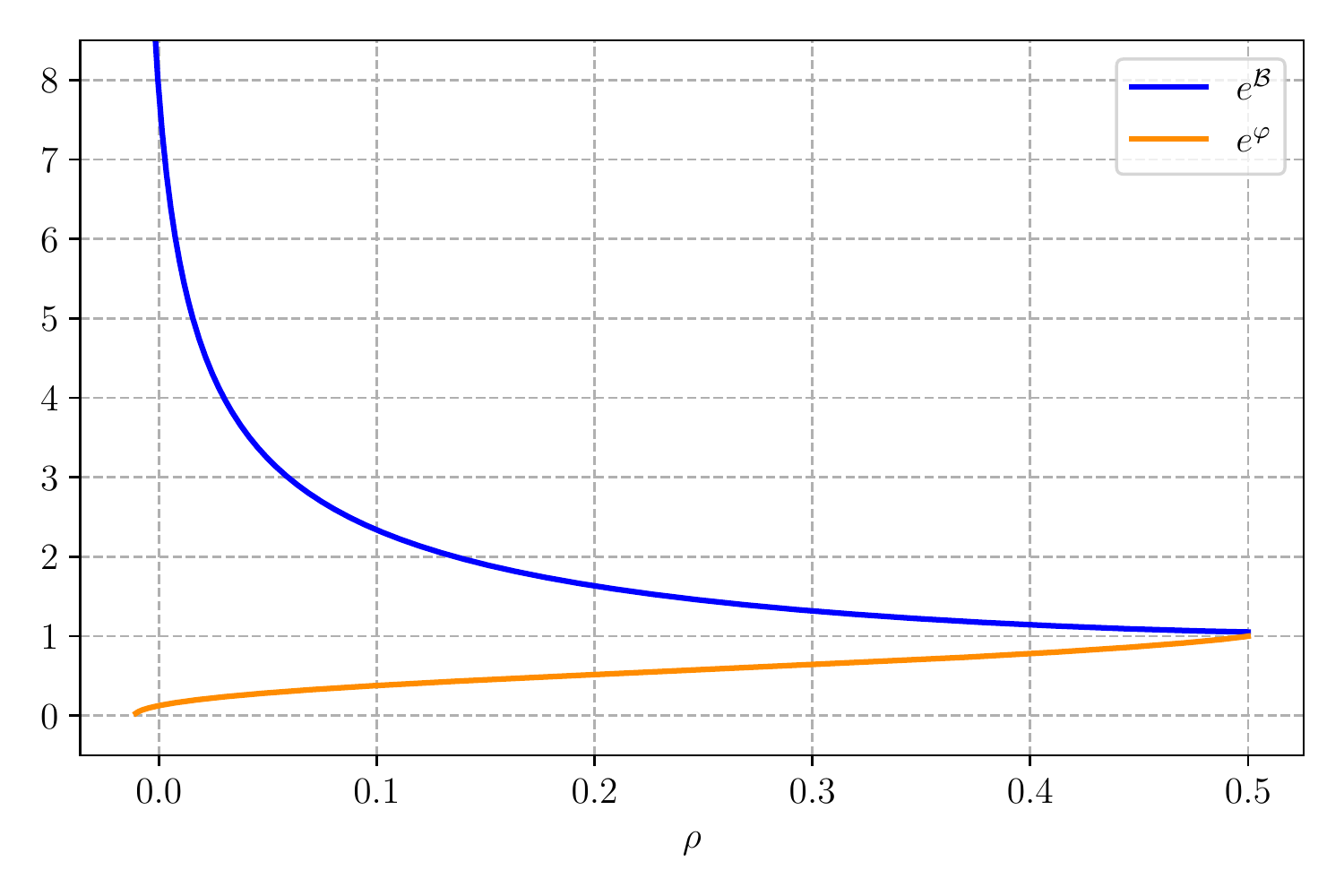} \\
\end{tabular}
\caption{ \small The upper--left panel displays the potential of eq.~\eqref{pot_35_2} for $C = 1$ and $D = - \log(2)$, so that $f = \frac{1}{2}$, while the upper--right panel displays the inverted potential. The lower panels display $e^{\cB}$ and $e^{\vf}$ for $\alpha = \frac{1}{4}$ and $\xi_0 = 2$, so that $ \alpha  < 1 - f$ and $f< 1$, and for $\alpha =  \frac34$ and $\xi_0 = 2$, so that $ 1 - f < \alpha < 1 + f$ and $f< 1$.}
\label{fmin1alphadiffzero1}
\end{figure}

\noindent Here it will suffice to confine the attention to the range $\alpha > 0$. A closer look at the function
 \begin{equation}
 	h(\rho, \alpha) \ = \  1 - f^2 \exp \left\{ 2 \, \bigg[ \erf^{-1} (\rho -\alpha) \bigg]^2 -  2 \, \bigg[ \erf^{-1} \bigg(\frac{\rho}{f} \bigg) \bigg]^2  \right\} \ ,
 \end{equation}
 indicates that one must distinguish the two cases $\alpha < 1- f$ and $1-f < \alpha <  1 + f$. The former is similar to the previous one: $e^{\cB} $ is always bounded and the string coupling is always perturbative. Fig.~\ref{fmin1alphadiffzero1} displays an example of this type, which closely resembles the case $\alpha =  0$ but is somewhat deformed. For the same reason, the $r$--direction is compact, since $e^{\cB + \frac{\vf}{3}}$ is integrable, and the reduced Planck mass and gauge coupling are finite. Once more, however, the scalar curvature in string frame diverges at the boundaries.
If $1-f < \alpha <  1 + f$, the functions are defined for $\rho \in ( \bar \rho, f)$, where $h(\bar \rho, \alpha) = 0$, since for $\rho \longrightarrow (-1 + \alpha)^+$ the argument of the square root becomes negative. The $r$--direction is again compact, with finite reduced Planck mass and gauge coupling, while the scalar curvature is singular at $\rho  =  f$, for the same reason as in previous cases, and at $\rho = \bar \rho$ where $h(\rho, \alpha)$ changes sign.

\begin{figure}[ht]
\centering
\begin{tabular}{cc}
\includegraphics[width=45mm]{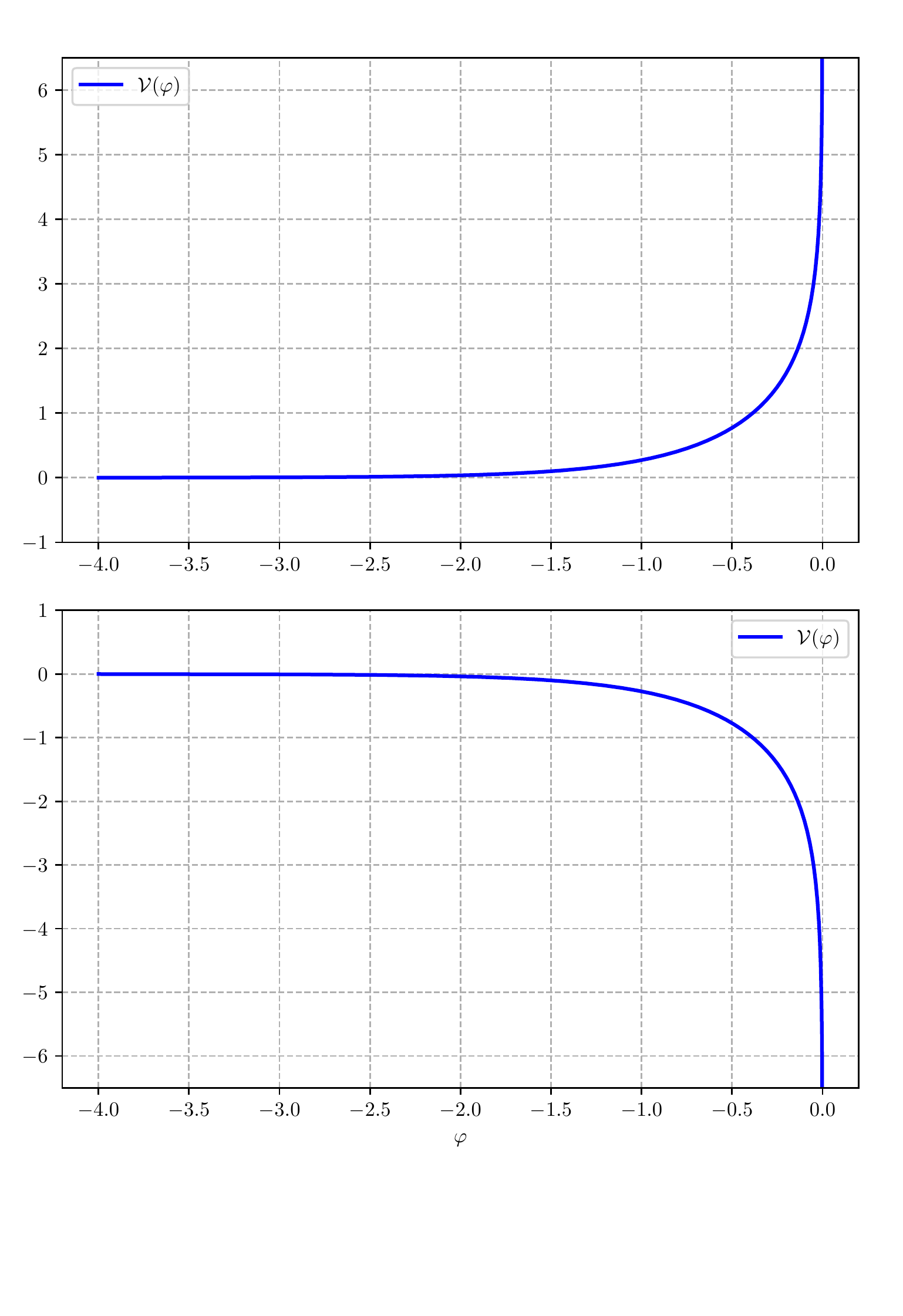} &
\includegraphics[width=0.57\textwidth]{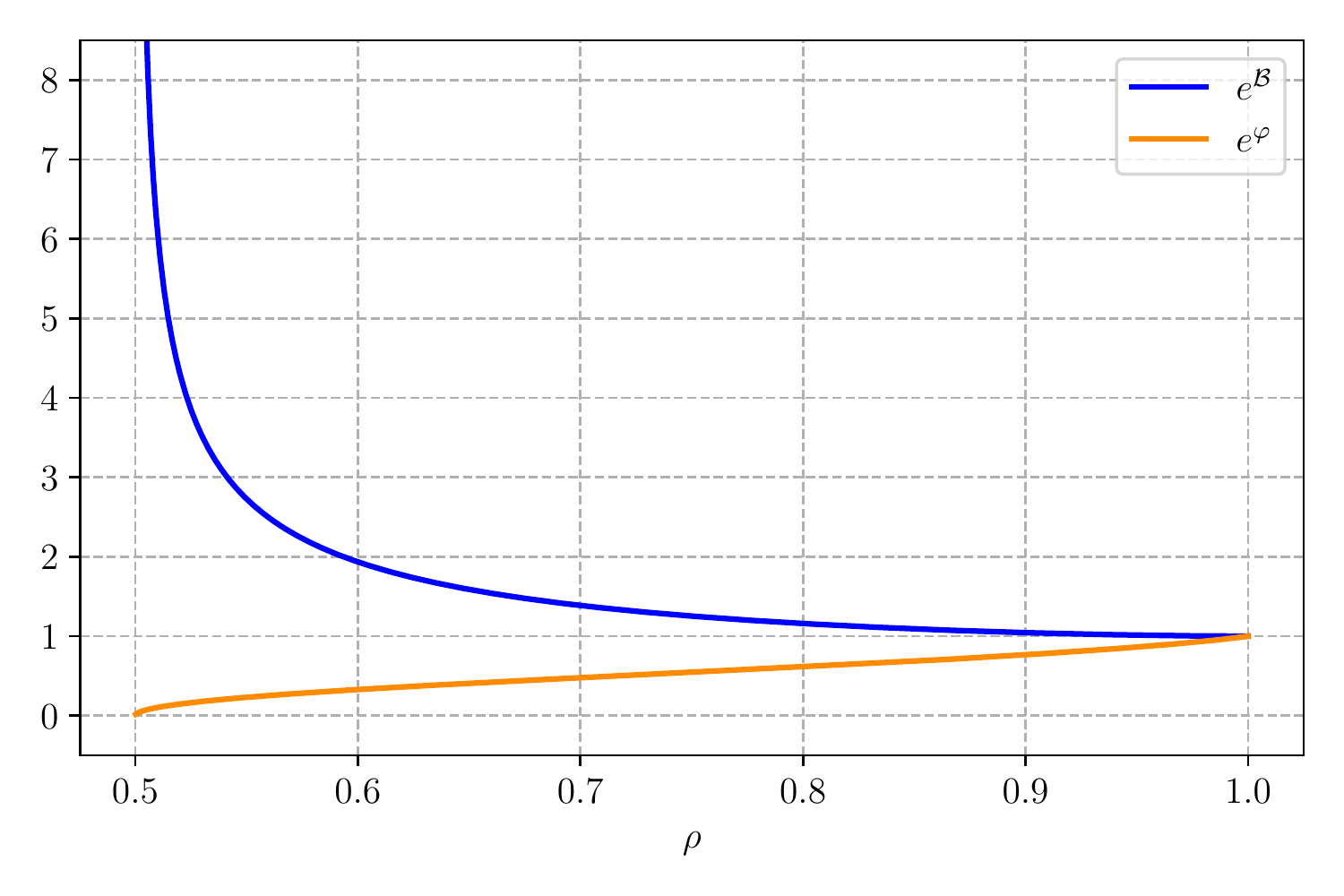}
\end{tabular}
\caption{ \small The upper--left panel displays the potential of eq.~\eqref{pot_35_2} for $C = 1$ and $D = 0$, so that $f = 1$, while the lower--left panel displays the inverted potential. The right panel displays $e^{\cB}$ and $e^{\vf}$ for $\alpha = 1$ and $\xi_0 = 2$. As expected, there is a divergence in $\rho  = \tilde \rho$, and in this particular case $\tilde \rho \simeq \frac{1}{2}$.}
\label{f1alphadiffzero2}
\end{figure}

\begin{figure}[ht]
\centering
\begin{tabular}{cc}
\includegraphics[width=45mm]{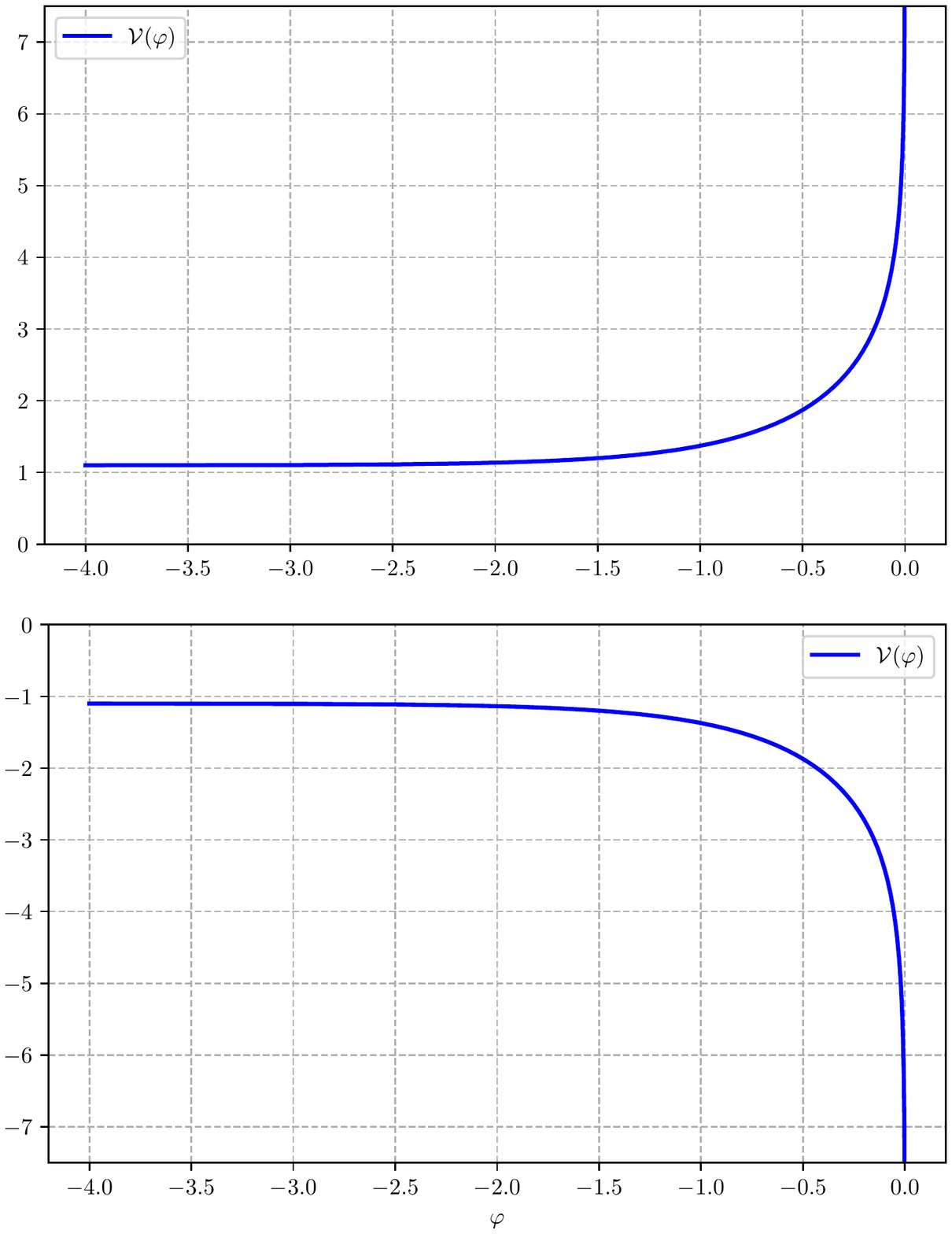} &
\includegraphics[width=0.57\textwidth]{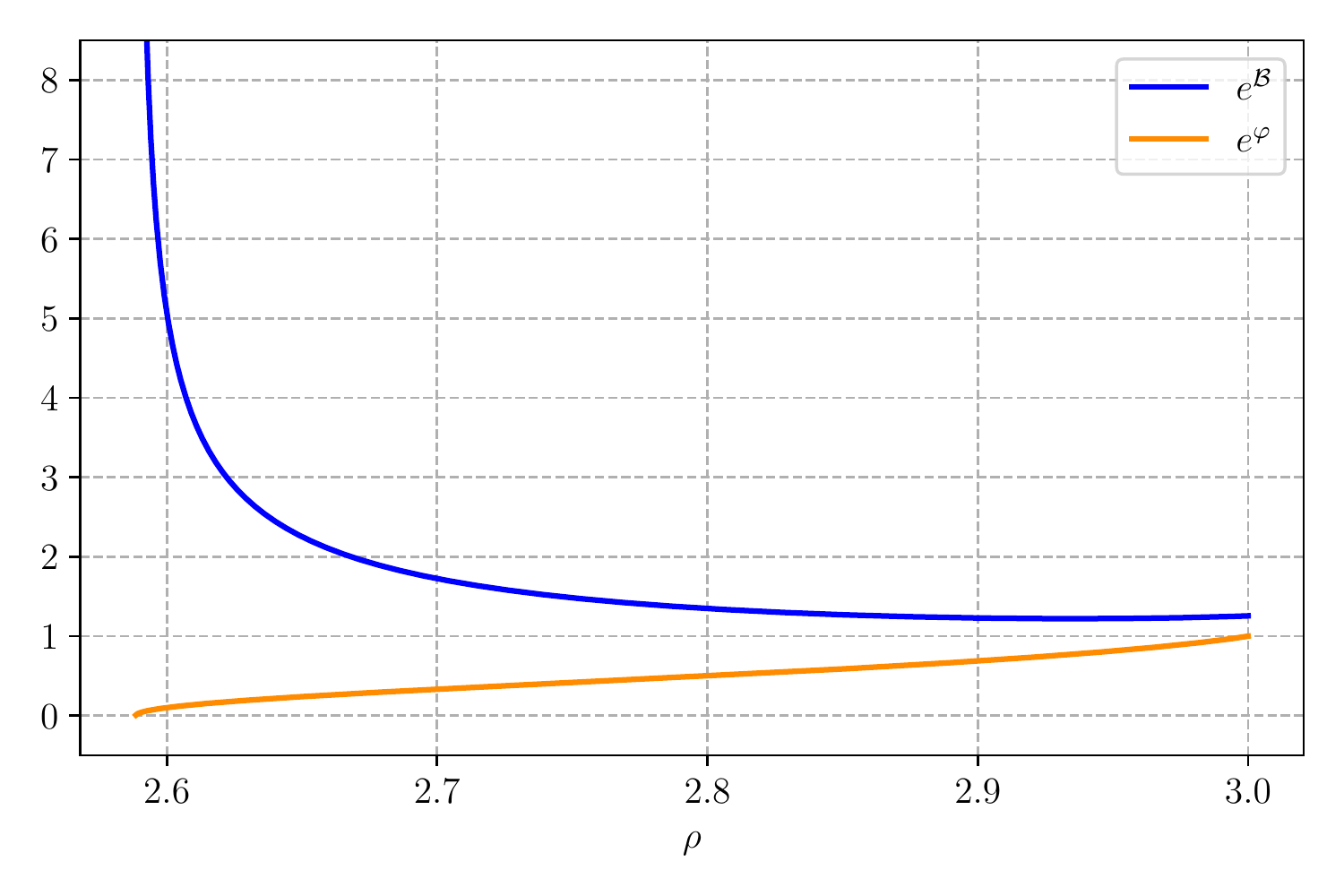}
\end{tabular}
\caption{ \small The upper--left panel displays the potential of eq.~\eqref{pot_35_2} for $C = 1$ and $D = \log(3)$, so that $f = 3$, while the lower--left panel displays the inverted potential. The right panel displays $e^{\cB}$ and $e^{\vf}$ for $\alpha = \frac{5}{2}$ and $\xi_0 = 2$, so that $ f - 1< \alpha  < f+1$ and $f> 1$.}
\label{fmagg1alphadiffzero1}
\end{figure}

The other cases, \framebox{$f =  1$ , $\alpha \neq 0$} and \framebox{$f > 1$, $f -1 < |\alpha| < f +1 $}, can be treated in a similar fashion, and the reader can verify that they also result in a bounded string coupling, an internal interval of finite length and finite values for the reduced Planck mass and gauge coupling. Once more, however, the scalar curvature is singular at the ends of the range of $r$. Figs.~\ref{f1alphadiffzero2} and~\ref{fmagg1alphadiffzero1} display examples of these types.

\noindent It is also interesting to analyze the case $C < 0$, so that the potential of eq.~\eqref{pot_35} is unbounded from below. The equations of motion~\eqref{logcothcharges} become
\begin{equation}
	\dot \xi^2  \ = \  \log \left( \frac{\xi}{\xi_0} \right) \ , \qquad \qquad \dot \eta^2  \ = \   \log \left( \frac{\eta}{\eta_0} \right) \ ,
\label{logcothcharges2}
\end{equation}
and the Hamiltonian constraint demands again that
\begin{equation}
    f \ \equiv \  e^{\frac{D}{C}} \ = \ \frac{\eta_0}{\xi_0} \ .
\end{equation}
In this case then, solving eq.~\eqref{logcothcharges2} and returning to the original variables,
\begin{align}
    e^{\mathcal A} \, = & \; \frac{\xi_0}{2} \ \sqrt{ 1 - f^2 \exp \Bigg\{  2 \bigg[-i \, \erf^{-1} \left( \displaystyle i \, \frac{\rho}{f} \right) \bigg]^2 - 2 \bigg[ - i \, \erf^{-1} \big(i(\rho - \alpha)\big)\bigg]^2   \Bigg\}} \; \; \times \nonumber  \\
	& \qquad \qquad \qquad \qquad \qquad \qquad \qquad \qquad \qquad \exp\left\{\left[ - i \,  \erf^{-1}\big(i(\rho - \alpha)\big) \right]^2 \right\} \ , \nonumber \\
	e^{\vf} \, = & \; \sqrt{\frac{1 - f \exp \Bigg\{  \bigg[-i \, \erf^{-1} \left( \displaystyle i \, \frac{\rho}{f} \right)\bigg]^2 - \bigg[ - i \, \erf^{-1} \big(i(\rho - \alpha)\big)\bigg]^2  \Bigg\}}{1 + f \exp \Bigg\{  \bigg[-i \, \erf^{-1} \left( \displaystyle i \, \frac{\rho}{f} \right) \bigg]^2 - \bigg[ - i \, \erf^{-1} \big(i(\rho - \alpha)\big)\bigg]^2   \Bigg\}}} \ , \nonumber \\
	\mathcal B \, = & \;  -\mathcal A \ .
\end{align}
In general, the error function $\erf(z)$ has no single--valued inverse in the complex plane, but here we are confined to the imaginary axis and $\erf(iy)$, with $y \in \mathbb R$, is invertible. Moreover, one can show that, in the large--$x$ limit, the behavior of the inverse error function is well captured by the simple expression
\begin{equation}
    -i \, \erf^{-1}(\, i \, x\, ) \ \sim \ \sqrt{\log(\sqrt{\pi} \, x)} \ + \ \frac{1}{4} \, \frac{\log(\log(\sqrt{\pi} \, x))}{\sqrt{\log(\sqrt{\pi} \, x)}} \ .
\end{equation}
Consequently, for large values of $\rho$,
\begin{equation}
    e^{\vf} \ \sim \ \sqrt{\frac{1 -  \sqrt{1 - \frac{\log(f)}{\log{(\sqrt{\pi} \rho)}} }}{2}}
\end{equation}
and
\begin{equation}
    e^{\cA} \ \sim \ \sqrt{\pi} \, \rho \  \sqrt{\frac{\log(f)}{\log{(\sqrt{\pi} \rho)}}} \ .
\end{equation}
From these behaviors one can deduce the following results.
\begin{itemize}
    \item For $f > 1$ $e^{\vf}$ vanishes for large values of $\rho$, but rather slowly, since
\begin{equation}
    e^{\vf} \ \sim \ \frac{1}{2} \ \sqrt{\frac{\log(f)}{\log(\sqrt{\pi} \rho)}}\ .
\end{equation}
Moreover $e^{\cA}$ has essentially a linear behavior, so that for large values of $\rho$
\begin{equation}
    e^{\cB} \ \sim \ \frac{1}{\rho} \ .
\end{equation}
These results imply that the internal $r$--direction is not compact.
    \item For $f < 1$ the argument of the square root in $e^{\vf}$ becomes negative for large--enough values of $\rho$, and therefore $e^{\vf}$ is defined only on an interval, in which it is always bounded. At the boundaries of this interval both $e^{\vf}$ and $e^{\cA}$ vanish with a square--root behavior. Consequently, $e^{\cB}$ has merely an inverse square--root divergence. As a result, this solution also combines a bonded string coupling, a compact $r$--direction, finite reduced 9D Planck mass and gauge coupling, but the string--frame scalar curvature~\eqref{string_curvature} is again unbounded, according to eq.~\eqref{string_curvature}.
    \item  For $f = 1$, the nature of the solution depends crucially on $\alpha$. For $\alpha > 0$, the solution is defined for $\rho <  \rho^*$, where the arguments of the square roots change sign, but the solutions in this class also combine a bounded string coupling, a compact $r$--direction and finite reduced 9D Planck mass and gauge coupling, while the string--frame scalar curvature is once more unbounded.
\end{itemize}

\noindent {\sc 6). The Potential of eq.~\eqref{pot_36}}

The next potential is
\begin{equation}
	\cV(\vf) \ = \  C \cosh(\vf) + \Lambda \ .
\end{equation}
The equations of motion~\eqref{EOM_five_potential} can be directly integrated, but the nature of the resulting solutions depends strongly on the values of  $\Lambda$ and $C$. Therefore, it is now convenient to let
\begin{equation}
	\lambda_1 \ = \  \sqrt{\left| \frac{\Lambda + C}{2}   \right|} \ , \qquad \qquad \lambda_2 \ = \  \sqrt{\left| \frac{\Lambda - C}{2}   \right|} \ ,
\end{equation}
while also considering separately the following ranges:
\begin{enumerate}
	\item \qquad $\Lambda \ >  \  |C|$ ,
	\item \qquad $\Lambda \ =  \  C$ ,
	\item \qquad $|C| \ >  \  \Lambda \ >\  -|C|$ ,
	\item \qquad $\Lambda \ = \  - |C|$ ,
	\item \qquad $\Lambda \ <  \  -|C|$ .
\end{enumerate}

As an illustration, we discuss explicitly the first two, which are more interesting.

\noindent \framebox{$1. \; \; \Lambda \ > \ |C|$}

\begin{figure}[ht]
\begin{tabular}{cc}
\includegraphics[width=45mm]{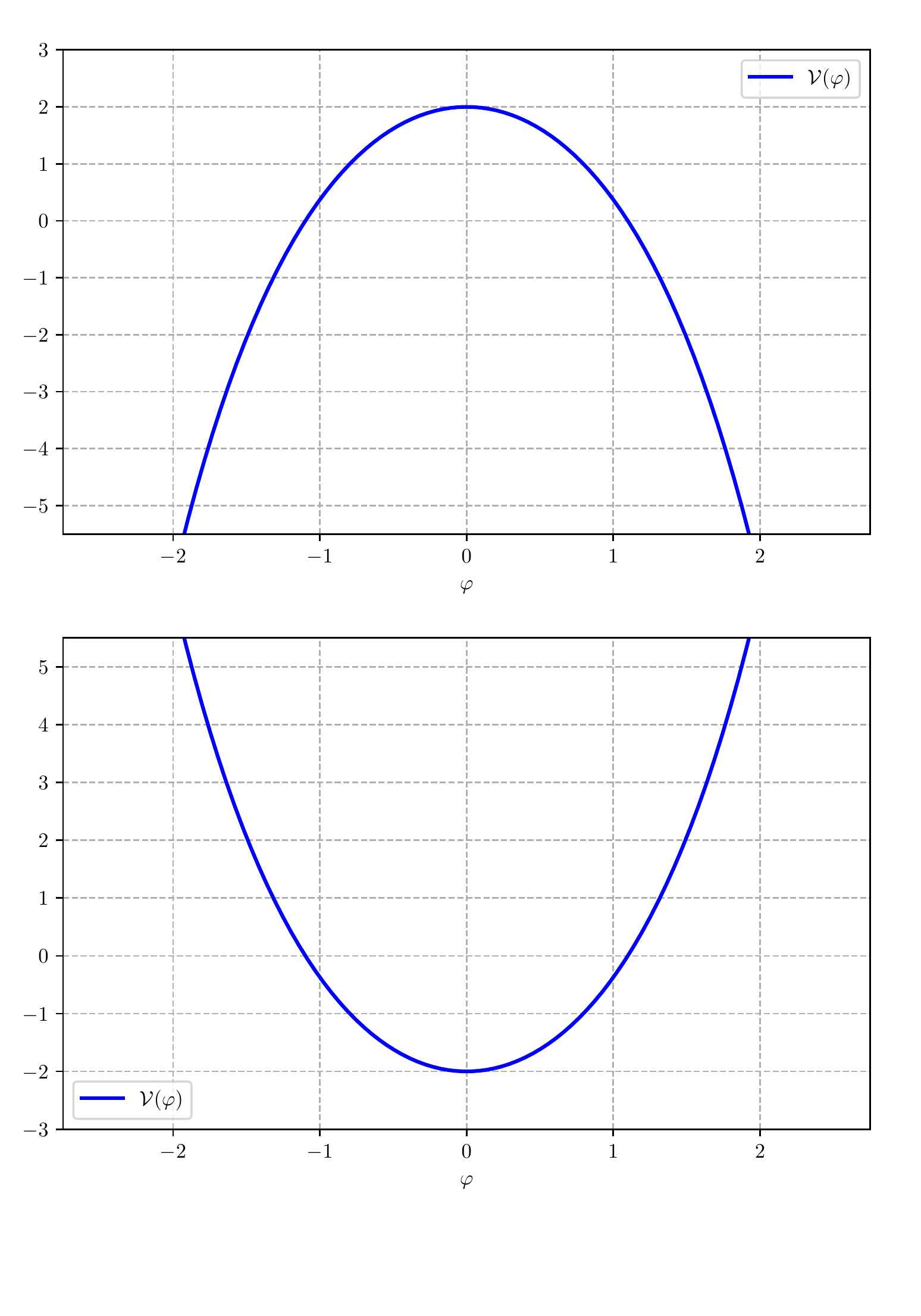} &
\includegraphics[width=0.57\textwidth]{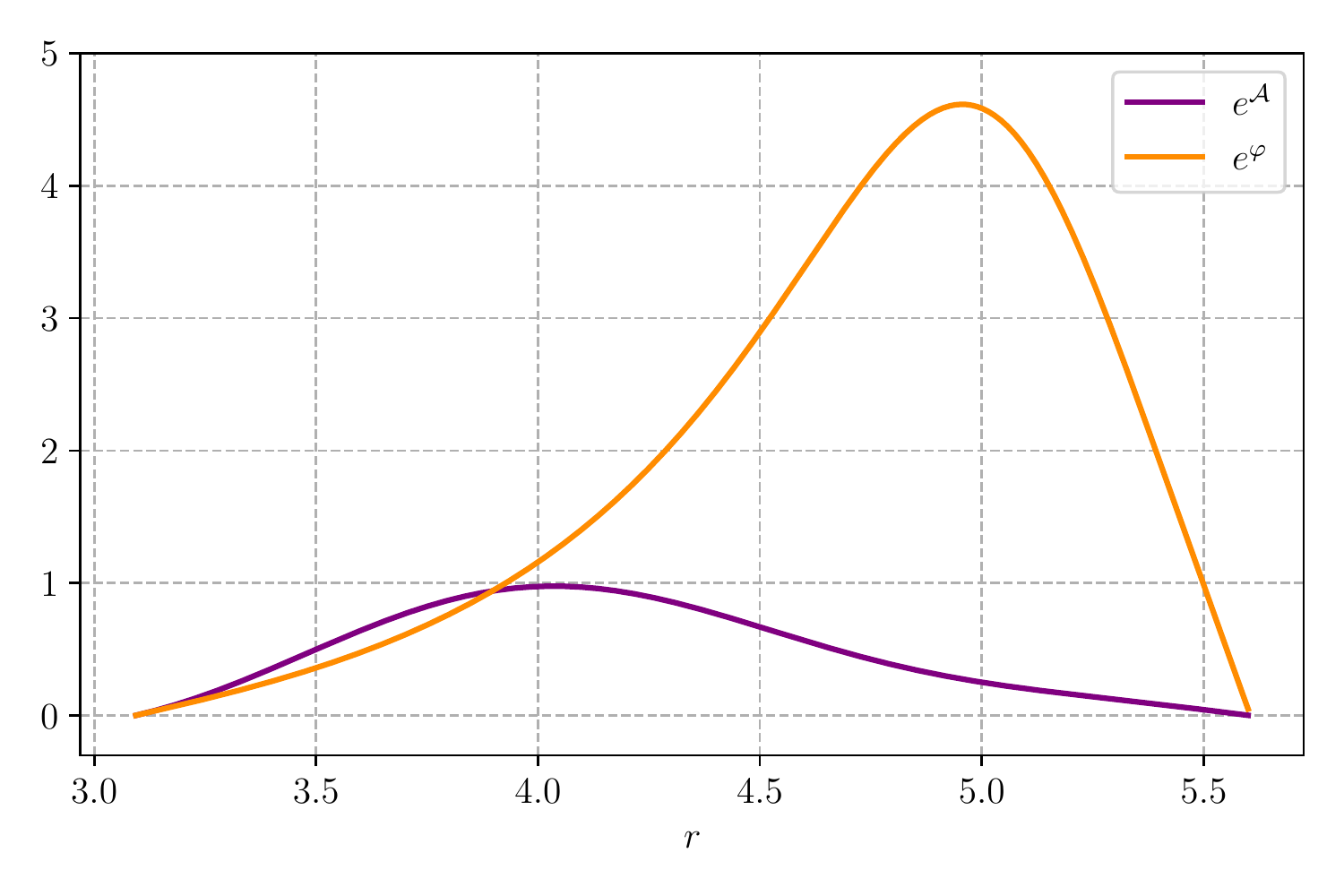}
\end{tabular}
\centering
\caption{ \small  The upper--left panel displays the potential of eq.~\eqref{pot_36} for $C = -3$ and $\Lambda = 5$, while the lower--left panel displays the inverted potential. In this case $\lambda_1 = 1$ and $\lambda_2 = 2$, so that $\lambda_2 > \lambda_1$. The right panel displays $e^{\cA}$ and $e^{\vf}$ ($\cB = 0$ for the gauge choice) for $r_\xi = 1$ and $r_\eta = 0$, in an example where the string coupling is bounded and the space is compact.}
\label{pot_6_sol_1}
\end{figure}

\noindent When both eigenvalues in eq.~\eqref{EOM_five_potential} are positive, the solution is
\begin{equation}
	\xi \ = \  \alpha \cos(\lambda_1 (r - r_{\xi}))  \ , \qquad \qquad \eta \ = \  \beta \cos(\lambda_2 (r - r_{\eta})) \ ,
\end{equation}
where $\alpha$, $\beta$, $r_{\xi}$ and $r_{\eta}$ are integration constants. The Hamiltonian constraint demands that
$\lambda_1^2\ \alpha^2 =  \lambda_2^2\ \beta^2$, and for simplicity we shall let $\alpha = \lambda_2$ and $\beta = \lambda_1$. In this case the solution is
\begin{align}
	e^{\mathcal A}  \ &= \    \frac{1}{4} \left(\lambda_2^2 \cos^2(\lambda_1 (r - r_{\xi})) - \lambda_1^2 \cos^2(\lambda_2 (r - r_{\eta})) \right) \ , \nonumber \\
	e^{\vf}  \ &= \    \frac{\lambda_2 \cos(\lambda_1 (r - r_{\xi})) + \lambda_1 \cos(\lambda_2 (r - r_{\eta}))}{\lambda_2 \cos(\lambda_1 (r - r_{\xi})) - \lambda_1 \cos(\lambda_2 (r - r_{\eta}))}  \ .
\end{align}
If $C > 0$, $\lambda_1 > \lambda_2$ and the solutions always have a strong coupling phase, since the potential is always positive. On the other hand, if $C< 0$ and thus $\lambda_2 > \lambda_1$, the considerations presented in Section~\ref{sec:compactness} imply the existence of solutions with a bounded string coupling, which are the counterparts of the cosmological climbing scalar for $\gamma<1$. Fig.~\ref{pot_6_sol_1} displays an example of this type, in which $\lambda_2 = 2$ and $\lambda_1 = 1$. However,  solutions with an unbounded string coupling also exist: they are the counterparts of the cosmological descending scalar present for $\gamma<1$.

\noindent \framebox{$2. \; \; \Lambda\  = \ C$}

\noindent In this case $\lambda_2 = 0$, and if $C > 0$ the solution, up to a rescaling and a shift of $r$, can be cast in the form
\begin{equation}
	\xi \ = \  \alpha \cos(\lambda_1 (r - r_{\xi}))  \ , \qquad \qquad  \eta \ = \  \lambda_1 r \ .
\end{equation}
The Hamiltonian constraint demands that $\alpha^2=1$, and setting for simplicity $\alpha = 1$,
\begin{equation}
	e^{\mathcal A} \ = \   \frac{1}{4} \left(\cos^2(\lambda_1 (r - r_{\xi}) -  \lambda_1^2 r^2 \right) \ , \qquad \qquad  e^{\vf} \ = \   \frac{ \cos(\lambda_1 (r - r_{\xi})) + \lambda_1 r}{\cos(\lambda_1 (r - r_{\xi})) -  \lambda_1 r} \ .
\end{equation}
However, there is surely a point where the denominator of $e^{\vf}$ vanishes and the string coupling diverges. On the other hand, if $C < 0$ the solution is
\begin{equation}
	\xi \ = \  \alpha \ \sinh(\lambda_1\,r) \ +\  \beta \cosh(\lambda_1\, r)  \ , \qquad \qquad  \eta \ = \  \lambda_1 r \ ,
\end{equation}
again up to a rescaling and a shift of $r$. The Hamiltonian constraint now demands that $\alpha^2 - \beta^2 =1$, which can be conveniently solved in terms of a rapidity $\zeta$
\begin{equation}
    \alpha \ = \  \cosh(\zeta) \ , \qquad \qquad \beta \ = \ \sinh(\zeta) \ ,
\end{equation}
so that
\begin{equation}
	\xi \ = \  \sinh(\lambda_1\,r + \zeta) \ .
\end{equation}
The complete solution in this case is therefore
\begin{equation}
	e^{\mathcal A} \ = \   \frac{1}{4} \left(\sinh^2(\lambda_1\,r + \zeta) -  \lambda_1^2 r^2 \right) \ , \qquad \qquad  e^{\vf} \ = \   \frac{ \sinh(\lambda_1 r + \zeta) + \lambda_1 r}{ \sinh(\lambda_1 r + \zeta) - \lambda_1 r} \ .
\end{equation}
For any choice of $\zeta$, the denominator of $e^{\vf}$ has a zero, where string coupling diverges. Similar considerations apply to all the other cases, so that this class of models is not particularly interesting for our purposes.

\noindent {\sc 7). The Potential of eq.~\eqref{pot_37}}

\begin{figure}[ht]
\centering
\begin{tabular}{cc}
\includegraphics[width=40mm]{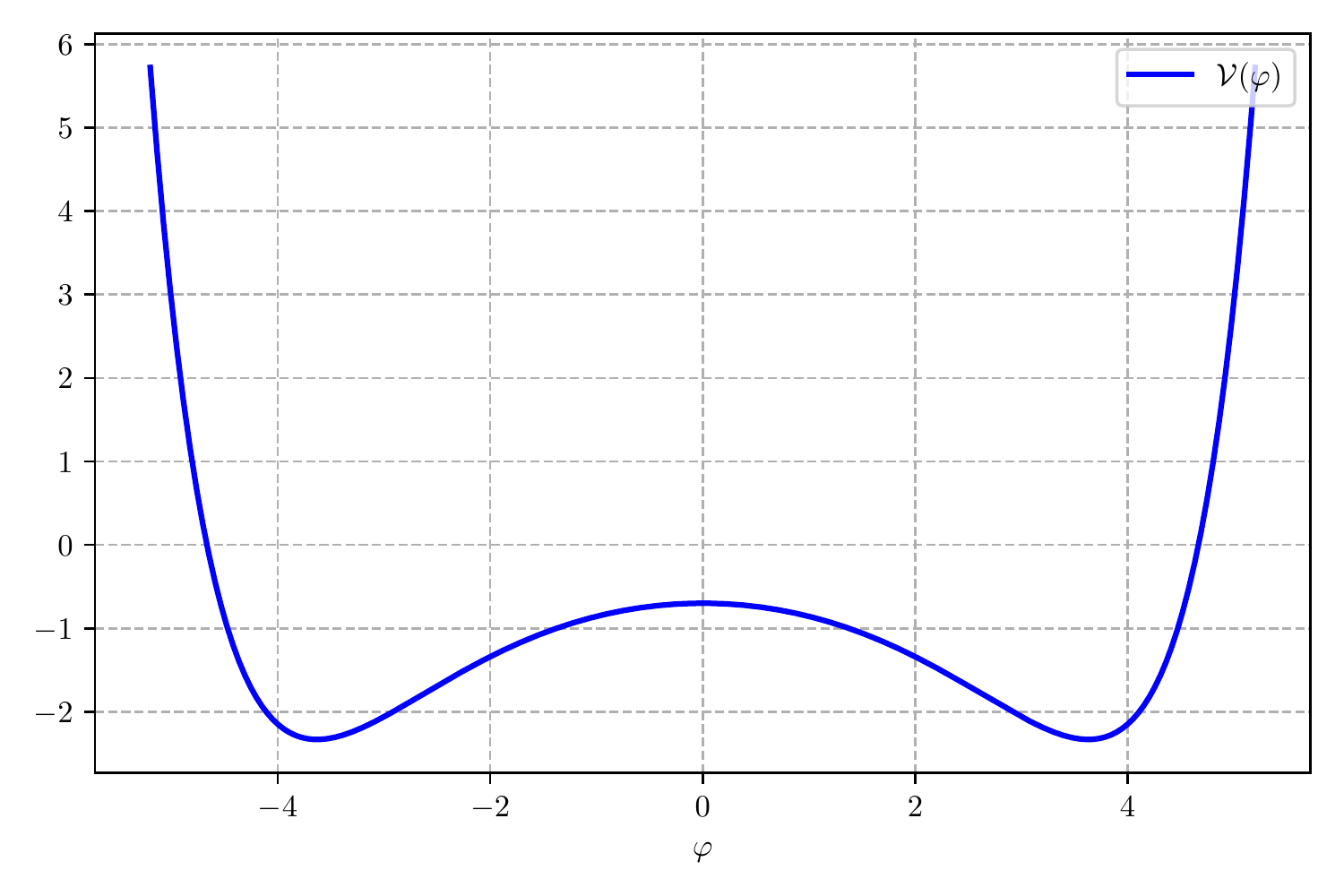} &
\includegraphics[width=40mm]{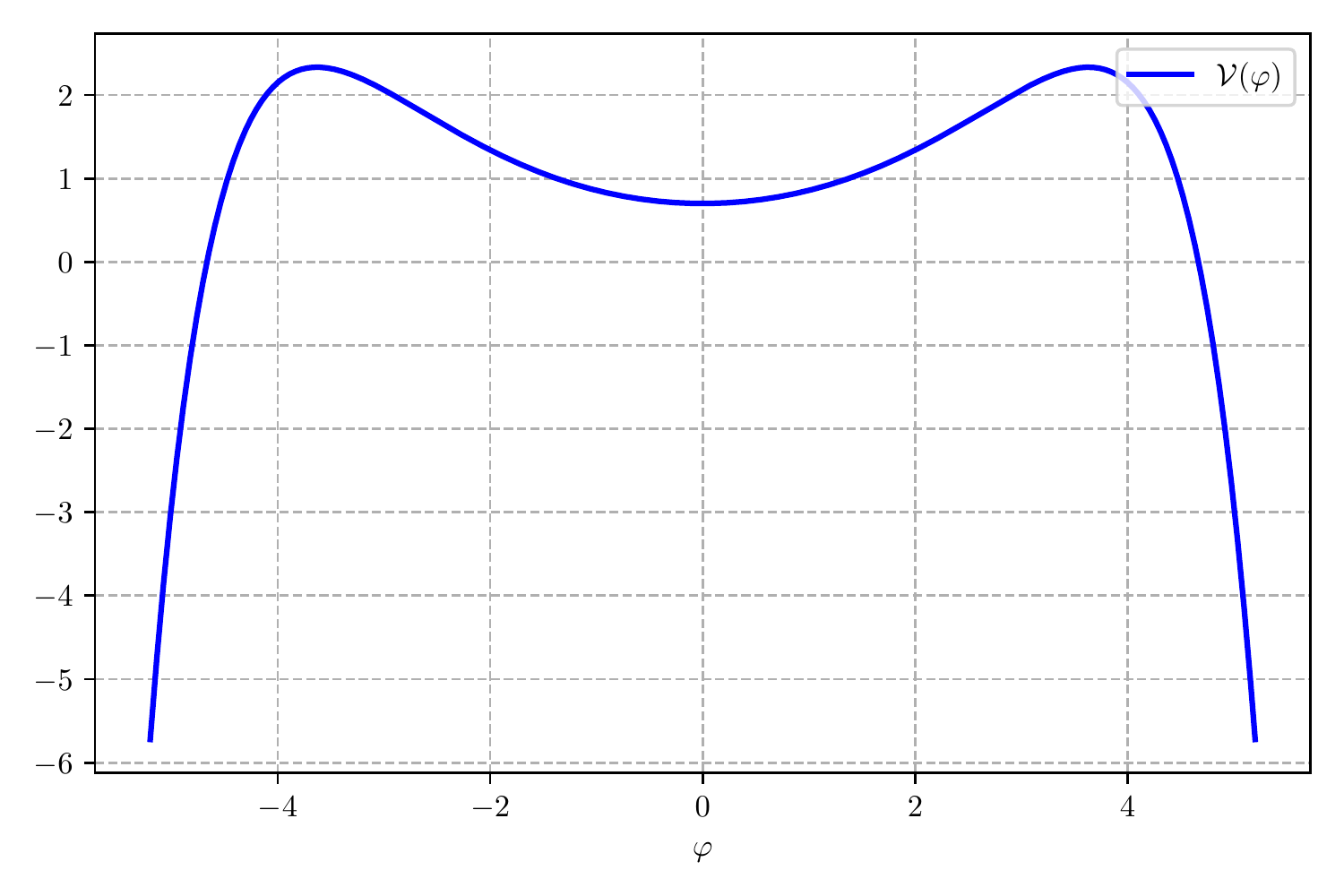} \\
\includegraphics[width=0.4\textwidth]{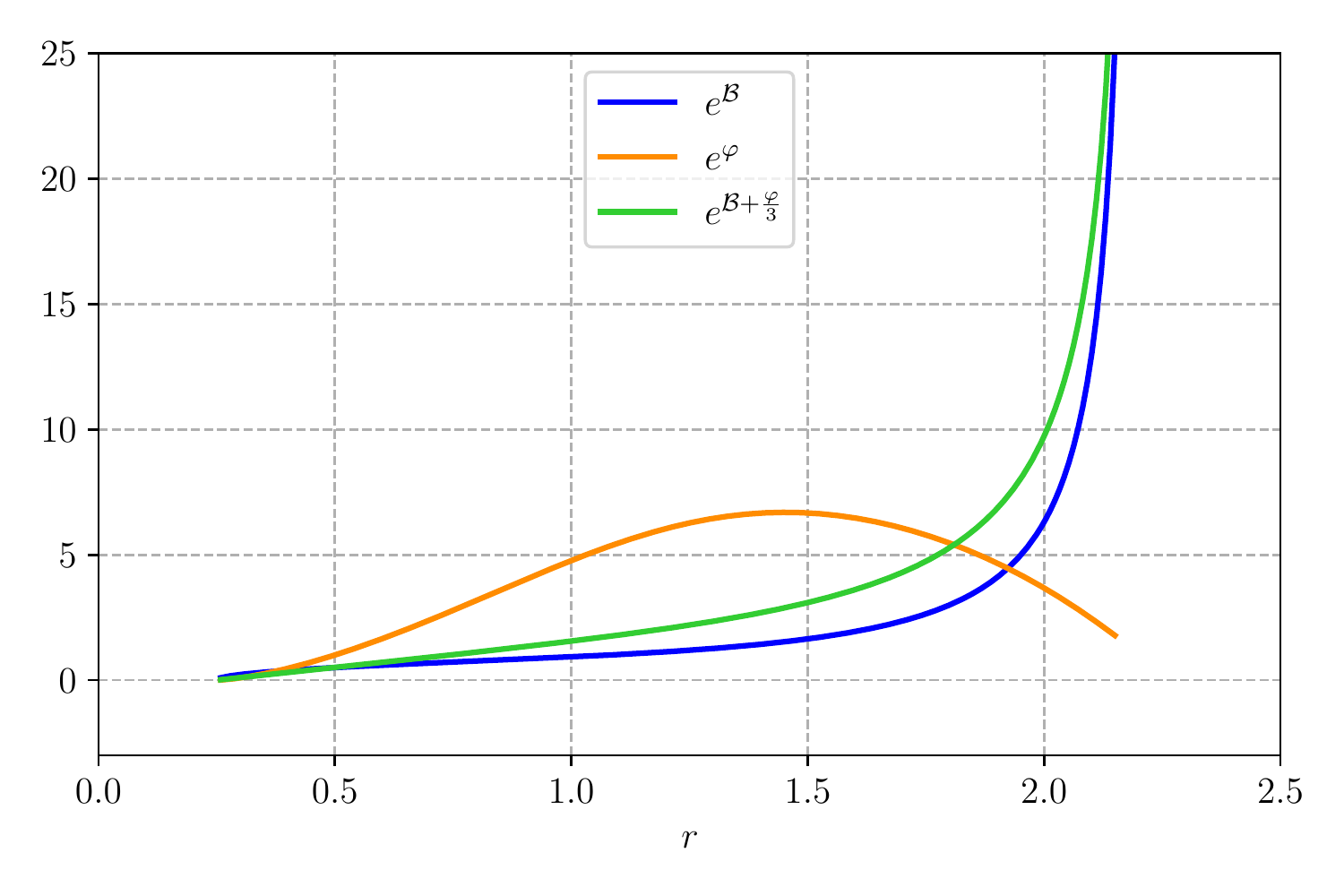} &
\includegraphics[width=0.4\textwidth]{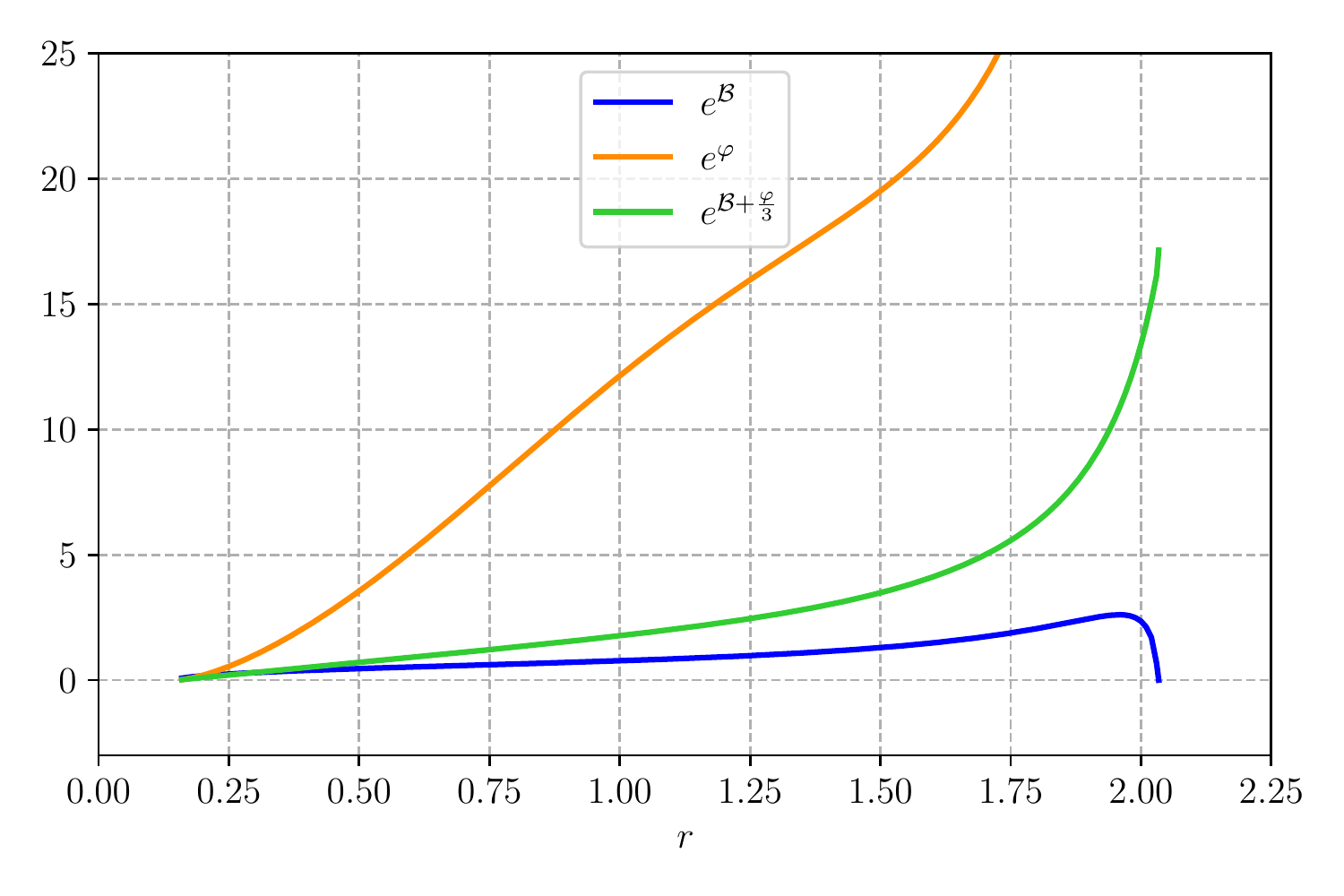} \\
\includegraphics[width=0.4\textwidth]{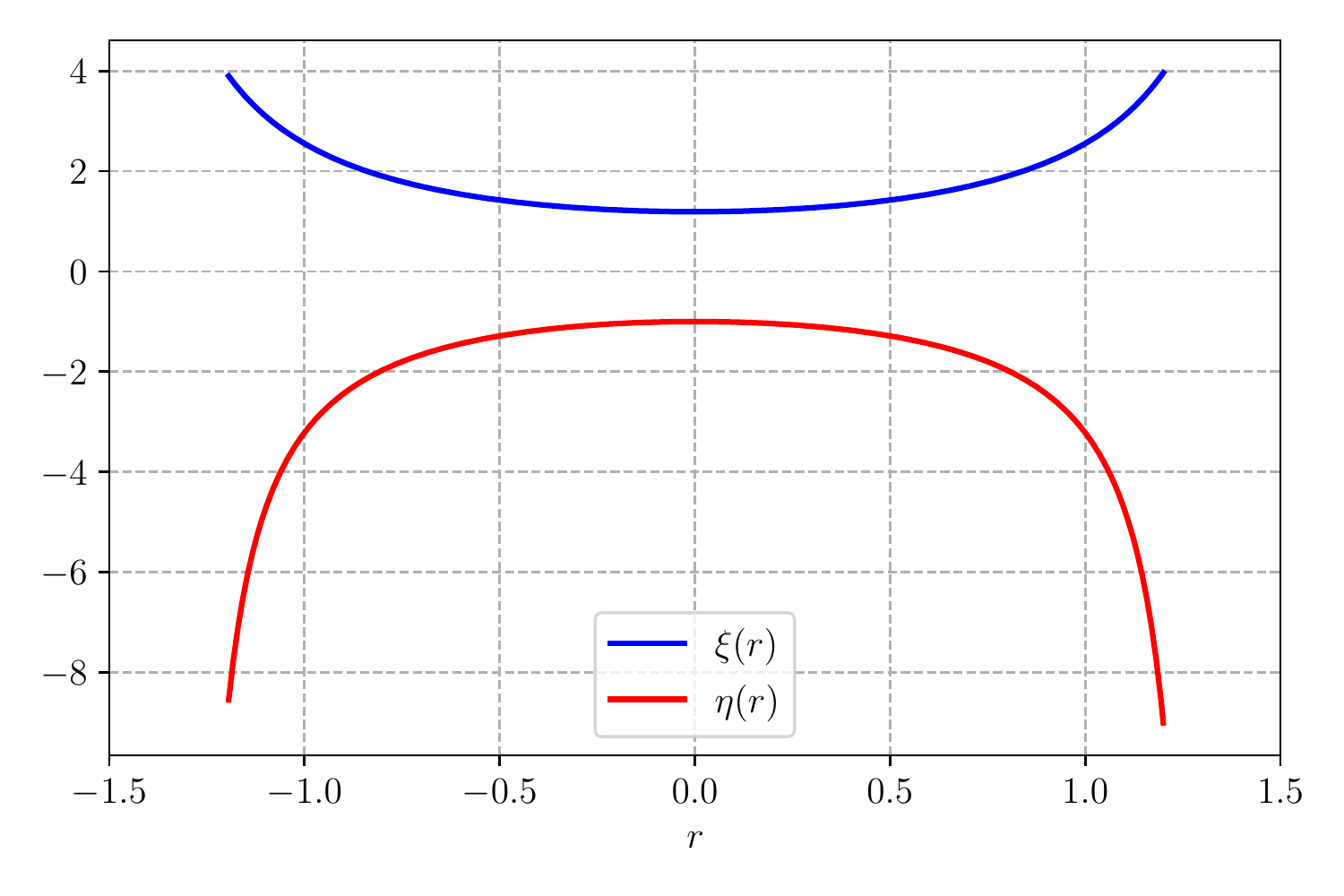} &
\includegraphics[width=0.4\textwidth]{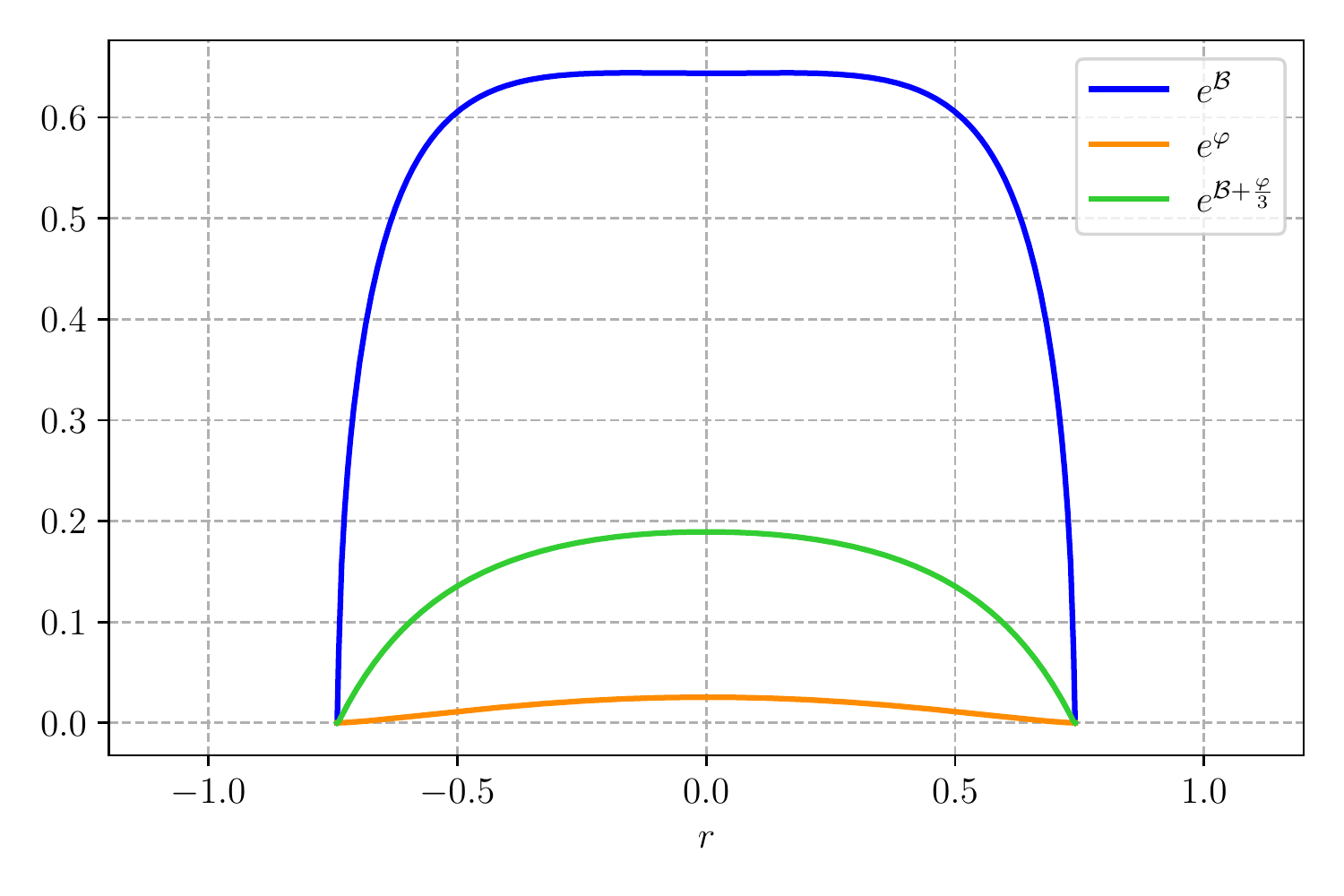} \\
\end{tabular}
\caption{ \small The upper panels display the potential of eq.~\eqref{pot_37_2} for $C_1 = - \frac{7}{10}$ and $C_2 = 1$ and the corresponding inverted potential. The two middle panels display solutions of eq.~\eqref{quadrat_seventh} with $\ve = 0.5$ and positive energy, but with different integration constants. In the first case the dilaton is ``trapped'' in the well of the inverted potential, while in the second case it overcomes the inverted potential. The two lower panels refer to a dilaton that does not overcome the inverted potential, but is confined to the left of it. In this case $\xi > \eta$, which is particularly interesting since the string coupling is then inevitably bounded. Moreover, the internal space has a finite length.  }
\label{cosh4sinh4_2}
\end{figure}

\noindent The next potential,
\begin{equation}
	\cV(\vf) \ = \  C_1 \cosh^4\left(\frac{\vf}{3}\right) \ + \ C_2 \sinh^4\left(\frac{\vf}{3}\right)  \ ,
\label{pot_37_2}
\end{equation}
is more interesting, and the solutions are particularly rich if $C_1<0$ and $C_2>0$ and $|C_1|<C_2$. The solution of eqs.~(\ref{EOM_six_potential}) can be expressed in terms of Jacobi elliptic functions, but the qualitative behavior is well captured recasting the system in the form
\beq
{\dot \xi}^2 \ = \ \pm 1\ + \ \ve \, \xi^4 \ , \qquad {\dot \eta}^2 \ = \ \pm 1\ + \ \eta^4 \ ,
\label{quadrat_seventh}
\eeq
up to a rescaling of the radial variable, where $0 < \varepsilon < 1$. In the Newtonian analogy, the upper signs correspond to particles whose positive total energy exceeds the peak value of the inverted quartic potentials, while the negative signs correspond to particles with negative total energy, which are reflected by them. The solutions $\xi(r)$ and $\eta(r)$ determine the original variables according to
\beq
    e^{\cA} \ = \ \left[ \xi^2 \ - \ \eta^2 \right]^{\frac{3}{2}} \ , \qquad e^{\vf} \ = \ \left[ \frac{\xi \ + \ \eta}{\xi \ - \ \eta} \right]^{\frac{3}{2}} \ , \qquad e^{\cB} \ = \ \left[ \xi^2 \ - \ \eta^2 \right]^{\frac{1}{2}} \ ,
\eeq
and $\xi$ and $\eta$ have simple poles. Three types of qualitative behavior can emerge at the ends of the range of $r$, which manifest themselves in the panels of fig.~\ref{cosh4sinh4_2}:
\begin{itemize}
    \item both $e^{\vf}$ and $e^{\cB}$ vanish, and this happens when $\xi +\eta = 0$ ;
    \item $e^{\cB}$ diverges and $e^{\vf}$ approaches a constant value, but $e^{\cB + \frac{\vf}{3}}$ has a simple pole there and thus the space is not compact;
    \item $e^{\cB}$ vanishes and $e^{\vf}$ goes to strong coupling, when $\xi - \eta = 0$.
\end{itemize}

These behaviors can also be understood from the Newtonian ``inverse potential'' picture, because different scenarios are possible:
\begin{itemize}
    \item the dilaton manages to climb the inverse potential barrier and attains arbitrarily large positive values, which translates into the presence of a strong coupling phase for $g_s$;
    \item the dilaton is trapped in the inner well of the inverted potential, and actually approaches its bottom, so that the space is asymptotically $AdS_{10}$;
    \item the dilaton does not manage to climb the inverse potential barrier, which reflects it. This, however, can occur in two different ways: if the dilaton lives to the left of the inverted barrier one finds a compact internal space and a bounded string coupling. On the other hand, if it lives to the right of the inverted barrier one finds again a compact internal space, but an unbounded string coupling.
\end{itemize}

For brevity, we have not considered a negative potential, but the corresponding behavior can be anticipated by the analysis of Section~\ref{sec:compactness}, which implies the existence of two classes of solutions with bounded and unbounded string coupling, since the dominant contribution to the potential for $\vf \to \infty$ is an exponential term with $\gamma<1$.

\noindent {\sc 8). The Potential of eq.~\eqref{pot_38}}

\noindent The following potential is
\begin{equation}
	\cV (\vf)  \ = \   \Im \left[  C \log \left( \frac{e^{-2 \vf} + i}{e^{-2 \vf} - i} \right)  + i \Lambda \right] \ ,
\end{equation}
and since the $\log$ is purely imaginary $C$ can be assumed to be real.
The corresponding equation of motion can be turned into
\begin{equation}
	\dot z^2 \ = \  - \, 8 \, C \, \log(z) \ + \ E \ + \ i \, F \ ,
\end{equation}
where $z$ was defined in eq.~\eqref{eight_z} and $E$ and $F$ are two real integration constants. The Hamiltonian constraint demands that
\begin{equation}
	\Im \Big[  E + iF + 4i \Lambda   \Big]  \ = \   0 \ ,
\end{equation}
and therefore $F = - 4 \Lambda$ . Consequently, one must solve
\begin{equation}
	\dot z^2 \ = \  - \, 8 \, C \, \log(z) \ + \ E \ - \ 4 \, i \, \Lambda
\end{equation}
in the complex plane, which leads, along the lines of previous examples, to
\begin{equation}
	\int_{w_0}^{\sqrt{- 8 C \log(z) + E - 4 i \Lambda}} e^{- \frac{1}{8C} w^2} \; \td w \ = \  4C\, e^{\frac{i \Lambda}{2C}} e^{-\frac{E}{8C}} r \ .
\end{equation}
As a result, The solution can be expressed in terms of the error function in the complex plane,
\begin{equation}
	\erf \left[ \sqrt{- \log(z) + \frac{E - 4 i \Lambda}{8C}} \right] \ = \   \frac{2}{\sqrt{\pi}} \, e^{-\frac{E}{8C}} \, e^{\frac{i \Lambda}{2C}} (4C \, r + i R) \ ,
\label{z_erf}
\end{equation}
where $i R$ is an imaginary integration constant, up to a shift of $r$ that can absorb the corresponding real part. $z$ is obtained inverting this relation, and the result can be cast in the form
\begin{equation}
	z \ = \  e^{\frac{E - 4 i \Lambda}{8C}} \exp \left\{ - \left[ \erf^{-1} \left( \frac{2}{\sqrt{\pi}} e^{-\frac{E- 4i\Lambda}{4C}} (4C \, r + i R) \right) \right]^2 \right\} \ .
\label{z_erfinv}
\end{equation}

This solution would require some discussion, since the function $\erf^{-1}$ is not single--valued in the complex plane. However, this example concerns a potential that is essentially piece-wise constant, since
\beq
{\cal V}(\varphi) \ \simeq \ \Lambda \ + \ C\, \theta(\varphi) \ ,
\eeq
where $\theta$ is a Heaviside step function, and one can obtain simple approximate solutions that apply in regions where the dilaton has a given sign.

Let us therefore consider a constant potential equal to a generic value $V_0$. In this case, the equation of motion for $\vf$ of eq.~\eqref{staticequations} leads to
\begin{equation}
    \dot \vf \ = \ \k \, e^{- 2 \cA} \ ,
\label{vf_and_A}
\end{equation}
where $\k$ is an integration constant, which we shall take to be non negative. In the gauge $\cB = - \cA$, the Hamiltonian constraint then leads to
\begin{equation}
    \k^2 \, e^{-4\cA} \ - \ 2\, e^{-2\cA} \, V_0 \ = \ \dot \cA^2 \ .
\end{equation}
The solution of this equation is
\begin{equation}
    e^{\cA} = \sqrt{\frac{\k^2}{2V_0} \ -\ 2\, V_0 \, r^2 }
\end{equation}
if $V_0 \neq 0$, and then
\begin{equation}
    e^{\vf} \ = \ \sqrt{\left| \frac{r \, + \, \frac{\k}{2\,V_0}}{r \, - \, \frac{\k}{2\,V_0}} \right|} \ .
\end{equation}
If $V_0>0$ this type of solution applies in the region $\left(-\,\frac{\k}{2 V_0},0\right)$, or in the region $\left(0,\frac{\k}{2 V_0}\right)$, where $\varphi$ has a given sign. In this case there is a strong--coupling region and the contribution to the internal length is finite. Alternatively, if $V_0<0$ this type of solution applies in the region $\left(\frac{\k}{2 |V_0|},+\infty\right)$, or in the region $\left(- \infty,-\frac{\k}{2 |V_0|}\right)$, where $\varphi$ has a given sign. In this case there is a strong--coupling region and the contribution to the internal length is infinite. The two simple approximate solutions are to be smoothly connected in the middle, but the singularities are already manifest. Finally, if $V_0=0$,
\begin{equation}
    e^{\cA} \ = \ \sqrt{2 \, \k \, r}
\end{equation}
where $r>0$, and then
\begin{equation}
    e^{\vf} \ = \ \sqrt{\left|\frac{r}{r_0}\right|} \ ,
\end{equation}
so that there is a region of strong coupling and the contribution to the internal length is infinite.
Finally, if $V_0 = 0$ and $\k = 0$, both $e^{\vf}$ and $e^{\cA}$ are constant, one recovers flat space with a constant string coupling.

\noindent {\sc 9). The Potential of eq.~\eqref{pot_39}}

\noindent Finally, let us consider the potential
\begin{equation}
	\cV (\vf) \ = \  \Im \left[ \,C\, \Big( i + \sinh (2 \gamma \vf) \Big)^{\frac{1}{\gamma} - 1} \right] \ ,
\label{pot_39_2}
\end{equation}

\begin{figure}[ht]
\centering
\begin{tabular}{cc}
\includegraphics[width=0.4\textwidth]{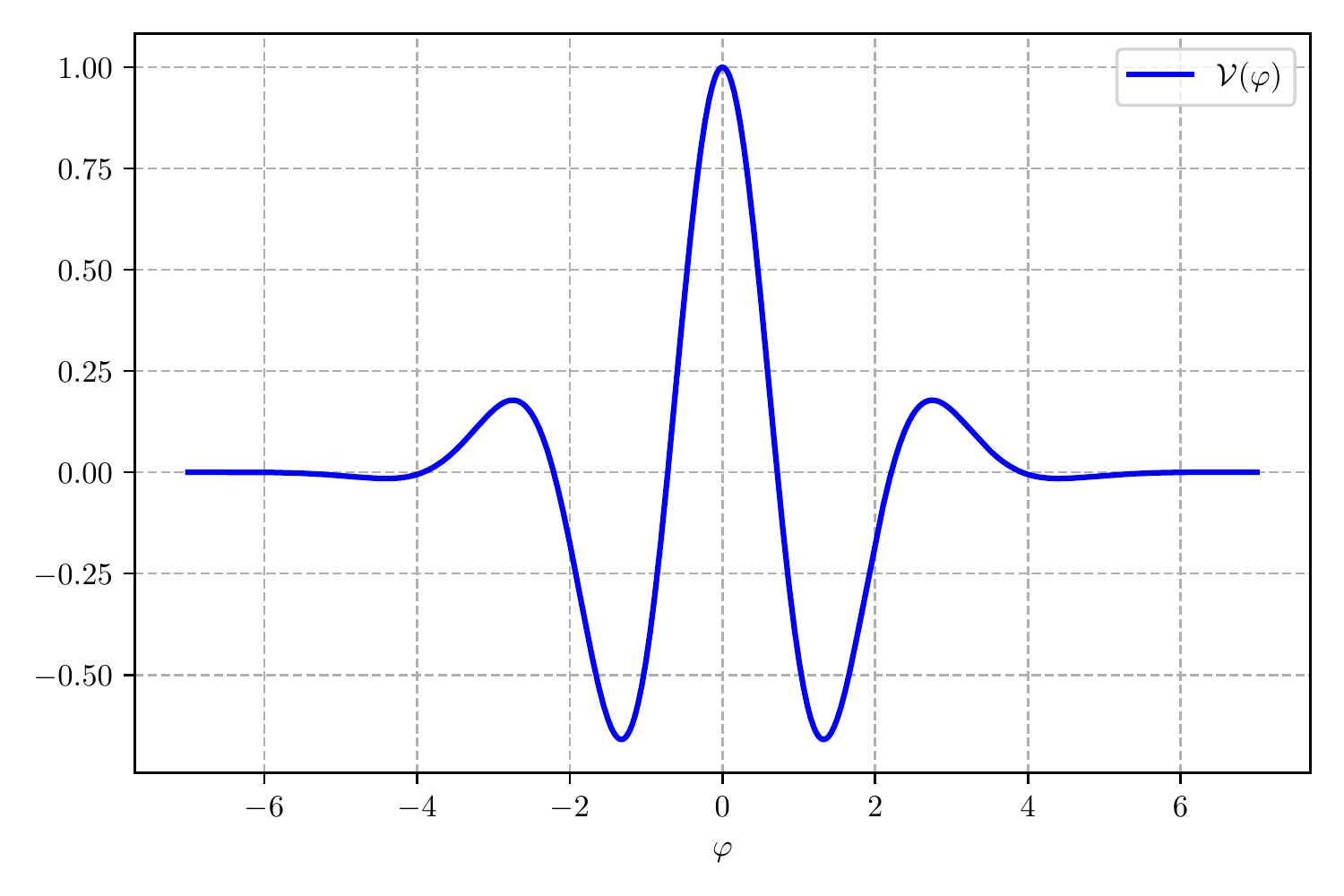} &
\includegraphics[width=0.4\textwidth]{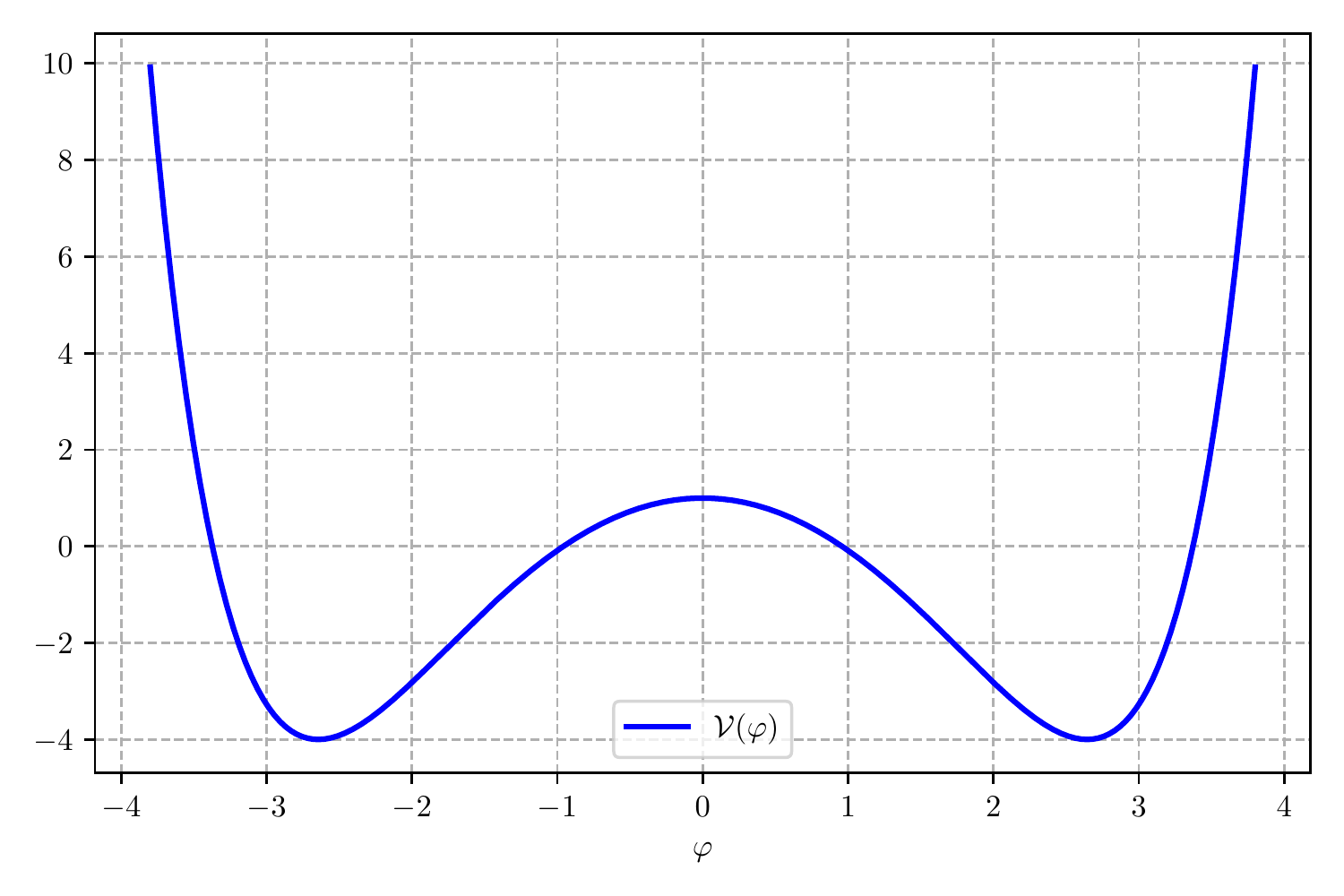} \\
\end{tabular}
\includegraphics[width=0.4\textwidth]{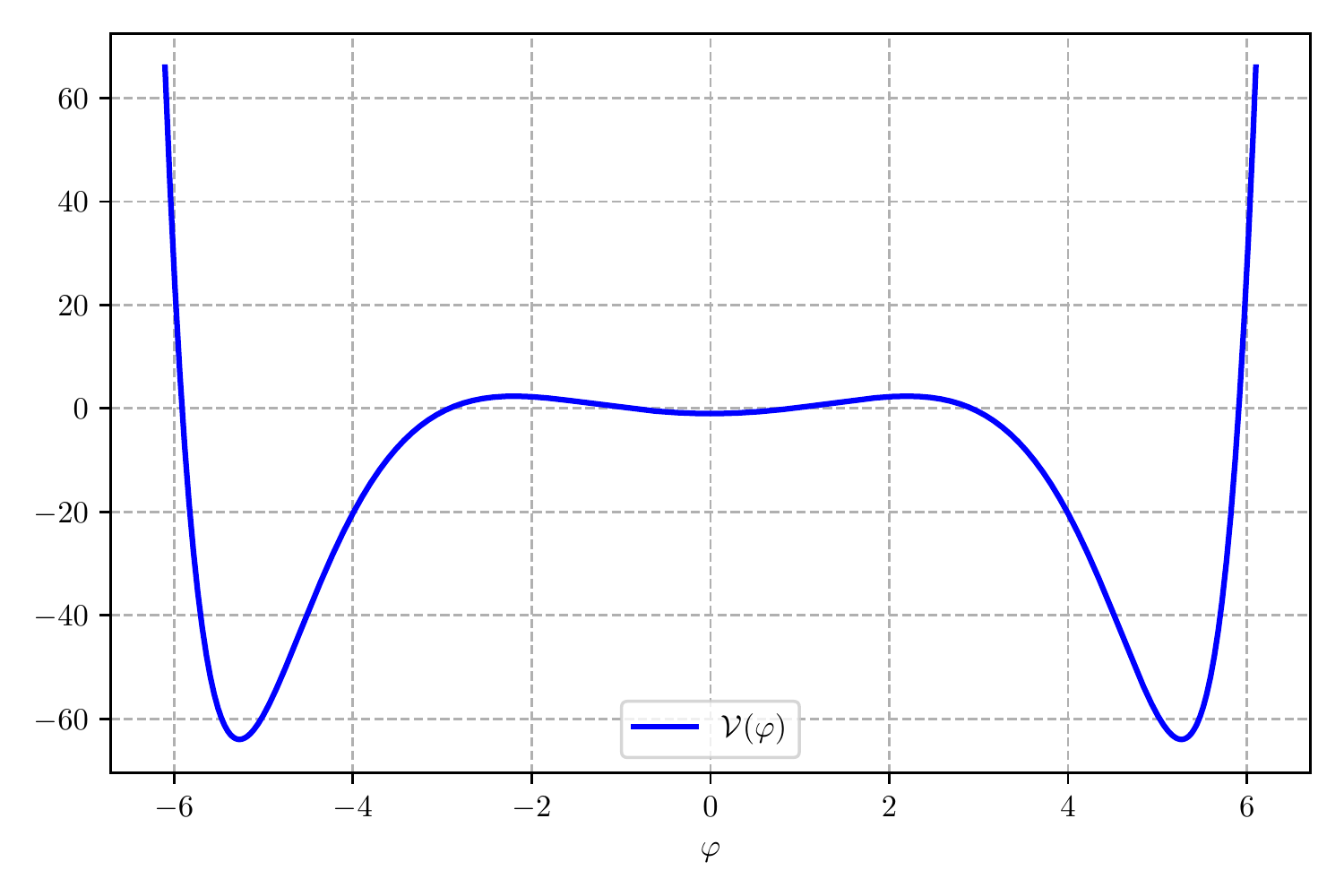}
\caption{ \small The potential of eq.\eqref{pot_39_2} with $C =1$. The upper--left panel shows $\gamma = - \frac{1}{10}$, the upper--right panel shows $\gamma = \frac{1}{6}$ and the lower panel shows $\gamma = \frac{1}{8}$. These potentials were also considered in \cite{fss}.}
\label{nine_pot}
\end{figure}

Multiplying both sides by $\dot z$, the equations of motion are equivalent to
\begin{equation}
	\dot z^2 \ = \  - \gamma^2 \tilde C z^{\frac{2}{\gamma} - 2} + E + i F \ ,
\end{equation}
where $E, F \in \mathbb R$ are two integration constants. The Hamiltonian constraint imposes $F = 0$, so that one has to solve
\begin{equation}
	\dot z^2 \ = \  -\, \gamma^2 \, \tilde C z^{\frac{2}{\gamma} - 2} + E \ .
\label{ediffzero}
\end{equation}

Equations of this type can be solved by quadratures in the complex plane, considering paths compatible with the reality of $r$. In general, an equation of the form
\beq
\frac{\td x + i \,\td y}{A+ i B} \ =\ \td r  \ ,
\eeq
where in our case
\beq
A(x,y) \ = \ \Re\,\sqrt{-\, \gamma^2 \, \tilde C z^{\frac{2}{\gamma} - 2} + E} \ , \qquad  B(x,y) \ = \ \Im\sqrt{-\, \gamma^2 \, \tilde C z^{\frac{2}{\gamma} - 2} + E } \ ,
\eeq
is equivalent to
\beq
\frac{\td y}{\td x} \ = \ \frac{B(x,y)}{A(x,y)} \ , \qquad \frac{\td x}{A(x,y(x))} \ = \ \td r \ .
\eeq
The first determines the paths in the complex plane that are compatible with the reality of $r$, while the second defines $x$ implicitly as function of $r$. The solutions can be investigated numerically, in particular for this class of potentials. However, we shall refrain from adding further details, since the intuition fostered by the preceding examples should allow an attentive reader to anticipate the key features of the resulting dynamics, in particular for the three interesting examples displayed in fig.~\ref{nine_pot}.

\vskip 12pt
\section{\sc  Conclusions }\label{sec:conclusions}

In this paper we have elaborated on the behavior of scalar--gravity systems in backgrounds of the type
\beq
	\td s^2 \ = \  e^{\frac{2}{9} \, \cA(r)} \, \eta_{\mu\nu} \, \td x^{\mu} \,  \td x^{\nu} \, + \, e^{2 \,  {\cal B}(r)} \, \td r^2  \ ,
\eeq
which are meant to describe compactifications from ten to nine dimensions.
We were inspired by the original result of Dudas and Mourad~\cite{dm_vacuum}, who showed how the leading tadpole potentials arising in non--tachyonic non--supersymmetric ten--dimensional strings~\cite{susy95,sugimoto,so1616} can induce the \emph{spontaneous compactification of an internal dimension down to a scale determined by their strength}. Their solutions lead to the emergence of nine--dimensional Minkowski spaces where finite values for the Planck mass and the gauge coupling are inherited, so that the gravitational and gauge interactions are effective there. These striking results, however, are accompanied by the presence of regions where the string coupling and/or the spacetime curvature in string units become large, which raises legitimate questions on what String Theory ought to add to this picture.

Here we have relied on a detailed analysis of scalar--gravity models with exponential potentials and on an elegant set of additional potentials that in several cases combine the leading perturbative contribution of orientifold models, which in the present notation takes the form in eq.~\eqref{V_orientifold}, and in some cases the leading contribution~\eqref{V_heterotic} for the heterotic model of~\cite{so1616}, with additional terms that can mimic possible perturbative or non--perturbative corrections. The resulting models are integrable, and mostly in an elementary fashion. The Hamiltonian constraint is particularly helpful, since it suggests a ``particle analogy'' that links the problem to intuitions that were widely developed for instanton methods. In brief, the inverted potential $- \cV(\vf)$, rather the potential $\cV(\vf)$ itself, drives the ``dynamical'' evolution in the internal coordinate, with some additional subtleties brought along by gravity. Relying on different techniques, we have explored under what circumstances solutions with a compact internal space, as in~\cite{dm_vacuum}, but devoid of strong--coupling regions, can exist.

We have thus collected a few general lessons:
\begin{itemize}
    \item Obtaining a bounded string coupling is generally possible when the potential $\cV(\vf)$ has a local well, which translates into a local bump in the inverted potential that the ``particle'' cannot overcome for a certain range of integration constants. Remarks of this type apply to the potentials n. 1, 3, and 4, and also to the step--like potential n. 8 and to special cases within the family of potentials n. 9. An interesting issue, which we have not touched upon, is the possibly metastable nature of these types of vacua.
    \item Infinite wells provide additional types of interesting scenarios. If the dilaton cannot overcome the corresponding bump of the inverted potential, it will be confined to one of its sides, and in one of them it is bounded from above. This type of situation presents itself in the potentials n. 6 and 7 and in cases belonging to the family of potentials n. 9.
    \item Potentials that become unbounded from below can naturally place upper bounds on $\vf$. Actually, our potentials are often dominated in some regions by a single exponential, and the corresponding exact solutions, discussed in detail in Section~\ref{sec:compactness}, provide detailed indications on the corresponding dynamics. The main mechanism at work is the Euclidean counterpart of the ``climbing scalar'': it grants, for instance, that \emph{an inverted exponential $- e^{2\gamma\varphi}$ with $\gamma \geq 1$ places inevitably upper bounds on the string coupling}. This setting and the presence of dips are both favored by corrections involving exponential terms of negative sign.
    \item Potentials that are not defined beyond a certain value of $\vf$ provide another very interesting option to place upper bounds on the string coupling. This situation presents itself in the potential n. 5, and more generally is favored by series of corrections involving exponential terms with identical signs.
    \item There is generally a tension between upper bounds on the scalar curvature and finite sizes for the internal space. This can be foreseen comparing eq.~\eqref{compact_cond}, which defines the internal length in the string frame, and eq~\eqref{string_curvature}, which defines the scalar curvature in the string frame. Our examples suggest that this be true in general, and in Section~\ref{sec:compactness} we saw clearly that the two options are incompatible whenever the potential is dominated by an exponential term.
    \item We have not found a simple way to anticipate whether or not the internal space is compact. This is simple, however, if the dilaton stops at a finite value $\varphi_0$, which can occur if the potential has a negative critical point there. The resulting space is then non--compact and is asymptotically $AdS_{10}$. If the potential is dominated by an exponential term, the discussion in Section~\ref{sec:compactness}, summarized in the Tables~\ref{tab:dilaton_dynamics_1} and~\ref{tab:dilaton_dynamics_2}, indicates that a number of windows exist that can grant a finite length for the $r$--direction, but these depend in a complicated way on the nature of the potential.
\end{itemize}
Consistently with the preceding considerations, \emph{we have not found any example that combines an interval of finite length with bounded values for both string coupling and curvature}. Higher--derivative corrections to the low--energy effective theory might prove crucial to bypass this limitation, as was the case in other contexts~\cite{small_black_holes}. However, the analysis of~\cite{cdud} indicates that quadratic curvature corrections do not suffice: one would need higher--order terms, and presumably resummations thereof. Table~\ref{Interesting_results} collects some properties of the most interesting models that we have analyzed.

\begin{table}[ht]
\centering
\begin{tabular}{|c|c|c|c|c|c|}
\hline
    $n$ & Potential & Parameters & Int. Const. & $M_P^7$ & $\frac{1}{g_{YM}^2}$ \\
\hline\hline
    1. & $C \vf  e^{2\vf}$ &  $C \, < \, 0$ & $D \, = \, - \,  a^2$ & finite & finite \\
\hline
    2. &$C_1 e^{2 \vf} + C_2$ & $C_1 \, < 0$,  $C_2 \, > 0$ & all & finite & finite \\
\hline
    3. & $\ve_1 e^{2\gamma \vf} + \ve_2 e^{(\gamma+1) \vf} $ & $0 < \gamma < \frac{1}{3}$,  & \eqref{b_condit} & infinite & infinite \\
    & & $\ve_1 = -1$, $\ve_2 = 1$ & &  & \\
    & & & &  & \\
    & & $-1 < \gamma < 1$ & \eqref{b_bounded_gs} & finite & finite \\
    & & $\ve_1 = 1$, $\ve_2 = -1$ & &  & \\
    & & & &  & \\
    & & $0 < \gamma < \frac{1}{3}$ & \eqref{ineq_2} & infinite & infinite \\
    & & $\ve_1 = -1$, $\ve_2 = -1$ & &  & \\
    & & & & & \\
    & & $\gamma > 1$ & \eqref{cond_b} & finite & finite \\
    & & $\ve_1 = -1$, $\ve_2 = 1$ & & & \\
\hline
    4. & $\lambda \left( e^{\frac{2}{\gamma} \vf} - e^{2 \gamma \vf} \right)$ & $0 < \gamma < \frac{1}{3}$ & $\eta = 1$ & infinite & infinite \\
    & & $\lambda > 0$ & & & \\
    & & & & & \\
    & & $0 < \gamma < 1$ & $\eta = - 1$ & finite & finite \\
    & & $\lambda < 0$ & &  & \\
\hline
    5. & $C \log(- \coth(\vf)) + D$ & $C > 0$ & & finite & finite \\
     &  & $C < 0$, $D \geq 0$ & & finite & finite \\
\hline
    6. & $C \cosh(\vf) + \lambda$ & $C < 0$, $\Lambda > 0$ & & finite & finite \\
\hline
    7. & $C_1 \cosh^4\left(\frac{\vf}{3}\right) + C_2 \sinh^4\left(\frac{\vf}{3}\right)$ & $C_1 < 0$, $|C_1| < C_2$  & & finite & finite \\
\hline
\end{tabular}
\caption{\small A brief summary of most of our results.}
\label{Interesting_results}
\end{table}

\begin{figure}[ht]
\includegraphics[width=0.57\textwidth]{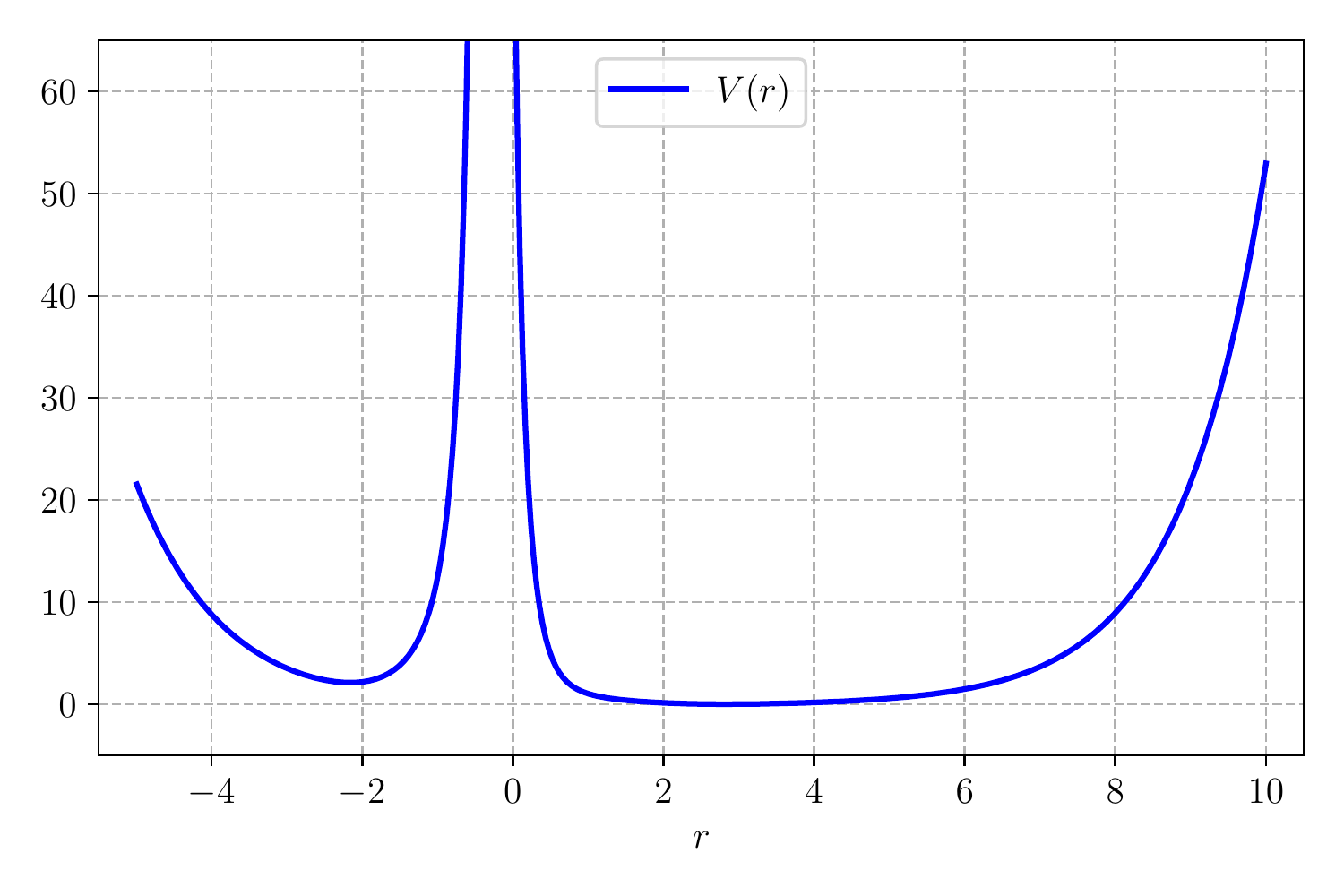}
\centering
\caption{ \small The potential of eq.~\eqref{pot_schrod} for the solution of fig.~\ref{fourth_inverted} is always positive. Consequently, the eigenvalues of the corresponding Schr\"odinger problem, which lie above its minimum, are also positive, and the scalar perturbations contain no tachyonic modes.}
\label{stability_fourth}
\end{figure}

As we have stressed, several interesting examples, albeit not of all of them, rest on potentials that are not bounded from below. Moreover, around the turning point for the ``particle'' they look like inverted harmonic potentials. This raises some legitimate concerns about the possible emergence of tachyonic modes in nine dimensions. On the other hand, the Breitenlohner--Freedman~\cite{bf_bound} bounds in AdS spaces raise some hope that these systems be stable. Before concluding, we would like to elaborate briefly on this point. According to~\cite{bms}, the squared masses of scalar perturbations within the class of metrics of eq.~\eqref{einstein_metric} are determined by the eigenvalues of Schr\"odinger--like operators involving the potentials
\begin{equation}
    V(r) \ = \frac{ e^{- \left( \cB - \frac{1}{9} \cA \right)} \, \dot{a}(r)}{2} \ + \ \frac{a(r)^2}{4} \ + \ b(r) \ , \label{pot_schrod}
\end{equation}
where, in our conventions,
\bea
a(r) &=&   \frac{8}{3}\, \dot{{\cal A}}(r)\,e^{-\left({\cal B} - \frac{1}{9}\cA\right)} \ - \ \frac{2}{\dot \vf(r)}\, e^{{\cal B} + \frac{1}{9} \cA}\, \cV'(\vf) \ , \nonumber \\
b(r) &=& \frac{28}{9} \; e^{\frac{2 \cA}{9}} \, \left[ \cV (\vf) \, + \, \frac{1}{2} \, \frac{\dot \cA}{\dot \vf} \; \cV'(\vf)\right] \ . \label{pot_st}
\eea
A positive $b$ would imply perturbative stability, since the remaining terms combine with the ``kinetic operator'' into manifestly non--negative contributions, and this is precisely what happens for the Dudas--Mourad vacua of~\cite{dm_vacuum}. This condition, however, is sufficient but not necessary, and in fact is not fulfilled in the cases of interest. Therefore, in principle one should study the ground state of the Schr\"odinger--like systems
\beq
\left[ - \left(e^{- \left(\cB -\frac{1}{9} \, \cA \right)} \, \frac{\td}{\td r}\right)^2  \ + \ V(r) \right] \Psi \ = \ m^2 \, \Psi \ ,
\eeq
in order to ascertain whether or not the lowest possible value of $m^2$ is positive in the different toy models.
Actually, a closer look reveals that matters seem to conspire in interesting ways. This is true, in particular, for the example of eq.~\eqref{fourth_cosh}, which belongs to the family n. 4 of Sections~\ref{sec:models} and~\ref{sec:profiles}. As shown in fig.~\ref{fourth_inverted}, the string coupling has a desired upper bound, and its potential $V$, displayed in fig.~\ref{stability_fourth}, is indeed everywhere positive, despite a negative $b(r)$! This clearly suffices to conclude that $m^2>0$ in this case, so that, surprisingly, no tachyons would emerge in nine dimensions. It is natural to suspect that similar results hold for all cases in Table~\ref{Interesting_results}, but we do not have a general argument to this effect. We shall stop here for the moment, leaving a detailed stability analysis for the future.

\vskip 18pt
\section*{Acknowledgments}
This work was supported in part by Scuola Normale, by INFN (IS GSS-Pi) and by the MIUR-PRIN contract 2017CC72MK\_003. P.P. was partly supported by the SNF through Project Grants 200020 and 182513, and by the NCCR 51NF40-141869 ``The Mathematics of Physics'' (SwissMAP). A.S. is grateful to P.~Fr\'e and A.~Sorin, who introduced him to the techniques used extensively in~\cite{fss} and reconsidered here in a different context, and to the Alexander von Humboldt Foundation for its kind and generous support. We are both grateful to E.~Dudas and J.~Mourad for useful discussions.

\begin{appendices}
\section{\sc Some Useful Results}
In this Appendix we collect some useful formulas for the Einstein--frame curvatures corresponding to the class of metrics in eq.~\eqref{einstein_metric}, and for their string--frame counterparts.

The Einstein--frame scalar curvature is
\begin{equation}
    R \ = \ 4\; \left[ \, \frac{2}{9}\, e^{-2{\cal B}}\,\dot{\cal A}^2 \ + \ {\cal V}\left(\varphi\right)  \right] \ ,
\label{einstein_curvature2}
\end{equation}
while the independent components of the Einstein--frame Riemann tensor are
\begin{align}
	{R^r}_{rrr} & \; = \; 0 \ , \nonumber \\
	{R^r}_{\mu\nu r} & \; = \; e^{\frac{2}{9} \, \cA \, - \, 2 \, \mathcal B} \, \left[ \frac{\ddot{\cA}}{9}  \ + \ \frac{\dot{\cA}}{9}\,  \left(\frac{\dot {\cA}}{9}\,  - \,  \dot {\mathcal B}\right)\right] \eta_{\mu\nu}\  ,  \nonumber \\
	{R^\mu}_{r\nu r} & \; = \; - \left[ \frac{\ddot{\cA}}{9}  \ + \ \frac{\dot{\cA}}{9}\,  \left(\frac{\dot {\cA}}{9}\,  - \,  \dot {\mathcal B}\right)\right] \delta^\mu_\nu \ , \nonumber \\
	{R^\mu}_{\nu\rho \sigma} & \; = \; \frac{\dot{\cA}^2}{81} \,  e^{ \frac{2}{9} \, \cA \, - \, 2 \,\mathcal B } (\delta^\mu_\sigma \, \eta_{\nu\rho} \,  - \, \delta^\mu_\rho \, \eta_{\nu\sigma}) \ .
\end{align}
Consequently the Einstein--frame Ricci tensor has, in the cases of interest, the non--vanishing components
\begin{align}
    R_{\mu\nu} &=\ - \ e^{ \frac{2}{9} \, \cA \, - \, 2 \, \mathcal B}\left[ \frac{\ddot{\cA}}{9} \,+\, \frac{\dot{\cA}}{9}\,  \left(\dot {\cA} \,- \,\dot {\cB}\right) \right] \eta_{\mu\nu} \ ,\nonumber \\
    R_{rr} &= \ - \ \left[ \ddot{\cA} \, +\, \dot{\cA}\, \left(\frac{\dot {\cA}}{9}  \,- \,\dot {\cB}\right) \right] \ .
\end{align}

The counterparts of these expressions in the string frame read
\begin{align}
	{{R_{(s)}}^r}{}_{rrr} & \; = \; 0 \ , \nonumber \\
	{{R_{(s)}}^r}{}_{\mu\nu r} & \; = \; e^{\frac{2}{9} \, \cA \, - \, 2 \,\mathcal B} \, \left[\frac{\ddot{\cA}}{9} \ + \  \frac{\ddot \vf}{3}  \ + \ \left(\frac{\dot{\cA}}{9} + \frac{\dot \vf}{3} \right) \, \left(\frac{\dot {\cA}}{9} - \dot {\mathcal B}\right)\right] \eta_{\mu\nu}\  ,  \nonumber\\
	{{R_{(s)}}^\mu}{}_{r\nu r} & \; = \; - \left[\frac{\ddot{\cA}}{9} \ + \  \frac{\ddot \vf}{3}  \ + \ \left(\frac{\dot{\cA}}{9} + \frac{\dot \vf}{3} \right) \, \left(\frac{\dot {\cA}}{9} - \dot {\mathcal B}\right)\right] \delta^\mu_\nu \ , \nonumber \\
	{{R_{(s)}}^\mu}{}_{\nu\rho \sigma} & \; = \; \left( \frac{\dot{\cA}}{9}\  +\ \frac{\dot \vf}{3} \right)^2 e^{\frac{2}{9} \, \cA \, - \, 2 \, \mathcal B} \,  \Big(\delta^\mu_\sigma \, \eta_{\nu\rho} \,  - \, \delta^\mu_\rho \, \eta_{\nu\sigma} \Big) \ ,
\end{align}
and therefore
\begin{align}
    {R_{(s)}}{}_{\mu\nu} &=\ - \ e^{\frac{2}{9} \, \cA \, - \,2 \, \mathcal B}\left[ \frac{\ddot{\cA}}{9} \,+\, \frac{\ddot \vf}{3} \, + \,  \left( \frac{\dot{\cA}}{9} \, + \, \frac{\dot \vf}{3} \right) \left(\dot {\cA} \,- \,\dot {\cB} \, + \, \frac{8}{3} \, \dot \vf \right) \right] \eta_{\mu\nu} \ , \nonumber \\
    {R_{(s)}}{}_{rr} &= \ - \ \left[ \ddot{\cA} \,+\, 3\, \ddot \vf \,  +\, \left( \dot{\cA} \, + \, 3\, \dot \vf \right) \left(\frac{\dot {\cA}}{9} \,- \,\dot {\mathcal B}\right) \right] \ .
\end{align}
\end{appendices}


\vskip 18pt
\begin{thebibliography}{99}

\bibitem{stringtheory}
For reviews see: M.~B.~Green, J.~H.~Schwarz and E.~Witten, ``Superstring Theory'', 2 vols., Cambridge, UK: Cambridge Univ. Press (1987); J.~Polchinski, ``String theory'', 2 vols. Cambridge, UK: Cambridge Univ. Press (1998);  C.~V.~Johnson, ``D-branes,'' USA: Cambridge Univ. Press (2003) 548 p; B.~Zwiebach, ``A first course in string theory''
Cambridge, UK: Cambridge Univ. Press (2004); K.~Becker, M.~Becker and J.~H.~Schwarz,
``String theory and M-theory: A modern introduction'' Cambridge, UK: Cambridge Univ.
Press (2007); E.~Kiritsis, ``String theory in a nutshell'', Princeton, NJ: Princeton Univ. Press (2007);
P.~West, ``Introduction to strings and branes,'' Cambridge: Cambridge Univ. Press (2012).

\bibitem{supersymmetry} For a review of early results see: P.~Fayet and S.~Ferrara,
Phys. Rept. \textbf{32} 249 (1977).
More recent reviews are: P.C.~West,
``Introduction to Supersymmetry and Supergravity'' (World Scientific, Singapore, 1990);
J.~Wess and J.~Bagger, ``Supersymmetry and supergravity,'' (Princeton University Press,
1992); S.~Weinberg,``The quantum theory of fields. Vol. 3: Supersymmetry,'' (Cambridge
University Press, 2005).

\bibitem{supergravity}
D.~Z.~Freedman, P.~van Nieuwenhuizen and S.~Ferrara,
Phys.\ Rev.\ {\bf D 13} (1976) 3214;
S.~Deser and B.~Zumino,
Phys.\ Lett.\ {\bf B 62} (1976) 335.
For a recent review see:
D.~Z.~Freedman and A.~Van Proeyen,
(Cambridge Univ. Press, 2012).  A quick survey of many developments can be found in: S.~Ferrara and A.~Sagnotti,
Riv. Nuovo Cim. \textbf{40} (2017) no.6, 279
[arXiv:1702.00743 [hep-th]].

\bibitem{CJS}
E.~Cremmer, B.~Julia and J.~Scherk,
Phys.\ Lett.\  {\bf 76B} (1978) 409.

\bibitem{witten1011}
E.~Witten,
Nucl. Phys. B \textbf{443} (1995), 85
[arXiv:hep-th/9503124 [hep-th]].

\bibitem{susy95}
A.~Sagnotti,
[arXiv:hep-th/9509080 [hep-th]],
Nucl. Phys. B Proc. Suppl. \textbf{56} (1997) 332
[arXiv:hep-th/9702093 [hep-th]].

\bibitem{sugimoto}
S.~Sugimoto,
Prog.\ Theor.\ Phys.\  {\bf 102} (1999) 685 [arXiv:hep-th/9905159].

\bibitem{so1616}
L.~J.~Dixon and J.~A.~Harvey,
Nucl.\ Phys.\ B {\bf 274} (1986) 93;
L.~Alvarez-Gaume, P.~H.~Ginsparg, G.~W.~Moore and C.~Vafa,
Phys.\ Lett.\ B {\bf 171} (1986) 155.

\bibitem{va}
D.~V.~Volkov and V.~P.~Akulov,
Phys.\ Lett.\ B {\bf 46} (1973) 109.

\bibitem{nonlinearsusy}
E.~Dudas and J.~Mourad,
Phys.\ Lett.\ B {\bf 514} (2001) 173
[hep-th/0012071];
G.~Pradisi and F.~Riccioni,
Nucl.\ Phys.\ B {\bf 615} (2001) 33
[hep-th/0107090].

\bibitem{orientifolds}
A.~Sagnotti, in Cargese '87, ``Non-Perturbative Quantum Field
Theory'', eds. G. Mack et al (Pergamon Press, 1988), p. 521,
[arXiv:hep-th/0208020];
G.~Pradisi and A.~Sagnotti,
Phys.\ Lett.\ {\bf B 216} (1989) 59;
P.~Horava,
Nucl.\ Phys.\ {\bf B 327} (1989) 461,
Phys.\ Lett.\ {\bf B 231} (1989) 251;
M.~Bianchi and A.~Sagnotti,
Phys.\ Lett.\ {\bf B 247} (1990) 517;
M.~Bianchi and A.~Sagnotti,
Nucl.\ Phys.\ {\bf B 361} (1991) 519;
M.~Bianchi, G.~Pradisi and A.~Sagnotti,
Nucl.\ Phys.\ {\bf B 376} (1992) 365;
A.~Sagnotti,
Phys.\ Lett.\  {\bf B 294} (1992) 196
[arXiv:hep-th/9210127].
For reviews see: E.~Dudas,
Class.\ Quant.\ Grav.\  {\bf 17} (2000) R41 [arXiv:hep-ph/0006190];
C.~Angelantonj and A.~Sagnotti,
Phys.\ Rept.\  {\bf 371} (2002) 1 [Erratum-ibid.\  {\bf 376} (2003)
339]
[arXiv:hep-th/0204089].

\bibitem{bsb}
I.~Antoniadis, E.~Dudas and A.~Sagnotti,
Phys.\ Lett.\ {\bf B 464} (1999) 38 [arXiv:hep-th/9908023];
C.~Angelantonj,
Nucl.\ Phys.\ {\bf B 566} (2000) 126
[arXiv:hep-th/9908064];
G.~Aldazabal and A.~M.~Uranga,
JHEP {\bf 9910} (1999) 024
[arXiv:hep-th/9908072];
C.~Angelantonj, I.~Antoniadis, G.~D'Appollonio, E.~Dudas and
A.~Sagnotti,
Nucl.\ Phys.\ {\bf B 572} (2000) 36
[arXiv:hep-th/9911081].

\bibitem{dm_vacuum}
E.~Dudas and J.~Mourad,
Phys.\ Lett.\  {\bf B 486} (2000) 172
[arXiv:hep-th/0004165].

\bibitem{bms}
I.~Basile, J.~Mourad and A.~Sagnotti,
JHEP \textbf{01} (2019) 174
[arXiv:1811.11448 [hep-th]].

\bibitem{fss}
P.~Fr\'e, A.~Sagnotti and A.~S.~Sorin,
Nucl. Phys. B \textbf{877} (2013) 1028
[arXiv:1307.1910 [hep-th]];
V.~V.~Sokolov and A.~S.~Sorin,
Lett. Math. Phys. \textbf{107} (2017) no.9, 1741-1768
[arXiv:1608.08511 [hep-th]].

\bibitem{russo}
J.~G.~Russo,
Phys. Lett. B \textbf{600} (2004), 185-190
[arXiv:hep-th/0403010 [hep-th]].

\bibitem{dks2010} E.~Dudas, N.~Kitazawa and A.~Sagnotti,
Phys. Lett. B \textbf{694} (2011), 80-88
[arXiv:1009.0874 [hep-th]];
E.~Dudas, N.~Kitazawa, S.~P.~Patil and A.~Sagnotti,
JCAP \textbf{05} (2012) 012
[arXiv:1202.6630 [hep-th]];
N.~Kitazawa and A.~Sagnotti,
JCAP \textbf{04} (2014) 017
[arXiv:1402.1418 [hep-th]];
A.~Sagnotti,
Phys. Part. Nucl. Lett. \textbf{11} (2014), 836-843
[arXiv:1303.6685 [hep-th]].

\bibitem{gm}
S.~S.~Gubser and I.~Mitra,
JHEP \textbf{07} (2002), 044
[arXiv:hep-th/0108239 [hep-th]].

\bibitem{ms}
J.~Mourad and A.~Sagnotti,
Phys. Lett. B \textbf{768} (2017), 92
[arXiv:1612.08566 [hep-th]].
For a review see J.~Mourad and A.~Sagnotti,
[arXiv:1711.11494 [hep-th]].

\bibitem{swampland} C.~Vafa,
[arXiv:hep-th/0509212 [hep-th]].
For an extensive review, see:
E.~Palti,
Fortsch. Phys. \textbf{67} (2019) no.6, 1900037
[arXiv:1903.06239 [hep-th]].


\bibitem{chsw}
P.~Candelas, G.~T.~Horowitz, A.~Strominger and E.~Witten,
Nucl. Phys. B \textbf{258} (1985) 46.

\bibitem{lm}
F.~Lucchin and S.~Matarrese,
Phys. Rev. D \textbf{32} (1985) 1316.

\bibitem{small_black_holes}
G.~Lopes Cardoso, B.~de Wit and T.~Mohaupt,
Phys. Lett. B \textbf{451} (1999) 309
[arXiv:hep-th/9812082 [hep-th]];
A.~Sen,
JHEP \textbf{05} (2005) 059
[arXiv:hep-th/0411255 [hep-th]].

\bibitem{cdud}
C.~Condeescu and E.~Dudas,
JCAP \textbf{08} (2013) 013
[arXiv:1306.0911 [hep-th]].


\bibitem{bf_bound}
P.~Breitenlohner and D.Z.~Freedman,
Phys. Lett. B {\bf 115} 3 (1982) 197;
P.~Breitenlohner and D.Z.~Freedman,
Annals Phys. \textbf{144} (1982) 249.



\end{thebibliography}
\end{document}